\documentclass{jfm}

\usepackage{graphicx}
\usepackage{epstopdf, epsfig}
\usepackage{amsmath,rotating}
\usepackage{dashrule}
\usepackage{psfrag}
\usepackage{latexsym}
\usepackage{mathrsfs}
\usepackage{pifont}
\usepackage{amsmath}
\usepackage{subfigure}
\usepackage{mathtools}
\usepackage{mathcomp}
\usepackage{mathdots}
\usepackage{rotating}
\usepackage{array}
\usepackage{color}
\usepackage{algorithm}
\usepackage{algorithmic}
\usepackage[normalem]{ulem}
\usepackage{stackengine}
\usepackage{booktabs}
\usepackage{hyperref}
\usepackage[utf8]{inputenc}
\usepackage{multirow}
\usepackage{arydshln}

\usepackage{gensymb}
\usepackage{amssymb}
\usepackage[dvipsnames]{xcolor}
\usepackage{tikz}
\usepackage{natbib}
\usepackage{lineno}
\usetikzlibrary{arrows,shapes,trees,calc,positioning}
\usetikzlibrary{matrix}
\usepackage{cancel}

\newcommand{\cA}{{\cal A}}
\newcommand{\cB}{{\cal B}}
\newcommand{\cC}{{\cal C}}
\newcommand{\cM}{{\cal M}}
\newcommand{\cX}{{\cal X}}

\newlength\myheight
\newlength\mydepth
\settototalheight\myheight{Xygp}
\makeatletter
\def\underbracex#1#2{\mathop{\vtop{\m@th\ialign{##\crcr
   $\hfil\displaystyle{#2}\hfil$\crcr
   \noalign{\kern3\p@\nointerlineskip}%
   #1\crcr\noalign{\kern3\p@}}}}\limits}

\def\underbracea{\underbracex\upbracefilla}

\def\upbracefilla{$\m@th \setbox\z@\hbox{$\braceld$}%
  \bracelu\leaders\vrule \@height\ht\z@ \@depth\z@\hfill 
\kern\p@\vrule \@width\p@\kern\p@\vrule \@width\p@\kern\p@\vrule \@width\p@
$}

\def\upbracefillb{$\m@th \setbox\z@\hbox{$\braceld$}%
\vrule \@width\p@\kern\p@\vrule \@width\p@\kern\p@\vrule \@width\p@\kern\p@
 \leaders\vrule \@height\ht\z@ \@depth\z@\hfill\bracerd
  \braceld\leaders\vrule \@height\ht\z@ \@depth\z@\hfill
\kern\p@\vrule \@width\p@\kern\p@\vrule \@width\p@\kern\p@\vrule \@width\p@
$}

\def\upbracefillc{$\m@th \setbox\z@\hbox{$\braceld$}%
\vrule \@width\p@\kern\p@\vrule \@width\p@\kern\p@\vrule \@width\p@\kern\p@
\leaders\vrule \@height\ht\z@ \@depth\z@\hfill
\kern\p@\vrule \@width\p@\kern\p@\vrule \@width\p@\kern\p@\vrule \@width\p@
$}

\def\underbraced{\underbracex\upbracefilld}
\def\upbracefilld{$\m@th \setbox\z@\hbox{$\braceld$}%
\vrule \@width\p@\kern\p@\vrule \@width\p@\kern\p@\vrule \@width\p@\kern\p@
 \leaders\vrule \@height\ht\z@ \@depth\z@\hfill\braceru$}

\def\upbracefillbd{$\m@th \setbox\z@\hbox{$\braceld$}%
\vrule \@width\p@\kern\p@\vrule \@width\p@\kern\p@\vrule \@width\p@\kern\p@
\bracerd\braceld
 \leaders\vrule \@height\ht\z@ \@depth\z@\hfill\braceru$}
 \newcommand*\bigcdot{\mathpalette\bigcdot@{.4}}
\newcommand*\bigcdot@[2]{\mathbin{\vcenter{\hbox{\scalebox{#2}{$\m@th#1\bullet$}}}}}
 \makeatother

\newcommand{\enma}[1]   {\ensuremath{#1}}

\newcommand{\beq}{\begin{equation}}
\newcommand{\eeq}{\end{equation}}
\newcommand{\bseq}{\begin{subequations}}
\newcommand{\eseq}{\end{subequations}}
\newcommand{\beqn}{\begin{eqnarray}}
\newcommand{\eeqn}{\end{eqnarray}}
\newcommand{\ba}{\begin{array}}
\newcommand{\ea}{\end{array}}
\newcommand{\bct}{\begin{center}}
\newcommand{\ect}{\end{center}}
\newcommand{\btmz}{\begin{itemize}}
\newcommand{\etmz}{\end{itemize}}
\newcommand{\benum}{\begin{enumerate}}
\newcommand{\eenum}{\end{enumerate}}

\newcommand{\cH}{\enma{\mathcal H}}

\newcommand{\diag}      {\enma{\mathrm{diag}}}

\newcommand{\trace}     {\enma{\mathrm{trace}}}

\newcommand{\col}       {\enma{\mathrm{col}}}

\newcommand{\bv}{{\bf v}}

\newcommand{\matbegin}{
        \left[
}
\newcommand{\matend}{
        \right]
}

\newcommand{\tbo}[2]{
  \matbegin \begin{array}{c}
       #1 \\ #2
       \end{array} \matend }
\newcommand{\thbo}[3]{
  \matbegin \begin{array}{c}
       #1 \\ #2 \\ #3
       \end{array} \matend }

\newcommand{\tbt}[4]{
  \matbegin \begin{array}{cc}
       #1 & #2 \\ #3 & #4
       \end{array} \matend }

\newcommand{\be}{\begin{equation}}
\newcommand{\ee}{\end{equation}}

\newcommand{\cplxs}{ C\kern -.35em \rule{0.03 em}{.7 ex}~   }

\def\complex{\hbox{C\kern -.45em \rule{0.03 em}{1.5 ex}}~}

\newcommand{\bi}{\begin{itemize}}
\newcommand{\ei}{\end{itemize}}

\newcommand{\bu}{{\bf u}}

\newcommand{\bbZ}{\mathbb{Z}}

\newcommand{\btab}{\begin{tabular}}
\newcommand{\etab}{\end{tabular}}
\newcommand{\bd}{{\bf d}}

\newcommand{\bpsi}{\mbox{\boldmath$\psi$}}

\newcommand{\bPhi}{\mbox{\boldmath$\Phi$}}

\newcommand{\non}{\nonumber}
\newcommand{\mrd}{\mathrm{d}}
\newcommand{\mre}{\mathrm{e}}
\newcommand{\mri}{\mathrm{i}}

\newcommand{\ds}{\displaystyle}

\newcommand{\eps}{{\epsilon}}

\newcommand{\DefinedAs}[0]{\mathrel{\mathop:}=}

\definecolor{bgblue}{rgb}{0.04,0.19,0.53}
\definecolor{dblue1}{rgb}{0,0.3,0.7}
\definecolor{dred}{rgb}{0.4,0.2,0}

\shorttitle{Modeling riblets via domain transformation}
\shortauthor{M. Naseri and A. Zare}

\title{Turbulence modeling over riblets via domain transformation}

\author{Mohammadamin Naseri and Armin Zare
 \corresp{\email{armin.zare@utdallas.edu}}}
\affiliation{Department of Mechanical Engineering, University of Texas at Dallas,
Richardson, TX 75080, USA}

\begin{document}


\maketitle

\begin{abstract}
Numerical and experimental studies have demonstrated the drag-reducing potential of carefully designed streamwise-elongated riblets in lowering skin-friction drag. To support the systematic design of such surface corrugations, recent efforts have integrated simplified versions of the governing equations with innovative methods for representing the effects of rough boundaries on flow dynamics. Notably, the statistical response of the eddy-viscosity-enhanced linearized Navier-Stokes equations has been shown to effectively capture the ability of riblets in suppressing turbulence, quantify the influence of background turbulence on the mean velocity, and reproduce established drag-reduction trends. In this paper, we enhance the flexibility and computational efficiency of this simulation-free approach by implementing a domain transformation for surface representation, along with a perturbation analysis on a small geometric parameter of the riblets. While domain transformation complicates the differential equations, it provides accurate boundary representations and facilitates the analysis of complex riblet shapes at high Reynolds numbers by enabling perturbation analysis to simplify the dimensional complexity of the governing equations. Our method successfully predicts drag reduction trends for triangular and scalloped riblets, consistent with existing literature. We further utilize our framework to investigate flow mechanisms influenced by riblets and extend our study to channel flows with friction Reynolds numbers up to 2003. Our findings reveal the emergence of K-H rollers over large and sharp scalloped riblets, contributing to the degradation of drag reduction in these geometries. Additionally, we examine the impact of riblets on near-wall flow structures, focusing on their suppression of streamwise-elongated structures in flows over large riblets.
\end{abstract}


\section{Introduction}
Skin friction accounts for approximately 45\% of the total drag in aircraft transportation systems~\citep{cou92}. 
{Consequently, reducing skin-friction drag presents a significant opportunity to lower energy consumption and operational costs in both} the energy and transportation sectors. To harness this potential, numerous experimental~\citetext{\citealp{wal82,wallin84,becbruhaghoehop97}; \citealp*{becbruhag00}} and numerical~\citep*{chomoikim93,golhansir95,garjim11a} studies have demonstrated the drag-reducing advantages of streamwise-aligned, spanwise-periodic surface corrugations, commonly known as riblets. These investigations have identified a broad spectrum of drag reduction trends associated with the size and shape of riblets, paving the way for further optimization. Drag reduction achieved through riblets is attributed to their ability to regulate the spanwise movement of streamwise vortices in the near-wall region and to impede the downward transfer of momentum toward the wall by elevating vortices away from it~\citep{chomoikim93,suzkas94,golhansir95}. This mechanism aligns with observations that riblets with spacing smaller than the diameter of near-wall streamwise vortices can prevent these vortices from settling into the riblet grooves~\citep{leelee01}. Building on this understanding, earlier studies parameterize the drag-reducing behavior of riblets using metrics
such as streamwise and spanwise protrusion heights~\citetext{\citealp{becbar89}; \citealp*{lucmanpoz91}; \citealp{ibrgomchugar21}} and {the roughness function~\citep{orlleo06,spamcl11}. Furthermore, \cite{garjim11b} showed that expressing riblet size in terms of the square root of the groove cross-sectional area, $l_g^+ = \sqrt{A_g^+}$, provides a universal drag-reduction curve for riblets of various sizes and shapes, with the optimal size occurring at $l_g^+ \approx 11$ beyond which the drag-reducing performance of riblets would decline. This decline is linked to the breakdown of the viscous regime~\citep{garjim11b},} with numerous studies aiming to uncover geometric features of the surface that can describe the performance of riblets across both drag-reducing and drag-increasing regimes (e.g.,~\cite*{deygatfro22}).

\subsection{Performance decline in large riblets}

Several mechanisms have been proposed to explain the decline in drag reduction as riblet size increases. {\cite{chomoikim93} showed} that riblets with viscous spacings $s^+ \approx 40$ allow streamwise vortices to lodge within their grooves, thereby exposing a larger surface area to turbulent flow. In contrast, riblets with smaller spacings suppress cross-flow in the near-wall region and push streamwise vortices away from the wall. In a complementary study, \cite{suzkas94} conducted experiments demonstrating that a secondary flow develops near the tips of large riblets, enhancing {the downward transport of turbulent momentum and reducing the effectiveness of the riblets. Similarly, the numerical study by \cite{goltua98} attributed the drag increase over widely-spaced riblets to dispersive stresses carried by secondary motions that are absent in smooth-wall turbulent flows. On the other hand, \cite{garjim11b} offered an explanation that hinges on the formation of spanwise-coherent (Kelvin-Helmholtz-like) rollers above the tip of large riblets. This phenomenon has also been observed in flows over plant canopies~\citep{fin00,shagar20}, as well as permeable~\citep{jimuhlpinkaw01} and porous walls~\citep{breboeuit06}, and is attributed to localized transpiration or vertical momentum transport.}

{The coexistence of different mechanisms contributing to the reduced performance of large riblets was finally highlighted in~\cite{modendhutchu21}. By employing minimal-span channel simulations (e.g.,~\cite{macchuhutchaooigar17}), they recognized the viscous-scaled groove width at the riblet mean height as a reliable indicator of whether the flow within the groove is governed by viscous or inertial effects. Vortices were observed to penetrate grooves with a mean-height width of more than 20 viscous units, resulting in the emergence of secondary motions causing dispersive stresses. However, for certain geometries, such as large sharp triangular or blade riblets, the groove width required for the appearance of such flow mechanisms was shown to exceed the drag-reducing optimum ($l_g^+ \approx 11$), suggesting the coexistence of other mechanisms that can also cause the breakdown of the viscous regime. Specifically, enhanced vertical permeability and shear in the inertial flow surrounding the tips of such riblets can instigate a K-H instability that plays a similar role in drag reduction degradation~\citep{garjim11b,endmodgarhutchu21}. In other words, depending on whether turbulence can penetrate deeply into the riblet grooves or not, either K-H rollers or the dispersive stresses can lead to the breakdown of the drag reduction trend observed in riblets with $l_g^+ \lesssim 11$}.

\subsection{Prior model-based efforts in capturing the effects of riblets}

Solving the governing equations with boundary conditions that account for riblet geometry requires a stretched mesh conforming to the surface. However, such an approach demands a high number of discretization points, imposing a computational cost that hinders its feasibility for design optimization and real-time decision-making in engineering applications.  This challenge drives the development of low-complexity models that are capable of capturing the multi-scale nature of {high-Reynolds-number} turbulent flows over periodic surface geometries. In this vein, recent efforts have focused on creating models that accurately represent the dynamics of flow around drag-reducing riblets, aiding their design and providing insights into the mechanisms behind drag increase in off-design conditions. For instance,~\cite{vigcamwonluhgarchugay24} assessed the onset of drag increase using the restricted nonlinear (RNL) and augmented RNL (ARNL) models to identify the prominent nonlinear interactions contributing to drag and to capture their effect with limited wavenumber pairs. {\cite{woncamgarhutchu24} used a viscous vortex model that solves the two-dimensional Stokes-flow equations to predict drag reduction for small to optimally sized riblets. In solving these equations, the boundary conditions on the wall-normal and spanwise velocities were informed by smooth-wall direct numerical simulations (DNS) and correspond to a quasi-streamwise vortex that bounds the computational domain from above. Furthermore,~\citep{botinnahm25} used a variant of the homogenization technique~\citep{bot19} that accounts for advection in conjunction with a synthetic vortex model to capture the transverse flow over riblets and improve overall predictions of skin-friction drag beyond the viscous regime.}

Systems-theoretic tools have also been applied to quantify the stochastic and harmonic responses of turbulent flows over riblets. Notably, the $\cH_2$ norm of the linearized dynamics~\citep*{kasdunpap12} and resolvent analysis~\citep{chaluh19} have been utilized to examine the receptivity of channel flows to corrugated surface geometry. Temporally periodic forcing of the linearized Navier-Stokes (NS) equations has also been employed to model the effect of periodic surfaces~\citep{mormck18,huymck20}. Although these studies provide essential components for reduced-order models capable of receptivity analysis by creatively addressing rough boundary conditions, they fall short in accurately predicting skin-friction drag.~\cite*{ranzarjovJFM21} proposed a model-based framework for {predicting the effects} of riblets, which explicitly accounted for harmonic interactions induced by the spatially periodic geometry. By incorporating the second-order statistics of flow fluctuations around riblets to adjust the turbulent eddy-viscosity near the rough surface, this framework demonstrated reliable drag predictions, even beyond the viscous regime.

A key factor in the effectiveness of reduced-order models for capturing the influence of riblets on flow is their approach to handling rough boundary conditions.~\cite{chaluh19} and~\cite{ranzarjovJFM21} employed a volume penalization technique~\citep{khaangparcal00} to approximate the effects of spatially periodic surfaces on turbulent flow. This method relies on a resistive function that represents the surface geometry as a static feedback term, penalizing the momentum equations within the roughness structure. {Designing} such resistive functions often requires parametric tuning, leading to approximate solutions{, which} may fail to strictly adhere to no-penetration conditions at the riblet surface and do not provide a sharp representation of the immersed boundary~\citep{fadverorlmoh00}. An alternative approach uses discrete forcing to impose boundary conditions directly on the immersed boundary. In this method, each cell adjacent to the fluid domain employs an interpolation scheme to implicitly incorporate the boundary conditions~\citep{mitiac05}. Since the interpolation procedure uses linearization to enforce the appropriate velocity at the first cell outside the boundary, the discrete forcing method requires a sufficiently fine grid near the boundary to maintain accuracy. Unlike volume penalization, discrete forcing is heavily tied to the specifics of the discretization scheme, making its implementation more complex. Moreover, imposing pressure boundary conditions on the immersed boundary involves solving a Poisson equation, which can introduce spurious pressure oscillations at the wall~\citep{mitdonboznajvarloe08,ver23}.

An effective approach to capturing the effects of corrugated boundaries is to transform the physical domain into a computational domain that incorporates the boundary geometry into the differential operators~\citep*{cabszuflo02}. For spatially periodic riblets, this transformation results in spatially periodic differential operators. Although this method complicates the governing equations, it allows for an accurate representation of the surface geometry. Previous studies have employed this technique to analyze the stability of channel flow over longitudinal riblets and to investigate transition mechanisms in the presence of riblets~\citep{ehr96,kasdunpap12,morflo14}. More recently,~\cite*{jourobche24} employed the domain transformation method to perform both modal and non-modal stability analyses of transitional channel flow over riblets, demonstrating that riblets can induce an earlier laminar-turbulent transition through the formation of oblique waves.

\subsection{Preview of modeling framework and results} 

{In~\cite{ranzarjovJFM21}, volume penalization was combined with the turbulence modeling framework of~\cite{moajovJFM12} to investigate the influence of riblets on turbulent channel flow. It was demonstrated that the statistical response of the eddy-viscosity-enhanced linearized NS equations can effectively capture the impact of background turbulence on the mean velocity and aid the prediction of skin-friction drag. Specifically, a spatially period resistive function was introduced into the governing equations to model the effect of the riblet surface on the flow dynamics. However, this approach induces a broad range of harmonic interactions in the fluctuation field, requiring a prohibitively large state space for accurate analysis. These computational challenges become especially pronounced at high Reynolds numbers and for sharp riblet geometries.}

{In the present work, we build on this foundation and address the associated limitations by introducing techniques that offer  flexibility and computational efficiency in capturing the effects of sharp riblets in high-Reynolds-number flows. Unlike~\cite{ranzarjovJFM21}, we employ the domain transformation technique of~\cite{ehr96} which facilitates an accurate representation of riblet-induced geometric effects. In the transformed domain, we show that the height of riblets can be treated as a small parameter, enabling a perturbation-based analysis that significantly reduces the computational cost of evaluating turbulent flow statistics.}
{We use our simulation-free approach to study various riblet geometries, especially scalloped riblets, a geometry known for its robustness and superior drag-reducing performance relative to more commonly studied shapes such as blades and sawtooth riblets~\citep{becbruhaghoehop97}. Our approach not only predicts drag reduction trends consistent with existing experimental and numerical data~\citep{becbruhaghoehop97,garjim11b}, but also enables detailed investigation of underlying flow mechanisms—including Kelvin–Helmholtz (K–H) instabilities and the near-wall cycle—at high friction Reynolds numbers (up to $Re_\tau = 2003$ in this study), even in the presence of large and sharp riblets.}

\subsection{Paper outline}

The paper is organized as follows. In \S~\ref{sec.formulation}, we formulate the problem, introduce the domain transformation to capture the shape of riblets, evaluate the mean flow in the transformed coordinates, and describe the necessity for studying the dynamics of velocity fluctuations. In \S~\ref{sec.Velfluc}, we form the linearized eddy-viscosity-enhanced NS equations around the initial mean velocity profile and employ perturbation analysis to efficiently compute the second-order statistics of velocity fluctuations, which are then used to correct the turbulent viscosity and refine our predictions of the mean velocity and skin-friction drag. In \S~\ref{sec.result}, we use our approach to capture the drag-reducing trends of scalloped riblets in a turbulent channel flow. In \S~\ref{sec.structure}, we analyze the statistical response of the linearized dynamics to explain the degraded performance of large riblets at high Reynolds numbers. Finally, in \S~\ref{sec.conclusion}, we conclude with a summary of contributions and an outlook for future research directions.
 
\section{Problem formulation}
\label{sec.formulation}
\begin{figure}
        \begin{center}
        \begin{tabular}{cccc}
        \hspace{-.6cm}
        \subfigure[]{\label{fig.channelriblets}}
        &&
        \subfigure[]{\label{fig.ribshapes}}
        &
        \\[-.4cm]
	&
	\begin{tabular}{c}
       \includegraphics[width=0.4\textwidth]{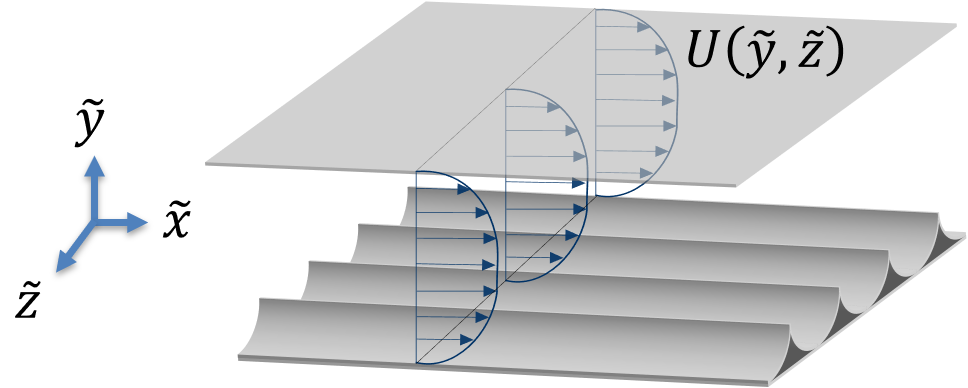}
       \end{tabular}
       &&
       \vspace{.1cm}
       \begin{tabular}{c}
       \includegraphics[width=0.4\textwidth]{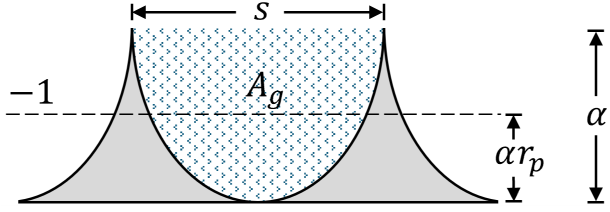}
       \end{tabular}
       \end{tabular}
       \end{center}
        \caption{(a) Configuration of a channel flow with streamwise-constant spanwise-periodic riblets on the lower wall together with turbulent mean velocity profiles. (b) scalloped riblets of height $\alpha$, peak to peak spacing $s=2\pi/\omega_z$, and groove cross-sectional area  $A_g$. The parameter $r_p$ represents the proportion of the riblet height {below $-1$}.}
\end{figure}

The incompressible NS and continuity equations governing the dynamics of turbulent flow within a channel that has longitudinal riblets mounted on its lower wall (figure~\ref{fig.channelriblets}) are given by
\begin{subequations}
\label{eq.NS-BC}
\begin{eqnarray}
\label{eq.NS}
    \ba{rcl}
        \partial_t \tilde{\bu} 
        &\!\!=\!\!&
         - \,
        (\tilde{\bu} \cdot  \tilde{\nabla}) \tilde{\bu}
                \,-\,
        \tilde{\nabla} \tilde{P}
                \,+\,
        \dfrac{1}{Re_\tau} \, \tilde{\Delta}\, \tilde{\bu}, 
        \\[0.15 cm]
        0 
        &\!\!=\!\!&
        \tilde{\nabla} \cdot \tilde{\bu},
    \ea
\end{eqnarray}
subject to no-slip and no-penetration boundary conditions that respect the shape of the spanwise-periodic surface corrugation dictated by the shape function $r(\tilde{z})>0$, i.e.,
\begin{align}
\label{eq.BC}
    \tilde{\bu}(\tilde{x},\tilde{y}=1,\tilde{z},t) \;=\; 0, \quad  \tilde{\bu}(\tilde{x},\tilde{y}=-1+r(\tilde{z}),\tilde{z},t) \;=\; 0.
\end{align}
\end{subequations}
Here, $\tilde{x}$, $\tilde{y}$, and $\tilde{z}$ denote the streamwise, wall-normal, and spanwise coordinates, respectively, $t$ is time, $\tilde{\bu}$ is the velocity vector, $\tilde{P}$ is the pressure, $\tilde{\nabla}$ is the gradient, $\tilde{\Delta}=\tilde{\nabla}\cdot\tilde{\nabla}$ is the Laplacian, and $Re_\tau=u_{\tau}h /\nu$ is the friction Reynolds number defined in terms of the friction velocity $u_{\tau} = \sqrt{\tau_w/\rho}$, where $\tau_w$ is the wall-shear stress (averaged over horizontal directions and time) and $\rho$ is the fluid density, $h$ is the channel half height, and $\nu$ is the kinematic viscosity. 
In the governing equations, space is non-dimensionalized by $h$, velocity by $u_\tau$, time by $h/u_{\tau}$, and pressure by $\rho u_{\tau}^{2}$. In this paper, we analyze the effect of riblets under constant-bulk conditions, in which the bulk flux remains constant via adjustment of the streamwise pressure gradient $P_x$. 

\begin{figure}
  \centerline{\includegraphics[width=10cm]{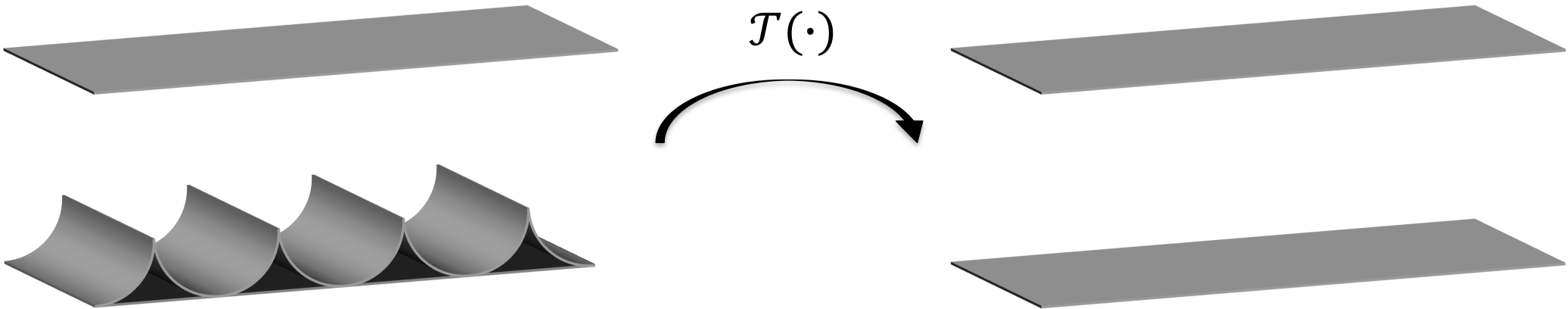}}
  \caption{Schematic of the domain transformation $\mathcal{T}(\cdot)$ to translate the effect of spanwise periodic surface roughness onto the differential operators.}
\label{fig.coc}
\end{figure}

In solving the NS equations subject to boundary conditions that capture the shape of riblets (equations~\eqref{eq.NS-BC}), multiple approaches have been offered to bypass the need for a stretched mesh that conforms to the surface geometry. Herein, we employ the domain transformation
\begin{align}
\label{eq.coor-transform}
    x \; = \; \tilde{x},
    \quad 
    y \; = \; F(\tilde{y},\tilde{z}),
    \quad 
    z \; = \; \tilde{z},
\end{align}
that maps the physical domain $(\tilde{x}, \tilde{y}, \tilde{z})$ where $\tilde{y}\in [-1+r(\tilde{z}),1]$ to a computational domain in which $y\in[-1,1]$
(cf.~figure~\ref{fig.coc}). This is achieved using the mapping function
\begin{align}
\label{eq.mapping}
            F(\tilde{y},\tilde{z})
            \; \DefinedAs \;
            \dfrac{2\tilde{y} - r(\tilde{z})}{2-r(\tilde{z})}.
\end{align}
{The shape function $r(\tilde{z})$ captures the spanwise dependence of the surface corrugation, e.g., for scalloped riblets with spanwise frequency $\omega_z$, the shape function is constructed by concatenating two sinusoidal segments over the spanwise interval $[-\pi/\omega_z,\pi/\omega_z]$.} 
{In addition to the aforementioned mapping,} the riblet base is lowered by a proportion of the roughness height {($\alpha$)}, i.e., $\alpha\, r_p$ with $r_p\in[0,1]$, to ensure that the bulk flux matches that of a smooth channel flow of height $2$; see figure~\ref{fig.ribshapes}. {This results in a stretched vertical extent for both} the original and transformed wall-normal domains to $\tilde{y}\in[-1-\alpha \,r_p,1]$ and $y\in[-1-\alpha \,r_p,1]$, respectively. {We note that in this study, the value of $r_p$ is determined in an iterative manner.}
{Following the chain rule,} the domain transformation reflects the surface geometry on the differential operators as
\begin{align}
\label{eq.diffOP}
    \ba{rclrclrcl}
        \partial_{\tilde{x}} 
        &\!\!\!=\!\!\!& 
        \partial_{x},
        &
        \partial_{\tilde{y}} 
        &\!\!\!=\!\!\!& 
        F_{\tilde{y}}\, \partial_{y},
        & 
        \partial_{\tilde{z}} 
        &\!\!\!=\!\!\!&  
        F_{\tilde{z}}\, \partial_{y} \,+\, \partial_{z},
        \\[0.15cm]
        \partial_{\tilde{x}\tilde{x}} 
        &\!\!\!=\!\!\!& 
        \partial_{xx},
        & 
        \partial_{\tilde{y}\tilde{y}} 
        &\!\!\!=\!\!\!&  
        F_{\tilde{y}}^{2}\, \partial_{yy},
        & 
        \partial_{\tilde{z}\tilde{z}} 
        &\!\!\!=\!\!\!& 
        F_{\tilde{z}}^{2}\, \partial_{yy} \,+\, 2 F_{\tilde{z}}\, \partial_{yz} \,+\, \partial_{zz},
    \ea
\end{align}
where $F_{\tilde{y}}$ and $F_{\tilde{z}}$ are the ${\tilde{y}}$ and ${\tilde{z}}$ derivatives of the mapping function~\eqref{eq.mapping}, respectively.

While carefully designed small-size riblets have been shown to reduce drag and suppress the energy of the flow, large riblets are known for converse effects. This results in an {\em optimal\/} parameterization for conventional riblets (e.g.,~\cite{garjim11a}) that corresponds to the maximum achievable reduction in drag. In this paper, we not only analyze the effect of riblets on skin-friction drag and the turbulent kinetic energy, but also conduct a model-based analysis of previously identified flow mechanisms that not only deteriorate the drag-reducing capabilities of larger riblets, but can lead to an increase in skin-friction drag.

\subsection{Mean flow equations}
\label{sec.Umean}

{Skin-friction drag depends on the gradient of the turbulent mean velocity at the wall. The mean flow equations can be derived by applying the Reynolds decomposition to the governing equations~\eqref{eq.NS} as}
\begin{align}
\label{eq.RANS}
    \ba{rcl}
        \partial_t \bar{\bu} 
        &\!\!=\!\!&
            - \,
        (\bar{\bu} \cdot  \nabla) \bar{\bu}
                \,-\,
        \nabla \bar{P}
                \,+\,
        \dfrac{1}{Re_\tau} \, \Delta \bar{\bu}
                \,-\,
        \nabla \cdot \langle\textbf{v}\textbf{v}^T\rangle,
        \\[0.15 cm]
        0 
        &\!\!=\!\!&
        \nabla \cdot \bar{\bu}.
    \ea
\end{align}
Here, $\bar{\bu}=[\,U \,~ V \,~ W\,]^T$ is the vector of mean velocity components, $\bv = [\,u ~\, v ~\, w\,]^T$ is the vector of {zero-mean velocity fluctuations around $\bar{\bu}$ ($\langle\textbf{v}\rangle=0$)}, $p$ is the {zero-mean fluctuating pressure around the mean $\bar{P}$ ($\langle p \rangle=0$)}, $\langle\cdot\rangle$ denotes the expected value,
\begin{align}
\label{eq.expect}
        \langle\bu(x,y,z,t)\rangle
        \;=\;
        \ds{\lim_{T\rightarrow \infty }}\, \dfrac{1}{T} \int_{0}^{T}\bu(x,y,z,t+\tau) \,\mrd\tau,
\end{align}
and $\nabla$ and $\Delta$ are the gradient and Laplacian that take the following form by virtue of the domain transformation~\eqref{eq.coor-transform}:
\begin{align*}
            \nabla
            \;=\;
            \left [\, \partial_{x},\,~ F_{\tilde{y}} \partial_{y},\,~ F_{\tilde{z}} \partial_{y} \,+\, \partial_{z}\, \right ]^T,
            \quad
            \Delta
            \;=\;
            \partial_{xx} \,+\, F_{\tilde{y}}^{2}\, \partial_{yy} \,+\, F_{\tilde{z}}^{2} \partial_{yy} \,+\, 2 F_{\tilde{z}}\, \partial_{yz} \,+\, \partial_{zz}. 
\end{align*}
In equations~\eqref{eq.RANS}, the Reynolds stress tensor $\langle\textbf{v}\textbf{v}^T\rangle$ captures the effect of background turbulence by quantifying momentum transport due to turbulent fluctuations~\citep{mcc91}, but is unknown. In the absence of a fully determined stress-tensor, the mean flow equations are not closed and cannot be solved without adopting a turbulence model. To overcome the closure problem, we employ the turbulent viscosity hypothesis~\citep{mcc91} and assume turbulent momentum to be transported in the direction of the mean rate of strain, i.e.,
\begin{align}
\label{eq.turbvischyp}
        \langle\overline{\textbf{v}\textbf{v}^T} \rangle
        \,-\,
        \dfrac{1}{3}\trace\left(  \langle\overline{\textbf{v}\textbf{v}^T} \rangle \right)I 
        \;=\;
        -\,
        \dfrac{\nu_T}{Re_\tau}\left( \nabla\bar{\bu} 
        \, + \,
        (\nabla\bar{\bu})^T \right).
\end{align}
Here, overline indicates averaging over horizontal dimensions, $I$ is the identity operator, and $\nu_T(y)$ is the turbulent eddy viscosity normalized by molecular viscosity. Incorporating the turbulent viscosity hypothesis~\eqref{eq.turbvischyp} into equations~\eqref{eq.RANS} yields
\begin{align}
\label{eq.RANSVisc}
    \ba{rcl}
        \partial_t \bar{\bu} 
        &\!\!=\!\!&
                - \,
        (\bar{\bu} \cdot  \nabla) \bar{\bu}
                \, - \,
        \nabla \bar{P}
                \, + \,
        \dfrac{1}{Re_\tau} \nabla \cdot \left ( \left(1+\nu_T\right) ( \nabla\bar{\bu} \, + \, (\nabla\bar{\bu})^T  ) \right ), 
        \\[0.15cm]
        0 
        &\!\!=\!\!&
        \nabla \cdot \bar{\bu}.
    \ea
\end{align}

After applying the domain transformation~\eqref{eq.coor-transform}, the steady-state solution to the nonlinear mean flow equations~\eqref{eq.RANSVisc} can be obtained using Newton's method to only contain a streamwise velocity component, i.e., $\bar{\bu}\:=\:[U(y,z) \;\; 0 \;\; 0]^T$, which solves the linear equation
\begin{align}
\label{eq.meanVelT}
            (1 \,+\, \nu_{T})\left [F_{\tilde{y}}^{2}\, U_{yy}
            \,+\,
            F_{\tilde{z}}^{2}\, U_{yy} 
            \,+\, 
            2F_{\tilde{z}}\, U_{yz} 
            \,+\, U_{zz} \right ] 
            \,+\, F_{\tilde{y}}^{2} \,\nu_T'\, U_y
            \; = \;
            Re_{\tau}\,\bar P_x 
\end{align}
Here, $\nu_{T}'$ is the wall-normal derivative of $\nu_T$, and the mean velocity $U(y,z)$ obeys no-slip boundary conditions on both walls by virtue of the domain transformation. Due to the periodic geometry of riblets, a harmonic expansion of the mapping function $F(\tilde{y},\tilde{z})$ (equation~\eqref{eq.mapping}) in the spanwise direction, i.e., 
\begin{align}
\label{eq.FFourier}
    	F(\tilde{y},\tilde{z})
            \;=
            \ds{\sum\limits_{m=-\infty}^{\infty} {F}_{m}(\tilde{y})\,\mre^{\mri m\, \omega_z\tilde{z}},}
\end{align}
warrants the parameterization of equation~\eqref{eq.meanVelT} and its solution $U$ over the spanwise frequency of riblets, $\omega_z$, i.e.,
\begin{align}
    \label{eq.UFourier}
    U(y,z)
    \;=
    \ds{\sum\limits_{k=-\infty}^{\infty} U_{k}(y) \,\mre^{\mri k\omega_zz}.}
\end{align}
Substituting expansions~\eqref{eq.FFourier} and~\eqref{eq.UFourier} into equation~\eqref{eq.meanVelT} yields the equation for the $k$th harmonic $U_k(y)$ as
\begin{align*}
    \hspace{-.2cm}
             \underbrace{\Big[(1+\nu_T)
            \left(({F}_{\tilde{y},0}^{2}+{F}_{\tilde{z},0}^{2})\,\partial_{yy} + 2 {F}_{\tilde{z},0}\, \partial_{yz} + \partial_{zz}\right)
             + 
             \nu_T'{F}_{\tilde{y},0}^{2}\partial_{y}\Big]}_{\mathbf{L}_{k,0}} U_{k} +\! \underbracea{ \sum\limits_{m=-\infty\backslash{\{0\}}}^{k} \!\left[ \nu_T' {F}_{\tilde{y},m}^{2}\,\partial_{y} \right.}
            \\[0.2cm]
            \hspace{1.2cm}
            \underbraced{\left. + \, (1+\nu_T)\!
            \left(({F}_{\tilde{y},m}^{2}+{F}_{\tilde{z},m}^{2})\,\partial_{yy}
            + 
            2
            (k-m){F}_{\tilde{z},m}\,\partial_{yz}\right) \right]}_{\mathbf{L}_{k,m}}\! U_{k-m}
            \, = \,
            \left\{\begin{matrix}
            Re_{\tau}\bar P_x,& k=0\\ 
            0,& k\neq 0
            \end{matrix}\right.
\end{align*}
which can be brought into the bi-infinite matrix form
\begin{align}
\label{eq.meanVelmateix}
    \ds{
         \begin{bmatrix}
        \ddots  & \vdots  &  \vdots & \vdots  & \iddots 
        \\[0.1cm] 
        \cdots  & {\bf L}_{-1,0} & {\bf L}_{-1,1} & {\bf L}_{-1,2} & \cdots 
        \\[0.1cm]
        \cdots & {\bf L}_{0,-1} & {\bf L}_{0,0} & {\bf L}_{0,1} & \cdots 
        \\[0.1cm] 
        \cdots & {\bf L}_{1,-1} & {\bf L}_{1,-1} & {\bf L}_{1,0} & \cdots
        \\[0.1cm] 
         \iddots & \vdots  & \vdots  & \vdots  &  \ddots
        \end{bmatrix}
        \,
        \begin{bmatrix}
        \vdots\\[0.1cm] 
        U_{-1} \\[0.1cm] 
        U_{0}\\[0.1cm] 
        U_{1} \\[0.1cm] 
        \vdots
        \end{bmatrix}
    }
    \; = \;
    \begin{bmatrix}
    \vdots\\[0.1cm] 
    0 \\[0.1cm] 
    Re_\tau \bar{P}_x\\[0.1cm] 
    0 \\[0.1cm] 
    \vdots
    \end{bmatrix}.
\end{align}
Depending on the significance of higher-order harmonics in the Fourier expansion of $F$ (equation~\eqref{eq.FFourier}), the bi-infinite matrices and vectors can be truncated to account for a finite number of harmonic that provide a good approximation to the solution of equation~\eqref{eq.meanVelT}.

{For small-size riblets,} the flow in the vicinity of the solid surface is dominated by viscosity, and can therefore, be assumed laminar within the grooves of riblets~\citep{garjim11a}.
Because of this, we consider $\nu_T = 0$ for {$y\leqslant -1$} (cf.~figure~\ref{fig.ribshapes}).
{Note that due to the inclusion of $r_p$, which controls the level of protrusion into the turbulent regime, our model remains valid even when turbulence penetrates into the riblet grooves.}
On the other hand, a well defined turbulent viscosity $\nu_T$ is needed for {$y> -1$}, which of course, is not easy to come by as it depends on the velocity fluctuations around the turbulent mean $U$. A good starting point may be provided by the turbulent viscosity profile {of a smooth channel flow. While such a viscosity profile can be directly computed from DNS data, in this study, we take the analytical expression given by~\citep{reytie67}
\begin{align}
\label{eq.Cessmodel}
        \nu_{T_0}(y)
        \,=\,
        \dfrac{1}{2} \left(\left( 1 + \left( \dfrac{c_1}{3} Re_\tau\,(1-y^2)(1+ 2y^2)(1 - \mre^{-(1 - |y|)Re_\tau/c_2}) \right)^2\, \right)^{1/2} \!-\, 1\right),
\end{align}
which is obtained from extending the model introduced by Cess for pipe flow~\citep{ces58} to the channel flow. This model has been shown to reasonably approximate the turbulent mean velocity in channel flow, especially at high Reynolds numbers. In equation~\eqref{eq.Cessmodel},} parameters $c_1$ and $c_2$ are selected to minimize the least squares deviation between the mean streamwise velocity obtained in experiments and simulations and the steady-state solution to the mean flow equations using the averaged wall-shear stress $\tau_w=1$. {For example, at $Re_\tau = 186$, $c_1 =0.61$ and $c_2 = 46.2$ were reported by \cite{moajovJFM12} to provide the best fit to the turbulent mean velocity resulting from the DNS of~\cite{deljim03}; see table~\ref{tab.ribconf} for the optimal choices of $c_1$ and $c_2$ at other Reynolds numbers.} Given the aforementioned parameterization for $\nu_T$, we can solve equation~\eqref{eq.meanVelmateix} using block operators~\citep{aurtre17} that divide the wall-normal extent of the computational domain into upper turbulent and lower laminar regions. In this case, smoothness would be enforced at the intersection of these regions via interface conditions,
\begin{align*}
    \ba{rcl}
        U_k(y=-1^+,z) 
        &\!\!=\!\!& 
        U_k(y=-1^-,z),
        \\[0.2cm]
        \dfrac{\partial U_k}{\partial y} (y=-1^+,z) 
        &\!\!=\!\!& 
        \dfrac{\partial U_k}{\partial y} (y=-1^-,z)
    \ea
\end{align*}
for all $k$. Figure~\ref{fig.MeanVelocity} shows the solution to equation~\eqref{eq.meanVelT} for a turbulent channel flow with $Re_{\tau} = 186$ subject to $\bar P_x = -1$ over scalloped riblets with $\alpha/s = 0.87$, $\omega_z = 60$, and $r_p=0.487$. Here, we use a pseudospectral scheme with Chebyshev polynomials~\citep{Reddy2000} with $N_t = 140$ and $N_b = 30$ collocation points to discretize the top ($y \in [-1,1]$) and bottom ($y\in[-1-\alpha\, r_p,-1]$) portions of the wall-normal dimension, respectively, and $25$ harmonics were used to capture the spanwise-periodic shape of the riblets, i.e., $m \in [-12, 12]$ in the Fourier expansion of $F(\tilde{y},\tilde{z})$ (equation~\eqref{eq.FFourier}).

\subsection{Skin-friction drag reduction}
\label{sec.DRnocontrol}
\begin{figure}
        \begin{center}
        \begin{tabular}{cccc}
        \hspace{-.6cm}
        \subfigure[]{\label{fig.MeanVelocity}}
        &&
        \hspace{-1cm}
        \subfigure[]{\label{fig.DRnocontrol}}
        &
        \\[-.6cm]
        \hspace{-.1cm}
	\begin{tabular}{c}
        \vspace{.6cm}
        \rotatebox{90}{{$\tilde{y}$}}
       \end{tabular}
       &\hspace{-.3cm}
	\begin{tabular}{c}
       \includegraphics[width=0.46\textwidth]{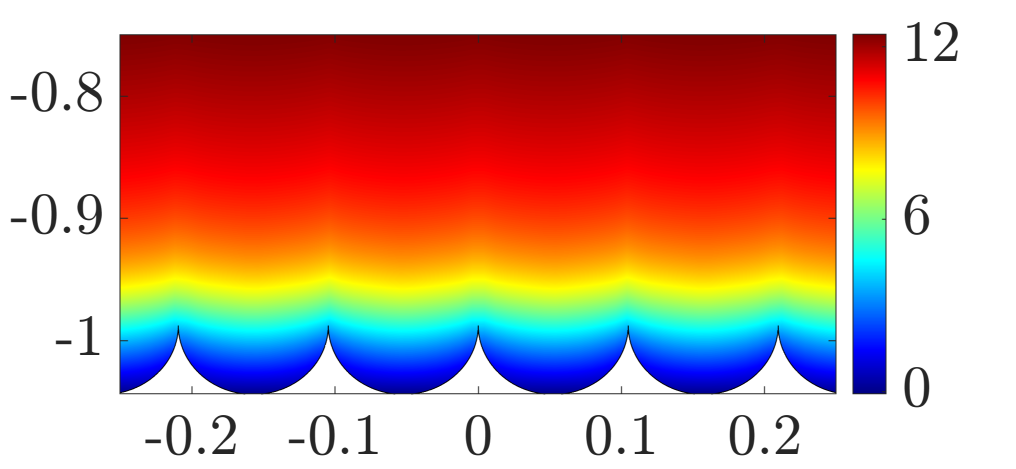}
        \\[-.05cm]
            \hspace{-.45cm}
            {$\tilde{z}$}
       \end{tabular}
       &\hspace{-.4cm}
       \begin{tabular}{c}
        \vspace{.2cm}
        {\small \rotatebox{90}{{$\Delta D$}}}
       \end{tabular}
       &\hspace{-.5cm}
    \begin{tabular}{c}
       \includegraphics[width=0.45\textwidth]{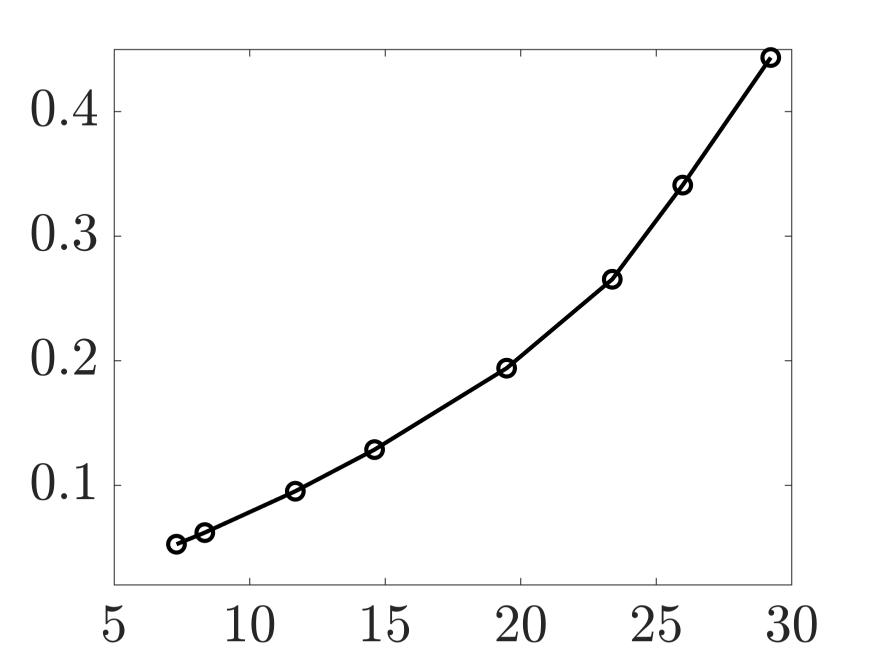}
       \\[-.1cm]
            \hspace{.2cm}
            $s^+$
       \end{tabular}
       \end{tabular}
       \end{center}
        \caption{{(a) The streamwise mean velocity $U(\tilde{y},\tilde{z})$ for turbulent channel flow with $Re_{\tau} = 186$ over scalloped riblets of $\alpha/s=0.87$ and $\omega_z = 60$; (b) The relative drag reduction $\Delta D \DefinedAs (D - D_s)/D_s$ computed using the solution to equation~\eqref{eq.meanVelT} with $\nu_T=\nu_{T_0}$.}}
        \label{fig.velocitydragnocontrol}
\end{figure}

In the presence of riblets, skin-friction drag at the lower wall, $D$, can be computed using the slope of the mean velocity at the upper wall,
\begin{align}
    \label{eq.drag}
    D 
    \;=\;
    \ds{\bar{P}_x \,-\, \dfrac{\omega_z}{2\pi}\, \int_{0}^{2\pi/\omega_z} \dfrac{\partial U}{\partial y}(y=1,z)\, \mrd z}
\end{align}
where $\bar{P}_x = -D_s/Re_{\tau}$, with $D_s$ denoting the slope of the mean velocity at the wall in the absence of riblets. 
It is evident that the mean velocity in figure~\ref{fig.MeanVelocity} respects the shape of the riblets. However, a parametric study over riblets with $\alpha/s = 0.87$ but with different frequencies $\omega_z$ fails to capture any drag reduction and provides no optimal spacing that maximizes drag reduction (figure~\ref{fig.DRnocontrol}). What's more, the resulting drag reduction does not follow commonly reported trends from prior numerical and experimental studies (see, e.g.,~\cite{becbruhaghoehop97,garjim11b}). While the turbulent viscosity hypothesis provides an opportunity for closing the mean flow equations, the choice of an appropriate turbulent eddy-viscosity $\nu_T$ that captures the effect of riblets on the background turbulence challenges our analysis. To address this issue, we adopt the iterative procedure of~\cite{moajovJFM12,ranzarjovJFM21} in starting from the turbulent viscosity over smooth surfaces and utilizing the dynamics of velocity fluctuations $\bv$ to correct this initial profile.

\section{Turbulence modeling in channel flow over riblets}
\label{sec.Velfluc}
{In this section, we adopt the turbulence modeling framework of~\cite{moajovJFM12} to determine a turbulent viscosity, $\nu_T$, that enables an accurate prediction of skin-friction drag in channel flow over riblets. This is achieved through a sequence of steps that introduce modifications to $\nu_{T_0}$, i.e., the turbulent viscosity of a smooth channel flow. First, an initial estimate of the turbulent mean velocity is obtained by solving equation~\eqref{eq.meanVelT} with $\nu_{T_0}$. Although this estimate does not achieve the correct skin-friction drag (cf.~figure~\ref{fig.DRnocontrol}), it serves as a base state for linearizing the NS equations and analyzing velocity fluctuations. The second-order statistics of the velocity field obtained from the linearized model are then used to modify $\nu_{T_0}$ via chosen turbulence model.
A key advantage of the coordinate transformation method employed here is its ability to compute these modifications through perturbation analysis in the riblet height, $\alpha$, which significantly enhances scalability of our approach compared to~\cite{ranzarjovJFM21}. The updated turbulent viscosity is subsequently used to refine predictions of the mean velocity and skin-friction drag. In \S~\ref{sec.result}, we validate this framework using triangular riblets and analyze the effects of scalloped riblets on skin-friction drag and turbulent stresses. Additionally, in \S~\ref{sec.structure}, we use flow statistics obtained from the linearized Navier–Stokes equations to investigate the physical mechanisms influencing drag in the presence of sharp riblets of varying size.}

\subsection{Turbulent viscosity model}
\label{sec.nuT}
If $k/\eps$ and $k^{3/2}/\eps$ are chosen as time and length scales, turbulent viscosity can be expressed as~\citep[Chapter 10]{pop00}
\begin{align}
\label{eq.KEpsilon}
            \nu_T\;=\;c\,Re_{\tau}^{2}\,\dfrac{{k}^2}{\epsilon},
\end{align}
where $k$ and $\eps$ are the turbulent kinetic energy and its rate of dissipation, respectively, and {$c=0.09$ is an established empirical constant in the absence of riblets}~\citep{pop00}. The kinetic energy and its rate of dissipation can be determined from second-order velocity statistics as
\begin{align}
\label{eq.kepsmodel}
    \ba{rcl}
        k(y)
        &\!\!=\!\!&
        \dfrac{1}{2}\left( \langle \overline{uu} \rangle+\langle \overline{vv} \rangle+\langle \overline{ww} \rangle \right), 
        \\[0.25 cm]
        \epsilon (y)
        &\!\!=\!\!&
        2\left( \langle \overline{u_xu_x} \rangle+\langle \overline{v_yv_y} \rangle+\langle \overline{w_zw_z} \rangle+\langle \overline{u_yv_x} \rangle+\langle \overline{u_zw_x} \rangle+\langle \overline{v_zw_y} \rangle\right)
        \\[0.25cm]
        && ~ + \;
        \langle \overline{u_yu_y} \rangle+\langle \overline{w_yw_y} \rangle+\langle \overline{v_xv_x} \rangle+\langle \overline{w_xw_x} \rangle+\langle \overline{u_zu_z} \rangle+\langle \overline{v_zv_z} \rangle.
    \ea
\end{align}
The widely used $k-\epsilon$ model~\citep{jonlau72,lausha74} provides differential transport equations for $k$ and $\epsilon$, but is computationally demanding. An alternative simulation-free way of approximating these quantities is to compute them from the second-order statistics of the linearized flow dynamics around the mean velocity.

\subsection{Stochastically forced linearized Navier-Stokes equations}

By linearizing the NS equations~\eqref{eq.NS} around the mean velocity $\bar{\bu}= [\,U(y,z) \,~ 0 \,~ 0\,]^T$ obtained from solving equation~\eqref{eq.meanVelT} and $\bar{P}$, we arrive at the dynamics of velocity fluctuations
\begin{align}
\label{eq.linfluct}
	\ba{rcl}
    		\partial_t \bv
    		&\!\!=\!\!&
    		-
    		\left(  \nabla  \cdot  \bar{\bu}  \right) \bv
     		\, - \,
            \left( \nabla  \cdot  \bv  \right) \bar{\bu}
                \, - \,
    		{\nabla} {p}
    		  \, + \,
    		\dfrac{1}{Re_\tau} {\nabla} \cdot ((1 \,+\, \nu_{T})(\nabla \bv \,+\, (\nabla \bv)^T))
                 \, + \,
            {\mathbf{f}}
    		\\[.15cm]
    		0
    		&\!\!=\!\!&
    		\nabla \cdot \bv .
    \ea
\end{align}
In these so called eddy-viscosity-enhanced linearized NS equations, molecular viscosity is augmented by turbulent viscosity $\nu_T$ to compensate for the nonlinear terms that are dropped through linearization~\citep{Reyhus72,deljimjfm06,pujgarcosdep09,hwacosJFM10b,jovARFM21,abozarJFM23} and ${\mathbf{f}}$ is a zero-mean white-in-time stochastic forcing that excites the stochastic response of the linearized dynamics. Due to the domain transformation~\eqref{eq.coor-transform}, the boundary conditions for equations~\eqref{eq.linfluct} are given by $\bv(x,y=\pm1,z,t) = 0$ and the $2\pi/\omega_z$ periodicity of the surface is instead reflected onto the differential operators. As a result, in contrast to the smooth channel flow, the normal mode in the spanwise direction is no longer given by $\mre^{\mri k_z z}$; the normal modes in the spatially periodic direction are given by Bloch waves~\citep*{odekel64,benliopap78}, resulting in the wavenumber parameterization 
\begin{align*}
    \bv(x,y,z,t)\;=\;\mre^{\mri k_xx}\mre^{\mri\theta z}\,\hat{\bv}_\theta(k_x,y,z,t)
\end{align*}
with $\theta \in [0,\omega_z)$ for velocity fluctuations and all other quantities in equation~\eqref{eq.linfluct}. Here, $k_x \in \mathbb{R}$ is the streamwise wavenumber and $\hat{\bv}_\theta(k_x,y,z,t)$ is a $2\pi/\omega_z$-periodic function in the spanwise direction with Fourier series expansion 
\begin{align}
\label{eq.offsetFourier}
    \hat{\bv}_\theta(k_x,y,z,t) \;=\; \sum_{n \in \mathbb{Z}} \hat{\bv}_n(k_x,y,\theta,t)\,\mre^{\mri n \omega_z z}.
\end{align}
In this expansion, $\{\hat{\bv}_n(k_x,y,\theta,t)\}_{n \in \mathbb{Z}}$ are the coefficients of the Fourier series expansion of $\hat{\bv}_\theta(k_x,y,z,t)$. Based on this, 
\begin{align}
\label{eq.bloch}
    \bv(x,y,z,t)\;=\;\sum_{n \in \mathbb{Z}} \hat{\bv}_n(k_x,y,\theta,t)\,\mre^{\mri \left(k_xx + \theta_n z\right)},
\end{align}
where $\theta_n=\theta + n\,\omega_z$ is the spanwise wavenumber. By substituting~\eqref{eq.bloch} into the linearized dynamics~\eqref{eq.linfluct} and eliminating pressure through a standard conversion~\citep{schhen01}, the differential equations for the dynamics of $\bv$ can be brought into the evolution form
\begin{align}
\label{eq.lnse}
	\ba{rcl}
    		\partial_t \bpsi_\theta(k_x,y,t)
    		&\!\!=\!\!&
    		\left [ \cA_\theta(k_x)\bpsi_\theta(k_x,\;\cdot\;,t)\right ](y) 
            \,+\,
            \cB_\theta(k_x)\mathbf{f}_\theta(k_x,y,t),
    		\\[.15cm]
    		\bv_\theta(k_x,y,t)
    		&\!\!=\!\!&
    		\left [\cC_\theta(k_x)\bpsi_\theta(k_x,\;\cdot\;,t)\right ](y),
    \ea
\end{align}
in which the state vector $\bpsi$ consists of the wall-normal velocity $v$ and vorticity $\eta \,=\, \partial_z u \,-\, \partial_x w$. Here, $\bpsi_\theta$, $\mathbf{v}_\theta$, and $\mathbf{f}_\theta$ are bi-infinite column vectors parameterized by the streamwise wavenumber $k_x$ and the spanwise wavenumber offset $\theta$, e.g., for each $(k_x,\theta)$ pair, $\bpsi_\theta(k_x,y,t)=\col\left \{ \hat{\bpsi}_n(k_x,y,\theta,t) \right \}_{n\in\mathbb{Z}}$ with $\hat{\bpsi}_n = [\,\hat{v}_n \,~ \hat{\eta}_n\,]^T$ for any integer $n$, and the state  $\cA_\theta(k_x)$, input $\cB_\theta(k_x)$, and output  $\cC_\theta(k_x)$ matrices are bi-infinite with operator-valued elements in $y$; see appendix~\ref{app.opABC} for details. 
We note that the input matrix $\cB_\theta(k_x)$ results from the conversion of equation~\eqref{eq.linfluct} into the evolution form~\eqref{eq.lnse} and the output matrix $\cC_\theta(k_x)$  establishes a kinematic relation between the state $\bpsi_\theta$ and the velocity vector $\mathbf{v}_\theta$. At both the top and bottom walls of the channel, homogeneous Dirichlet and Neumann boundary conditions are imposed on $\hat{v}_n$, and homogeneous Dirichlet boundary conditions are imposed on $\hat{\eta}_n$. Moreover, smoothness of the solution at the intersection of the top and bottom wall-normal regions, i.e., $y=-1$, is ensured by enforcing the following conditions:
\begin{align*}
    \ba{rclrcl}
        \hat{v}_n(y=-1^+,z) 
        &\!\!\! = \!\!\!& 
        \hat{v}_n(y=-1^-,z), \quad 
        &
        \dfrac{\partial^i \hat{v}_n}{\partial y^i} (y=-1^+,z) 
        &\!\!\! = \!\!\!& 
        \dfrac{\partial^i \hat{v}_n}{\partial y^i} (y=-1^-,z),
        \\[0.35cm]
        \hat{\eta}_n(y=-1^+,z) 
        &\!\!\! = \!\!\!& 
        \hat{\eta}_n(y=-1^-,z), \quad 
        &
        \dfrac{\partial \hat{\eta}_n}{\partial y} (y=-1^+,z) 
        &\!\!\! = \!\!\!& 
        \dfrac{\partial \hat{\eta}_n}{\partial y} (y=-1^-,z),
    \ea
\end{align*}
with $i=\{1$, $2$, $3\}$.

\subsection{Second-order statistics of the velocity fluctuations}
\label{sec.lyap}
The second-order statistics of the velocity fluctuations $\bv_\theta(k_x,y,t)$ {in model~\eqref{eq.lnse}} can be obtained from the solution $\cX_\theta(k_x)$ of the operator Lyapunov equation~\citep{farjovbam08},
\begin{align}
\label{eq.lyap}
    		\cA_\theta(k_x) \cX_\theta(k_x) + \cX_\theta(k_x) \cA^*_\theta(k_x)
            \; = \;
            - \cM_\theta(k_x).
\end{align}
Here, $*$ denotes the adjoint of an operator, $\cX_\theta(k_x) \DefinedAs \left< \bpsi_\theta(k_x,\;\cdot\;,t) \otimes \bpsi_\theta(k_x,\;\cdot\;,t)\right>$ is the steady-state covariance matrix of the state $\bpsi_\theta(k_x,y,t)$, $\otimes$ is the tensor product, and $\cM_\theta(k_x)=\cM^*_\theta(k_x)\succeq 0$ is the covariance matrix of the zero-mean white-in-time stochastic forcing $\mathbf{d}_\theta := \cB_\theta\mathbf{f}_\theta$, i.e., 
\begin{align}
\label{eq.cov-d}
    \langle \bd_\theta(k_x,\;\cdot\;,t_1) \otimes \bd^*_\theta(k_x,\;\cdot\;,t_2)\rangle = \cM_\theta(k_x)\,\delta (t_1-t_2),
\end{align}
where $\delta$ is the Dirac delta function. Following the bi-infinite structure of $\bd_\theta(k_x,\;\cdot\;,t)$, matrix $\cM_\theta(k_x)$ takes the bi-infinite block-diagonal form $\cM_\theta(k_x) = \diag\left \{ \cM(k_x,\theta_n) \right \}_{n\in\bbZ}$, where each block represents the covariance matrix of one of the harmonics of the forcing. Having obtained $\cX_\theta(k_x)$, the covariance matrix of the velocity field $\bv_\theta(k_x,\;\cdot\;,t)$ can be computed as $\Phi_\theta(k_x) = \cC_\theta (k_x) \cX_\theta (k_x) \cC_\theta^* (k_x)$. Note that operator adjoints appearing in equation~\eqref{eq.lyap} for generator $\cA_\theta$ or the expression of $\cM_\theta$ (appendix~\ref{app.Xpert}) should be determined with respect to the inner product that induces kinetic energy of flow fluctuations~\citep{jovbamJFM05}; see appendix A of~\cite{zarjovgeoJFM17} for a change of coordinates that provides a treatment by bringing equations~\eqref{eq.lnse} to a state-space in which the kinetic energy is determined by the Euclidean norm of the state vector.

Given its bi-infinite structure, solving Lyapunov equation~\eqref{eq.lyap} as done in~\cite{ranzarjovJFM21} and~\cite{naszarAIAA24} can become arduous, especially when considering sharp riblets (e.g., scalloped riblets). This is because the long tails of the Fourier expansions (cf.~equation\eqref{eq.FFourier}) give rise to a large number of significant harmonic interactions, and thereby, large dense matrices of dimension $2 m N_y \times 2 m N_y$ (for $N_y$ collocation points in $y$ and $m$ harmonics in~\eqref{eq.FFourier} and~\eqref{eq.UFourier}) after discretization. We address this issue using a perturbation analysis of flow quantities in the height of riblets $\alpha$, which allows us to break down equation~\eqref{eq.lyap}, and thereby the analysis of all riblet-induced effects, over different perturbation levels. As we demonstrate, this technique, which exploits the structure of the block operator matrices in model~\eqref{eq.lnse}, can provide sufficiently accurate solutions to equation~\eqref{eq.lyap} in a computationally efficient manner that facilitates analysis at high Reynolds numbers.

\subsection{Perturbation analysis of flow statistics}
\label{sec.pert_stats}

In order to reduce the computational complexity of solving equation~\eqref{eq.lyap}, we revisit \S~\ref{sec.formulation} and identify the height of riblets $\alpha$ 
as a small parameter around which the mapping function $F(\tilde{y},\tilde{z})$ can be expanded, i.e.,
\begin{align}
\label{eq.Fpert}
            F(\tilde{y},\tilde{z})
            \;=\;
            F_0(\tilde{y}) \,+\, \alpha\, F_1(\tilde{y},\tilde{z}) \,+\, \alpha^2\, F_2(\tilde{y},\tilde{z}) \,+\, O(\alpha^3).
\end{align}
Here, $F_0(\tilde{y}) = \tilde{y}$, $F_1(\tilde{y},\tilde{z}) = \bar{r}(\tilde{z})\,(\tilde{y}-1)/2$, $F_2(\tilde{y},\tilde{z}) = \bar{r}^2(\tilde{z})\,(\tilde{y}-1)/4$ are obtained from the Neumann series expansion of $F(\tilde{y},\tilde{z})$ given in equation~\eqref{eq.mapping} with $\bar{r}(\tilde{z})=r(\tilde{z})/\alpha$. Following the structure of equation~\eqref{eq.meanVelT}, a similar perturbation series can be considered for $U(y,z)$ as
\begin{align}
\label{eq.Upert}
    	U(y,z)
            \;=\;
            U_0(y) \,+\, \alpha\, U_1(y,z) \,+\, \alpha^2\, U_2(y,z)\,+\, O(\alpha^3),
\end{align}
where $U_0(y)$, $U_1(y,z)$, and $U_2(y,z)$ can be consecutively
obtained from the following sequence of linear equations:
\begin{align*}
            \alpha^0:~~ &
            \partial_y\, \big((1 \,+\, \nu_{T}) \, \partial_y\, U_0 \big)
            \;=\;
            Re_{\tau}\bar P_x
            \\[0.15cm]
            \alpha^1:~~ &
             (1 \,+\, \nu_{T})\left [ \partial_{yy} \,+\, \partial_{zz}  \right ] U_1
            \,+\, \nu_{T}'\, \partial_y\, U_1
            \;=\;
            -2 F_{\tilde{y}_1} \partial_y\, \big((1 \,+\, \nu_{T}) \, \partial_y\, U_0 \big)
            \\[0.15cm]
            \alpha^2:~~ & 
             (1 \,+\, \nu_{T})\left [\partial_{yy} \,+\, \partial_{zz}  \right ] U_2
            \,+\, \nu_{T}' \,\partial_y\, U_2
            \;=\;
            -(1 \,+\, \nu_{T}) \big[F_{\tilde{z}_1}^{2} \partial_{yy} U_0
            \,+\, 2 F_{\tilde{z}_1} \partial_{yz} U_1 \big ]
            \\
            &
            \hspace{4cm} -2 F_{\tilde{y}_1} \partial_y\, \big((1 \,+\, \nu_{T}) \, \partial_y\, U_1 \big)
            \,-\,
            F_{\tilde{y}_1}^{2} \partial_y\, \big((1 \,+\, \nu_{T}) \, \partial_y\, U_0 \big).
\end{align*}
Note that $U_0(y)$ is the mean velocity profile for turbulent channel flow with smooth walls and $U_1(y,z)$ and $U_2(y,z)$ capture riblet-induced perturbations at the levels of $\alpha^1$ and $\alpha^2$, respectively.
Following the perturbation expansion for $F(\tilde{y},\tilde{z})$ and $U(y,z)$ (equations~\eqref{eq.Fpert} and~\eqref{eq.Upert}) in the linearized dynamics, the dynamic generator $\cA_\theta(k_x)$ can be decomposed over various levels of $\alpha$ as
\begin{align}
\label{eq.Apert}
    \cA_\theta (k_x)
        \;=\;
        \cA_{0,\theta}(k_x) \,+\, \alpha\,\cA_{1,\theta}(k_x) \,+\, \alpha^2\,\cA_{2,\theta}(k_x) \,+\, O(\alpha^3),
\end{align}
where $\cA_{0,\theta}(k_x)$ corresponds to turbulent channel flow over smooth walls and $\cA_{1,\theta}(k_x)$ and $\cA_{2,\theta}(k_x)$ captures the effect of riblets on the flow dynamics at $\alpha^1$ and $\alpha^2$ levels, respectively; see appendix~\ref{app.Apert} for the structure of block operator matrices $\cA_{l,\theta}(k_x)$. As equation~\eqref{eq.lyap} is linear, its solution, $\cX_\theta(k_x)$, inherits a similar perturbation series, i.e., 
\begin{align}
\label{eq.Xpert}
    \cX_\theta (k_x)
        \;=\;
        \cX_{0,\theta}(k_x) \,+\, \alpha\,\cX_{1,\theta}(k_x) \,+\, \alpha^2\,\cX_{2,\theta}(k_x) \,+\, O(\alpha^3).
\end{align}
In appendix~\ref{app.Xpert}, we show that $\cX_{l,\theta}(k_x)$ are computed {from smaller-size Lyapunov equations compared to equation~\eqref{eq.lyap}} ($2N_y \times 2N_y$ vs $2mN_y \times 2mN_y$, for $N_y$ collocation points in $y$ and $m$ harmonics in~\eqref{eq.FFourier} and~\eqref{eq.UFourier}). {Given the cubic computational complexity of solving unstructured Lyapunov equations, an $m$-fold reduction in the size of the operator matrices leads to a substantial decrease in both memory usage and computation time required to obtain the second-order flow statistics over riblets. Specifically, perturbation analysis facilitates model-based predictions of the effects of small riblets on the flow using computations comparable in complexity to those for smooth channel flow. This demonstrates the computational efficiency of our approach relative to the analysis presented in~\cite{ranzarjovJFM21}.} 
It is also noteworthy that the appearance of odd-powered perturbation terms in~\eqref{eq.Apert} and~\eqref{eq.Xpert} is due to the offset introduced to $r(\tilde{z})$ in the mapping function $F(\tilde{y},\tilde{z})$, which gives rise to $0$th-order harmonics in~\eqref{eq.FFourier}.
Finally, the energy spectrum of the flow can be computed as $\bar{E}_\theta(k_x) = {\sum_{n\in\bbZ} \trace \left(X_d(k_x,\theta_n)\right)}$, where $X_d(k_x,\theta_n)$ are blocks on the main diagonal of $\cX_\theta(k_x)$ that have been confined to $y \in [-1,1]$. Based on this, the perturbation series for $\cX_\theta(k_x)$ yields the perturbation expansion
\begin{align}
\label{eq.Epert}
    \bar{E}_\theta (k_x)
        \;=\;
        \bar{E}_{0,\theta}(k_x) \,+\, \alpha\,\bar{E}_{1,\theta}(k_x) \,+\, \alpha^2\,\bar{E}_{2,\theta}(k_x) \,+\, O(\alpha^3)
\end{align}
for the energy spectrum of velocity fluctuations, where the energy at the level of $\alpha^l$ is computed from the trace of the corresponding covariance matrix in~\eqref{eq.Xpert}.

\subsection{Perturbation analysis of turbulent viscosity}
\label{sec.pert_visc}
Based on the model adopted in \S~\ref{sec.nuT}, the turbulent viscosity is determined by second-order statistics of the flow (cf.~equations~\eqref{eq.kepsmodel}), which can be computed from $X_d(k_x,\theta_n)$. Following~\eqref{eq.Xpert}, the perturbation analysis can be extended to the turbulent kinetic energy $k$ and its rate of dissipation $\eps$ as
\begin{align}
\label{eq.Pertexpansionkeps}
    \ba{rclclclcl}
        k(y)
            &\!\!=\!\!&
        k_0(y) 
        &\!\!\!\!+\!\!\!\!&
        \alpha\, k_1(y) &\!\!\!\!+\!\!\!\!& \alpha^2\, k_2(y) &\!\!\!\!+\!\!\!\!& O(\alpha^3),
        \\[0.15cm]
        \eps(y)
            &\!\!\!\!=\!\!\!\!&
        \eps_0(y) &\!\!\!\!+\!\!\!\!& \alpha\, \eps_1(y) &\!\!\!\!+\!\!\!\!& \alpha^2\, \eps_2(y) &\!\!\!\!+\!\!\!\!& O(\alpha^3),
    \ea
\end{align}
where the subscript $0$ denotes quantities in the absence of riblets and subscripts $1$ and $2$ denote changes due to the effect of riblets, which can be computed from the corresponding terms in the perturbation series of $\cX_\theta(k_x)$; see appendix~\ref{app.corrKEps} for details. Substituting~\eqref{eq.Pertexpansionkeps} into~\eqref{eq.KEpsilon} and employing the Neumann series expansion yields
\begin{align}
\label{eq.nuTpert}
    \nu_T(y) \;=\; \nu_{T_0}(y) \,+\, \alpha\,\nu_{T_1}(y) \,+\, \alpha^2\,\nu_{T_2}(y) \,+\, O(\alpha^3),
\end{align}
where $\nu_{T_0}$ is the turbulent viscosity for flows over smooth walls (equation~\eqref{eq.Cessmodel}) and
\begin{align}
    \label{eq.nuT12}
        \nu_{T_1}(y) 
        &\;=\;
        \nu_{T_0}(y)\! \left( 2\,\dfrac{k_1(y)}{k_0(y)} - \dfrac{\eps_1(y)}{\eps_0(y)} \right),
        \\[0.15cm]
        \non
        \nu_{T_2}(y) 
        &\;=\;
        \nu_{T_0}(y) \!\left( 2\, \dfrac{k_2(y)}{k_0(y)} - \dfrac{\eps_2(y)}{\eps_0(y)} - 2\,\dfrac{\eps_1(y)k_1(y)}{\eps_0(y) k_0(y)} + \dfrac{{\eps_1(y)}^2}{{\eps_0(y)}^2} + \dfrac{{k_1(y)}^2}{{k_0(y)}^2} \right)\!.
\end{align}
Here, $k_0(y)$ captures the wall-normal dependence of turbulent kinetic energy in channel flow over smooth walls and can be computed from DNS-generated datasets (see, e.g.,~\url{https://torroja.dmt.upm.es/channels/data}). On the other hand, $\eps_0(y)$ is computed using $\eps_0(y) = c Re_\tau^2 k_0^2(y)/\nu_{T_0}(y)$.
The influence of fluctuations on the turbulent mean velocity and, consequently, skin-friction drag can be quantified by substituting the perturbation series for $F$, $U$, and $\nu_T$ (equations~\eqref{eq.Fpert},~\eqref{eq.Upert}, and~\eqref{eq.nuTpert}, respectively) into equation~\eqref{eq.meanVelT}; see appendix~\ref{app.Upert-Dpert}.

\section{Turbulent drag reduction and {stress modulation}}
\label{sec.result}

We utilize the perturbation analysis presented in the previous section to study the effects of {triangular and} scalloped riblets with different geometric configurations (table~\ref{tab.ribconf}) on the statistical signature {of the velocity field as predicted by our reduced-order model. While most of the results in this section are presented for turbulent channel flow with $Re_\tau=186$, the case of $Re_\tau=547$ is used to for comparison with an experimental study.} 
In addition to analyzing changes to the energy spectrum, we use riblet-induced perturbations to the second-order statistics to compute changes to the mean velocity and skin-friction-drag. While the small size of riblets ($\alpha \ll 1$) enables our perturbation analysis, it also implies that the Reynolds number remains unchanged over various case studies as the influence of riblets on the channel height and shear velocity is negligible. We use $25$ harmonics in $z$ to capture the spanwise-periodic shape of the riblets via the domain transformation technique, i.e., $m \in [-12, 12]$ in the Fourier expansion~\eqref{eq.FFourier}. We use a DNS-generated dataset of flow statistics in channel flow over smooth walls~\citep{deljim03,deljimzanmos04,hoyjim06} to compute $k_0$ and $\eps_0$ in equations~\eqref{eq.nuT12} and to shape the energy spectrum of stochastic forcing $\bd_\theta(k_x)$ in equation~\eqref{eq.cov-d} (appendix~\ref{app.Xpert}). In obtaining a finite-dimensional approximation of the evolution model~\eqref{eq.lnse}, we use a total of $N_y$ Chebyshev collocation points made up of $N_t$ and $N_b$ points in the top and bottom regions (i.e., $N_y = N_b + N_t$) to discretize the differential operators in the wall-normal direction. We also use $N_x$ logarithmically spaced streamwise wavenumbers with $0.03 < k_x<k_{x,\mathrm{max}}${, where $k_{x,\mathrm{max}}$ is the largest streamwise wavenumber covered in the DNS-generated database,} together with 3 harmonics of $\omega_z$ ($n = 1$, $2$, and $3$) and $N_\theta$ logarithmically spaced offset points $0.01<\theta<\omega_z$ to parameterize the governing equations over the horizontal dimensions; see table~\eqref{tab.ribconf}. {The finest streamwise and spanwise scales that are refined by this spectral parameterization are given by $\lambda_x^+ = 27.49$ and $\lambda_z^+ = 5.5$ for flow with $Re_\tau = 186$, where $\lambda^+_x = 2\pi Re_\tau/k_x$ and $\lambda^+_z = 2\pi Re_\tau/\theta_n$ are the streamwise and spanwise wavelengths in viscous units, respectively.} 
We ensure grid convergence by doubling the number of collocation points used for discretization in the wall-normal dimension and the number of wavenumbers used for parameterization in the wall-parallel dimensions.
{In this section, we validate our model for flow over triangular riblets at $Re_\tau=186$ by comparing the profiles of mean velocity, drag reduction values, root-mean-square (rms) of velocity fluctuations, and Reynolds shear stress with DNS of~\cite{chomoikim93} and experiments of~\cite{becbruhaghoehop97}. Following this validation, we present our model-based predictions for flow over scalloped riblets at $Re_\tau=186$. While the presentation of results in either the computational or physical domain remains unchanged, the results in~\S\S~\ref{sec.result} and~\ref{sec.structure} are presented in the original physical coordinates ($\tilde{x}$, $\tilde{y}$, $\tilde{z}$) to avoid confusion.}

\begin{table}
 \begin{center}
    \setlength{\tabcolsep}{3.6pt}
    {\scriptsize \begin{tabular}{lcccccccccc}
    {\small $Re_\tau$}&& {\small $\alpha/s$}&{\small $\omega_z$}& {\small $N_x$}&{\small $N_t$}&{\small $N_b$}&{\small $N_\theta$}& {\small $k_{x,\mathrm{max}}$} &{\small $c_1$}& {\small $c_2$}\\[.2cm]
    186 & \includegraphics[height=.02\textwidth]{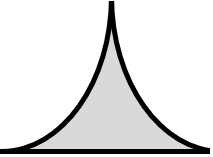} & 0.55 & 35, 40, 45, 60, 80, 100, 160 & 50 & 140 & 30 & 51 & 42.5 & 0.61 & 46.2\\
    & \includegraphics[height=.02\textwidth]{scalogo}& 0.65 & 45, 50, 60, 80, 100, 160\\
    & \includegraphics[height=.02\textwidth]{scalogo}& 0.87 & 45, 50, 55, 60, 80, 100, 140, 160 \\
    &\includegraphics[height=.02\textwidth]{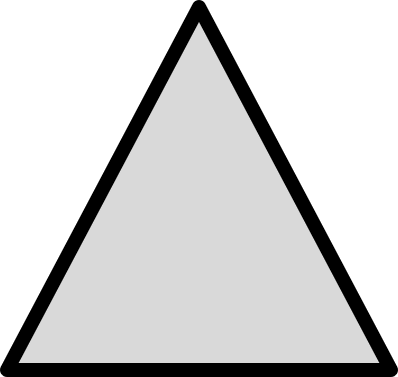}& {0.87} & 57\\
    & \includegraphics[height=.02\textwidth]{scalogo}& 1.2 &  60, 80, 100, 120, 160, 210 \\ [.2cm]
    547 & \includegraphics[height=.02\textwidth]{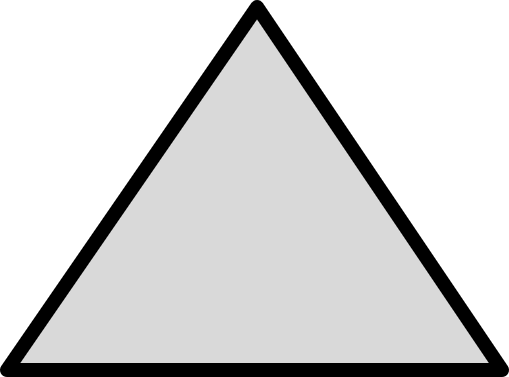}& {0.5} & 69, 179 & 96 & 170 & 30 & 101 & 128 & 0.45 & 29.4\\
    & \includegraphics[height=.02\textwidth]{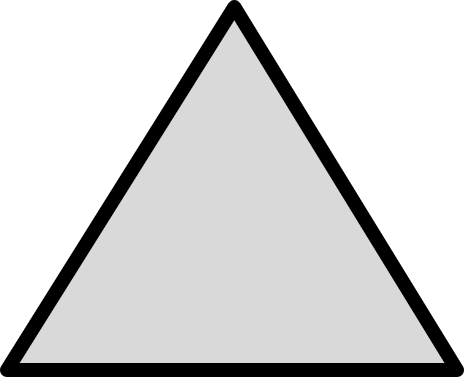} & {0.55} & 115\\
    & \includegraphics[height=.02\textwidth]{scalogo}& 0.55 & 115, 130, 175, 230, 290, 460\\
    & \includegraphics[height=.02\textwidth]{tri60logo}& {0.87} & 119,	129, 173, 194, 226, 293, 572\\
    & \includegraphics[height=.02\textwidth]{scalogo}& 0.65, 0.87, 1.2 & 115\\
    & \includegraphics[height=.02\textwidth]{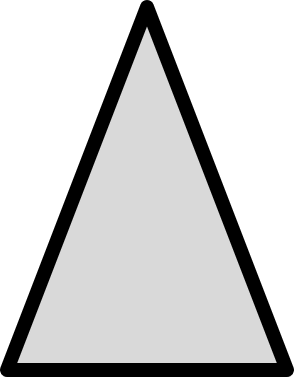}& {1.87} & 163\\ [.2cm]
    934 & \includegraphics[height=.02\textwidth]{scalogo}& 0.55 & 200, 220, 300, 390, 490, 780 & 192 & 200 & 30 & 151 & 255 & 0.43 & 27\\[.05cm]
    & \includegraphics[height=.02\textwidth]{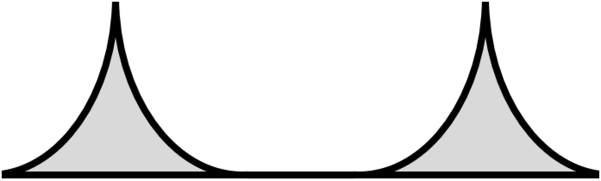} & {0.55} & 100, 110, 150, 195, 245, 390\\
    & \includegraphics[height=.02\textwidth]{scalogo}& 0.65, 0.87, 1.2 & 200\\ [.2cm]
    2003 & \includegraphics[height=.02\textwidth]{scalogo}& 0.55 & 420, 470, 640, 840, 1050, 1700 & 384 & 270 & 10 & 201 & 511 & 0.42 & 25.4\\
    & \includegraphics[height=.02\textwidth]{scalogo}& 0.65, 0.87, 1.2 & 420 
    \end{tabular}}
    \caption{Characteristic parameters corresponding to {various riblet} configurations examined in this study along with the number of discretization points used in different dimensions. Here, $\alpha /s$ and $\omega_z$ are the height-to-spacing ratio and spanwise frequency of riblets, $N_t$ is the number of collocation points in the top wall-normal region (between $-1$ and $+1$), $N_b$ is the number of collocation points in the bottom wall-normal region (between $-1-\alpha r_p$ and $-1$), $N_x$ is the number of wavenumbers in the streamwise direction, and $N_\theta$ is the number of logarithmically spaced offset wavenumbers $\theta < \omega_z$. {Parameters $c_1$ and $c_2$ are the constants for the turbulent viscosity profile in equation~\eqref{eq.Cessmodel}.}}
  \label{tab.ribconf}
 \end{center}
\end{table}

\subsection{Turbulent viscosity and turbulent mean velocity}
\label{sec.visvel}
\begin{figure}
    \begin{tabular}{cccc}
        \hspace{-.4cm}
        \subfigure[]{\label{fig.Ucompare}}
        &&
        \hspace{-0.8cm}
        \subfigure[]{\label{fig.vel12}}
        &
        \\[-.5cm]
        \hspace{.2cm}
        \begin{tabular}{c}
                \vspace{.8cm}
                {\normalsize \rotatebox{90}{$U$}}
            \end{tabular}
            &
            \hspace{-.5cm}
        \begin{tabular}{c}
                \includegraphics[width=0.5\textwidth]{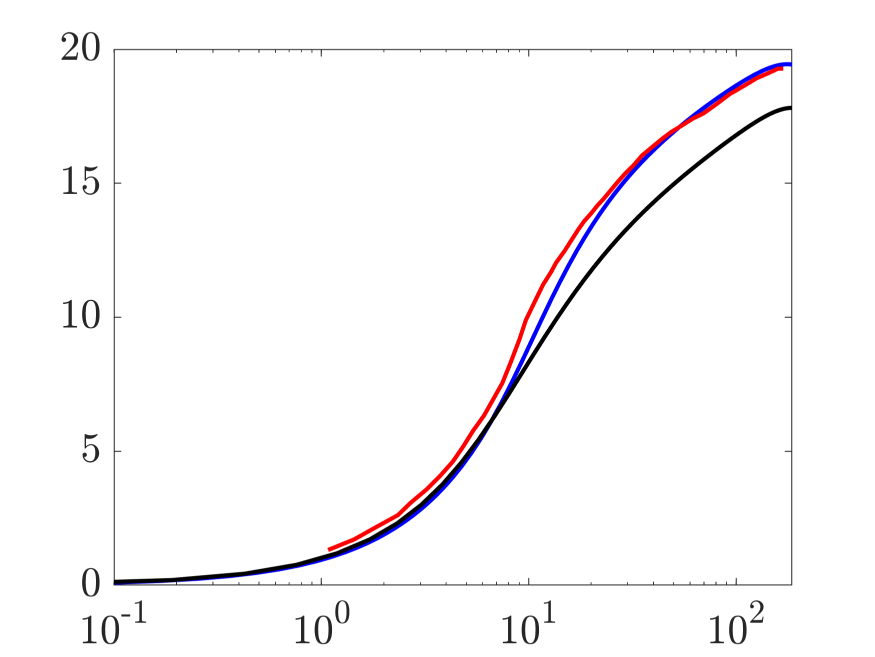}
                \\[.01cm]
                \hspace{-.1cm}
                {\normalsize $\tilde{y}^+$}
            \end{tabular}
            &
            \hspace{-0.1cm}
            \begin{tabular}{c}
                \vspace{.8cm}
                {\normalsize \rotatebox{90}{${\tilde{y}}$}}
            \end{tabular}
            &\hspace{-.4cm}
            \begin{tabular}{c}
                \includegraphics[width=0.35\textwidth]{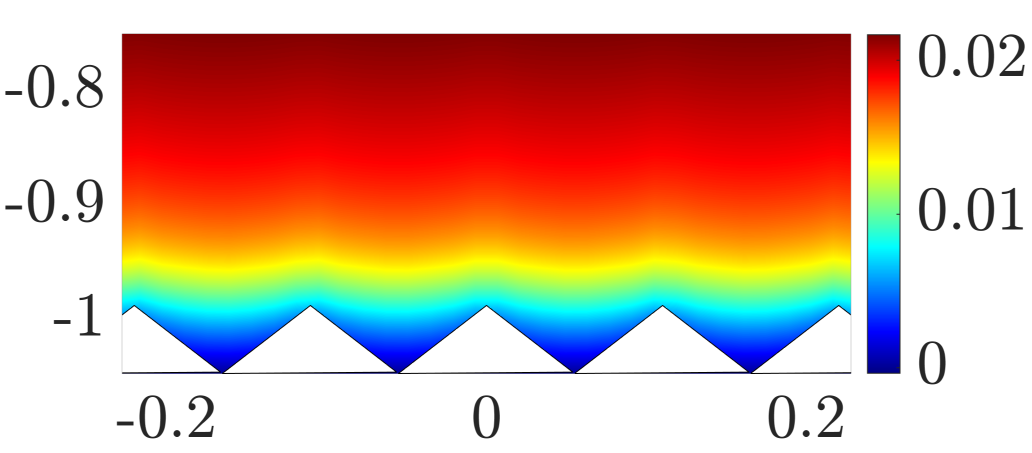}
                \\[-.38cm]
                \includegraphics[width=0.35\textwidth]{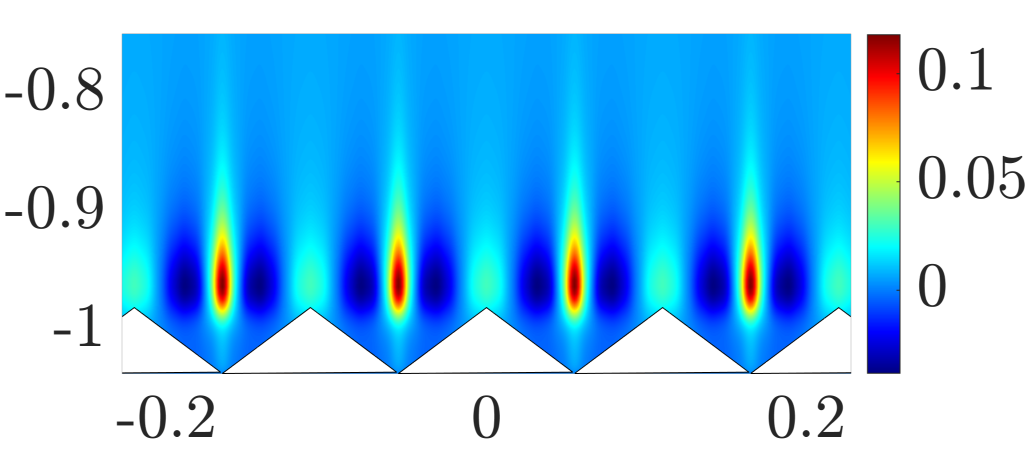}
                \\[-.38cm]
                \includegraphics[width=0.35\textwidth]{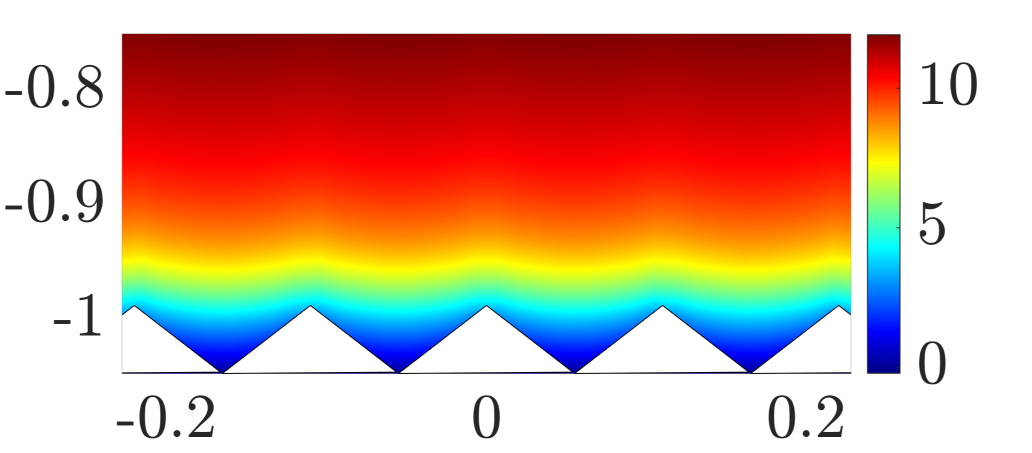}
                \\[-.1cm]
                \hspace{-.4cm}
                {\normalsize ${\tilde{z}}$}
            \end{tabular}
   \end{tabular}
\caption{{Mean velocity profiles of turbulent channel flow with $Re_\tau=186$ over triangular riblets with $\alpha/s = 0.87$ and $\omega_z = 57$. (a) Spanwise-averaged profiles from our model (blue) and the  DNS of~\cite{chomoikim93} (red) compared to the mean velocity in the absence of riblets (black); (b) Color plots of the $\tilde{y}$-$\tilde{z}$ dependence of the riblet-induced modifications $\alpha\,U_1(\tilde{y},\tilde{z})$ (top), $\alpha^2\,U_2(\tilde{y},\tilde{z})$ (middle) in addition to the total mean velocity $U(\tilde{y},\tilde{z}) = U_0(\tilde{y}) + \alpha\,U_1(\tilde{y},\tilde{z}) + \alpha^2\,U_2(\tilde{y},\tilde{z})$ up to $\alpha^2$ (bottom).}}
\label{fig.Uprofile}
\end{figure}

In order to provide a point of comparison with the result of numerical simulations, figure~\ref{fig.Ucompare} shows the mean velocity over {triangular riblets with $\alpha/s = 0.87$ and $\omega_z = 57$}, which corresponds to the configuration considered in~\cite{chomoikim93}.
While the mean velocity profiles from the DNS and our study show a good match for $\tilde{y}^+ \gtrsim 30$, they slightly deviate closer to the wall. {Here, $\tilde{y}^+=Re_\tau(1+\tilde{y})$ is the wall-normal coordinate in viscous units.}
The spanwise variation of the mean velocity is depicted in figure~\ref{fig.vel12}. While the first-order modification {$\alpha\,U_1$} shows a concentration of riblet-induced effects in the vicinity of the wall {($\tilde{y}<-0.9$)}, the second-order modification {$\alpha^2U_2$} shows high and low speed attributes alternating within the grooves and tips of the riblets that extend farther away from the wall. 

\begin{figure}
        \begin{center}
	\begin{tabular}{cc}
        \begin{tabular}{cc}
        \subfigure[]{\label{fig.nuTs}}
        &
        \\[-.4cm]
        \hspace{.3cm}
        \begin{tabular}{c}
            \vspace{2.1cm}
            {\scriptsize \rotatebox{90}{$\nu_{T_0}$}}
        \end{tabular}
        &
        \includegraphics[width=.21\textwidth]{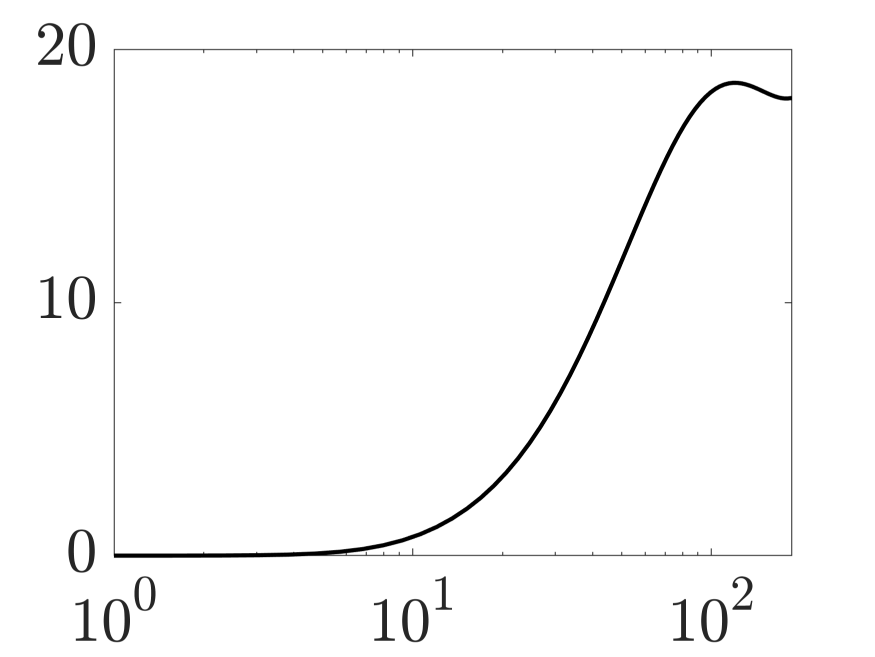}
        \\[-1.2cm]
        &
        {\scriptsize $\tilde{y}^+$}
       \end{tabular}
       &
      \begin{tabular}{cc}
        \subfigure[]{\label{fig.Usmooth}}
        &
        \\ [-.4cm]
        \hspace{.3cm}
        \begin{tabular}{c}
            \vspace{2.1cm}
            {\scriptsize \rotatebox{90}{$U_0$}}
        \end{tabular}
        &
        \includegraphics[width=.21\textwidth]{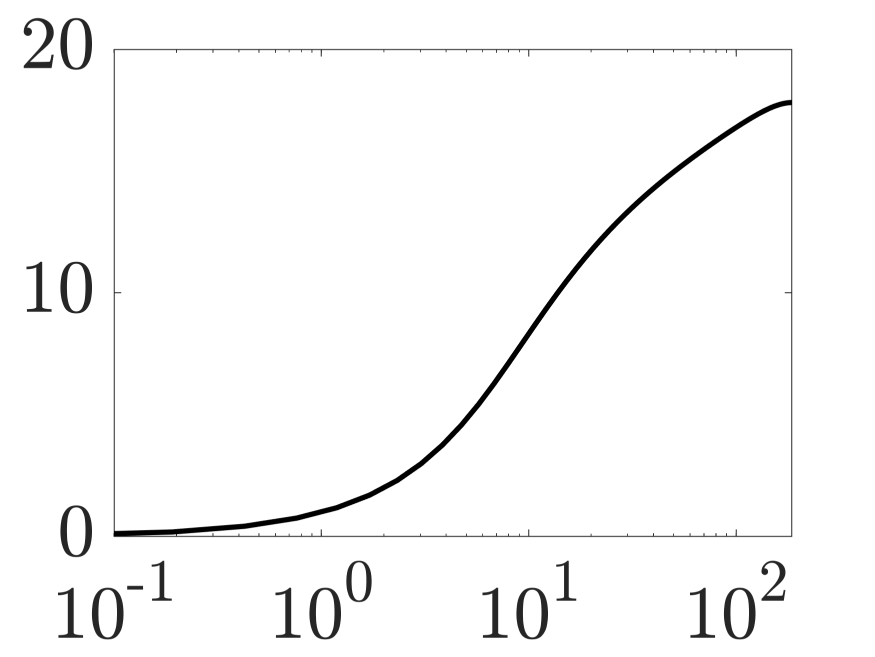}
        \\[-1.1cm]
        &
        {\scriptsize $\tilde{y}^+$}
       \end{tabular}
    \\
    \hspace{-.6cm}
    \begin{tabular}{ccc}
        \subfigure[]{\label{fig.nuT1lgpso}}
        &&
        \\[-.5cm]
        & {\scriptsize $l^+_g \in (6,12)$}
        &
        {\scriptsize $l^+_g \in (12,22)$}
        \\[.1cm]
        \hspace{.4cm}
        \begin{tabular}{c}
            \vspace{2.1cm}
            {\scriptsize \rotatebox{90}{$\alpha\, \nu_{T_1}$}}
        \end{tabular}
        &
        \includegraphics[width=.21\textwidth]{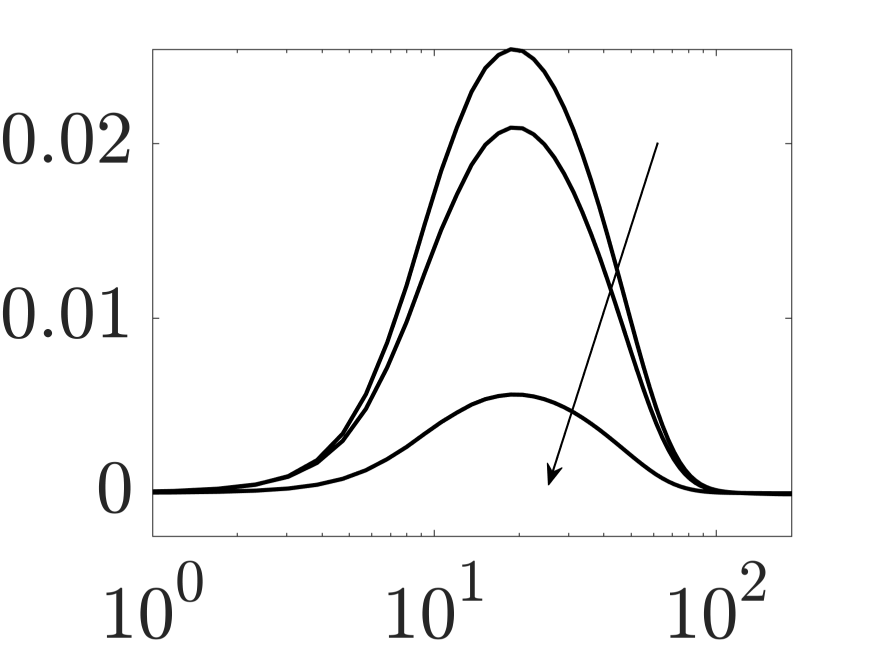}
        &
        \includegraphics[width=.21\textwidth]{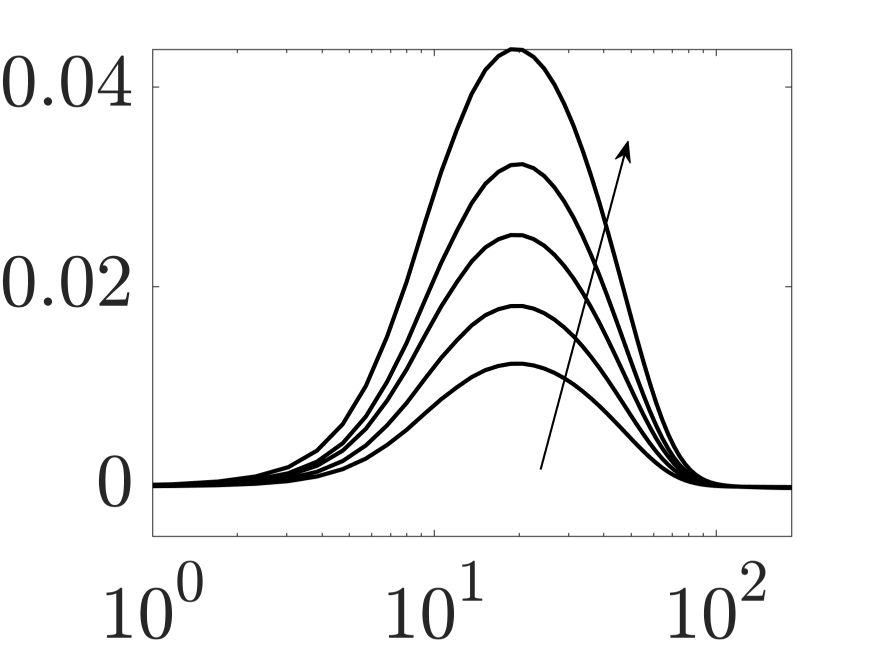}
    \end{tabular}
    &
    \hspace{-1cm}
    \begin{tabular}{ccc}
    \hspace{.1cm}
        \subfigure[]{\label{fig.U1as87so}}
        &&
        \\[-.5cm]
        & {\scriptsize $l^+_g \in (6,12)$}
        &
        {\scriptsize $l^+_g \in (12,22)$}
        \\[.1cm]
        \hspace{.4cm}
        \begin{tabular}{c}
            \vspace{2.1cm}
            {\scriptsize \rotatebox{90}{$\alpha\, U_1$}}
        \end{tabular}
        &
        \includegraphics[width=.21\textwidth]{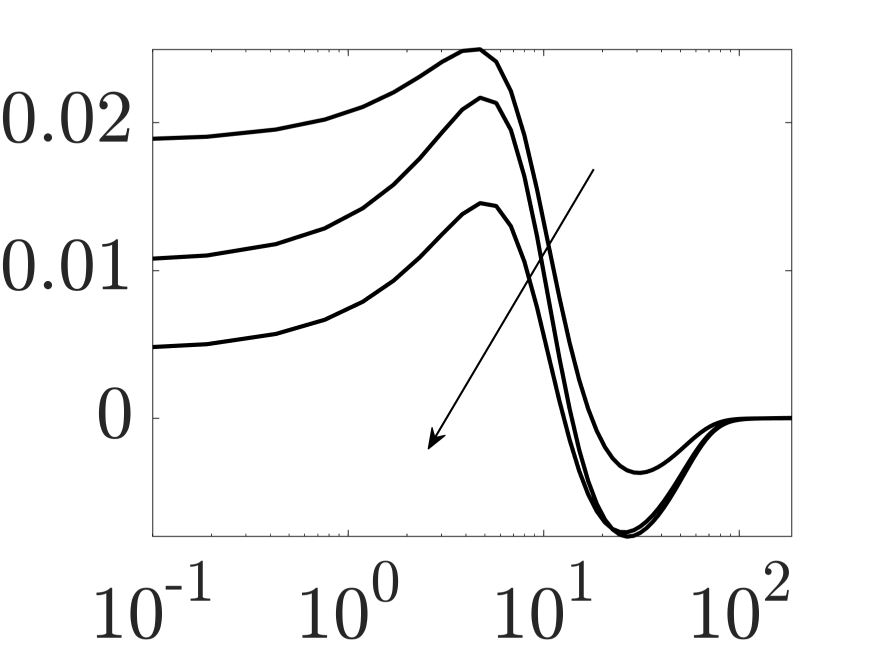}
        &
        \includegraphics[width=.21\textwidth]{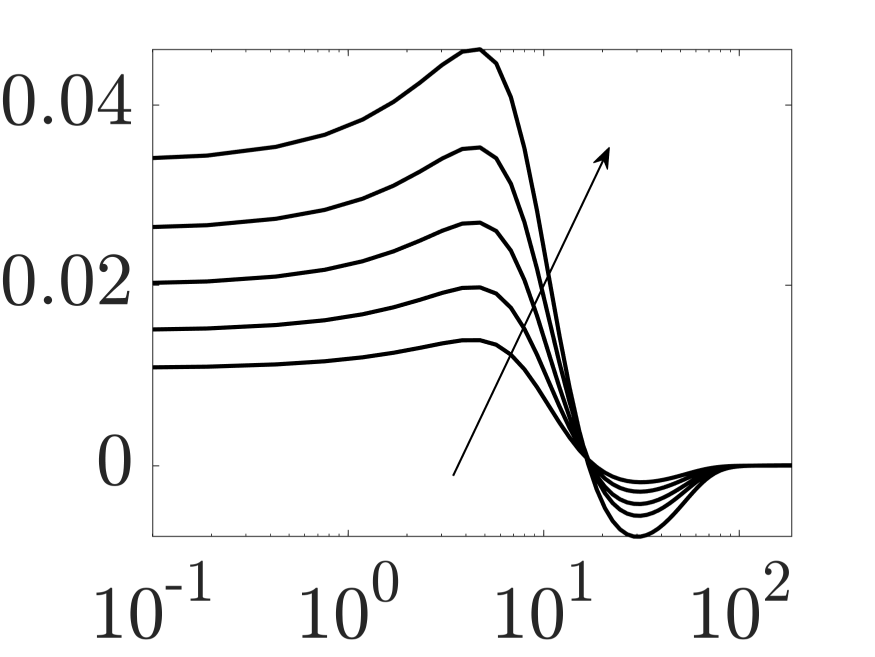}
    \end{tabular}
        \\[-1.4cm] 
    \begin{tabular}{ccc}
    \hspace{-.6cm}
        \subfigure[]{\label{fig.nuT2lgpso}}
        &&
        \\[-.4cm]
        \begin{tabular}{c}
            \vspace{2cm}
            {\scriptsize \rotatebox{90}{$\alpha^2 \nu_{T_2}$}}
        \end{tabular}
        &
        \includegraphics[width=.21\textwidth]{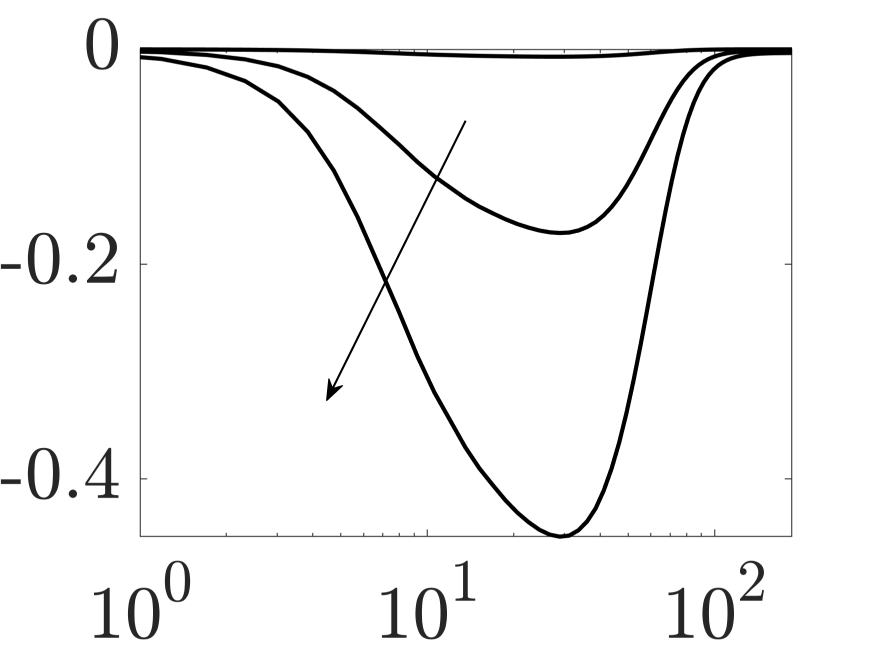}
        &
        \includegraphics[width=.21\textwidth]{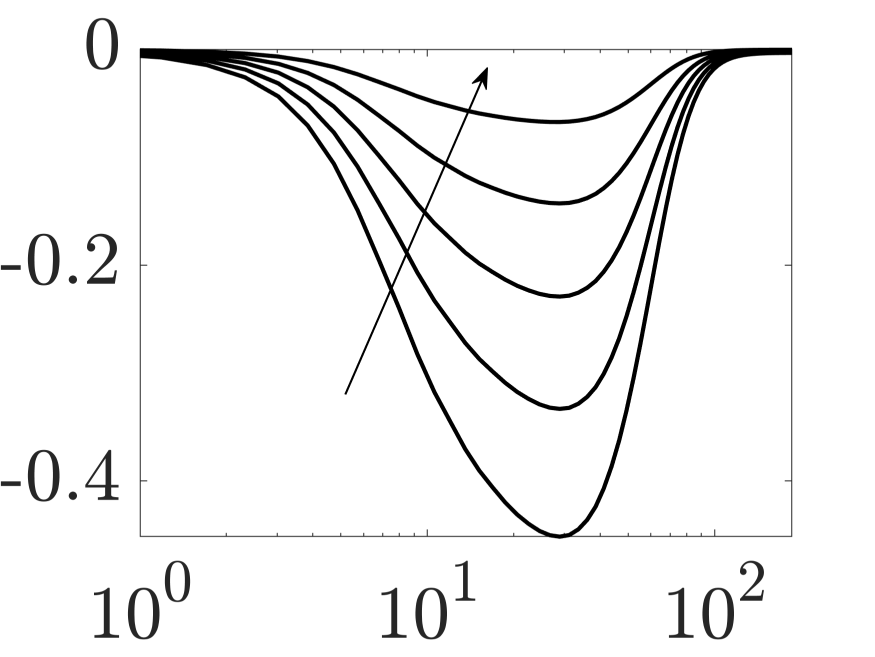}
    \end{tabular}
    &
    \hspace{-.5cm}
    \begin{tabular}{ccc}
    \hspace{-.3cm}
        \subfigure[]{\label{fig.U2as87so}}
        &&
        \\[-.4cm]
        \begin{tabular}{c}
            \vspace{2cm}
            {\scriptsize \rotatebox{90}{$\alpha^2 U_2$}}
        \end{tabular}
        &
        \includegraphics[width=.21\textwidth]{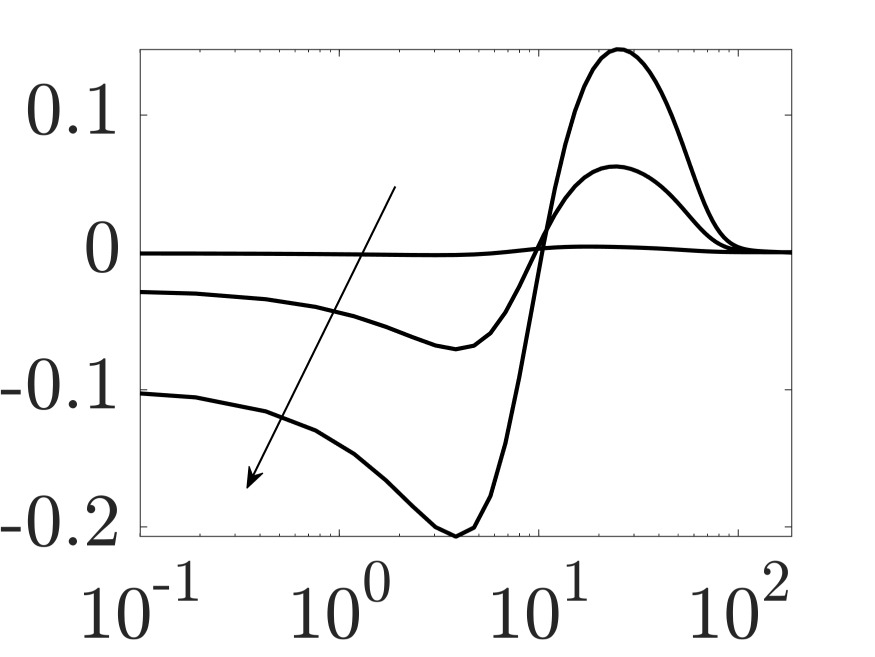}
        &
        \includegraphics[width=.21\textwidth]{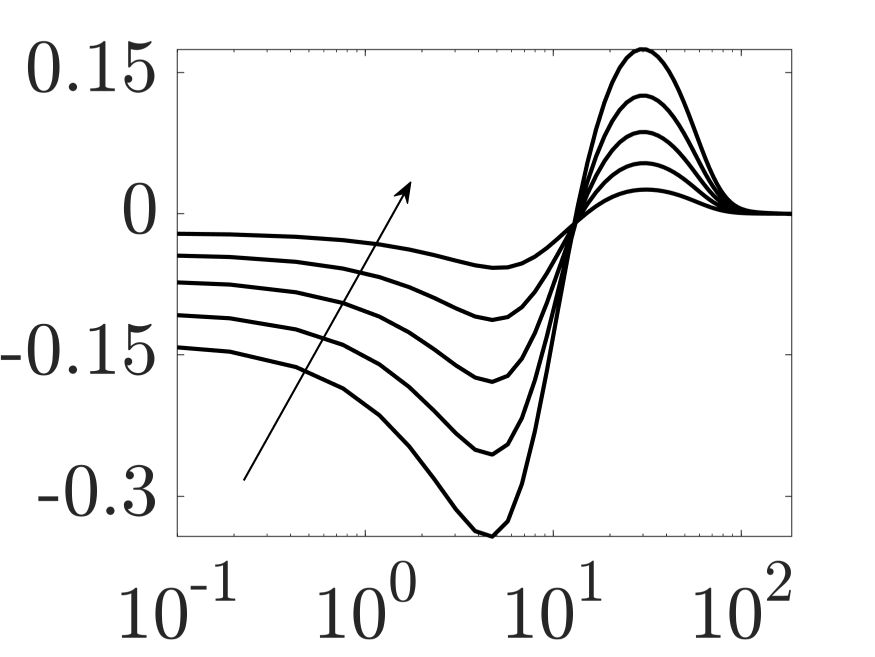}
    \end{tabular}
        \\[-1.4cm]
    \begin{tabular}{ccc}
    \hspace{-.6cm}
        \subfigure[]{\label{fig.nuT12lgpso}}
        &
        &
        \\[-.4cm]
        \begin{tabular}{c}
            \vspace{2.1cm}
            {\scriptsize \rotatebox{90}{$ \nu_{T_\mathrm{corr}}$}}
        \end{tabular}
        &
        \includegraphics[width=.21\textwidth]{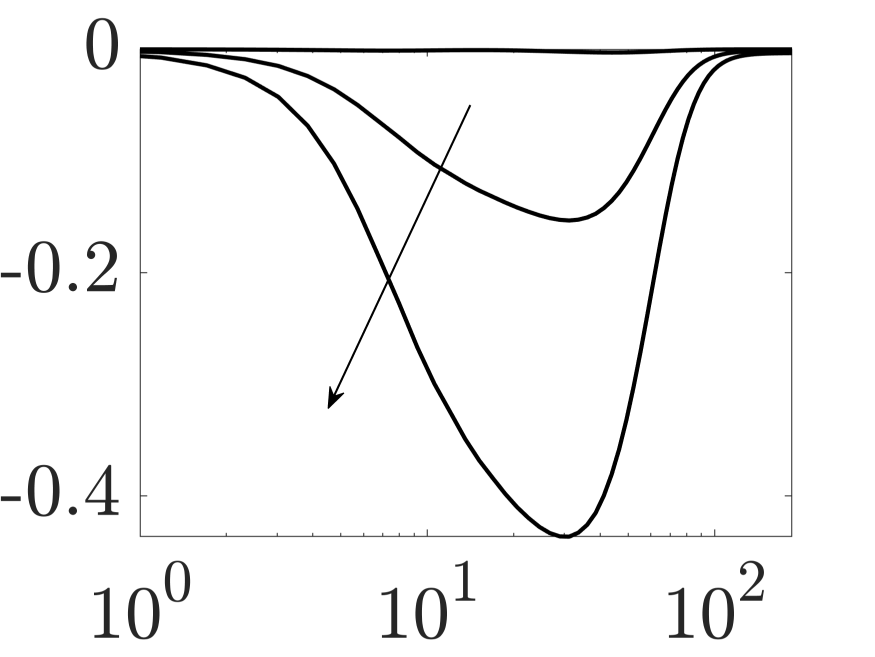}
        &
        \includegraphics[width=.21\textwidth]{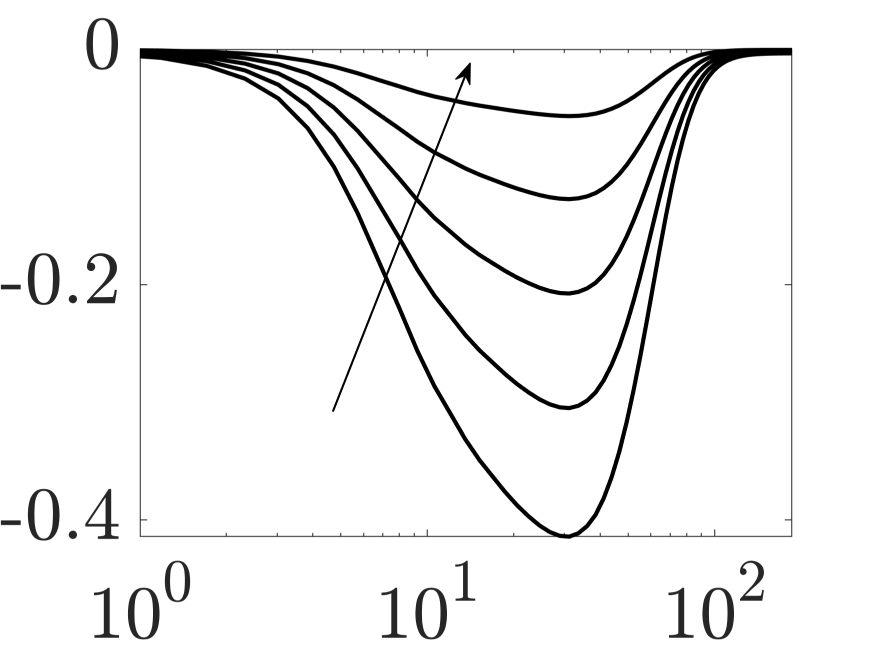}
        \\[-1.3cm]
        &
        {\scriptsize $\tilde{y}^+$}
        &
        {\scriptsize $\tilde{y}^+$}
    \end{tabular}
    &
    \hspace{-.5cm}
    \begin{tabular}{ccc}
    \hspace{-.3cm}
        \subfigure[]{\label{fig.U12as87so}}
        &
        &
        \\[-.4cm]
        \begin{tabular}{c}
            \vspace{2.1cm}
            {\scriptsize \rotatebox{90}{$U_\mathrm{corr}$}}
        \end{tabular}
        &
        \includegraphics[width=.21\textwidth]{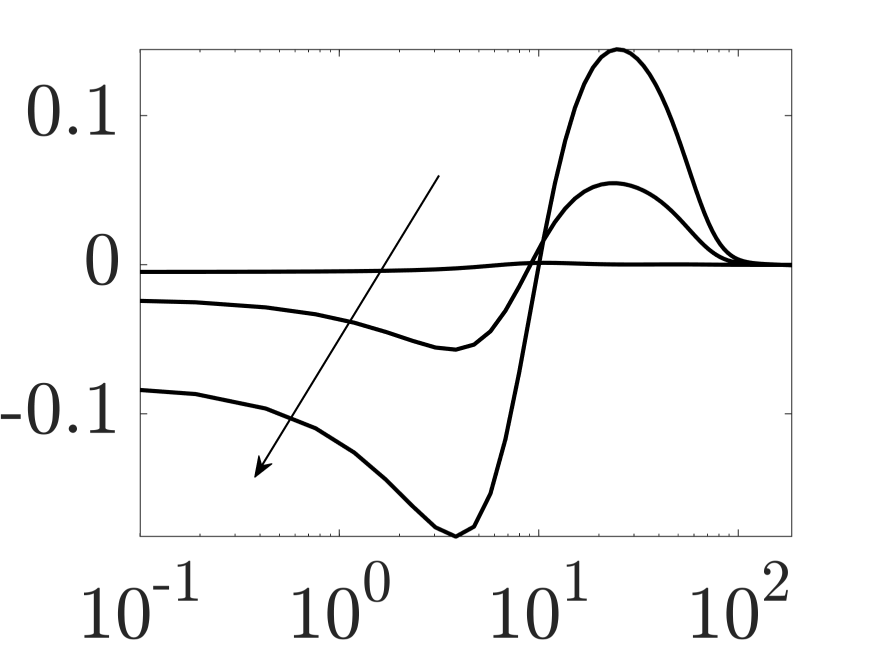}
        &
        \includegraphics[width=.21\textwidth]{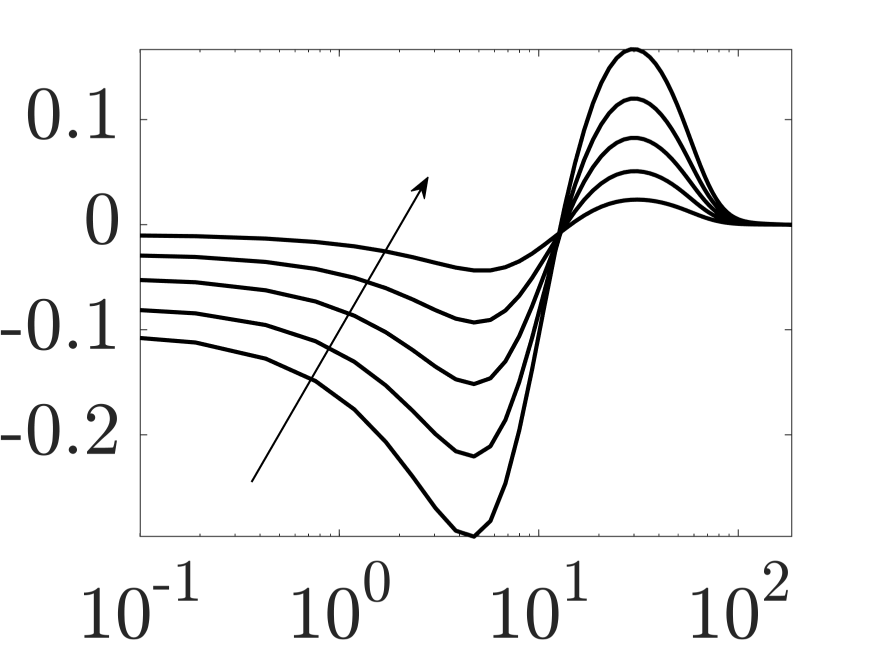}
        \\[-1.2cm]
        &
        {\scriptsize $\tilde{y}^+$}
        &
        {\scriptsize $\tilde{y}^+$}
    \end{tabular}
    \end{tabular}
\end{center}
        \caption{{(a) Turbulent viscosity $\nu_{T_0}$ and (b) mean velocity $U_0$ in smooth channel flow with $Re_\tau = 186$ along with first- (c,d) and second-order (e,f) modifications to these quantities due to scalloped riblets with $\alpha/s = 0.87$. The first and second columns in these subfigures correspond to small to optimal ($l^+_g \in (6,12)$) and optimal to large ($l^+_g \in (12,22)$) sized riblets, respectively. The final row shows the total modification (up to $\alpha^2$) to (g) turbulent viscosity, i.e., $\nu_{T_\mathrm{corr}} \DefinedAs \alpha \nu_{T_1} + \alpha^2 \nu_{T_2}$, and (h) the mean velocity, i.e., $U_\mathrm{corr} \DefinedAs \alpha U_{1} + \alpha^2 U_2$. In all figures, $l^+_g$ increases in the direction of the arrows.}}
		\label{fig.nuT012U012}
\end{figure}

We {next} examine the effect of scalloped riblets with $\alpha/s = 0.87$ but different size on the turbulent viscosity and mean velocity.
Figures~\ref{fig.nuTs} and~\ref{fig.Usmooth} show the turbulent viscosity $\nu_{T_0}$ and the mean velocity $U_0$ profiles in channel flow with smooth walls, respectively. 
Perturbation analysis allows us to separate the effect of riblets at $\alpha^1$ and $\alpha^2$ levels. As the spanwise variations (frequency) of riblet-induced perturbations to the turbulent viscosity and mean velocity are smaller than that of the surface roughness, all results are averaged over the spanwise dimension.
{We note that the modifications to the turbulent viscosity and mean velocity, as shown in figures~\ref{fig.nuT012U012}(c-h), are localized to the wall-normal region below $\tilde{y}^+ \approx 100$, which is in agreement with the results of prior numerical studies, e.g.~\citep{endmodgarhutchu21}, where an outer layer similarity is reported despite the presence of riblets.}
While $\alpha\,\nu_{T_1}$ shows a peak value located at {$\tilde{y}^+ \approx 20$} for various sizes of riblets (figure~\ref{fig.nuT1lgpso}), $\alpha^2\,\nu_{T_2}$ shows a significant trough at {$\tilde{y}^+ \approx 29$} (figure~\ref{fig.nuT2lgpso}). Given the typical values of $\alpha$ for which our perturbation analysis holds (e.g., $\alpha=0.05$), an overall suppression of turbulence is observed after summing the effect at $\alpha$ and $\alpha^2$ levels (cf.~equation~\eqref{eq.nuTpert} and figure~\ref{fig.nuT12lgpso}), especially in the near-wall region. As shown in \S~\ref{sec.rtke}, the near-wall suppression of $\nu_T$ results in an overall reduction in turbulent kinetic energy. We note, however, that roughness-induced turbulence suppression reduces for larger riblets. In a similar manner, figures~\ref{fig.U1as87so} and~\ref{fig.U2as87so} show the first- and second-order modifications to the mean velocity due to the presence of riblets. These corrective terms are obtained by substituting $\nu_{T_1}$ and $\nu_{T_2}$ into the mean flow equations (appendix~\ref{app.Upert-Dpert}). In the vicinity of the wall, the dominant second-order term ($\alpha^2 \nu_{T_2}$) decreases the mean velocity resulting in a reduction in the mean velocity gradient relative to the baseline (figure~\ref{fig.Usmooth}). At higher wall-normal locations, both $\alpha\, U_1$ and $\alpha^2 U_2$ show an increase in the mean velocity gradient (figures~\ref{fig.U1as87so} and~\ref{fig.U2as87so}). In spite of this, the effect of small riblets ($\alpha \ll 1$) on the mean velocity is concentrated in the vicinity of lower wall (figure~\ref{fig.U12as87so}).

\subsection{Skin-friction drag}
\label{sec.DR}

We next examine the effect of riblets on turbulent drag in flow with $Re_\tau=186$. Following~\cite{garjim11b}, we refer to the parameter space in which drag reduction is proportional to the size of riblets as the viscous regime. Due to inconsistencies in analyzing the height and spacing of riblets of different shape, we use the square root of the cross-sectional area of the riblet grooves, $l^+_g = \sqrt{A^+_g}$, for our parametric study. This geometric parameterization of riblets also achieves the best collapse of the breakdown dimensions for the linear viscous regime~\citep{garjim11a,garjim11b}. We normalize the drag reduction curves by their slope in the viscous regime, i.e., $m_l := \lim_{l^+_g\rightarrow 0} \Delta D/l^+_g$, to remove the effect of riblets' shape on the slope in this linear regime. 
{In this study, $\tilde{y} = -1$ serves as the virtual origin for drag calculations, which follows the lower bound of the wall-normal region used for computing riblet-induced corrections to turbulent statistics, and thereby, viscosity; see appendix~\ref{app.corrKEps} for details.}

\begin{figure}
        \begin{center}
        \begin{tabular}{cccc}
       \hspace{-.6cm}
        \subfigure[]{\label{fig.DRBechert}}
        &&
        \hspace{-0.8cm}
        \subfigure[]{\label{fig.DRmlBechert}}
        &
        \\[-.5cm]
	\begin{tabular}{c}
        \vspace{.2cm}
        {\rotatebox{90}{$\Delta D (\%)$}}
       \end{tabular}
       &\hspace{-.18cm}
	\begin{tabular}{c}
       \includegraphics[width=0.4\textwidth]{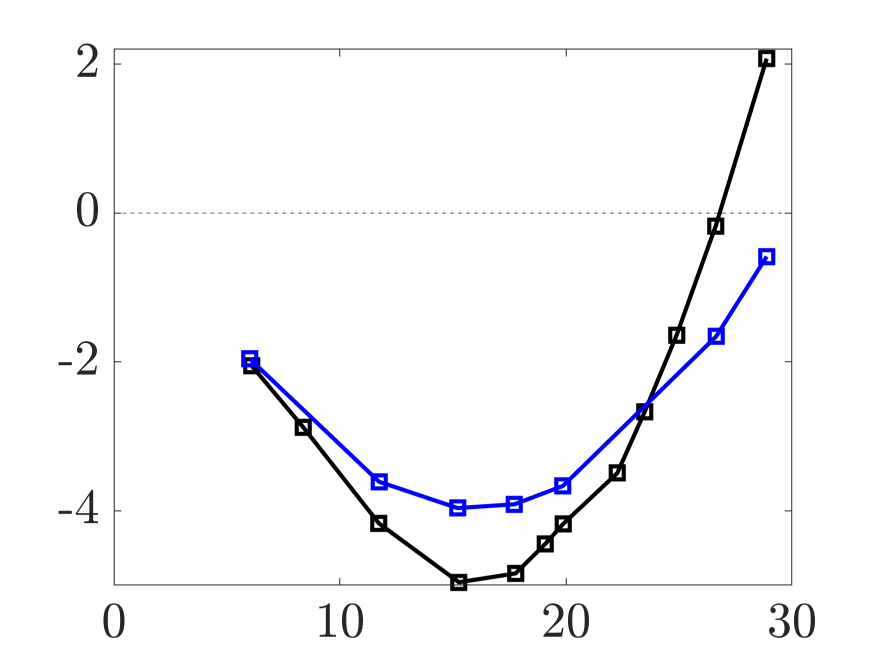}
        \\[-.1cm]
            \hspace{.2cm}
            $s^+$
       \end{tabular}
       &
       \hspace{-.3cm}
	\begin{tabular}{c}
        \vspace{.2cm}
        \rotatebox{90}{$-\Delta D/m_l$}
       \end{tabular}
       &\hspace{-.2cm}
    \begin{tabular}{c}
       \includegraphics[width=0.4\textwidth]{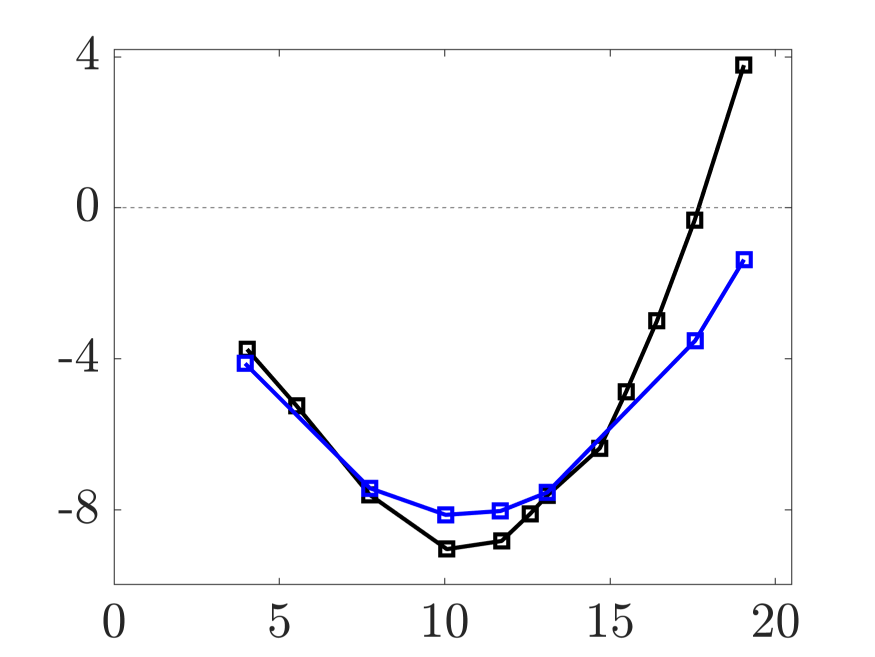}
       \\[-.1cm]
            \hspace{.2cm}
            $l_g^+$
       \end{tabular}
       \end{tabular}
       \end{center}
        \caption{{Comparison between the drag reduction predictions from our model (blue) of turbulent channel flow with $Re_\tau = 547$ over triangular riblets with $\alpha/s = 0.87$ and experimental results of~\cite{becbruhaghoehop97} (black). (a) Drag reduction as a function of riblet spacing in viscous units, i.e., $s^+ = Re_\tau s$; and (b) normalized drag reduction as a function of $l_g^+$. The groove area for triangular riblets is given by $A^+_g = s^+ \alpha^+/2$, where $\alpha^+ = Re_\tau \alpha$.}}
    \label{fig.DRBech}
\end{figure}

{For the triangular riblet configuration considered in figure~\ref{fig.Uprofile}, our model predicts a $4.94\%$ drag reduction, which is in close agreement with the $6\%$ drag reduction reported in~\cite{chomoikim93}. The difference between these predictions may stem from a host of simplifying assumptions adopted by our modeling framework, e.g., linearization and $\tilde{y} < -1$ being laminar. Nevertheless, as demonstrated in figure~\ref{fig.DRBech}, our model reliably predicts drag reduction trends and the optimal size of drag-reducing riblets when compared with the results of the experimental study of~\cite{becbruhaghoehop97}. We note that for an appropriate comparison, the height-to-spacing ratio and viscous-scaled spacing of triangular riblets were matched with the experimental study.}

\begin{figure}
\centering
    \begin{tabular}{ccc}
        \begin{tabular}{c}
            \vspace{0.5cm}
            \hspace{-1.35cm}
            \rotatebox{90}{$-\Delta D/m_l$}
        \end{tabular}
        &
        \hspace{-.9cm}
            \vspace{1.5cm}
        \begin{tabular}{c}
            \includegraphics[width=5.5cm]
            {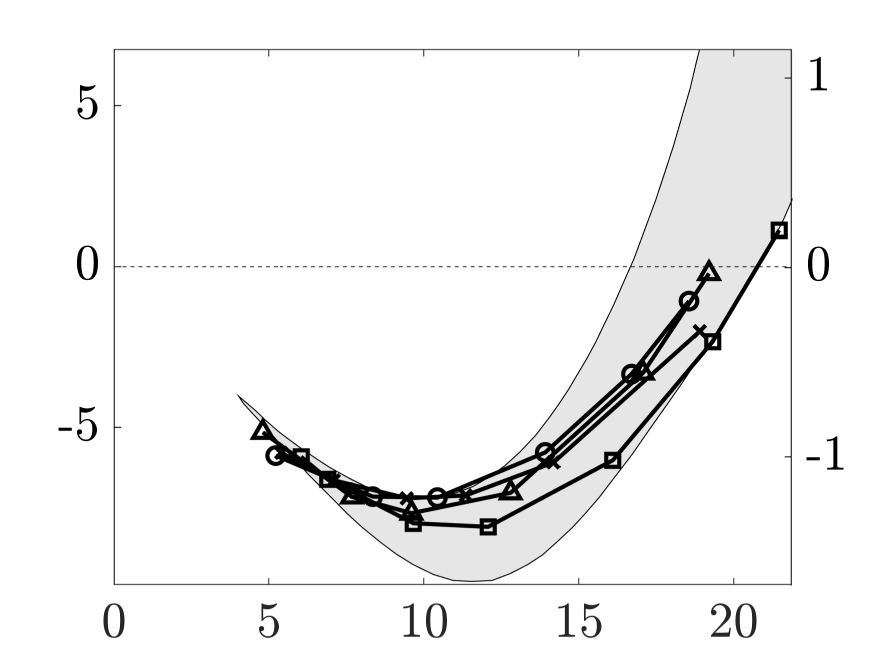}
            \\
            \hspace{0.01cm}
             $l^+_g$
        \end{tabular}
            &
        \begin{tabular}{c}
            \vspace{0.5cm}
            \hspace{-0.75cm}
            \rotatebox{90}{$\Delta U$}
        \end{tabular}
    \end{tabular}
\vspace{-1.5cm}
	\caption{{Normalized drag reduction (left axis) and roughness function (right axis) due to scalloped riblets on the lower wall of a turbulent channel flow with $Re_\tau = 186$ as a function of $l^+_g$. Different lines correspond to different riblet shapes: $\alpha/s=0.55$ ($\triangle $); $\alpha/s=0.65$ ($\bigcirc$); $\alpha/s=0.87$ ($\square$); and $\alpha/s=1.2$ ($\times$). The shaded region corresponds to the envelope of experimentally measured drag reduction levels from prior studies~\citep{becbruhaghoehop97,garjim11b}.}}
    \label{fig.DR}
\end{figure}

Figure~\ref{fig.DR} shows the $m_l$-normalized drag reduction {on the left axis and roughness function, i.e., riblet-induced variation in the mean velocity within the logarithmic layer, $\Delta U \DefinedAs U_{\mathrm{smooth}} - U$, on the right axis} as a function of $l^+_g$ for turbulent channel flow over scalloped riblets with height-to-spacing ratios, $\alpha/s=\{0.55$, $0.65$, $0.87$, $1.2\}$. 
For scalloped riblets, the groove cross-section area is given by $A^+_g = \pi/4\, s^+ \alpha^+$.
Our model-based predictions of the range of optimal riblet sizes ($l^+_g \in [9.6, 12.1]$) are in close agreement with $l^+_g \in [9.7, 11.7]$ reported in existing literature (e.g.,~\cite{garjim11b}), and fall within the envelope of the experimental and numerical results~\citep{becbruhaghoehop97,garjim11b} {corresponding to a variety of riblet shapes and sizes}. We identify $\alpha/s=0.87$ and {$l^+_g \approx 12$} ($\omega_z=80$) as the optimal configuration for drag-reducing scalloped riblets. Finally, we note that similar to prior studies {(e.g.,~\cite{garjim11b})}, we observe {an almost} perfect collapse of drag reduction curves in the linear viscous regime followed by a breakdown and scattering beyond the optimal size. Given that a second-order perturbation analysis effectively captures the drag reduction trends observed in previous experimental and numerical studies, we confine our analysis to corrections of this order throughout the remainder of the paper.

\subsection{{Turbulent stresses and energy spectrum}}
\label{sec.rtke}

{In this subsection, we further validate the results of our model by comparing the rms of velocity fluctuations and Reynolds shear stress profiles with those obtained from the DNS of~\cite{chomoikim93}. This comparison is conducted for turbulent channel flow with $Re_\tau = 186$ over triangular riblets. We then analyze the effect of scalloped riblets on the statistics of channel flow with $Re_\tau = 186$.}

\begin{figure}
        \begin{center}
        \begin{tabular}{cccc}
       \hspace{-.6cm}
        \subfigure[]{\label{fig.Euuvvww12rms}}
        &&
        \hspace{-0.8cm}
        \subfigure[]{\label{fig.Euv12rms}}
        &
        \\[-.5cm]\hspace{-.3cm}
	\begin{tabular}{c}
       \end{tabular}
       &\hspace{-.18cm}
	\begin{tabular}{c}
       \includegraphics[width=0.42\textwidth]{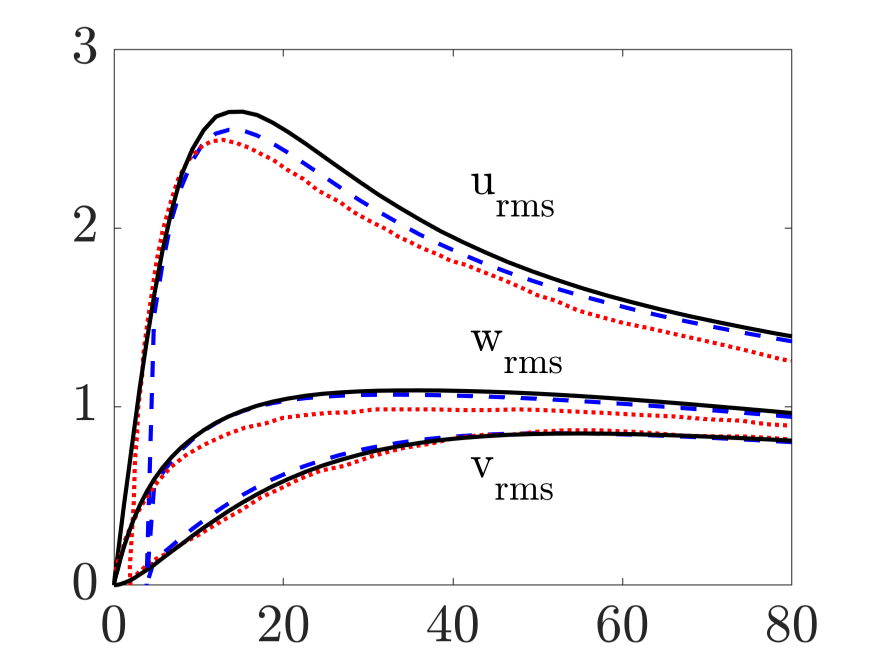}
       \end{tabular}
       &
       \hspace{-.3cm}
	\begin{tabular}{c}
        \vspace{.2cm}
        {\rotatebox{90}{$-uv$}}
       \end{tabular}
       &\hspace{-.2cm}
    \begin{tabular}{c}
       \includegraphics[width=0.42\textwidth]{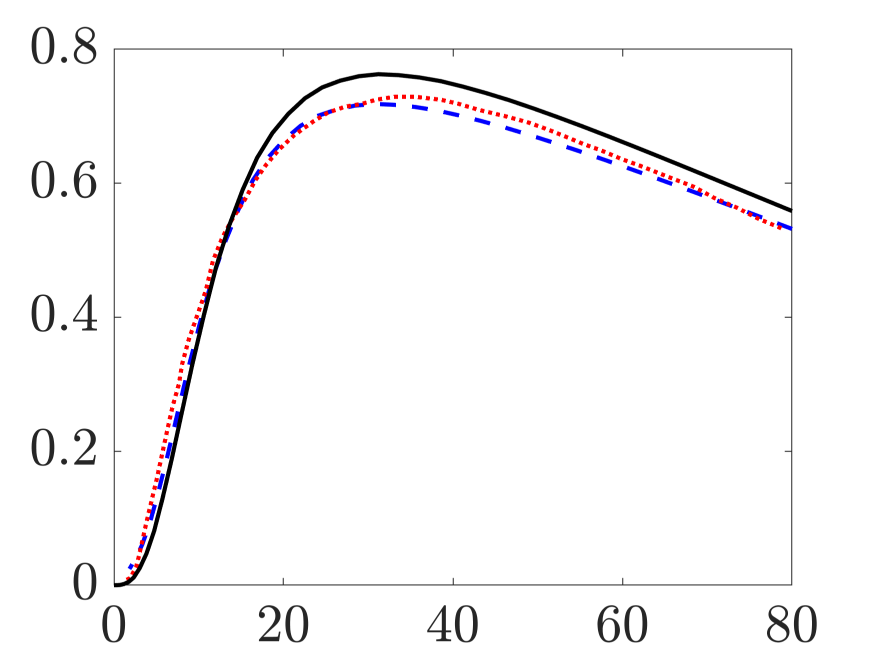}
       \end{tabular}
       \\
       \hspace{-.6cm}
        \subfigure[]{\label{fig.Euuvvww12rmsvalley}}
        &&
        \hspace{-0.8cm}
        \subfigure[]{\label{fig.Euv12rmsvalley}}
        &
        \\[-.5cm]\hspace{-.3cm}
	\begin{tabular}{c}
       \end{tabular}
       &\hspace{-.18cm}
	\begin{tabular}{c}
       \includegraphics[width=0.42\textwidth]{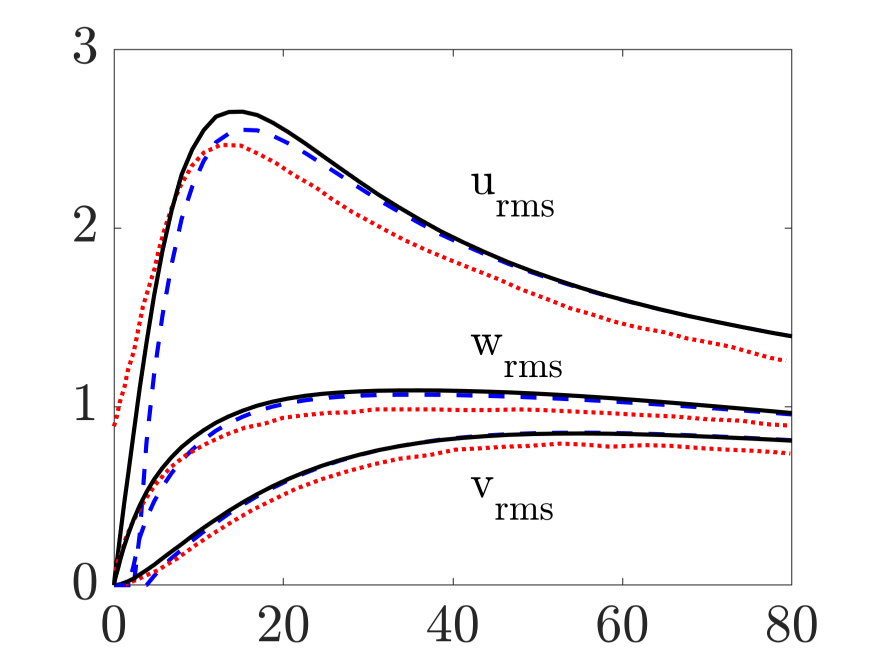}
       \end{tabular}
       &
       \hspace{-.3cm}
	\begin{tabular}{c}
        \vspace{.2cm}
        {\rotatebox{90}{$-uv$}}
       \end{tabular}
       &\hspace{-.2cm}
    \begin{tabular}{c}
       \includegraphics[width=0.42\textwidth]{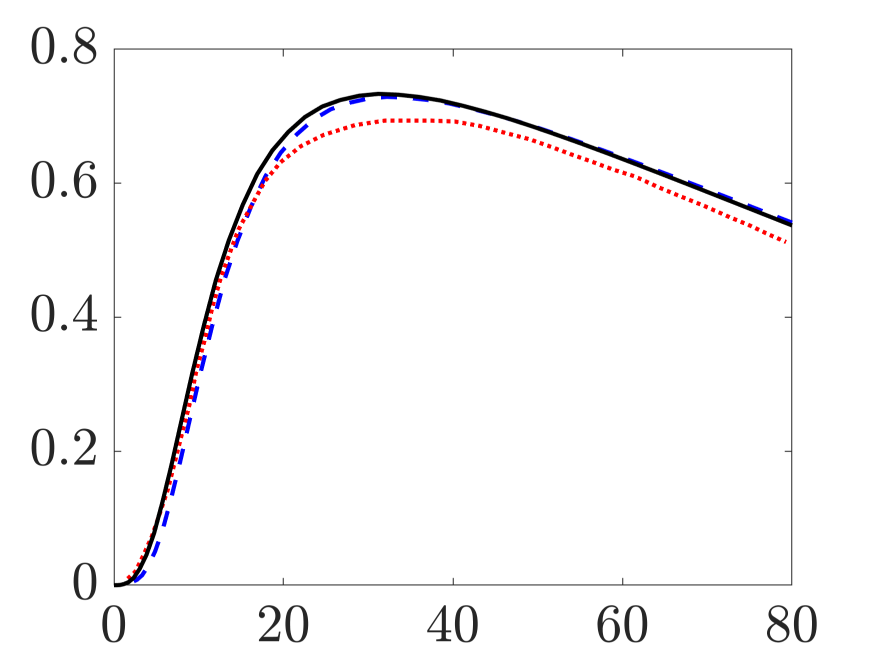}
       \end{tabular}
       \\
       \hspace{-.6cm}
        \subfigure[]{\label{fig.Euuvvww12rmsavg}}
        &&
        \hspace{-0.8cm}
        \subfigure[]{\label{fig.Euv12rmsavg}}
        &
        \\[-.5cm]\hspace{-.3cm}
	\begin{tabular}{c}
       \end{tabular}
       &\hspace{-.18cm}
	\begin{tabular}{c}
       \includegraphics[width=0.42\textwidth]{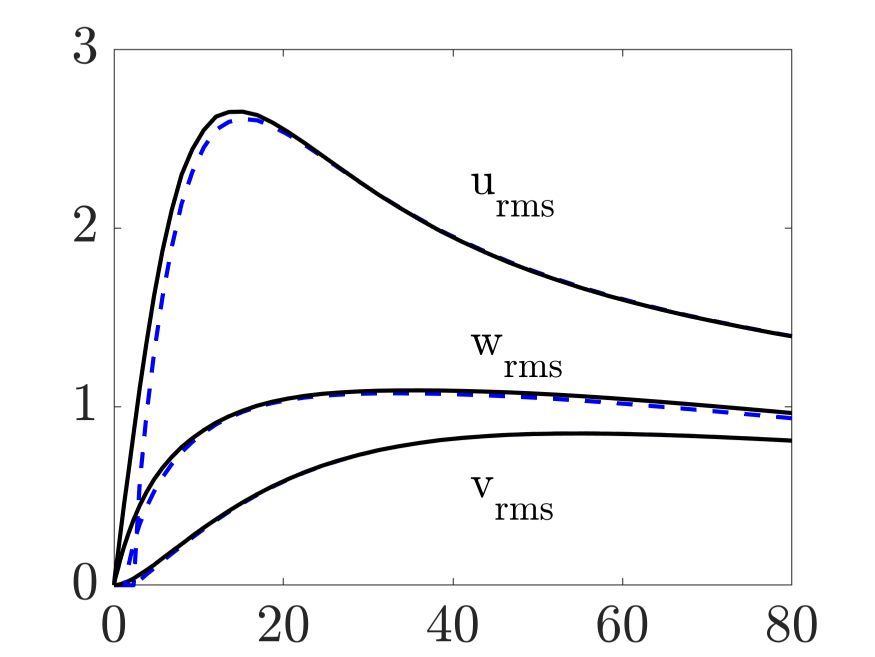}
        \\[-.1cm]
            \hspace{.2cm}
            $\tilde{y}^+$
       \end{tabular}
       &
       \hspace{-.3cm}
	\begin{tabular}{c}
        \vspace{.2cm}
        {\rotatebox{90}{$-uv$}}
       \end{tabular}
       &\hspace{-.2cm}
    \begin{tabular}{c}
       \includegraphics[width=0.42\textwidth]{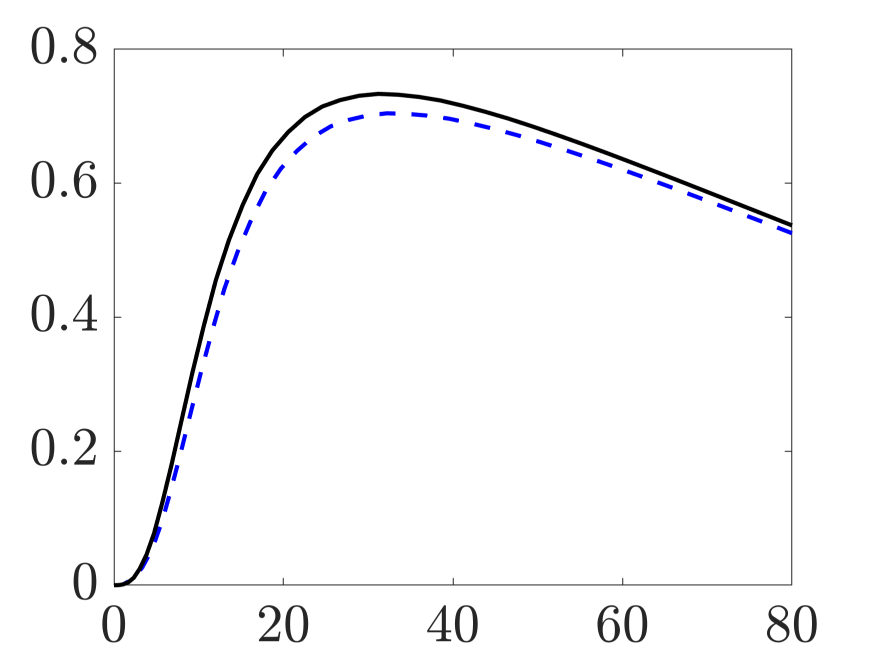}
       \\[-.1cm]
            \hspace{.2cm}
            $\tilde{y}^+$
       \end{tabular}
       \end{tabular}
       \end{center}
        \caption{{The rms of various velocity components (first column) and the Reynolds shear stress in turbulent channel flow with $Re_\tau = 186$ over triangular riblets with $\alpha/s = 0.87$ and $l^+_g \approx 13$ ($\omega_z = 57$). Predictions of the proposed model (dashed blue) are compared against the results of the DNS of~\cite{chomoikim93} (dotted red) and profiles from smooth channel flow (solid) at spanwise locations corresponding to the riblet tips (a,b) and trough center (c,d). The final row (e,f) shows the spanwise-averaged stresses in the presence of riblets.}}
	\label{fig.rmscompare}
\end{figure}

{Figures~\ref{fig.Euuvvww12rms} and~\ref{fig.Euuvvww12rmsvalley} compare the model-based predictions of the rms values of the velocity field with the results of DNS above the tip and within the valley of triangular riblets, respectively. The rms values are calculated from the steady-state covariances that solve Lyapunov equations~\eqref{eq.lyap-set}. For example, the dependence of the rms of streamwise velocity on the spanwise dimension can be obtained as
\begin{align*}
    \ba{rcl}
    u_\mathrm{rms}(y,z) 
        &\!\!=\!\!\!&
    \left (\ds{\int_{0}^\infty \int_{0}^{\omega_z} \sum_{n\in \bbZ}}\, \mathrm{Re}\!\left( \sum_{l=0}^{\infty} \alpha^l\, E_{uu,l}(y,k_x,\theta_n) \,\mre^{\mri k_x x}\right) \cos(\theta_n z)\, \mrd \theta\, \mrd k_x \right )^{1/2},
    \ea
\end{align*}
where, $\mathrm{Re}$ denotes the real part of a vector and  $E_{uu,l}(y,k_x,\theta_n)$ is the dominant eigenmode of the covariance matrix $\Phi_{u,l} (k_x,\theta_n) \,=\, C_u (k_x,\theta_n)\, X_{l,0} (k_x,\theta_n)\, C^*_u (k_x,\theta_n)$; see appendix~\ref{app.Xpert} for the perturbation analysis used to obtain covariance matrix $X_{l,0}$. A geometric parameterization of $\alpha/s = 0.87$ and $l^+_g \approx 13$ ($\omega_z = 57$) is selected to match that of the riblets considered in the DNS of~\cite{chomoikim93}. At the tips, our model captures the decrease in $u_\mathrm{rms}$, albeit slightly exaggerating changes below $\tilde{y}^+\approx 12$. It is notable that the wall-normal location of the peak streamwise velocity fluctuations is in good agreement with the result of DNS. Our model also closely follows the trend in 
$v_\mathrm{rms}$, while falling short of capturing the amount of reduction in $w_\mathrm{rms}$.  
At the center of the trough, our model captures the decrease in $v_\mathrm{rms}$ and $w_\mathrm{rms}$ below $\tilde{y}^+=6.7$, but fails to capture the increase in $u_\mathrm{rms}$ in the same region. The overall effect of the riblets on the rms values can be captured by averaging over the spanwise dimension (figure~\ref{fig.Euuvvww12rmsavg}). Our results show a decrease in $u_\mathrm{rms}$ below the inertial region and a slight decrease in $v_\mathrm{rms}$ and $w_\mathrm{rms}$ closer to the lower wall. 
For the same spanwise locations, figures~\ref{fig.Euv12rms} and~\ref{fig.Euv12rmsvalley} compare our model-based predictions of the Reynolds shear stress with the results of DNS. In agreement with~\cite{chomoikim93}, our model captures an elevated shear stress close to the riblet tips and up to $\tilde{y}^+ \approx 12$. While our model-based predictions of $-uv$ slightly deviate from the result of DNS in the riblet valleys, it captures the overall reduction. We note that the increased shear over the riblet tips is compensated by a decrease in the riblet valleys such that the spanwise-averaged profile shows an overall reduction for this drag-reducing configuration (figure~\ref{fig.Euv12rmsavg}). This is also in agreement with the observations of~\cite{chomoikim93}.}

\begin{figure}
        \begin{center}
        \begin{tabular}{cccccccc}
        \hspace{-.6cm}
        \subfigure[]{\label{fig.E0uu}}
        &&
        \hspace{-.8cm}
        \subfigure[]{\label{fig.E12uuas087O160}}
        &&
        \hspace{-.7cm}
        \subfigure[]{\label{fig.E12uuas087O80}}
        &&
        \hspace{-.7cm}
        \subfigure[]{\label{fig.E12uuas087O45}}
        &
        \\[-.5cm]\hspace{-.1cm}
	\begin{tabular}{c}
        {\small \rotatebox{0}{$\tilde{y}^+$}}
       \end{tabular}
       &\hspace{-.45cm}
	\begin{tabular}{c}
        \hspace{-.4cm}
        {\small smooth channel}
        \\
       \includegraphics[width=2.9cm]{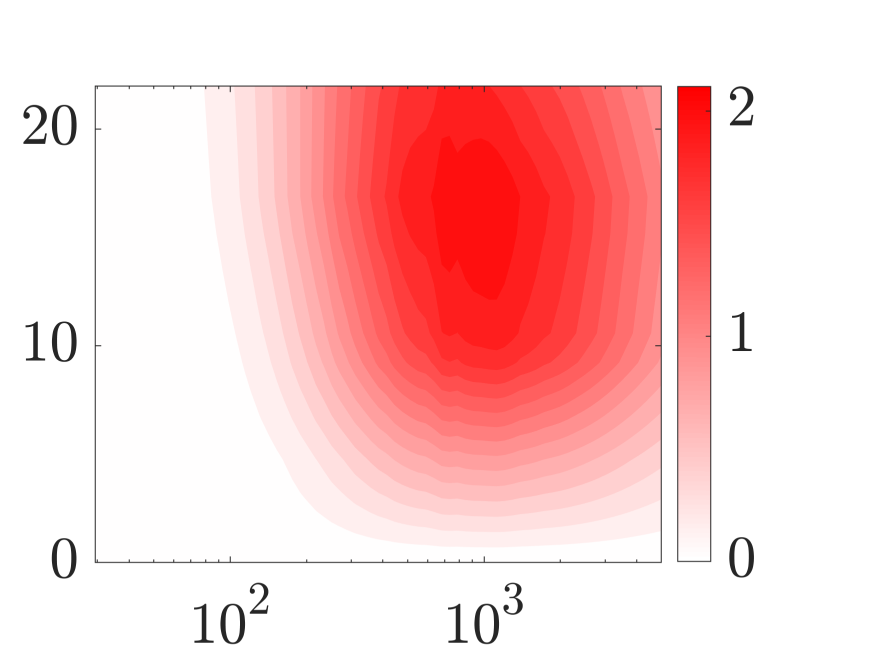}
        \end{tabular}
       &&\hspace{-.35cm}
    \begin{tabular}{c}
        {\small $l^+_g \approx 6$}
        \\
       \includegraphics[width=2.9cm]{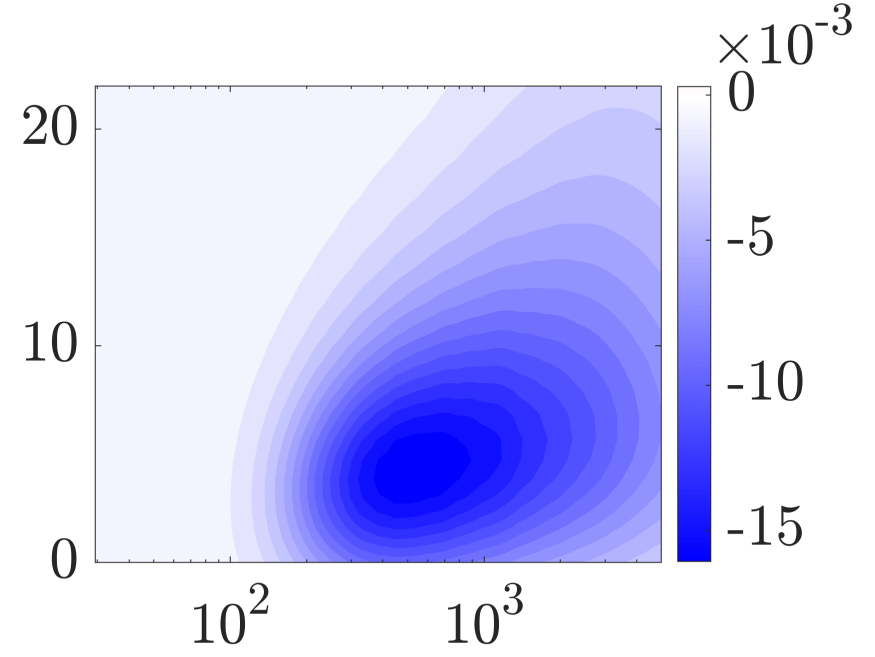}
       \end{tabular}
       &&\hspace{-.35cm}
    \begin{tabular}{c}
    {\small $l^+_g \approx 12$}
        \\
       \includegraphics[width=2.9cm]{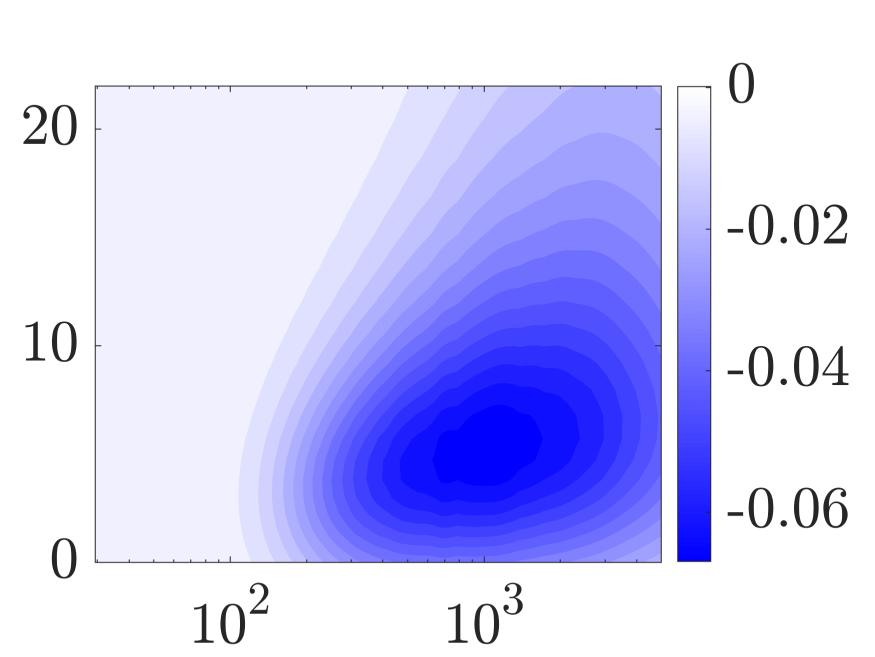}
       \end{tabular}
       &&\hspace{-.3cm}
    \begin{tabular}{c}
    {\small $l^+_g \approx 21$}
        \\
       \includegraphics[width=2.9cm]{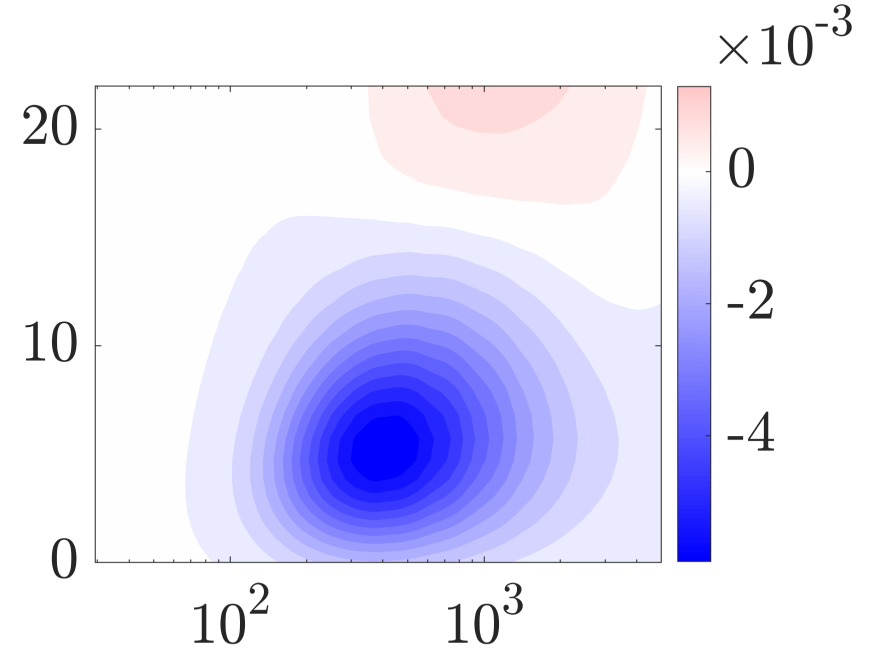}
       \end{tabular}
       \\
       \hspace{-.6cm}
        \subfigure[]{\label{fig.E0vv}}
        &&
        \hspace{-.8cm}
        \subfigure[]{\label{fig.E12vvas087O160}}
        &&
        \hspace{-.7cm}
        \subfigure[]{\label{fig.E12vvas087O80}}
        &&
        \hspace{-.7cm}
        \subfigure[]{\label{fig.E12vvas087O45}}
        &
        \\[-.5cm]\hspace{-.3cm}
	\begin{tabular}{c}
        \vspace{.3cm}
        {\small \rotatebox{0}{$\tilde{y}^+$}}
       \end{tabular}
       &\hspace{-.45cm}
	\begin{tabular}{c}
       \includegraphics[width=2.9cm]{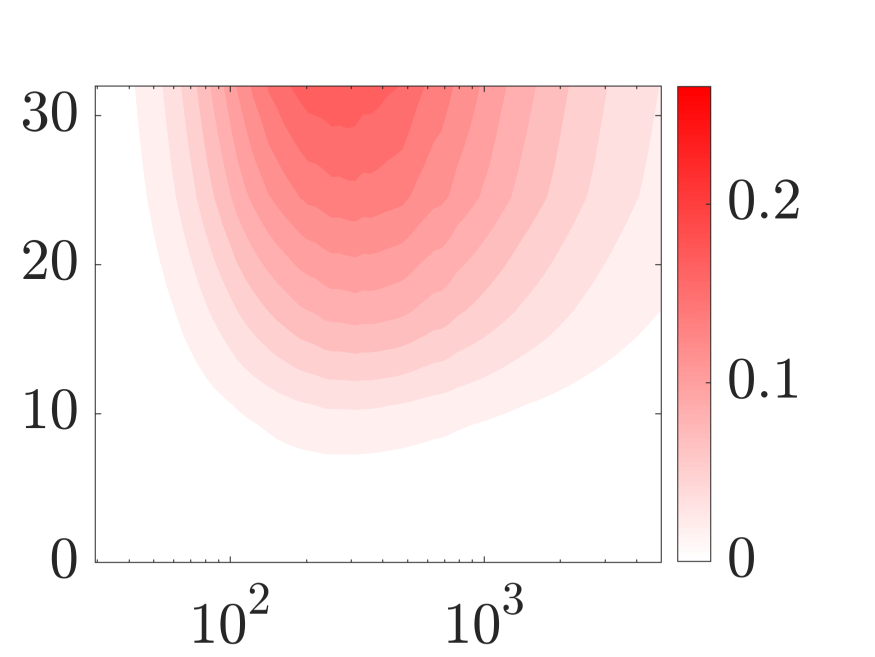}
        \end{tabular}
       &&\hspace{-.35cm}
    \begin{tabular}{c}
       \includegraphics[width=2.9cm]{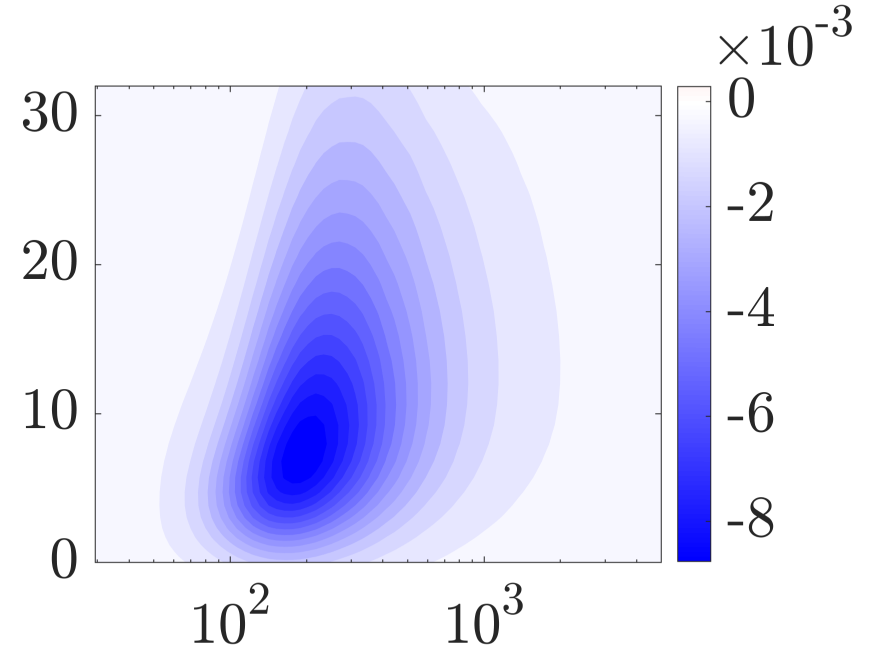}
       \end{tabular}
       &&\hspace{-.31cm}
    \begin{tabular}{c}
       \includegraphics[width=2.9cm]{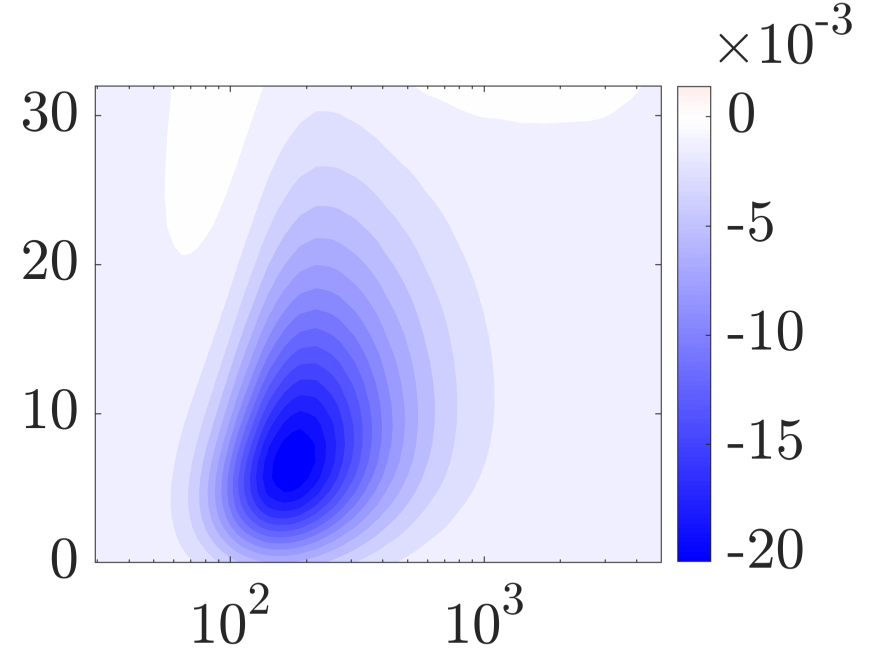}
       \end{tabular}
       &&\hspace{-.3cm}
    \begin{tabular}{c}
       \includegraphics[width=2.9cm]{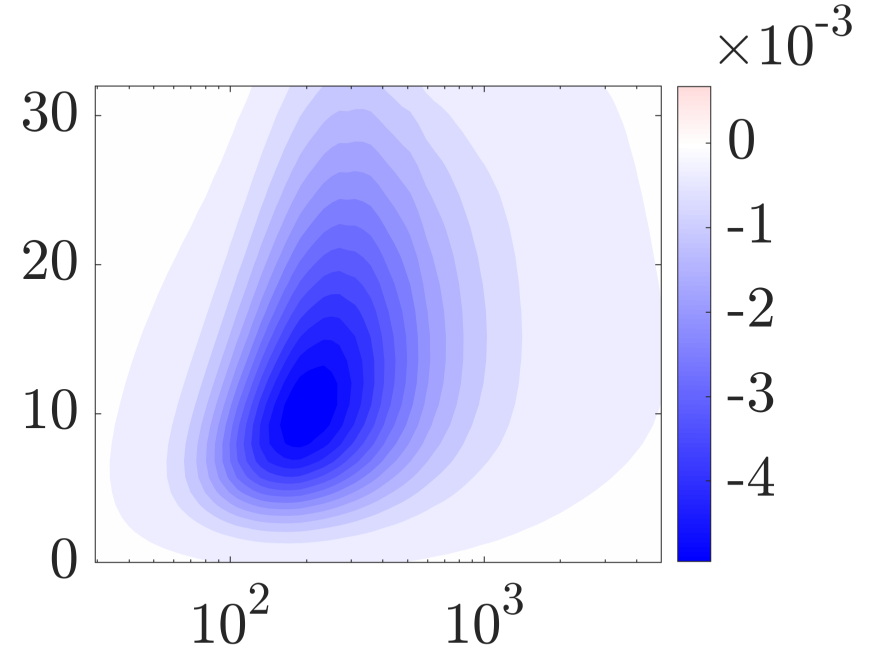}
       \end{tabular}
       \\
       \hspace{-.6cm}
        \subfigure[]{\label{fig.E0ww}}
        &&
        \hspace{-.8cm}
        \subfigure[]{\label{fig.E12wwas087O160}}
        &&
        \hspace{-.7cm}
        \subfigure[]{\label{fig.E12wwas087O80}}
        &&
        \hspace{-.7cm}
        \subfigure[]{\label{fig.E12wwas087O45}}
        &
        \\[-.5cm]\hspace{-.3cm}
	\begin{tabular}{c}
        \vspace{.3cm}
        {\small \rotatebox{0}{$\tilde{y}^+$}}
       \end{tabular}
       &\hspace{-.45cm}
	\begin{tabular}{c}
       \includegraphics[width=2.9cm]{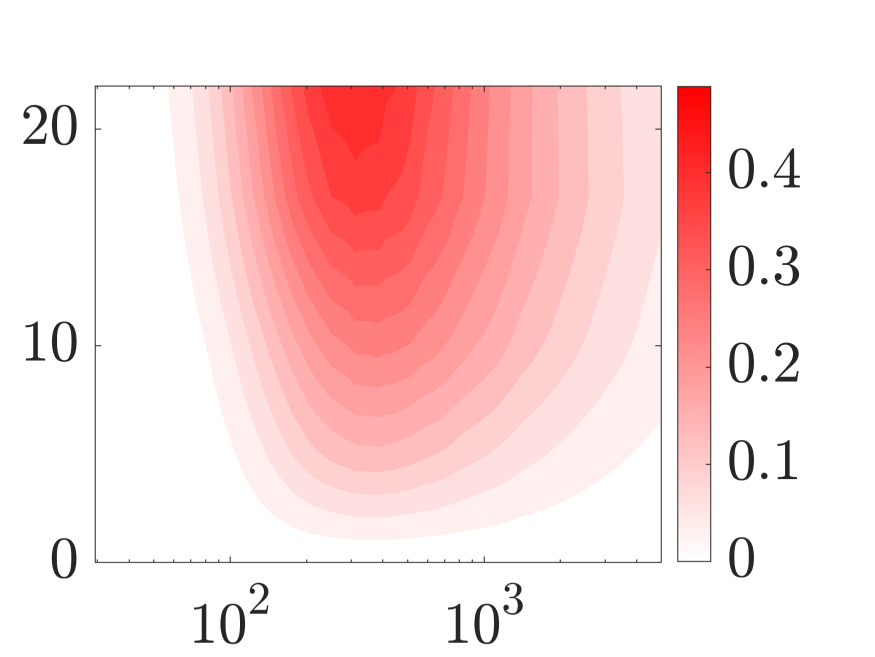}
        \end{tabular}
       &&\hspace{-.35cm}
    \begin{tabular}{c}
       \includegraphics[width=2.9cm]{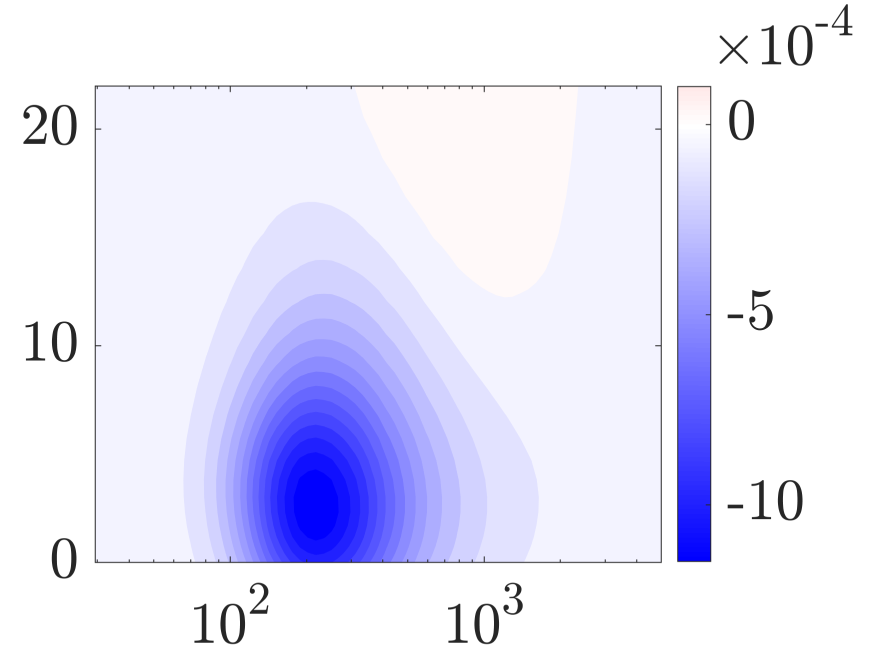}
       \end{tabular}
       &&\hspace{-.31cm}
    \begin{tabular}{c}
       \includegraphics[width=2.9cm]{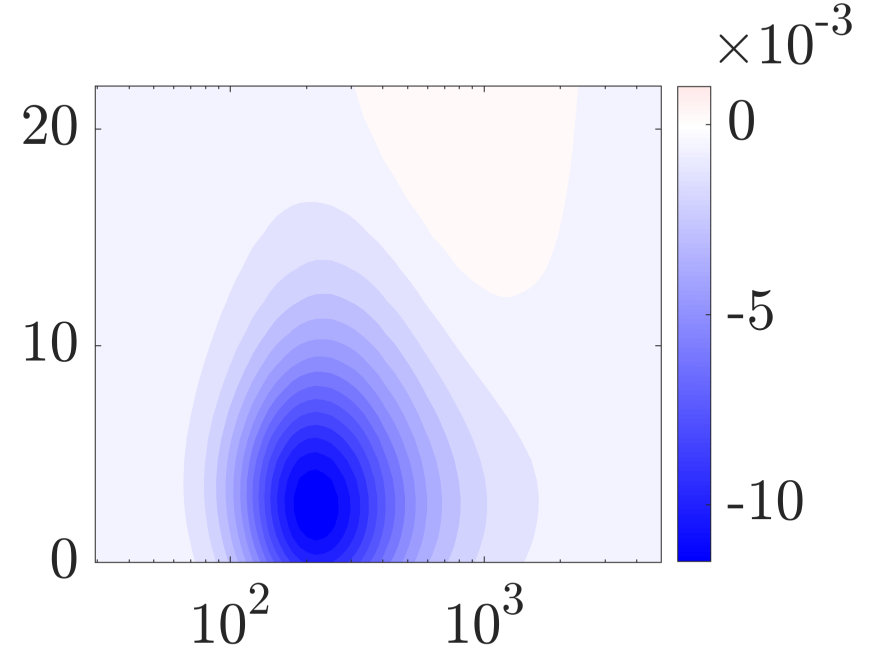}
       \end{tabular}
       &&\hspace{-.3cm}
    \begin{tabular}{c}
       \includegraphics[width=2.9cm]{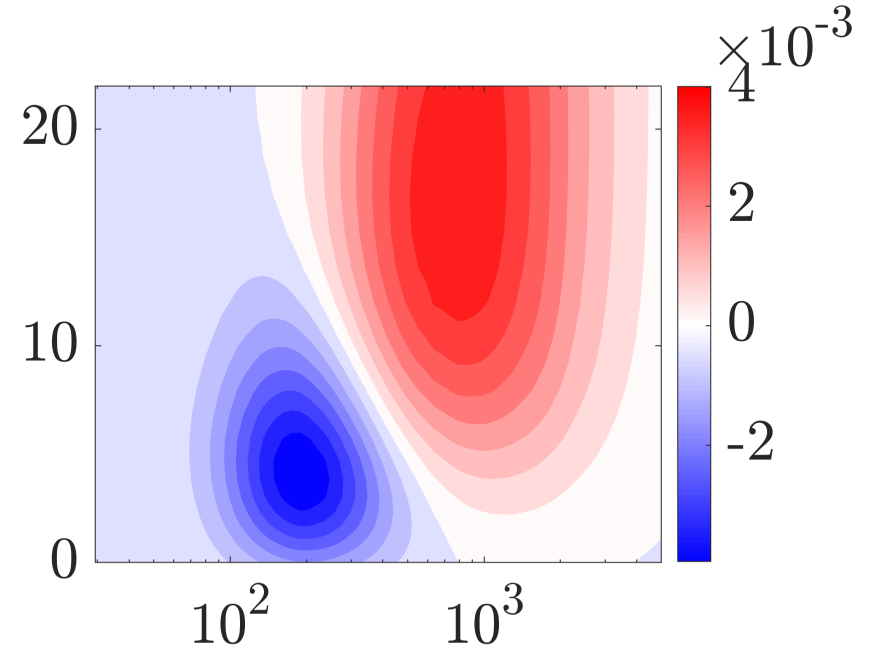}
       \end{tabular}
       \\
       \hspace{-.6cm}
        \subfigure[]{\label{fig.E0uv}}
        &&
        \hspace{-.8cm}
        \subfigure[]{\label{fig.E12uvas087O160}}
        &&
        \hspace{-.7cm}
        \subfigure[]{\label{fig.E12uvas087O80}}
        &&
        \hspace{-.7cm}
        \subfigure[]{\label{fig.E12uvas087O45}}
        &
        \\[-.5cm]\hspace{-.3cm}
	\begin{tabular}{c}
        \vspace{.3cm}
        {\small \rotatebox{0}{$\tilde{y}^+$}}
       \end{tabular}
       &\hspace{-.41cm}
	\begin{tabular}{c}
       \includegraphics[width=2.9cm]{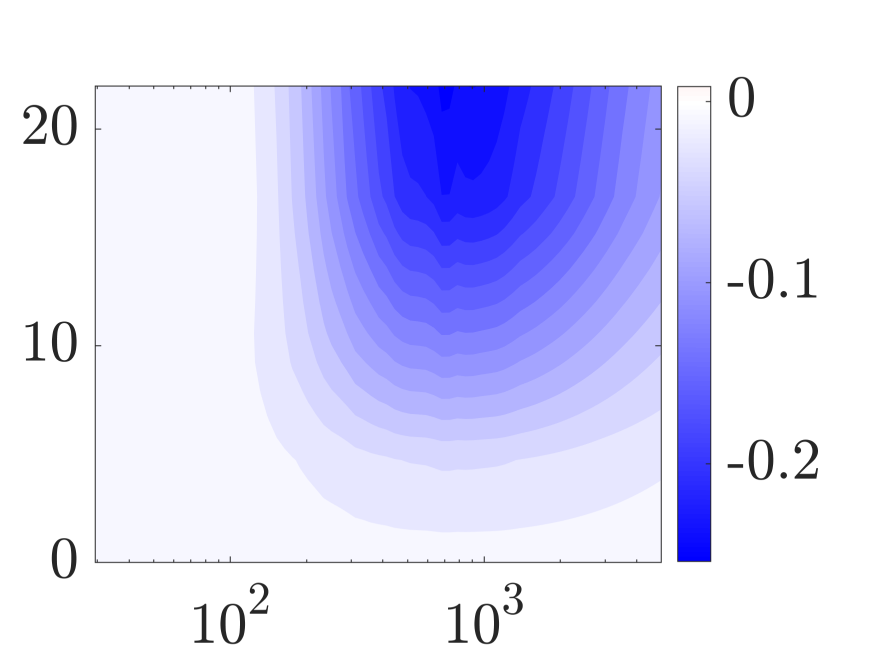}
        \\[-.1cm]
            \hspace{-.3cm}
            {\small $\lambda^+_x$}
       \end{tabular}
       &&\hspace{-.35cm}
    \begin{tabular}{c}
       \includegraphics[width=2.9cm]{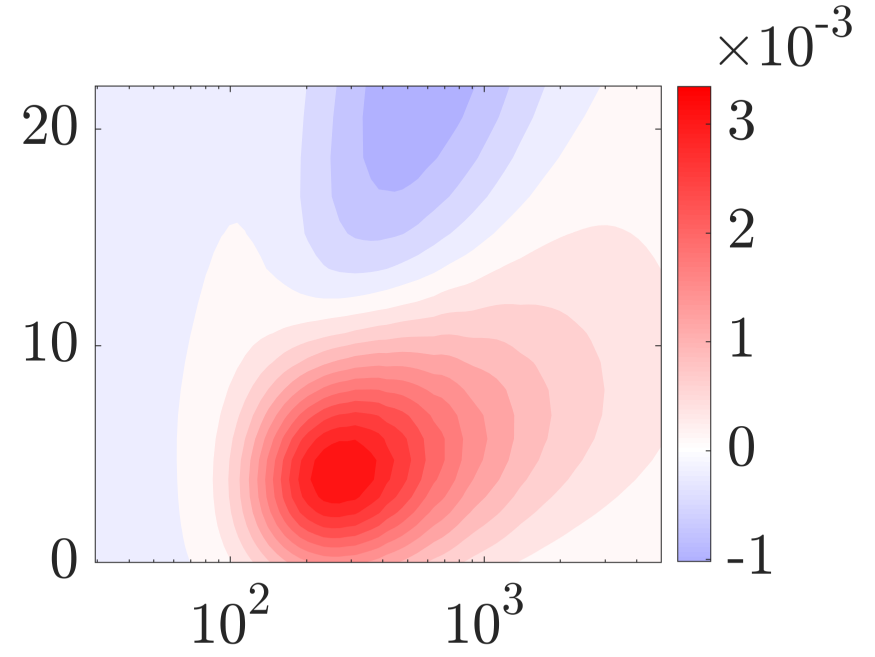}
       \\[-.1cm]
            \hspace{-.3cm}
            {\small $\lambda^+_x$}
       \end{tabular}
       &&\hspace{-.31cm}
    \begin{tabular}{c}
       \includegraphics[width=2.9cm]{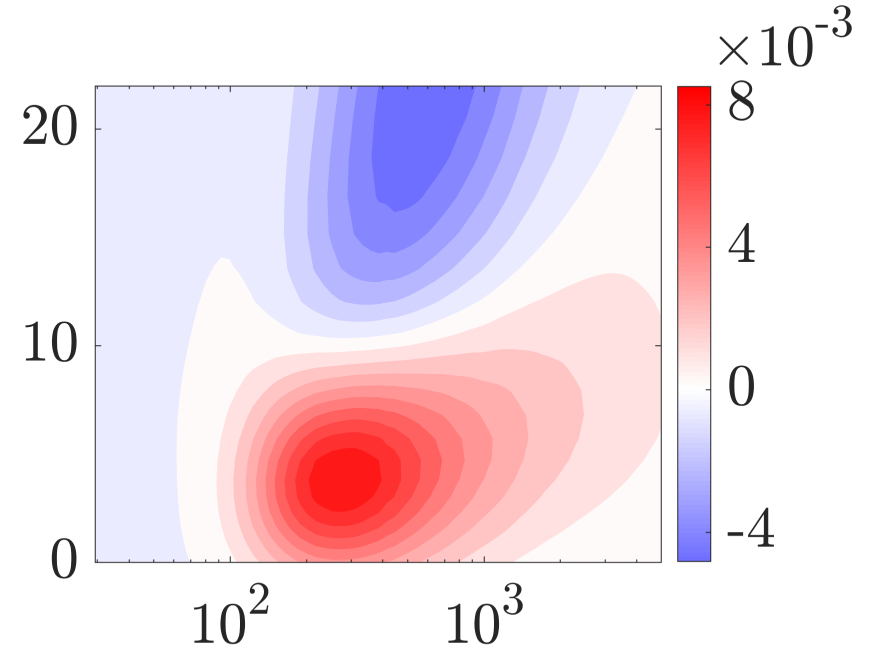}
       \\[-.1cm]
            \hspace{-.3cm}
            {\small $\lambda^+_x$}
       \end{tabular}
       &&\hspace{-.3cm}
    \begin{tabular}{c}
       \includegraphics[width=2.9cm]{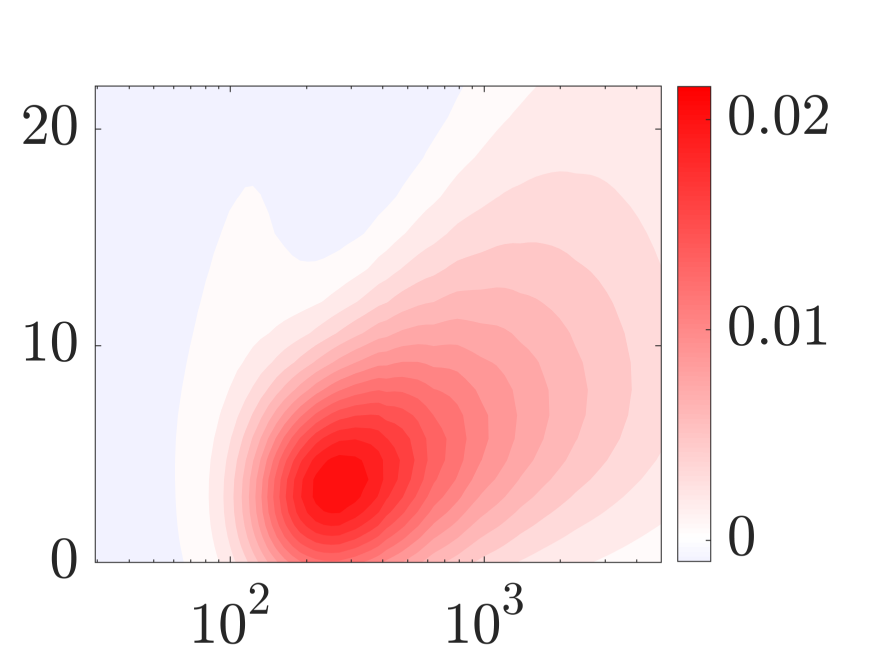}
       \\[-.1cm]
            \hspace{-.3cm}
            {\small $\lambda^+_x$}
       \end{tabular}
       \end{tabular}
       \end{center}
        \caption{{Premultiplied one-dimensional energy spectra of Reynolds stresses in channel flow with $Re_\tau=186$ resulting from the DNS of~\cite{deljim03} (first column) together with modifications (up to $\alpha^2$) induced by scalloped riblets with $\alpha/s=0.87$ and $l^+_g \approx 6$ ($\omega_z = 160$) (second column), $l^+_g \approx 12$ ($\omega_z = 80$) (third column), and $l^+_g \approx 21$ ($\omega_z = 45$) (fourth column). The energy spectra have been integrated over all spanwise wavelengths. (b-d) $k_x(\alpha\,E_{uu,1} + \alpha^2\, E_{uu,2})$; (f-h) $k_x(\alpha\,E_{vv,1} + \alpha^2\, E_{vv,2})$; (j-l) $k_x(\alpha\,E_{ww,1} + \alpha^2\, E_{ww,2})$; and (n-p) $-k_x (\alpha\,E_{uv,1} + \alpha^2\,E_{uv,2})$.}}
    \label{fig.Restresses}
\end{figure}

{For channel flow with $Re_\tau=186$ over scalloped riblets with $\alpha/s=0.87$, figure~\ref{fig.Restresses} analyzes the riblet-induced modifications to the premultiplied one-dimensional energy spectra of Reynolds stresses integrated over all spanwise wavelengths $\lambda^+_z$.
Hereafter, color plots use a red-white-blue colormap, where red denotes amplification, blue denotes suppression, and white denotes no change.
It is evident that the optimal drag-reducing scalloped riblets ($l^+_g \approx 12$) induce the largest suppression in the streamwise, wall-normal, and spanwise Reynolds stresses. Riblet-induced changes are concentrated close to the bottom wall of the channel and shift upwards as the riblet size increases. The relative impact of the scalloped riblets (compared to the smooth channel) is strongest in altering the wall-normal Reynolds stress and it is concentrated around $\lambda_x^+ \approx 200$ and $\tilde{y}^+ \approx 10$ for the largest riblets (figure~\ref{fig.E12vvas087O45}). We observe that larger-than-optimal riblets with $l^+_g \approx 21$ ($\omega_z = 45$) do not suppress the energy spectra as much as the optimal ones and that such large riblets can even result in an overall amplification of the spanwise energy spectrum (\ref{fig.E12wwas087O45}). 
Lastly, we observe a concentrated patch of added Reynolds shear stress at $\lambda_x^+ \approx 270$ and $\tilde{y}^+ \approx 4$ (figure~\ref{fig.E12uvas087O80}), which becomes stronger for larger riblets. While we over-predict the streamwise wavelength associated with peak amplification of the Reynolds shear stress ($\lambda_x^+ \approx 270$ in our model vs $\lambda_x^+ \approx 150$ in DNS), our observations of the energy spectra are generally in alignment with the DNS of~\cite{garjim11b}.} 

\begin{figure}
        \begin{center}
        \begin{tabular}{cccc}
       \hspace{-.6cm}
        \subfigure[]{\label{fig.kcorr}}
        &&
        \hspace{-0.8cm}
        \subfigure[]{\label{fig.NWhs087Re186YZ}}
        &
        \\[-.4cm]
        &
        \hspace{-.3cm}
	\begin{tabular}{cc}
        \vspace{.3cm}
        \begin{tabular}{c}
        \vspace{.2cm}
        {\rotatebox{90}{$k_0$}}
       \end{tabular}
       & 
       \hspace{-.4cm}
	\begin{tabular}{c}
       \includegraphics[width=0.38\textwidth]{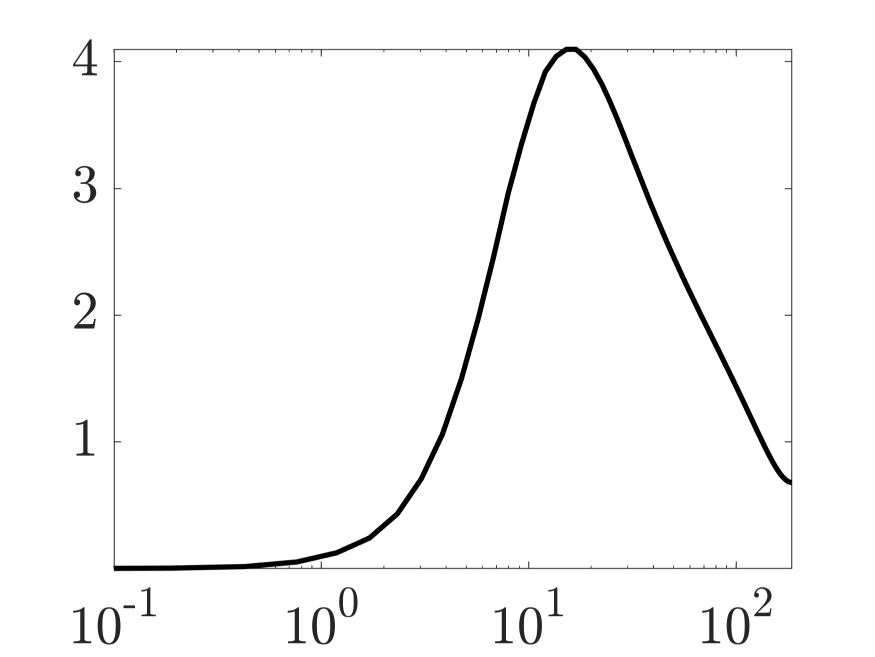}
       \end{tabular}
       \\[-.4cm]
        \begin{tabular}{c}
        \vspace{.2cm}
        {\rotatebox{90}{$k_\mathrm{corr}$}}
       \end{tabular}
       &
        \hspace{-.4cm}
	\begin{tabular}{c}
       \includegraphics[width=0.38\textwidth]{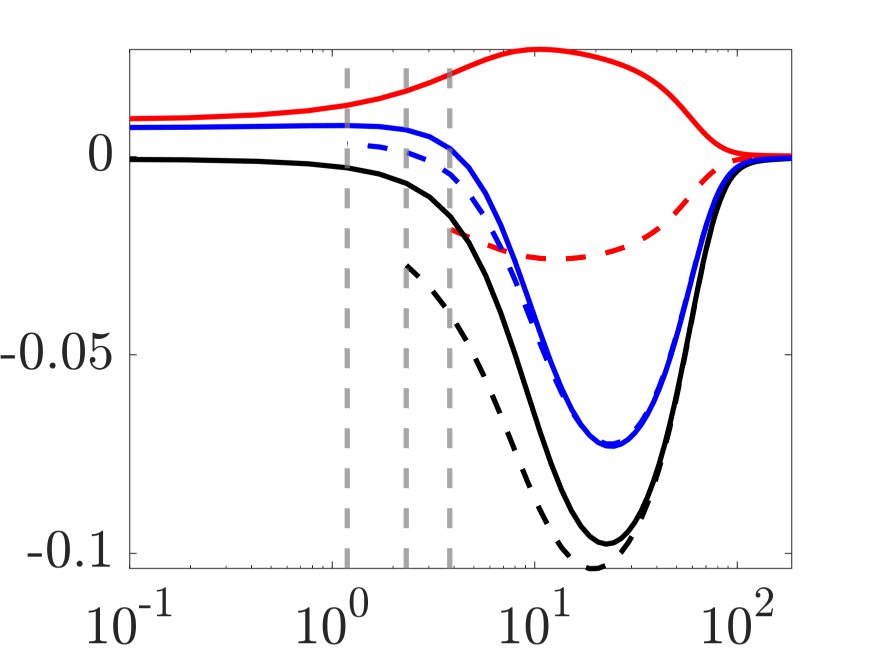}
        \\[-.1cm]
            \hspace{.2cm}
            $\tilde{y}^+$
       \end{tabular}
       \end{tabular}
       &&
       \hspace{-.6cm}
        \begin{tabular}{cc}
	\begin{tabular}{c}
        \vspace{.4cm}
        {\rotatebox{90}{$\tilde{y}^+$}}
       \end{tabular}
       &
       \begin{tabular}{c}
       \hspace{-.4cm}
       \includegraphics[width=0.45\textwidth]{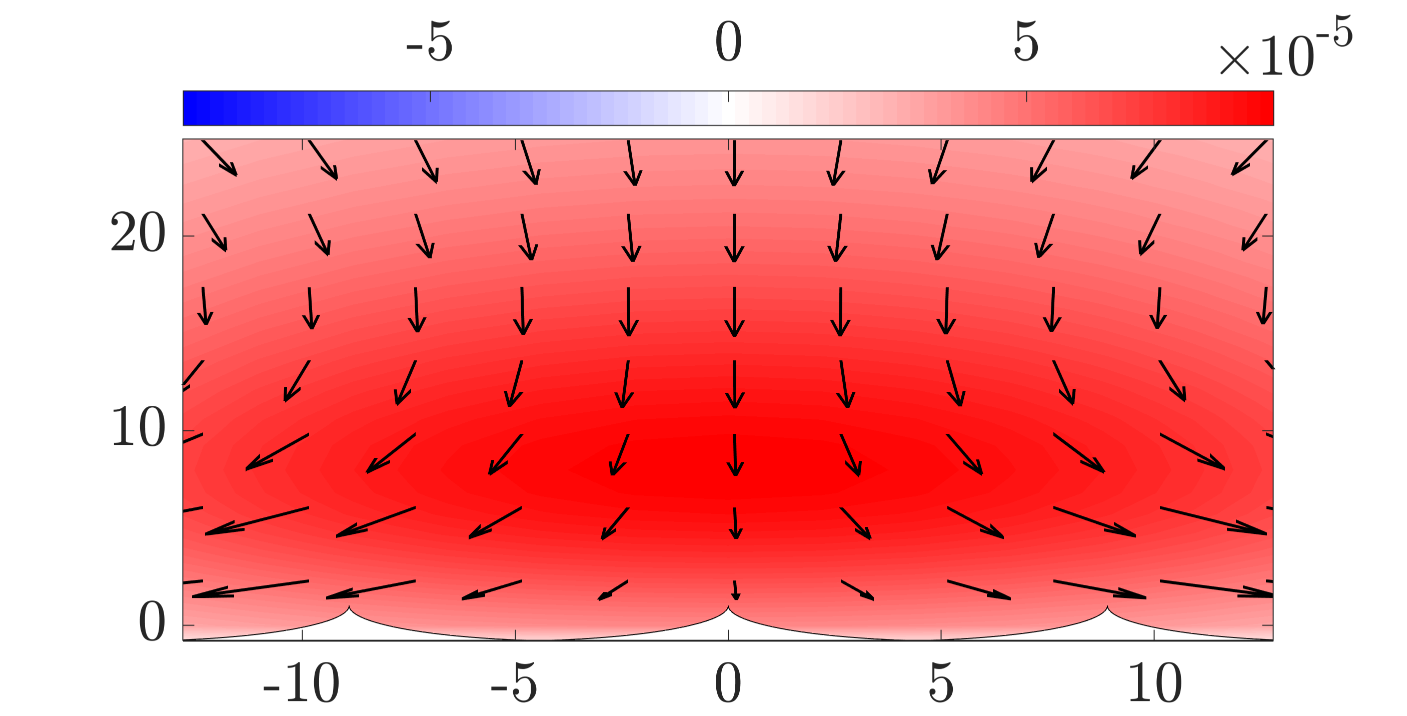}
       \end{tabular}
       \\
       \hspace{-.3cm}
	\begin{tabular}{c}
        \vspace{.4cm}
        {\rotatebox{90}{$\tilde{y}^+$}}
       \end{tabular}
       &
       \begin{tabular}{c}
       \hspace{-.4cm}
       \includegraphics[width=0.45\textwidth]{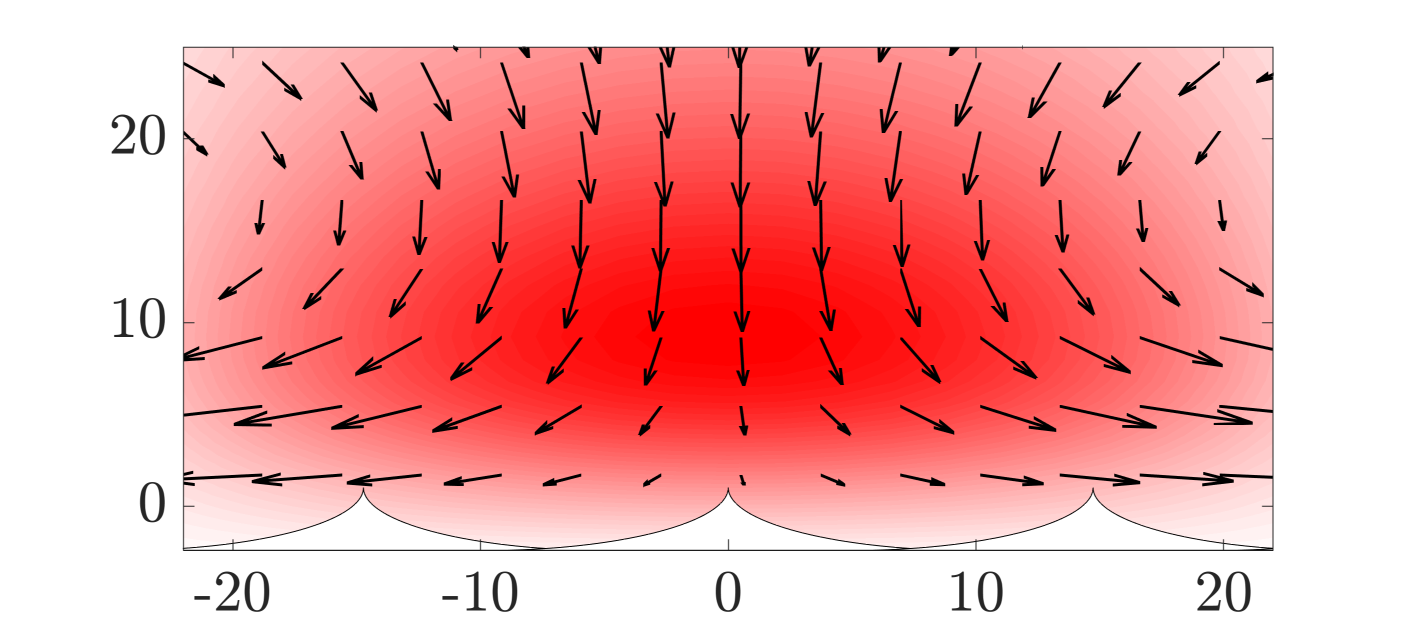}
       \end{tabular}
       \\
       \hspace{-.3cm}
	\begin{tabular}{c}
        \vspace{.4cm}
        {\rotatebox{90}{$\tilde{y}^+$}}
       \end{tabular}
       &
       \begin{tabular}{c}
       \hspace{-.4cm}
       \includegraphics[width=0.45\textwidth]{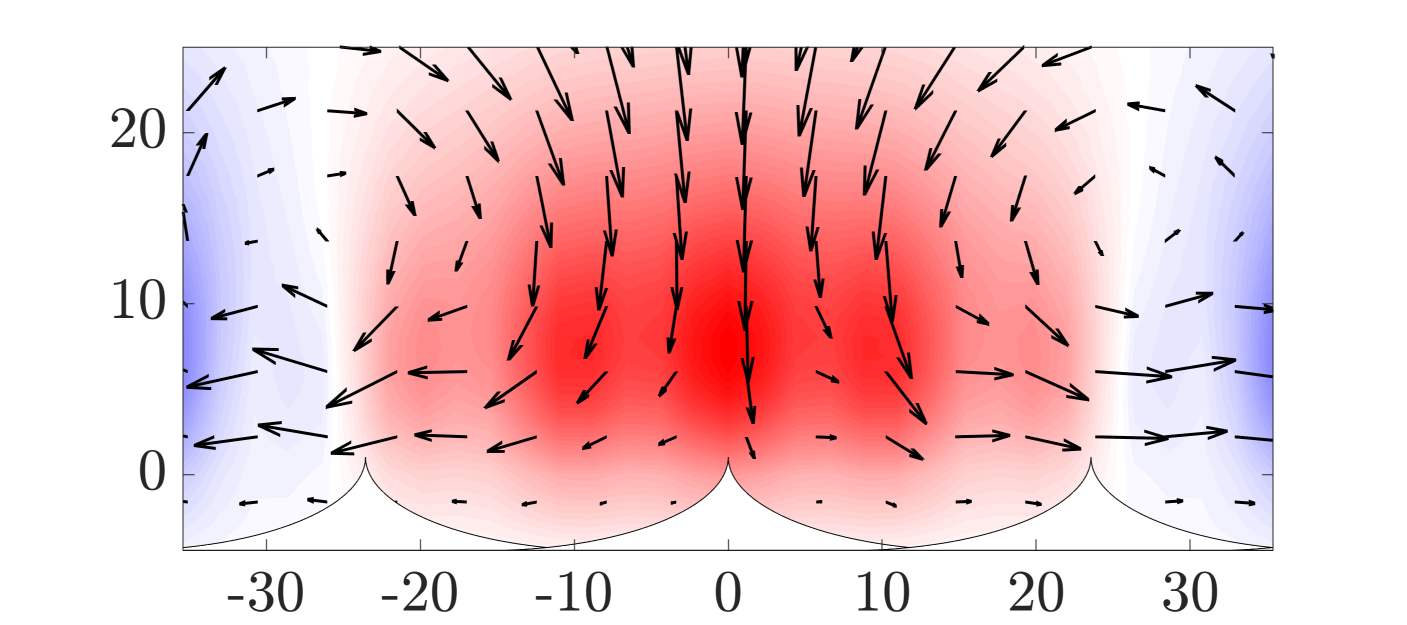}
        \end{tabular}
        \\
        &
        \hspace{.2cm}
        $\tilde{z}^+$
        \end{tabular}
    \end{tabular}
       \end{center}
        \caption{{(a) Turbulent kinetic energy in smooth channel flow with $Re_\tau = 186$ (top) together with modifications to turbulent kinetic energy $k_\mathrm{corr} \DefinedAs \alpha\,k_1 + \alpha^2\,k_2$, at the tip (dashed) and valley (solid) of scalloped riblets with $\alpha/s = 0.87$ for small ($l^+_g \approx 6$), optimal ($l^+_g \approx 12$), and large ($l^+_g \approx 21$) riblets (bottom) shown in blue, black, and red, respectively. The gray dashed lines mark the wall-normal location of the riblet tips. (b) Cross-plane view of the streamwise velocity $u$ at $\tilde{x}^+ = 500$ together with quiver lines corresponding to the wall-normal and spanwise velocity components.}} 
	\label{fig.keps}
\end{figure}

{To more thoroughly investigate the dependence of energetic modifications on the size of riblets, we study variations of the kinetic energy in the wall-normal direction after averaging over the translationally invariant streamwise dimension; see appendix~\ref{app.corrKEps} for details.
Comparing the kinetic energy at the riblet tips with the kinetic energy in smooth channel flow requires an appropriate baseline for the origin of the near-wall turbulence. As mentioned in \S~\ref{sec.DR}, $\tilde{y} = -1$ serves as such virtual origin for all cases considered in this study. For small to optimal-sized riblets, this origin would be located $\alpha (1-r_p)$ below their tips. For larger-than-optimal riblets, turbulence would protrude further down into the riblet grooves. We thus follow~\cite{endmodgarhutchu21} in determining the origin of turbulence in a way that maintains a constant ratio between the protrusion height and the riblet height ($1-r_p$) corresponding to that of optimal riblets. Figure~\ref{fig.kcorr} compares modifications ($k_{\mathrm{corr}}$) to the turbulent kinetic energy above the tip and within the valley of scalloped riblets of different size. In this figure, $k_0$ denotes the kinetic energy profile of smooth channel flow. Riblets of all sizes (small, optimal, and large) suppress the kinetic energy over their tips. However, it is the optimal drag-reducing riblets that achieve the most suppression, which is in agreement with the observations of figure~\ref{fig.Restresses}. While the energy is also suppressed in the valley of {small and optimal} drag-reducing riblets, it is amplified in the valley of {large} drag-increasing riblets. To analyze the reason behind this trend, we next analyze vorticity patterns in the proximity of the riblet-mounted surface.} 

{For three sizes of riblet, figure~\ref{fig.NWhs087Re186YZ} shows quiver lines of the cross-plane velocity fields ($v, w$) corresponding to streamwise vortices embedded on top of color plots of the streamwise velocity. Here, the choice of horizontal scales $(\lambda^+_x,\lambda^+_z) \approx (1100,110)$ corresponds to the dominant modes of the near-wall cycle.
As shown in this figure, besides an upward shift of the streamwise vortices, the flow over small- to optimally sized riblets remains similar to the flow over a smooth wall. In contrast, large riblets distort streamwise streaks and allow turbulence to penetrate the riblet grooves increasing the exposure of the surface to the faster stream above and increasing kinetic energy within the grooves of large riblets. The downward penetration of streamwise vortices captured by our model is consistent with the experimental observations of~\cite{leelee01}. We note that the steady-state flow structures shown in figure~\ref{fig.NWhs087Re186YZ} are constructed from the eigenvectors of the covariance matrix ${\bPhi}_\theta (k_x) \,=\, \cC_\theta (k_x) \cX_\theta (k_x)\, \cC^*_\theta (k_x)$, where $\cX_\theta (k_x)$ is the solution to Lyapunov equation~\eqref{eq.lyap}. The principal eigenvectors of this covariance matrix represent energetically dominant flow structures that reside in the vicinity of the upper and lower channel walls; see~\cite{moajovJFM12}[Appendix F] for additional details.}

\begin{figure}
        \begin{center}
        \begin{tabular}{cccccccc}
        \hspace{-.6cm}
        \subfigure[]{\label{fig.Esas87O80}}
        &&
        \hspace{-.8cm}
        \subfigure[]{\label{fig.E1as087O80}}
        &&
        \hspace{-.5cm}
        \subfigure[]{\label{fig.E2as087O80}}
        &&
        \hspace{-.6cm}
        \subfigure[]{\label{fig.E12as087O80}}
        &
        \\[-.5cm]\hspace{-.3cm}
	\begin{tabular}{c}
        \vspace{.2cm}
        {\small \rotatebox{90}{$k_x$}}
       \end{tabular}
       &\hspace{-.3cm}
	\begin{tabular}{c}
       \includegraphics[width=2.8cm]{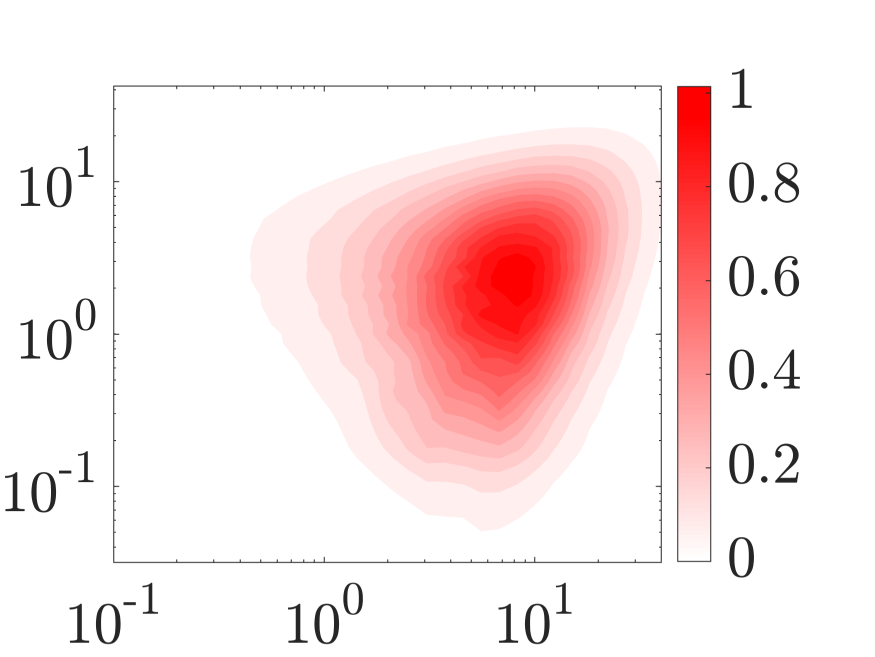}
        \\[-.1cm]
            \hspace{-.1cm}
            {\small $\theta$}
       \end{tabular}
       &&\hspace{-.35cm}
    \begin{tabular}{c}
       \includegraphics[width=2.8cm]{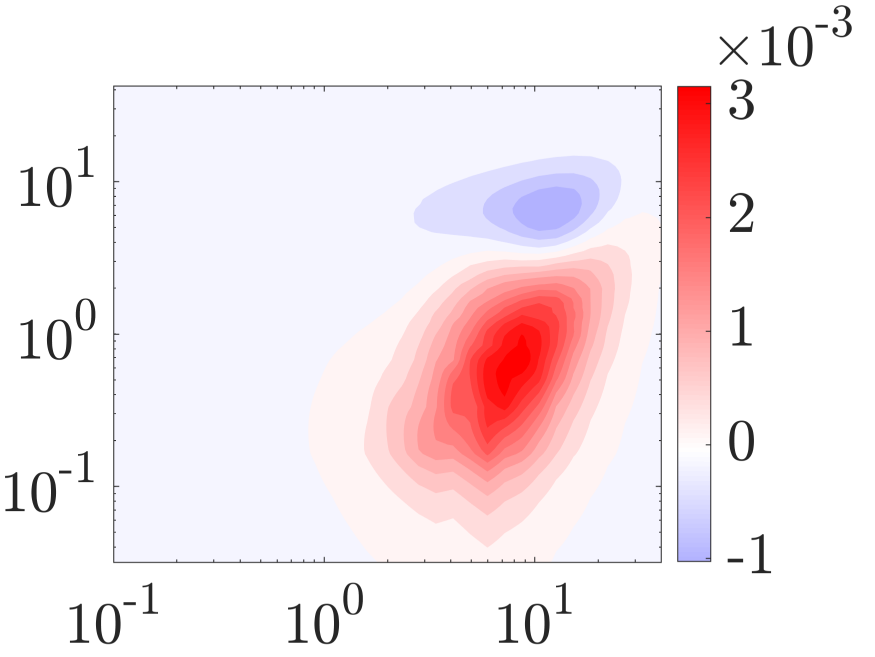}
       \\[-.1cm]
            \hspace{-.1cm}
            {\small $\theta$}
       \end{tabular}
       &&\hspace{-.25cm}
    \begin{tabular}{c}
       \includegraphics[width=2.8cm]{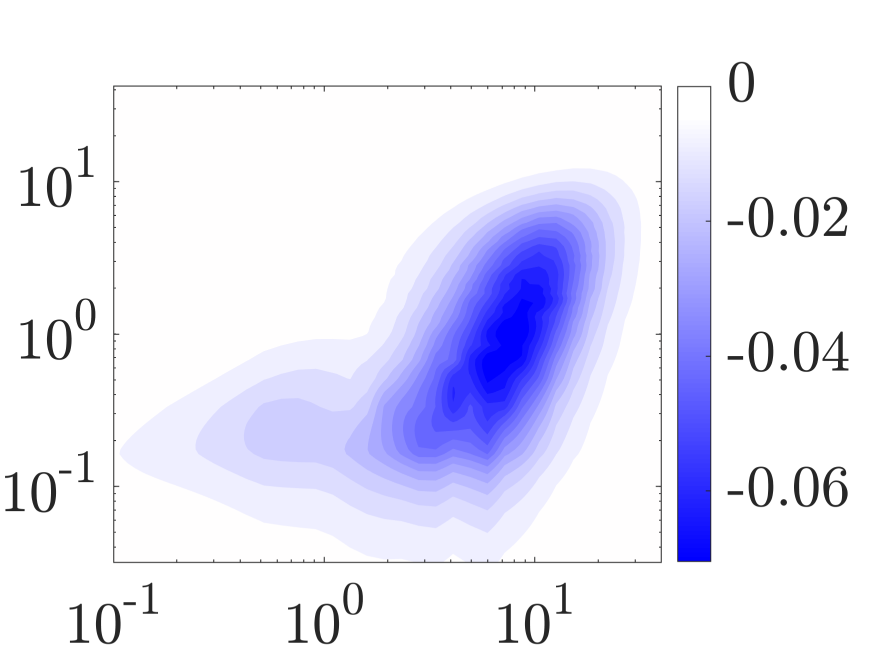}
       \\[-.1cm]
            \hspace{-.1cm}
            {\small $\theta$}
       \end{tabular}
       &&\hspace{-.15cm}
    \begin{tabular}{c}
       \includegraphics[width=2.8cm]{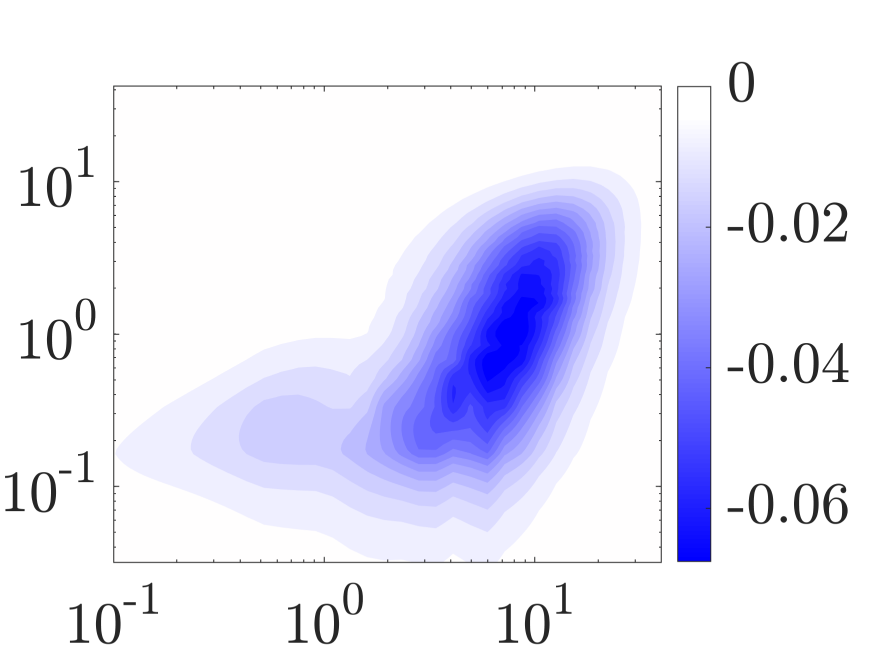}
       \\[-.1cm]
            \hspace{-.1cm}
            {\small $\theta$}
       \end{tabular}
       \end{tabular}
       \end{center}
        \caption{{(a) Premultiplied energy spectrum of the smooth channel flow, $k_x \theta \bar{E}_0(\kappa)$ at $Re_\tau=186$ from the DNS of~\cite{deljim03}; (b) first-order ($\alpha\,k_x \theta \bar{E}_1(\kappa)$); (c) second-order ($\alpha^2\,k_x \theta \bar{E}_2(\kappa)$); and (d) combined ($k_x \theta\,(\alpha\,\bar{E}_1(\kappa) + \alpha^2\, \bar{E}_2(\kappa))$) model-based modifications to the energy spectrum due to scalloped riblets with height-to-spacing ratio $\alpha/s = 0.87$ and {$l^+_g \approx 12$} ($\omega_z=80$).}}
	\label{fig.E012as87Omz}
\end{figure}

Finally, figure~\ref{fig.E012as87Omz} compares the premultiplied energy spectrum of smooth channel flow, $k_x \theta \bar{E}_0(\kappa)$, obtained from the DNS of~\cite{deljim03}, with riblet-induced modifications {predicted by our model}. Here, $\kappa = (k_x,\theta)$, where $\theta$ is {the} spanwise wavenumber offset. Following the parameterization in the spanwise direction, i.e., $\theta_n = \theta + n\,\omega_z$, the energy spectra have been summed over $n$ to integrate over the contributions of {all spanwise harmonics that are amplified in the fluctuation field}. The figures show the premultiplied spectra so that the areas under the log-log plots are equal to the total energy of fluctuations.
Energy modulations at the $\alpha^1$ and $\alpha^2$ perturbation levels due to riblets with $\alpha/s=0.87$ and {$l^+_g \approx 12$} are shown in figures~\ref{fig.E1as087O80} and~\ref{fig.E2as087O80}, respectively. The most energetic modes of the smooth channel flow take place at $(k_x,\theta)\approx (2.2,7.1)$. The largest suppression happens at $(k_x,\theta)\approx (6.7,12.1)$ and $(k_x,\theta)\approx (1.05,{7.2})$ for $\alpha^1$ and $\alpha^2$ levels, respectively. While figure~\ref{fig.E1as087O80} shows a slight energy amplification at $(k_x,\theta)\approx(0.73,8.85)$, the {total modification shown in figure~\ref{fig.E12as087O80}}, which is dominated by the second-order modification, does not {retain any amplification. Notably, the targeted modes are of the same spanwise but longer streamwise length-scale as the most amplified modes in the premultiplied energy spectrum of smooth channel flow.} The total effect of roughness (over all length-scales) on the turbulent kinetic energy can be determined as $\alpha^l \int_{\kappa } \bar{E}_l(\kappa) \mrd\kappa/\int_{\kappa } \bar{E}_0(\kappa) \mrd\kappa$ for different perturbation levels. For riblets with {$l^+_g \approx 12$} this quantity is $0.13\%$ and {$-4.72\%$} at $\alpha^1$ and $\alpha^2$ levels, respectively, and {$-4.59\%$} overall. This riblet configuration also yields the largest drag reduction (cf. figure~\ref{fig.DR}), which is suggestive of the {strong correlation and direct proportionality} between drag reduction and energy suppression demonstrated by~\cite{ranzarjovJFM21}.

\section{Analysis of flow mechanisms in the presence of riblets}
\label{sec.structure}

\begin{figure}
\centering
        \begin{tabular}{ccc}
            \begin{tabular}{c}
            \vspace{0.5cm}
            \hspace{-1.35cm}
            \rotatebox{90}{$-\Delta D/m_l$}
            \end{tabular}
		    &
	        \hspace{-.9cm}
                \vspace{1.5cm}
    		\begin{tabular}{c}
    		        \includegraphics[width=0.45\textwidth]{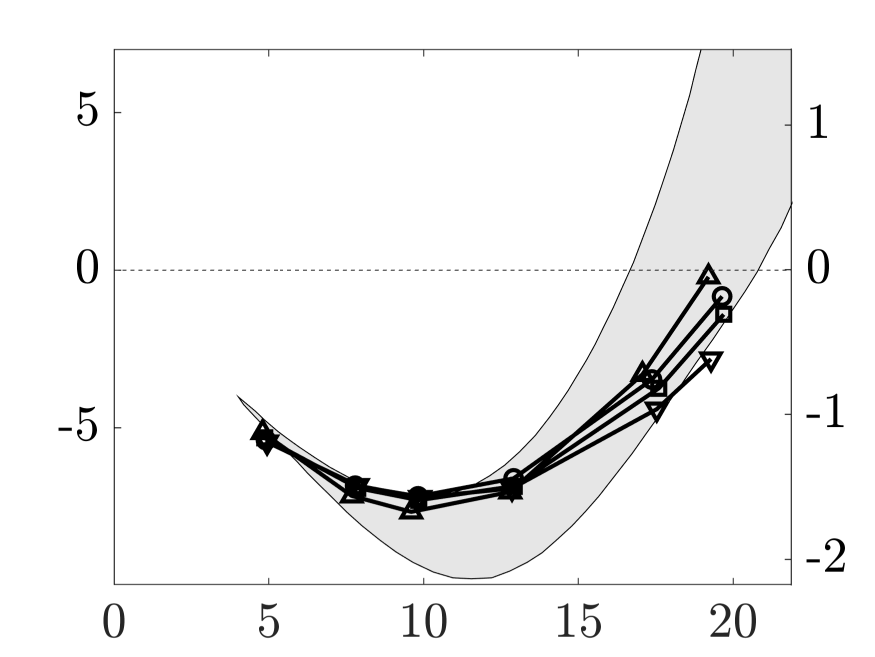}
    	        	\\
    		    \hspace{0.01cm}
    		     $l^+_g$
    	    \end{tabular}
                &
            \begin{tabular}{c}
                \vspace{0.5cm}
                \hspace{-0.75cm}
                \rotatebox{90}{$\Delta U$}
            \end{tabular}
            \end{tabular}
\vspace{-1.5cm}
	\caption{{Normalized drag reduction (left axis) and roughness function (right axis) due to the presence of scalloped riblets with $\alpha/s = 0.55$ mounted on the lower wall of a turbulent channel flow with $Re_\tau = 186 (\triangle)$, $547 (\bigcirc)$, $934 (\square)$, and $2003 (\bigtriangledown)$, as a function of $l^+_g$. The shaded region corresponds to the envelope of experimentally measured drag reduction levels from prior studies~\citep{becbruhaghoehop97,garjim11b}.}}
    \label{fig.DRRe}
\end{figure}

{The computational advantage afforded by the perturbation analysis of \S~\ref{sec.pert_stats} allows us to explore flow mechanisms over more complex geometries and at higher Reynolds numbers than would otherwise be feasible. Figure~\ref{fig.DRRe} shows the $m_l$-normalized drag reduction and roughness function for turbulent channel flow with $Re_\tau = 186$, $547$, $934$, and $2003$ over scalloped riblets with $\alpha/s = 0.55$; see table~\eqref{tab.ribconf} for the parameterization of the various cases considered in this figure. 
The degradation of drag reduction observed for riblets with larger than optimal $l_g^+$ has been associated with non-negligible inertial-flow mechanisms that lead to the eventual breakdown of the linear viscous regime~\citep{modendhutchu21}. Examples of such mechanisms that have been studied in the past include the lodging of near-wall vortices inside riblet grooves~\citep{chomoikim93, suzkas94, leelee01}, the generation of secondary flow~\citep{goltua98}, and the emergence of spanwise coherent rollers~\citep{garjim11b}. In this section, we exploit the computational efficiency of our framework to examine the emergence and prevalence of such flow mechanisms in high-Reynolds number channel flow over scalloped riblets.  Our analysis focuses on the occurrence of spanwise rollers associated with the K-H instability and the reorganization of turbulence near riblet-mounted surfaces--phenomena that have gained attention in recent numerical studies of large and sharp riblets~\citep{endmodgarhutchu21,endnewmodgarhutchu22}.}

\subsection{Kelvin-Helmholtz instability}
\label{sec.KH}

The emergence of long spanwise rollers induced by the K-H instability marks the onset of the breakdown of the linear viscous regime, {diminishing the effectiveness of riblets in reducing drag}~\citep{garjim11b}. This phenomenon enhances momentum exchange within and around the riblet grooves, leaving a distinct imprint on the wall-normal and shear stress energy spectra~\citep{garjim11b,gomgar19,shagar20}. In this subsection, {we first validate our model in generating a statistical flow signature that demonstrates the presence or absence of the K-H instability in accordance with prior literature. For this, we focus on turbulent channel flow with $Re_\tau = 547$ over triangular riblets of the same viscous spacing as~\cite{endmodgarhutchu21}. We then analyze model-based predictions of the spectral footprint of K-H rollers in flow over scalloped riblets.}

\subsubsection{Effect of Kelvin-Helmholtz instability on the wall-normal energy spectrum}
\label{sec.KH-vv}

\begin{figure}
        \begin{center}
        \begin{tabular}{cccccc}
        \hspace{-.6cm}
        \subfigure[]{\label{fig.E12vv2DT919o179Re547}}
        &&
        \hspace{-.7cm}
        \subfigure[]{\label{fig.E12vv2DT950o460Re547}}
        &&
        \hspace{-.6cm}
        \subfigure[]{\label{fig.E12vv2DT321o163Re547}}
        &
        \\[-.5cm]\hspace{-.3cm}
	\begin{tabular}{c}
        \vspace{.2cm}
        {\small \rotatebox{0}{$\lambda^+_z$}}
       \end{tabular}
       &\hspace{-.3cm}
	\begin{tabular}{c}
       \includegraphics[width=4cm]{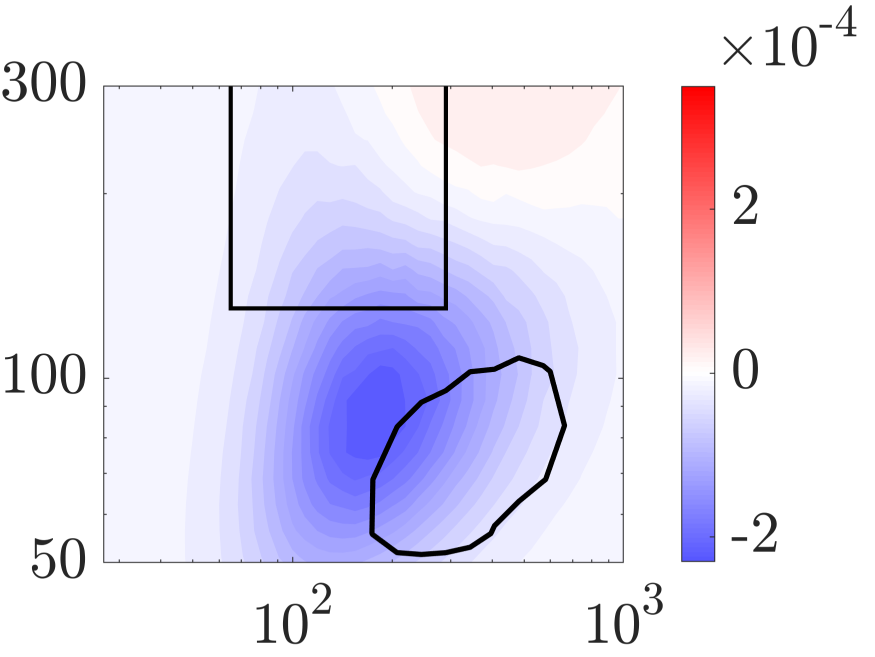}
       \\[-.1cm]
            \hspace{-.2cm}
            $\lambda^+_x$
       \end{tabular}
       &&\hspace{-.3cm}
    \begin{tabular}{c}
       \includegraphics[width=4cm]{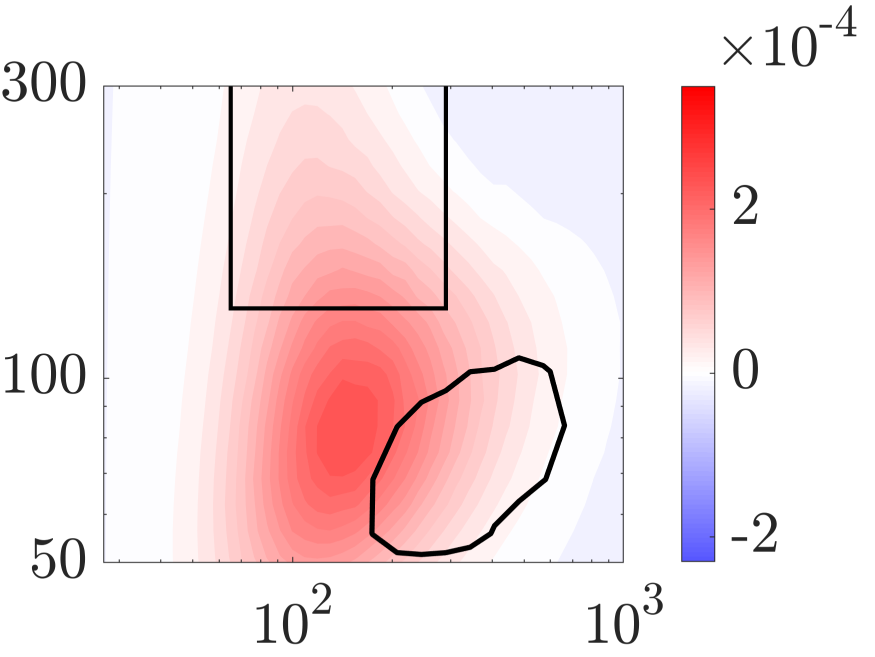}
       \\[-.1cm]
            \hspace{-.2cm}
            $\lambda^+_x$
       \end{tabular}
       &&\hspace{-.28cm}
    \begin{tabular}{c}
       \includegraphics[width=4cm]{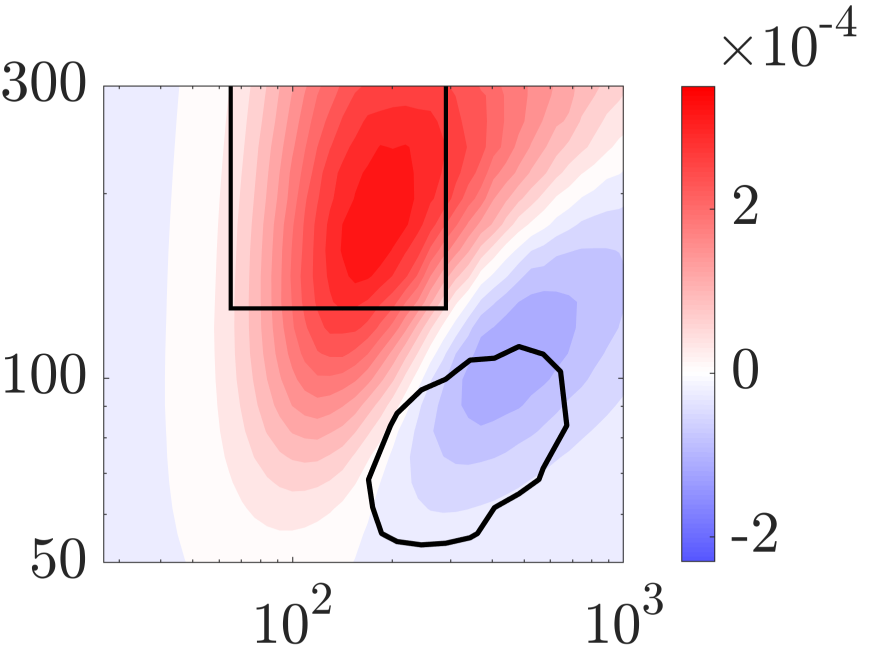}
       \\[-.1cm]
            \hspace{-.2cm}
            $\lambda^+_x$
       \end{tabular}
       \end{tabular}
       \end{center}
        \caption{{Premultiplied modifications to the wall-normal energy spectrum, approximated up to $\alpha^2$ as  $k_x \theta (\alpha\,E_{vv,1} + \alpha^2\,E_{vv,2})$, in turbulent channel flow with $Re_\tau=547$ one viscous unit above the crest of triangular riblets. (a) $\alpha/s=0.5$ and $l^+_g \approx 10$ ($\omega_z = 179$), (b) $\alpha/s=0.5$ and $l^+_g \approx 25$ ($\omega_z = 69$), and (c) $\alpha/s=1.87$ and $l^+_g \approx 20$ ($\omega_z = 163$). Black open boxes delimit the spectral window of K-H rollers according to~\cite{garjim11b} and black contour lines correspond to the $80\%$ contour level of the energy spectrum of smooth channel flow from the DNS of~\cite{deljimzanmos04}.}}
\label{fig.E12vvTvalid}
\end{figure}

{Figure~\ref{fig.E12vvTvalid} shows the premultiplied modifications to the wall-normal energy spectrum of turbulent channel flow with $Re_\tau = 547$ due to the presence of triangular riblets approximated up to a second order in $\alpha$, i.e., $k_x \theta (\alpha\,E_{vv,1} + \alpha^2\,E_{vv,2})$. The spectra are computed for a horizontal plane located 1 viscous unit above the riblets' crest. The geometric parameterization of riblets considered in this figure matches scenarios that have been previously associated with the absence or emergence of K-H instabilities that are known to give rise to spanwise rollers~\citep{endmodgarhutchu21}. Our model-based predictions show that riblets with $\alpha/s = 0.5$ and $l^+_g \approx 10$ ($\omega_z = 179$) suppress the energy spectrum over all horizontal wavenumbers. On the other hand, as shown in figures~\ref{fig.E12vvTvalid}(b,c), larger riblets amplify the wall-normal energy at this wall-normal location. However, for riblets with $\alpha/s = 0.5$, this amplification is quite weak in the spectral region associated with the K-H instability, i.e., $65 < \lambda^+_x < 290$ and $130 < \lambda^+_z$.  It is only for the sharper geometry with $\alpha/s=1.87$ that such modes are predominantly amplified (figure\ref{fig.E12vv2DT321o163Re547}). We therefore conclude that the spectral indicators of K-H instability onset are reserved to sufficiently large and sharp geometries, which is consistent with the observations of~\cite{endmodgarhutchu21}. We note that since the case of $\alpha/s=0.5$ and $l_g^+\approx 25$ corresponds to a drag-increasing riblet configuration~\citep{endmodgarhutchu21}, the weak footprint of K-H modes is indicative of an alternative destructive mechanism, e.g., dispersive stresses induced by secondary flows~\citep{goltua98,modendhutchu21}, which we, however, do not account for in this study.}

\begin{figure}
        \begin{center}
        \begin{tabular}{cccccc}
        \hspace{-.6cm}
        \subfigure[]{\label{fig.E12vvas55O115ypRe547Tri}}
        &&
        \hspace{-.7cm}
        \subfigure[]{\label{fig.E12vvas55O115ypRe547_3}}
        &&
        \hspace{-.6cm}
        \subfigure[]{\label{fig.ScallopedTriangular}}
        &
        \\[-.5cm]\hspace{-.3cm}
	\begin{tabular}{c}
        \vspace{.2cm}
        {\small \rotatebox{0}{$\lambda^+_z$}}
       \end{tabular}
       &\hspace{-.3cm}
	\begin{tabular}{c}
       \includegraphics[width=4cm]{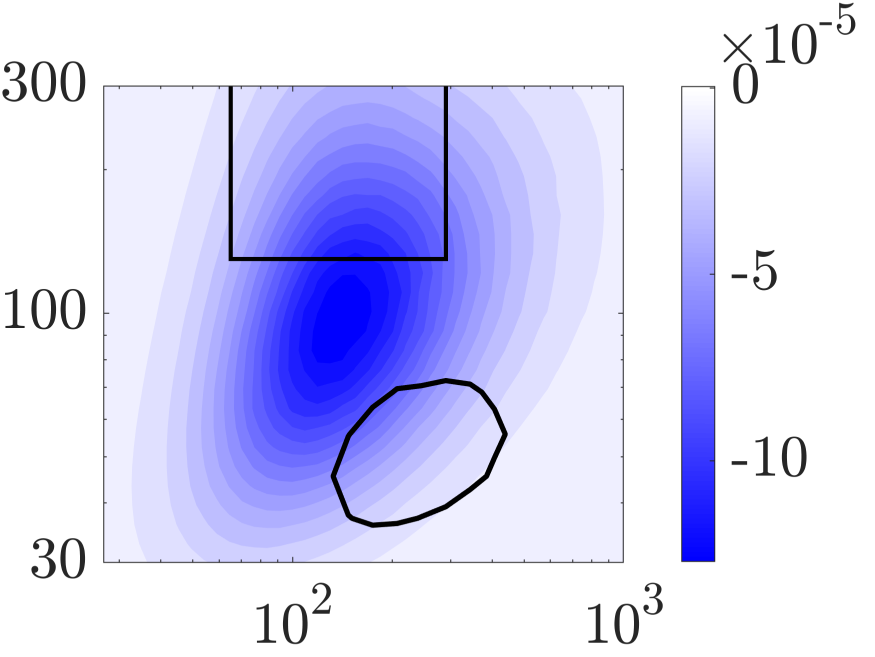}
       \\[-.1cm]
            \hspace{-.2cm}
            $\lambda^+_x$
       \end{tabular}
       &&\hspace{-.3cm}
    \begin{tabular}{c}
       \includegraphics[width=4cm]{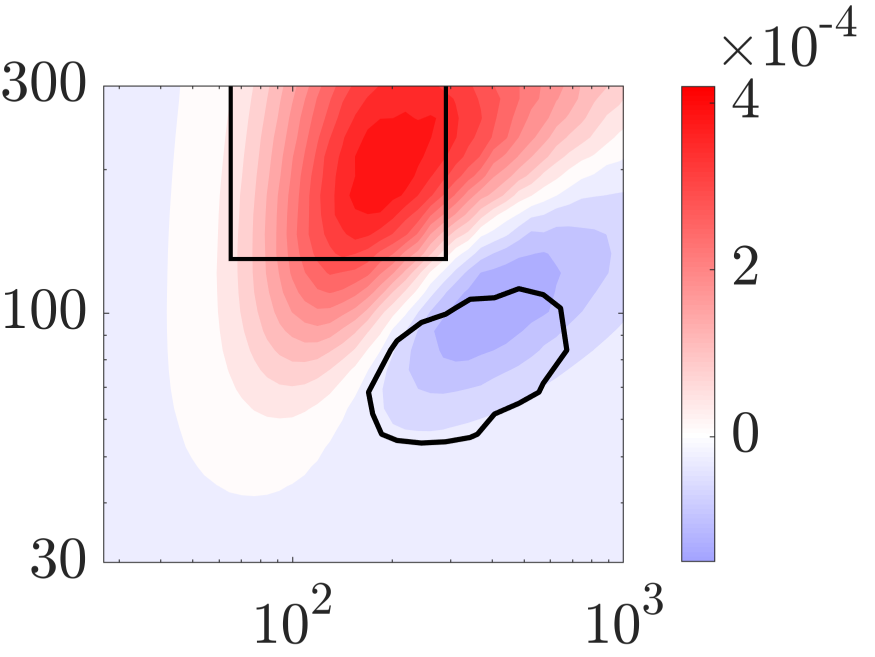}
       \\[-.1cm]
            \hspace{-.2cm}
            $\lambda^+_x$
       \end{tabular}
       &&\hspace{-.28cm}
    \begin{tabular}{c}
       \includegraphics[width=3.8cm]{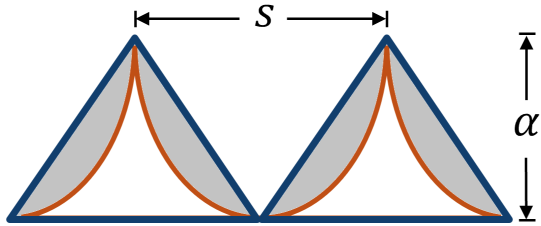}
       \end{tabular}
       \end{tabular}
       \end{center}
        \caption{{Premultiplied modifications to the wall-normal energy spectrum, approximated up to $\alpha^2$ as $k_x \theta (\alpha\,E_{vv,1} + \alpha^2\,E_{vv,2})$, in turbulent channel flow with $Re_\tau=547$ one viscous unit above the crest of (a) triangular ($l^+_g \approx 15.6$), and (b) scalloped ($l^+_g \approx 19.6$) riblets with $\alpha/s=0.55$ and $\omega_z = 115$.  Black open boxes delimit the spectral window of K-H rollers according to~\cite{garjim11b} and black contour lines correspond to the $80\%$ contour level of the energy spectrum of smooth channel flow from the DNS of~\cite{deljimzanmos04}. (c) A schematic of the two riblet geometries with the shaded regions highlighting the difference in groove area.}}
\label{fig.E12vvmixed}
\end{figure}

{An alternative approach to assess the role of sharpness in exciting the K–H modes is to compare the effects of triangular and scalloped riblets with identical height-to-spacing ratios and spanwise frequencies. Figure~\ref{fig.E12vvmixed} presents the premultiplied modifications to the wall-normal energy spectrum for flows over triangular and scalloped riblets with $\alpha/s=0.55$ and $\omega_z=115$. While the scalloped riblets show enhanced energy amplification in the spectral region associated with the K–H instability, the shaded surface addition illustrated in figure~\ref{fig.ScallopedTriangular} inhibits a similar amplification in the case of triangular riblets.}

\begin{figure}
        \begin{center}
        \begin{tabular}{cccccc}
        \hspace{-.6cm}
        \subfigure[]{\label{fig.E1vvas55O460ypRe547}}
        &&
        \hspace{-.7cm}
        \subfigure[]{\label{fig.E1vvas55O175ypRe547}}
        &&
        \hspace{-.7cm}
        \subfigure[]{\label{fig.E1vvas55O115ypRe547}}
        &
        \\[-.5cm]\hspace{-.3cm}
	\begin{tabular}{c}
        \vspace{-.2cm}
        {\small \rotatebox{0}{$\lambda^+_z$}}
       \end{tabular}
       &\hspace{-.3cm}
	\begin{tabular}{c}
        $l^+_g \approx 5$
        \\
       \includegraphics[width=4cm]{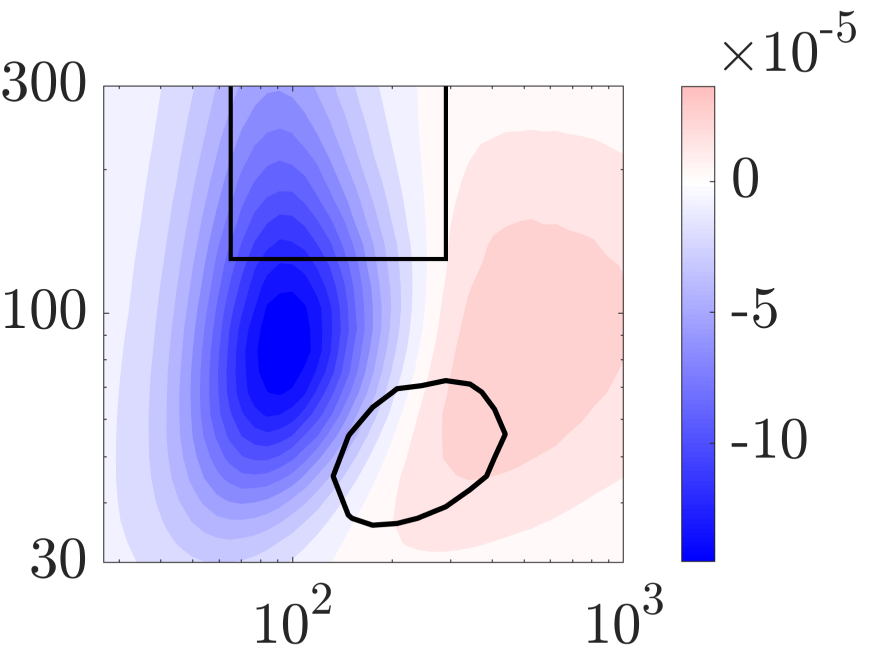}
       \end{tabular}
       &&\hspace{-.3cm}
    \begin{tabular}{c}
    $l^+_g \approx 10$
    \\
       \includegraphics[width=4cm]{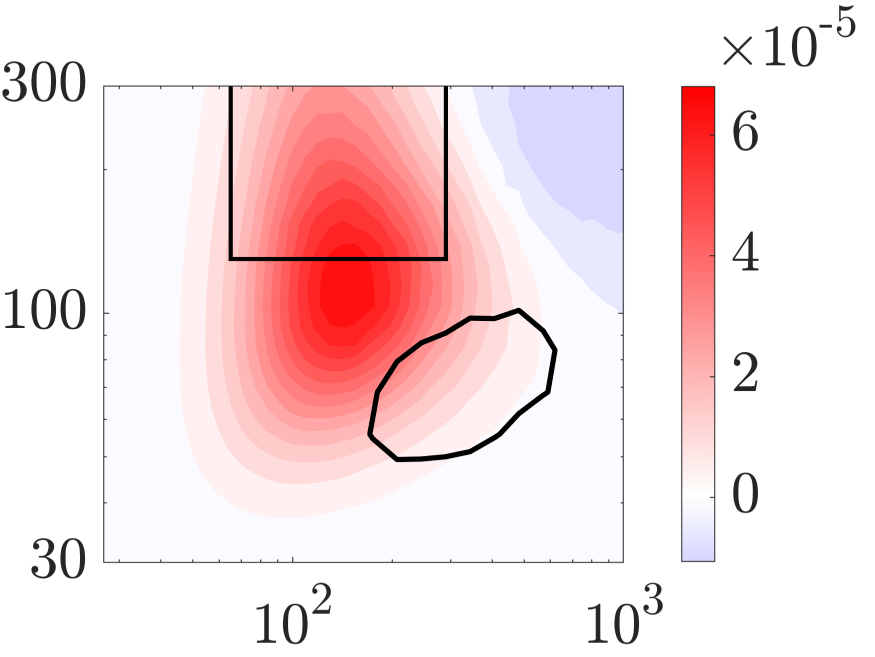}
       \end{tabular}
       &&\hspace{-.5cm}
    \begin{tabular}{c}
    $l^+_g \approx 20$
    \\
       \includegraphics[width=4cm]{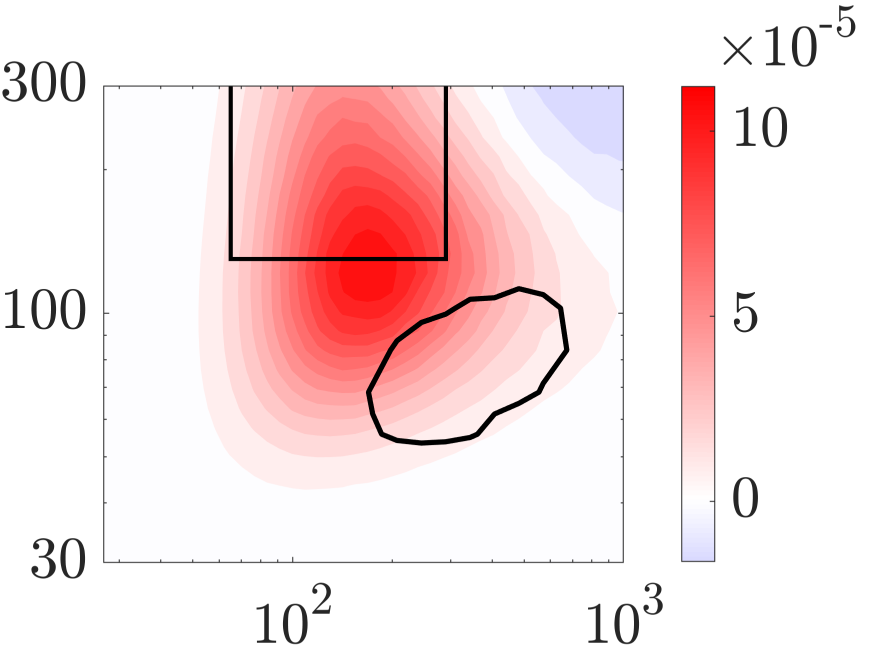}
       \end{tabular}
       \\[-0.1cm]
        \hspace{-.6cm}
        \subfigure[]{\label{fig.E2vvas55O460ypRe547}}
        &&
        \hspace{-.7cm}
        \subfigure[]{\label{fig.E2vvas55O175ypRe547}}
        &&
        \hspace{-.75cm}
        \subfigure[]{\label{fig.E2vvas55O115ypRe547}}
        &
        \\[-.5cm]\hspace{-.3cm}
	\begin{tabular}{c}
        \vspace{-.2cm}
        {\small \rotatebox{0}{$\lambda^+_z$}}
       \end{tabular}
       &\hspace{-.3cm}
	\begin{tabular}{c}
       \includegraphics[width=4cm]{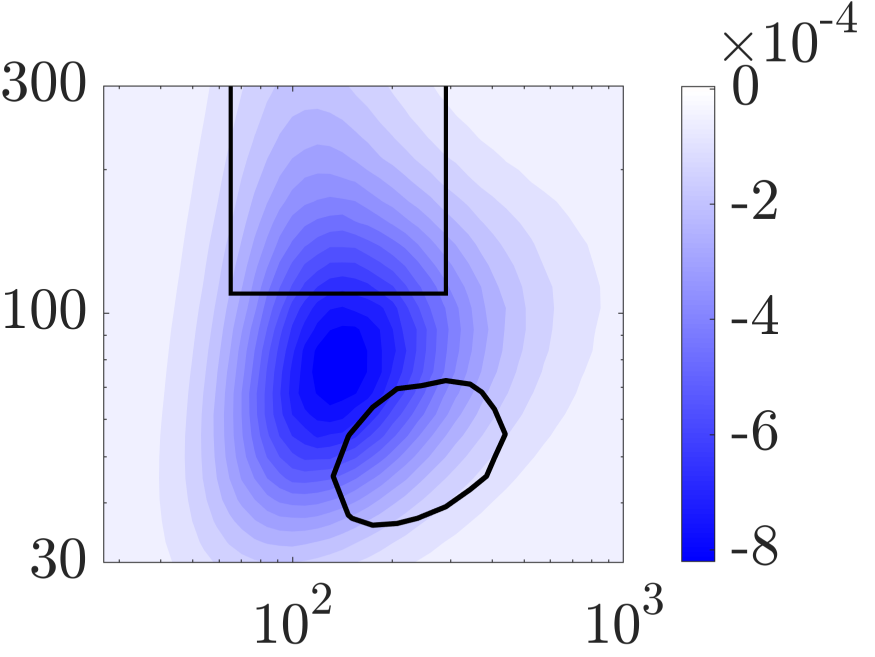}
       \end{tabular}
       &&\hspace{-.3cm}
    \begin{tabular}{c}
       \includegraphics[width=4cm]{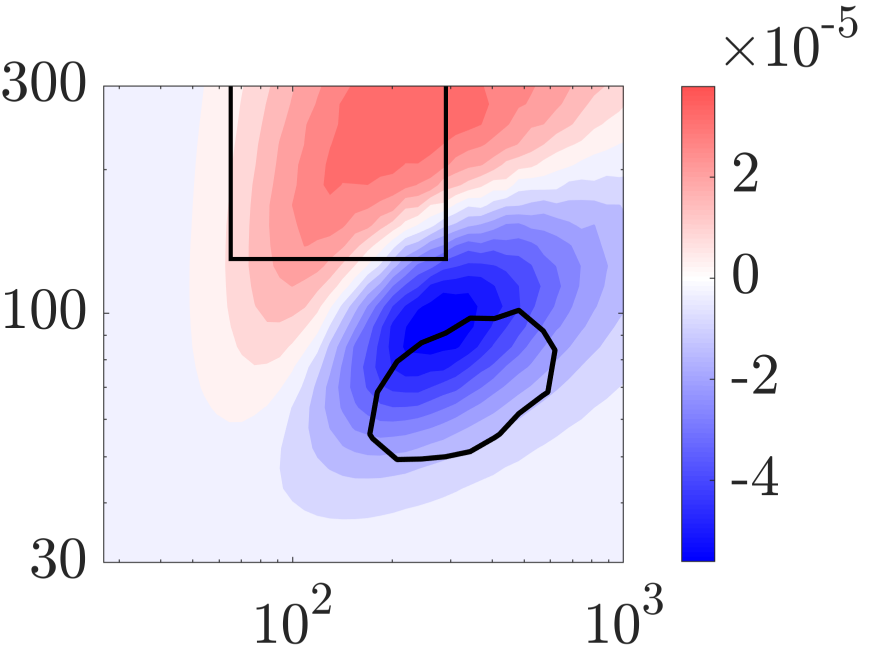}
       \end{tabular}
       &&\hspace{-.4cm}
    \begin{tabular}{c}
       \includegraphics[width=4cm]{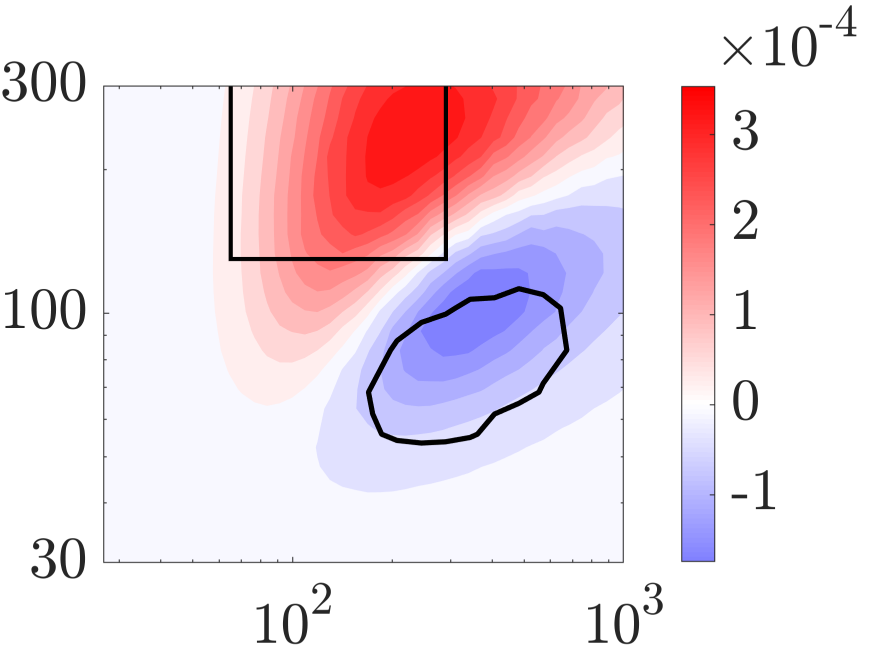}
       \end{tabular}
       \\[-0.1cm]
       \hspace{-.6cm}
        \subfigure[]{\label{fig.E12vvas55O460ypRe547}}
        &&
        \hspace{-.6cm}
        \subfigure[]{\label{fig.E12vvas55O175ypRe547}}
        &&
        \hspace{-.75cm}
        \subfigure[]{\label{fig.E12vvas55O115ypRe547}}
        &
        \\[-.5cm]\hspace{-.3cm}
	\begin{tabular}{c}
        \vspace{.2cm}
        {\small \rotatebox{0}{$\lambda^+_z$}}
       \end{tabular}
       &\hspace{-.3cm}
	\begin{tabular}{c}
       \includegraphics[width=4cm]{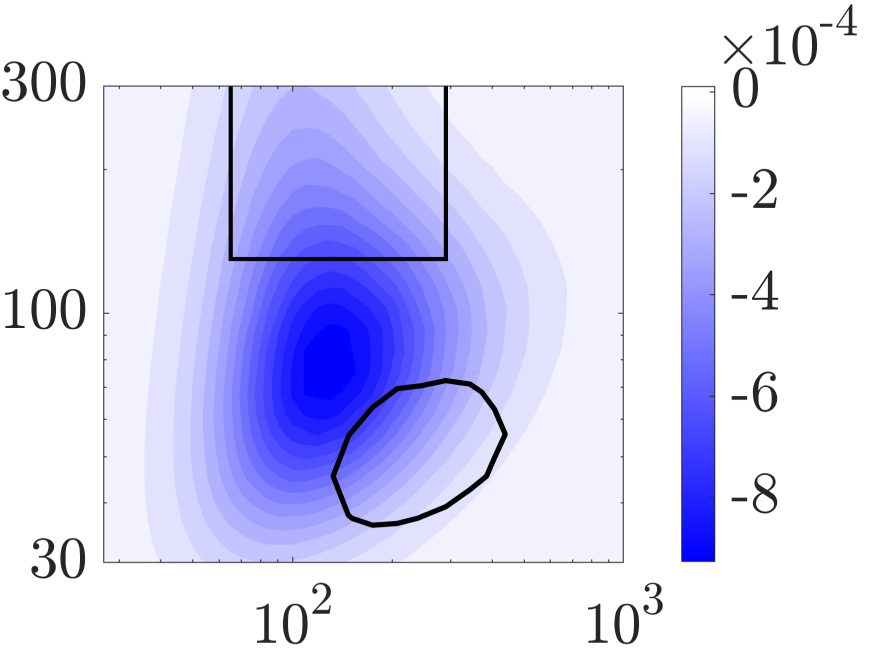}
        \\[-.2cm]
            \hspace{-.2cm}
            $\lambda^+_x$
       \end{tabular}
       &&\hspace{-.25cm}
    \begin{tabular}{c}
       \includegraphics[width=4cm]{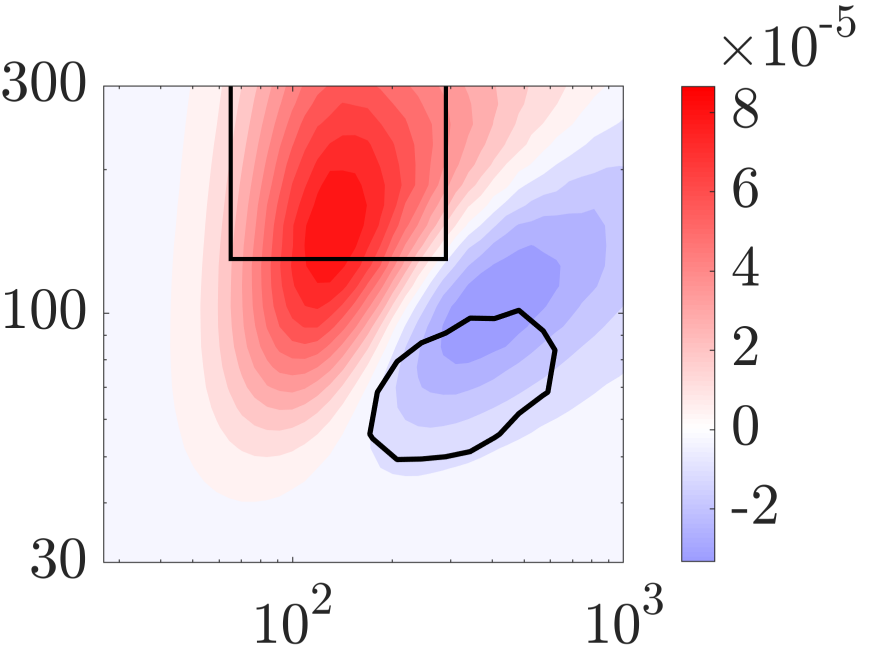}
       \\[-.2cm]
            \hspace{-.2cm}
            $\lambda^+_x$
       \end{tabular}
       &&\hspace{-.4cm}
    \begin{tabular}{c}
       \includegraphics[width=4cm]{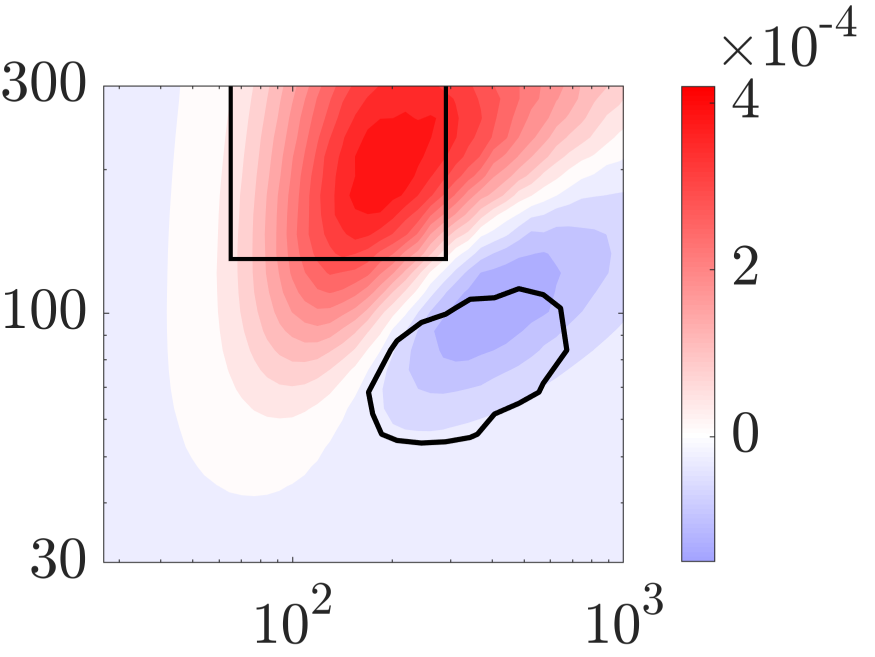}
       \\[-.2cm]
            \hspace{-.2cm}
            $\lambda^+_x$
       \end{tabular}
       \end{tabular}
       \end{center}
        \caption{{Premultiplied modifications to the wall-normal energy spectrum of turbulent channel flow with $Re_\tau=547$ one viscous unit above the crest of scalloped riblets with $\alpha/s = 0.55$ and $l^+_g \approx 5$ $(\omega_z = 460)$ (left column), $l^+_g \approx 10$ $(\omega_z = 230)$ (middle column), and $l^+_g \approx 20$ $(\omega_z = 115)$ (right column). (a-c) $\alpha\,k_x \theta E_{vv,1}$; (d-f) $\alpha^2\,k_x \theta E_{vv,2}$; and (g-i) $k_x \theta (\alpha\,E_{vv,1} + \alpha^2\, E_{vv,2})$. Black open boxes delimit the spectral window of K-H rollers according to~\cite{garjim11b} and black contour lines correspond to the $80\%$ contour level of the energy spectrum of smooth channel flow from the DNS of~\cite{deljimzanmos04}.}} 
        \label{fig.E2vvbeta75}
\end{figure}

\begin{figure}
        \begin{center}
        \begin{tabular}{cccccc}
        \hspace{-.6cm}
        \subfigure[]{\label{fig.E2vvas65O115ypRe547}}
        &&
        \hspace{-.78cm}
        \subfigure[]{\label{fig.E2vvas87O115ypRe547}}
        &&
        \hspace{-.78cm}
        \subfigure[]{\label{fig.E2vvas12O115ypRe547}}
        &
        \\[-.5cm]\hspace{-.3cm}
	\begin{tabular}{c}
        \vspace{.2cm}
        {\small \rotatebox{0}{$\lambda^+_z$}}
       \end{tabular}
       &\hspace{-.4cm}
	\begin{tabular}{c}
       \includegraphics[width=4cm]{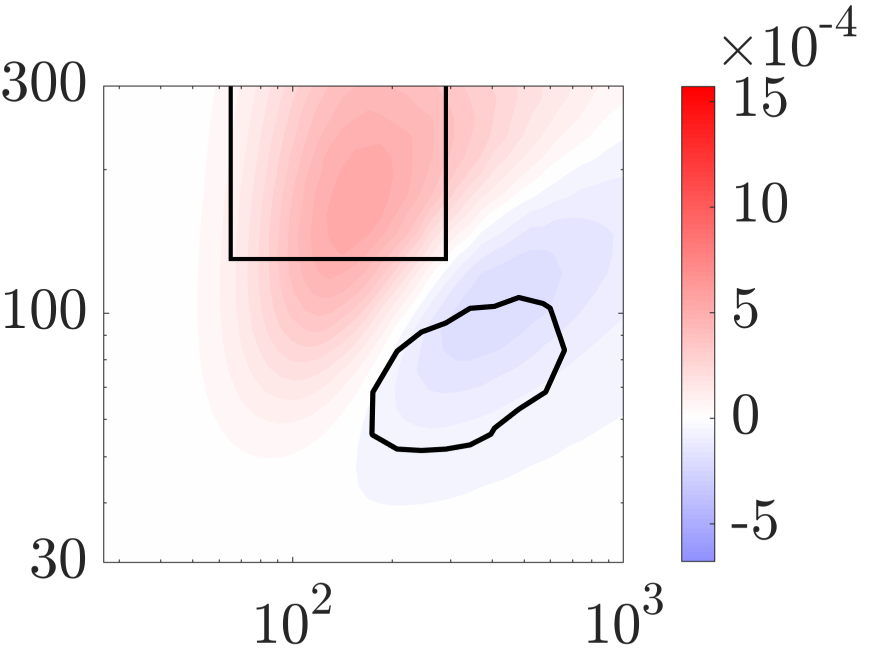}
       \\[-.2cm]
            \hspace{-.2cm}
            $\lambda^+_x$
       \end{tabular}
       &&\hspace{-.35cm}
    \begin{tabular}{c}
       \includegraphics[width=4cm]{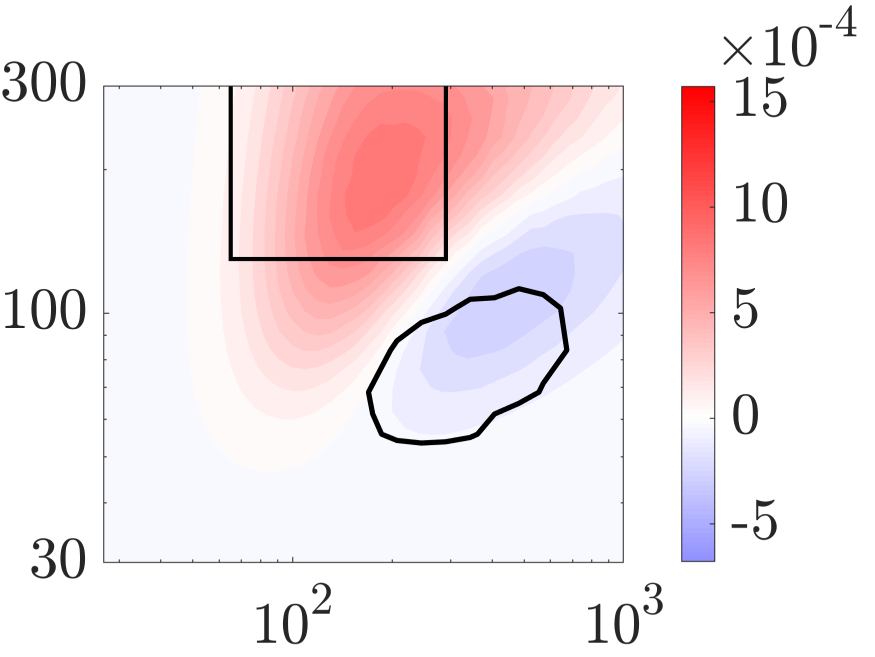}
       \\[-.2cm]
            \hspace{-.2cm}
            $\lambda^+_x$
       \end{tabular}
       &&\hspace{-.37cm}
    \begin{tabular}{c}
       \includegraphics[width=4cm]{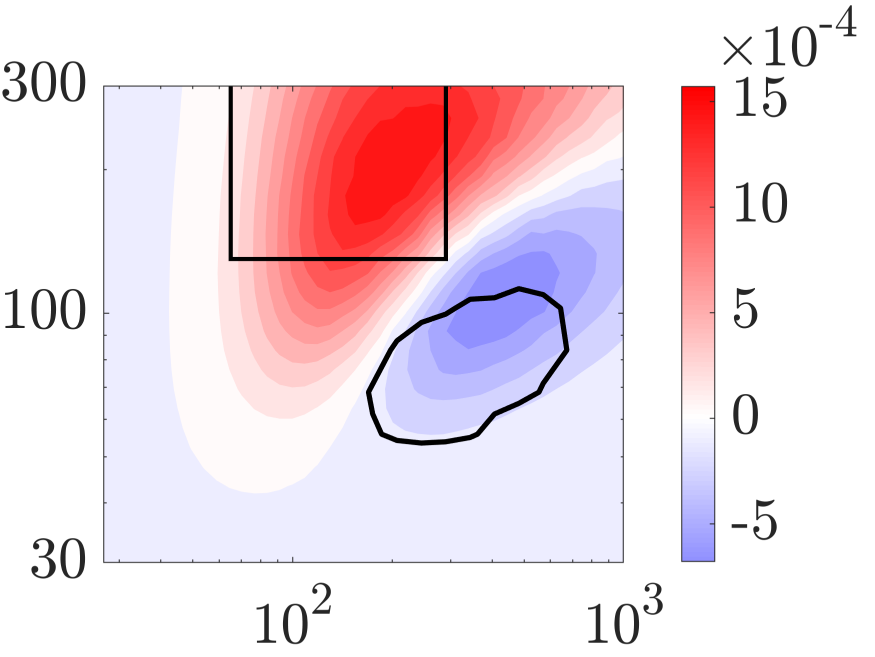}
       \\[-.2cm]
            \hspace{-.2cm}
            $\lambda^+_x$
       \end{tabular}
       \end{tabular}
       \end{center}
        \caption{{Premultiplied modifications to the wall-normal energy spectrum,  $k_x \theta (\alpha\,E_{vv,1} + \alpha^2\,E_{vv,2})$, of turbulent channel flow with $Re_\tau=547$ one viscous unit above the crest of scalloped riblets with the same viscous spacing but different viscous height. (a) $\alpha/s=0.65$ and $l^+_g \approx 22$ ($\omega_z = 115$); (b) $\alpha/s=0.87$ and $l^+_g \approx 25$ ($\omega_z = 115$); and (c) $\alpha/s=1.2$ and $l^+_g \approx 29$ ($\omega_z = 115$). Black open boxes delimit the spectral window of K-H rollers~\citep{garjim11b} and black contour lines correspond to the $80\%$ contour level of the energy spectrum of smooth channel flow from the DNS of~\cite{deljimzanmos04}.}}
    \label{fig.E2vvas}
\end{figure}

{Figure~\ref{fig.E2vvbeta75} shows spectral modifications at the same wall distance due to scalloped riblets of different size. To investigate the effect of riblet size on the emergence of K-H instability, the cases shown in this figure maintain a constant $\alpha/s = 0.55$ while varying $l_g^+$. Similar to triangular riblets (figure~\ref{fig.E12vvTvalid}), as the size of scalloped riblets increases, the modification to} the wall-normal energy spectrum transitions from damping to progressively amplifying, within a spectral range corresponding to the typical wavelengths of spanwise rollers.
This trend, which is uniformly observed at both $\alpha^1$ and $\alpha^2$ levels of our perturbation analysis, is in agreement with the observations of prior numerical studies~\citep{garjim11b,endmodgarhutchu21}. 
Interestingly, we also observe that larger riblets suppress the energy of the wall-normal velocity at the most energetic wavelengths of smooth channel flow; see regions delimited by {black contour} lines in {figures~\ref{fig.E12vvas55O175ypRe547} and~\ref{fig.E12vvas55O115ypRe547}.}
To investigate how things would change for taller riblets, figure~\ref{fig.E2vvas} extends the cases studied in figure~\ref{fig.E2vvbeta75} to $l^+_g \approx 22$, $25$, and $29$ {by solely increasing the height of riblets $\alpha$ ($\omega_z$ is kept constant at $115$).} The premultiplied modifications to the wall-normal energy spectra again show amplification for wavelengths delimited by the solid black lines. {This amplification is more pronounced for taller riblets, with the maximum amplification happening at larger streamwise wavelengths (from $160$ to $220$) and persisting} over the longest spanwise wavelengths ($\lambda^+_z \gtrsim 100$) corresponding to wide rollers excited by the K-H instability~\citep{garjim11b}. 
{Finally, despite a slight shift in amplification toward larger streamwise wavelengths, the spectral modifications shown in figure~\ref{fig.E2vvas} exhibit a consistent pattern across the horizontal wavenumber space, suggesting a potential geometric scaling of this quantity for riblets larger than the optimal size.}

\begin{figure}
        \begin{center}
        \begin{tabular}{cccccccc}
        \hspace{-.6cm}
        \subfigure[]{\label{fig.E2vvas55O45ypRe186}}
        &&
        \hspace{-.8cm}
        \subfigure[]{\label{fig.E12vvas55O115ypRe547_2}}
        &&
        \hspace{-.7cm}
        \subfigure[]{\label{fig.E2vvas55O200ypRe934}}
        &&
        \hspace{-.7cm}
        \subfigure[]{\label{fig.E2vvas55O420ypRe2003}}
        &
        \\[-.5cm]\hspace{-.3cm}
	\begin{tabular}{c}
        \vspace{.3cm}
        {\small \rotatebox{0}{$\lambda^+_z$}}
       \end{tabular}
       &\hspace{-.45cm}
	\begin{tabular}{c}
       \includegraphics[width=2.9cm]{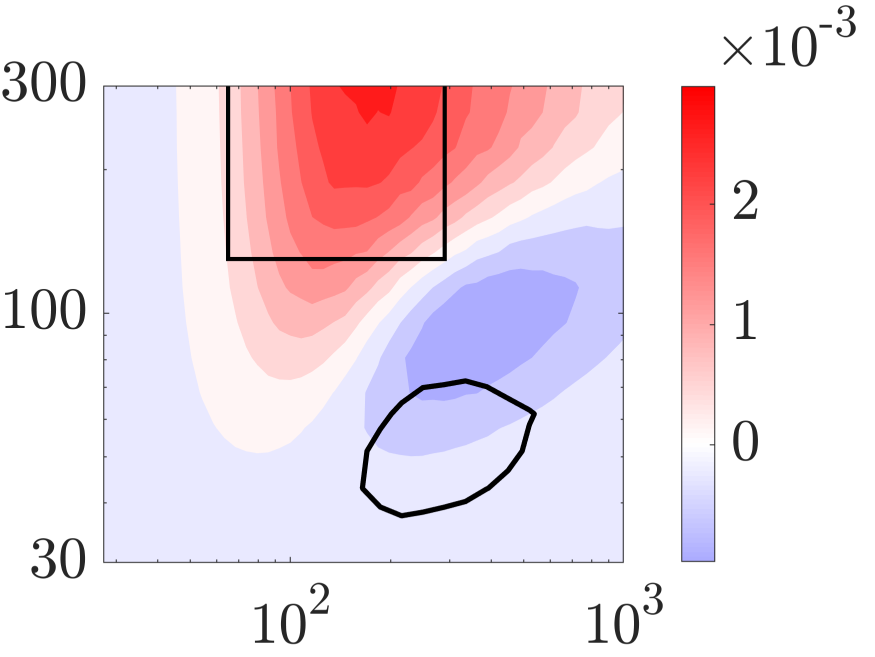}
        \\[-.1cm]
            \hspace{-.1cm}
            {\small $\lambda^+_x$}
       \end{tabular}
       &&\hspace{-.35cm}
    \begin{tabular}{c}
       \includegraphics[width=2.9cm]{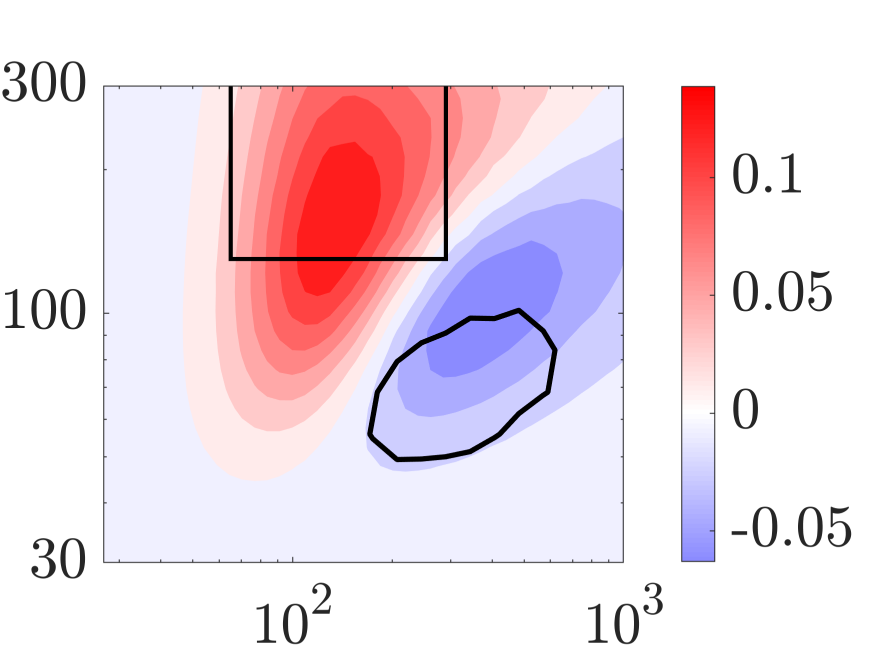}
       \\[-.1cm]
            \hspace{-.1cm}
            {\small $\lambda^+_x$}
       \end{tabular}
       &&\hspace{-.35cm}
    \begin{tabular}{c}
       \includegraphics[width=2.9cm]{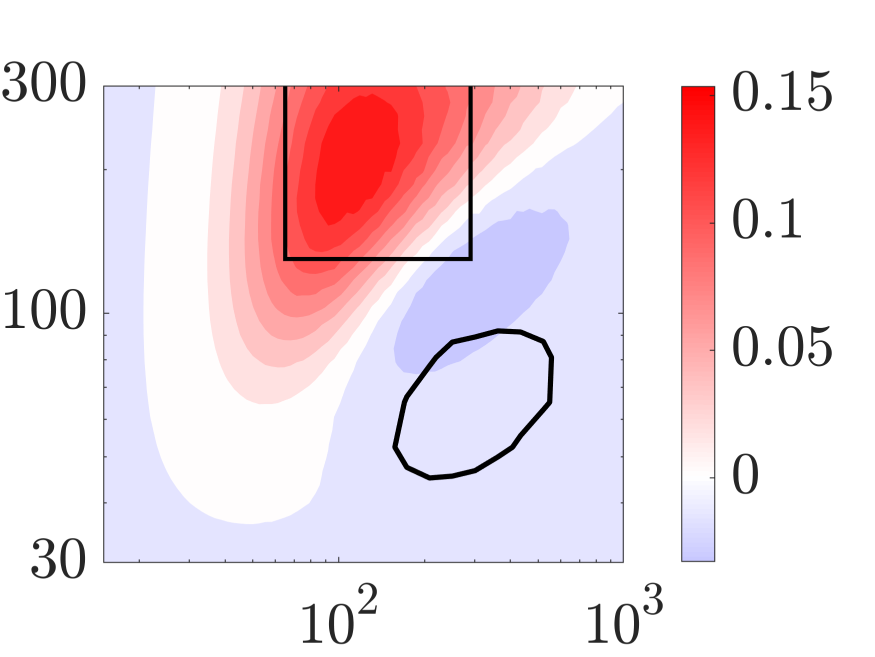}
       \\[-.1cm]
            \hspace{-.1cm}
            {\small $\lambda^+_x$}
       \end{tabular}
       &&\hspace{-.3cm}
    \begin{tabular}{c}
       \includegraphics[width=2.9cm]{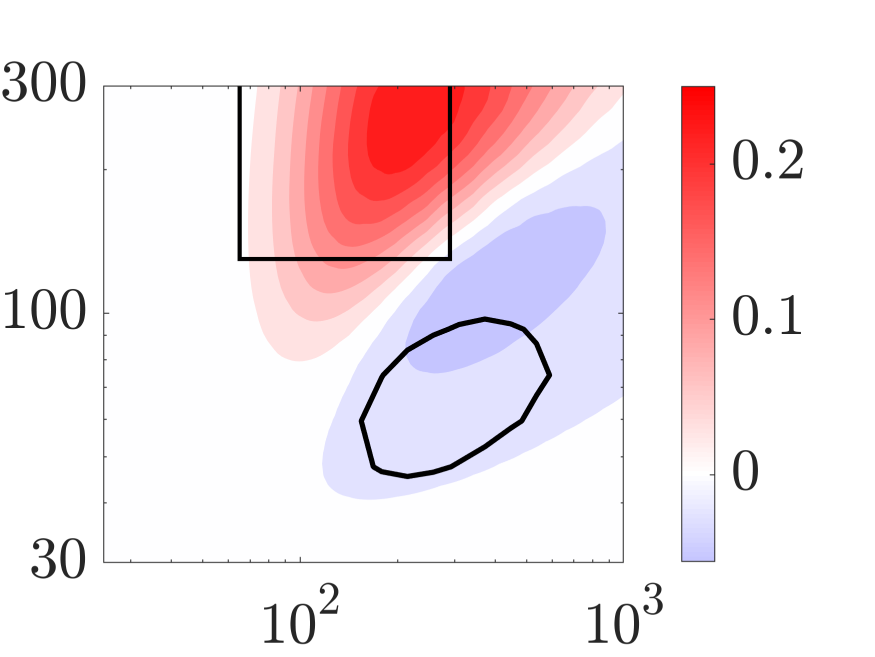}
       \\[-.1cm]
            \hspace{-.1cm}
            {\small $\lambda^+_x$}
       \end{tabular}
       \end{tabular}
       \end{center}
        \caption{{Premultiplied modifications to the wall-normal energy spectrum $k_x \theta (\alpha\,E_{vv,1} + \alpha^2\,E_{vv,2})$ for turbulent channel flow with (a) $Re_\tau=186$, (b) $Re_\tau=547$, (c) $Re_\tau=934$, and (d) $Re_\tau=2003$ one viscous unit above the crest of scalloped riblets with $\alpha/s=0.55$ and $l^+_g \approx 20$, which corresponds to spatial frequencies $\omega_z=40$, $115$, $200$, and $420$, respectively. Black open boxes delimit the spectral window of K-H rollers and black contour lines correspond to the $80\%$ contour level in the energy spectrum of smooth channel flow from the DNS of~\cite{deljim03,deljimzanmos04,hoyjim06}.}}
    \label{fig.E2vvRe}
\end{figure}

Figure~\ref{fig.E2vvRe} shows the effect of lower-wall scalloped riblets with $\alpha/s = 0.55$ and $l^+_g \approx 20$ on the wall-normal energy spectrum at different Reynolds numbers. {To better understand the contribution by riblets, the spectra have been normalized by the maximum wall-normal energy of smooth channel flow at the same wall-parallel plane.
It is evident from this figure that the energy amplification in the spectral window associated with the K-H instability is strengthened for higher Reynolds numbers. This trend is suggestive of the more pronounced footprint of K-H rollers and their earlier appearance over riblets with smaller $l^+_g$ at higher Reynolds numbers.}

\begin{figure}
        \begin{center}
        \begin{tabular}{cccc}
        \\[-.5cm]\hspace{-.3cm}
	\begin{tabular}{c}
        \vspace{.2cm}
        {\small \rotatebox{90}{${vv}_{\mathrm{KH}}$}}
       \end{tabular}
       &\hspace{-.3cm}
	\begin{tabular}{c}
       \includegraphics[width=0.45\textwidth]{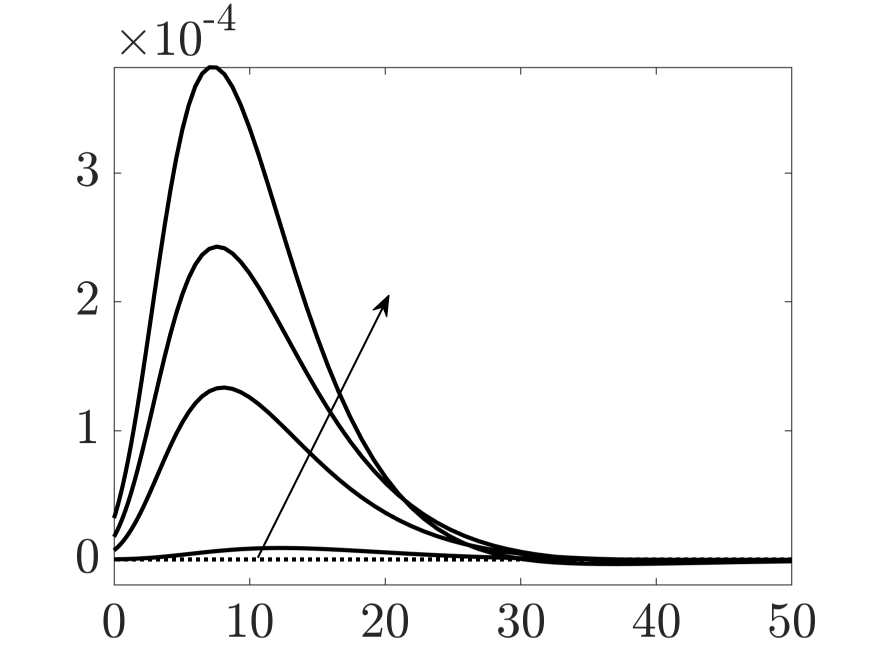}
        \\[-.1cm]
            \hspace{.2cm}
            $\tilde{y}^+$
       \end{tabular}
       \end{tabular}
       \end{center}
        \caption{{Modifications to the wall-normal stress ${vv}_\mathrm{KH}$ resulting from the K-H modes in channel flow with $Re_\tau = 547$ over scalloped riblets with $\alpha/s=0.55$ and $l^+_g \in (10,20)$ ($\omega_z \in [115,230]$). The black dotted line corresponds to the profile computed using equation~\eqref{eq.vv-KH} in the absence of riblets and $l^+_g$ increases in the direction of the arrow.}}
        \label{fig.EvvProfhs055Re547}
\end{figure}

For channel flow with $Re_\tau = 547$ over optimal to large-size scalloped riblets with $\alpha/s=0.55$, figure~\ref{fig.EvvProfhs055Re547} shows the wall-normal profiles of the modifications to the wall-normal stress given by
\begin{align}
\label{eq.vv-KH}
        {vv}_\mathrm{KH}
        \,=\,
        \ds{\int_{130}^{1000}\int_{65}^{290} \left(\alpha E_{vv,1} \,+\, \alpha^2 E_{vv,2}\right) \,\mrd \lambda^+_x \,\mrd \lambda^+_z},
\end{align}
{up to a second order in riblet height $\alpha$.}
Integration over $65<\lambda^+_x<290$ and $130<\lambda^+_z<1000$ ensures a separation of scales from the near-wall cycle at larger $\lambda^+_x$ and the association with the K-H instability and is achieved by integrating over the $k_x$ and $\theta_n$ values {corresponding to} the targeted wavelengths. 
Modifications to the wall-normal stress profile {peak at $\tilde{y}^+ \approx 9$.} Moreover, we observe the peak values to increase for larger riblets. The emergence of these peaks, driven by the K-H instability, indicates elevated turbulence levels on riblet-mounted surfaces compared to smooth walls, and is in agreement with the findings of~\cite{garjim11b}.

\begin{figure}
        \begin{center}
        \begin{tabular}{cccc}
        \hspace{-.8cm}
        \subfigure[]{\label{fig.sliceVhs055O45Re186all}}
        &&
        \hspace{-1.2cm}
        \subfigure[]{\label{fig.sliceVhs055O45Re547XYall}}
        &
        \\[-.3cm]
        \hspace{-.5cm}
	\begin{tabular}{c}
        \vspace{3.7cm}
        {\small \rotatebox{90}{$\tilde{y}^+$}}
        \\
        \vspace{2.3cm}
        {\small \rotatebox{90}{$\tilde{z}^+$}}
       \end{tabular}
       &\hspace{-.55cm}
	\begin{tabular}{c}
       \includegraphics[width=0.46\textwidth]{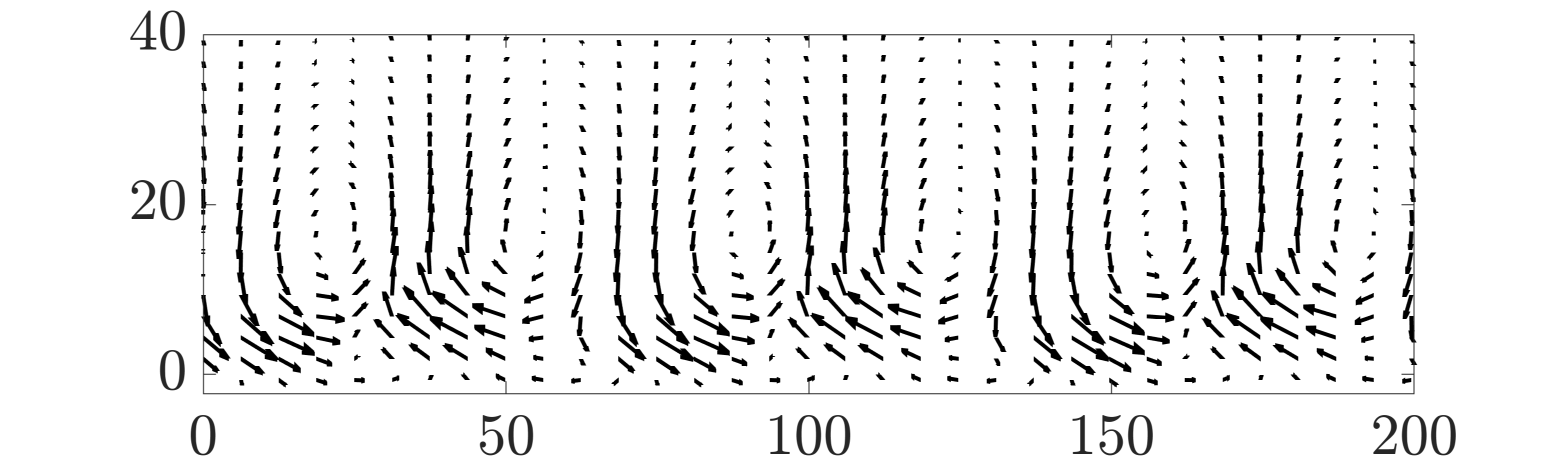}
       \\
       \includegraphics[width=0.47\textwidth]{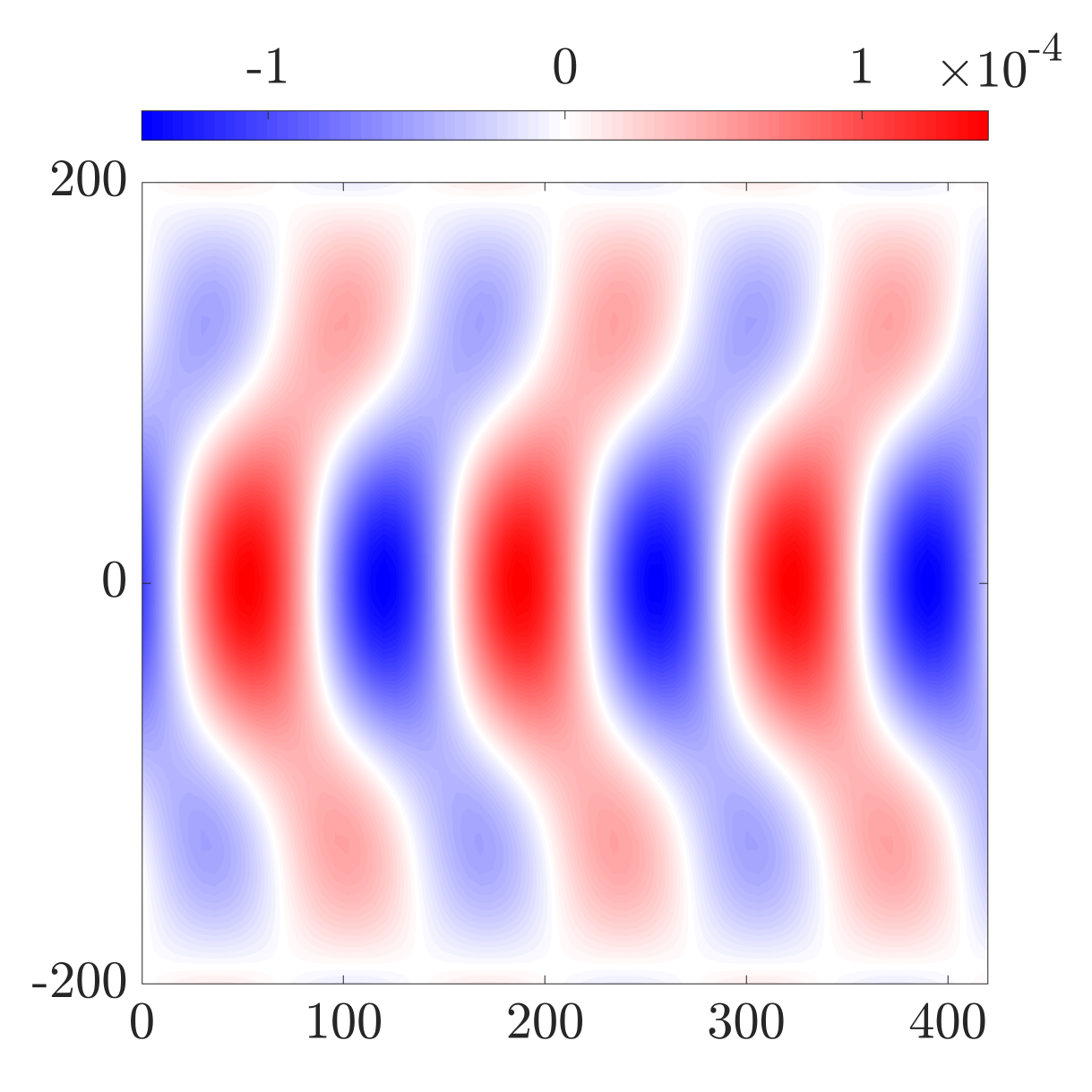}
       \end{tabular}
       &\hspace{.2cm}
       \begin{tabular}{c}
        \vspace{.25cm}
       \end{tabular}
       &\hspace{-.8cm}
    \begin{tabular}{c}
       \includegraphics[width=0.46\textwidth]{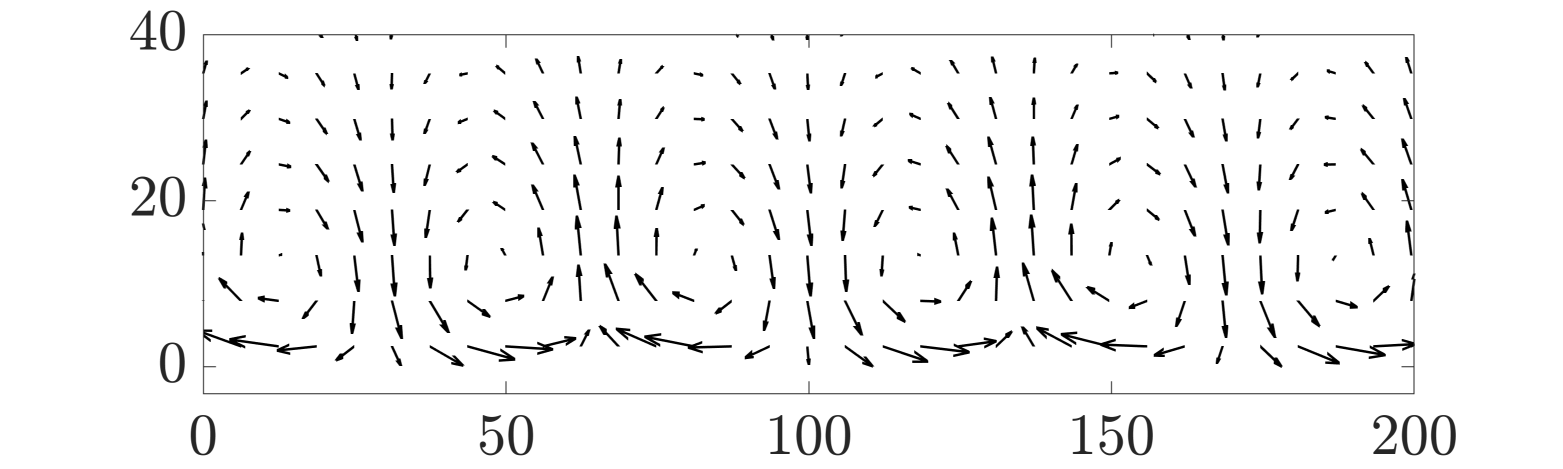}
       \\
       \includegraphics[width=0.47\textwidth]{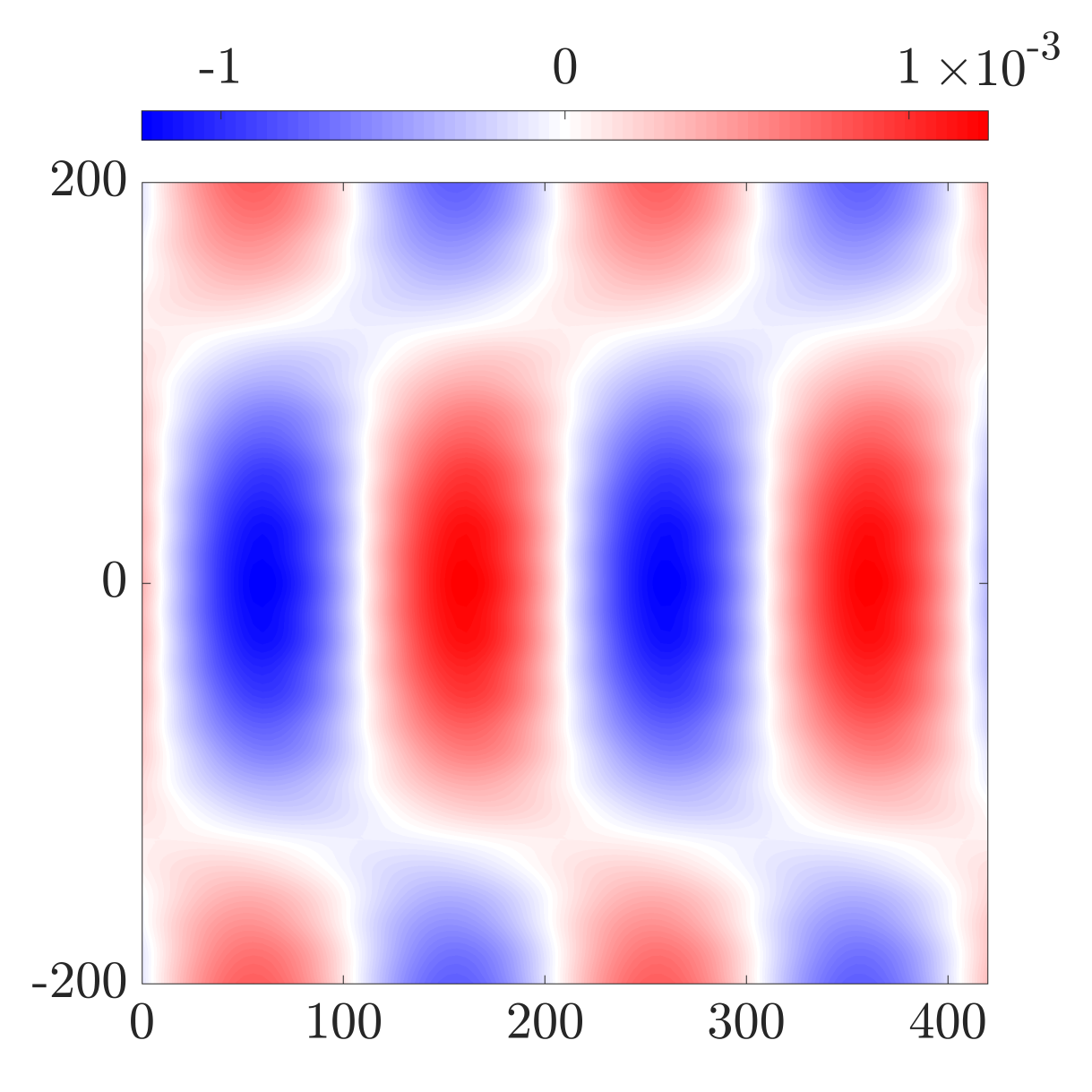}
       \end{tabular}
       \\
        \hspace{-.8cm}
        \subfigure[]{\label{fig.sliceVhs055O45Re934all}}
        &&
        \hspace{-1.2cm}
        \subfigure[]{\label{fig.sliceVhs055O45Re2003all}}
        &
        \\[-.3cm]
        \hspace{-.5cm}
	\begin{tabular}{c}
        \vspace{3.9cm}
        {\small \rotatebox{90}{$\tilde{y}^+$}}
        \\
        \vspace{2.55cm}
        {\small \rotatebox{90}{$\tilde{z}^+$}}
       \end{tabular}
       &\hspace{-.55cm}
	\begin{tabular}{c}
       \includegraphics[width=0.46\textwidth]{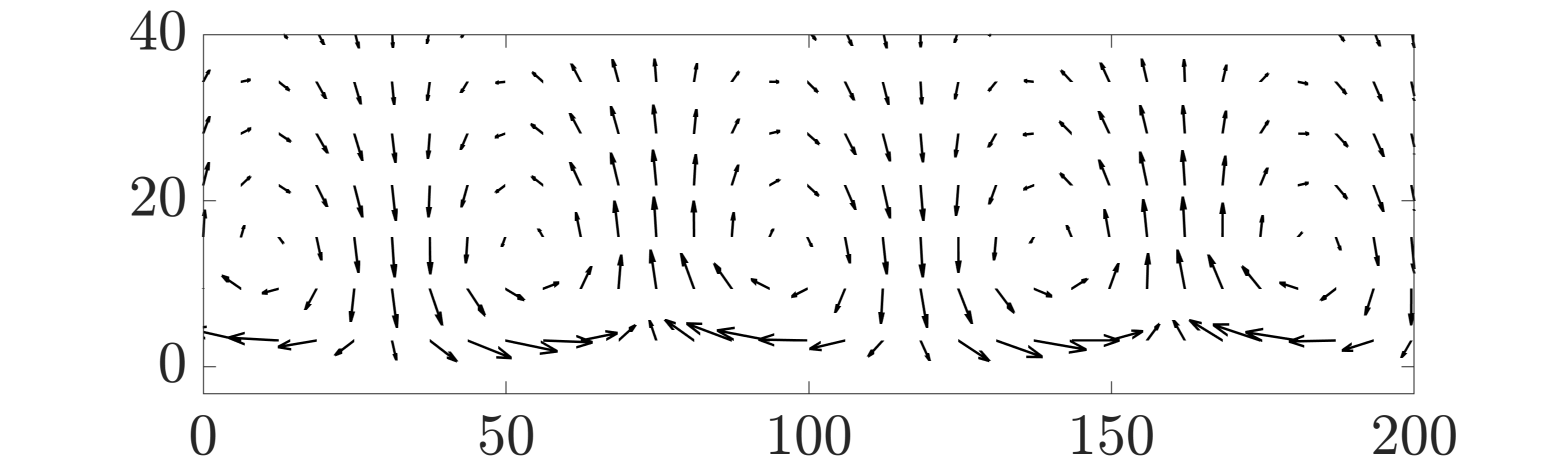}
       \\
       \includegraphics[width=0.47\textwidth]{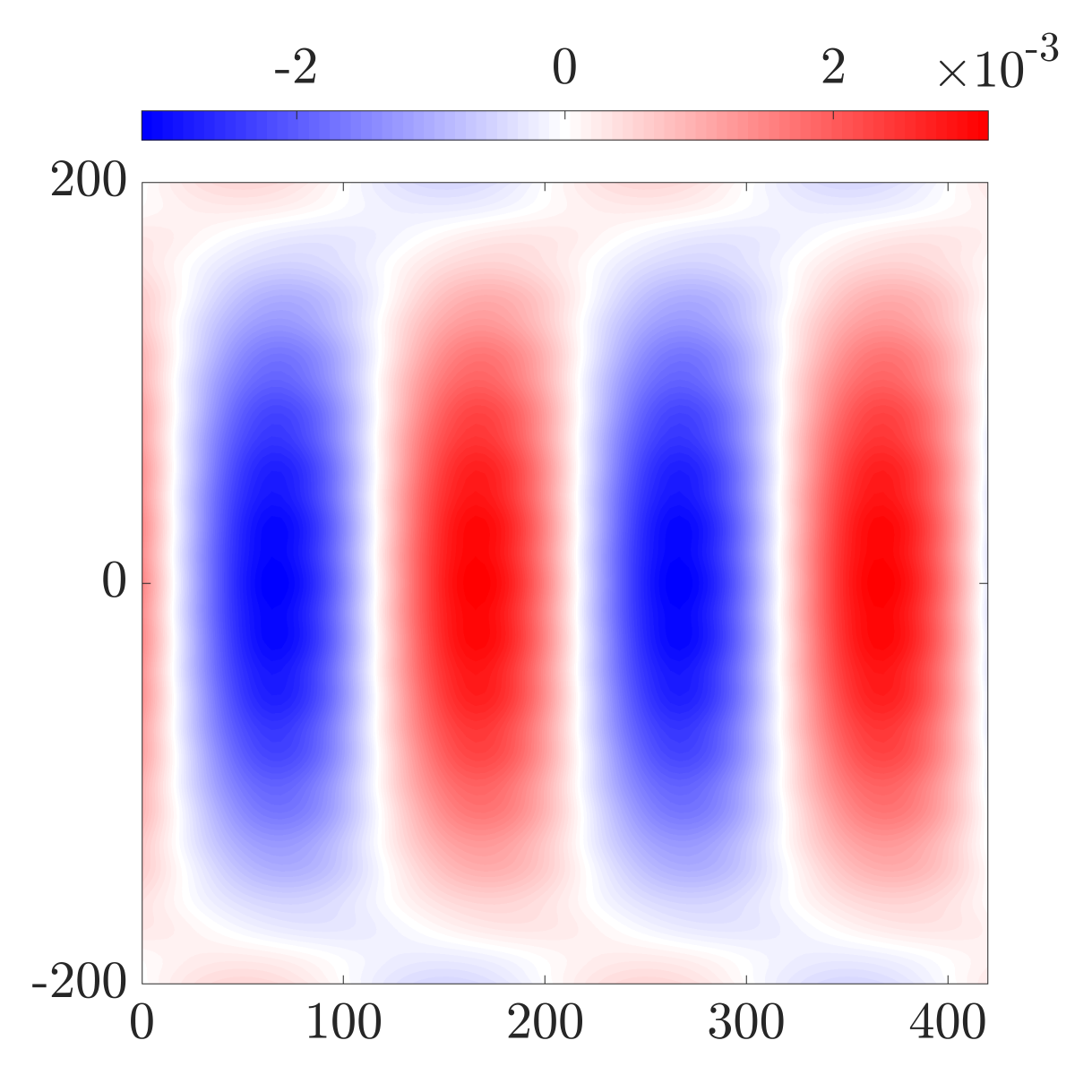}
       \\[-.1cm]
            \hspace{.2cm}
            $\tilde{x}^+$
       \end{tabular}
       &\hspace{.2cm}
       \begin{tabular}{c}
        \vspace{.2cm}
       \end{tabular}
       &\hspace{-.8cm}
    \begin{tabular}{c}
       \includegraphics[width=0.46\textwidth]{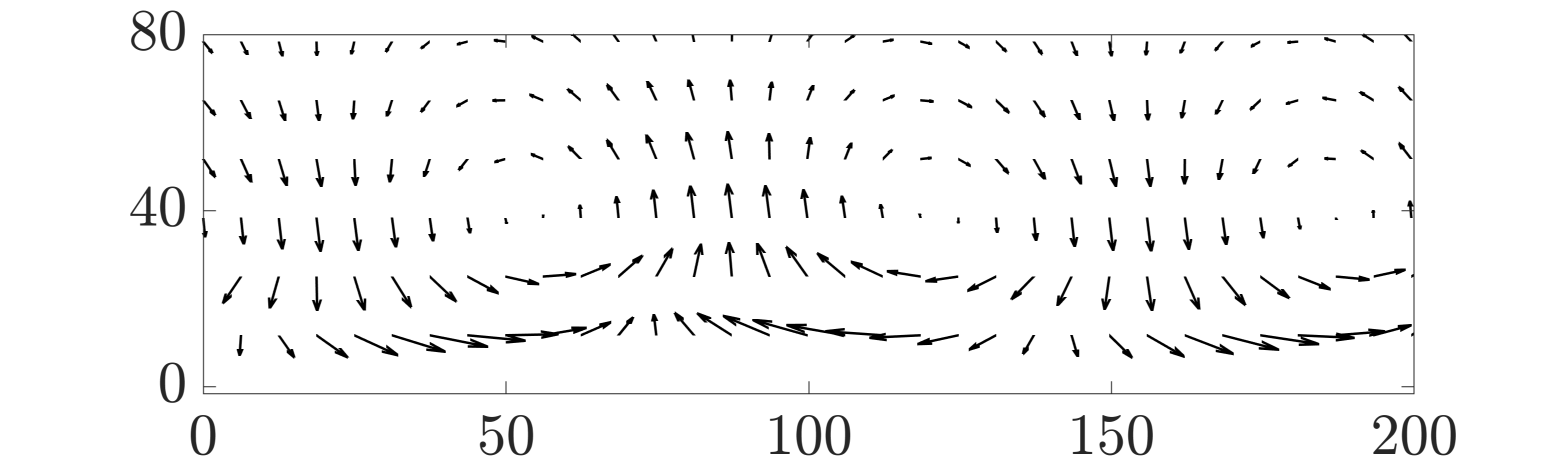}
       \\
       \includegraphics[width=0.47\textwidth]{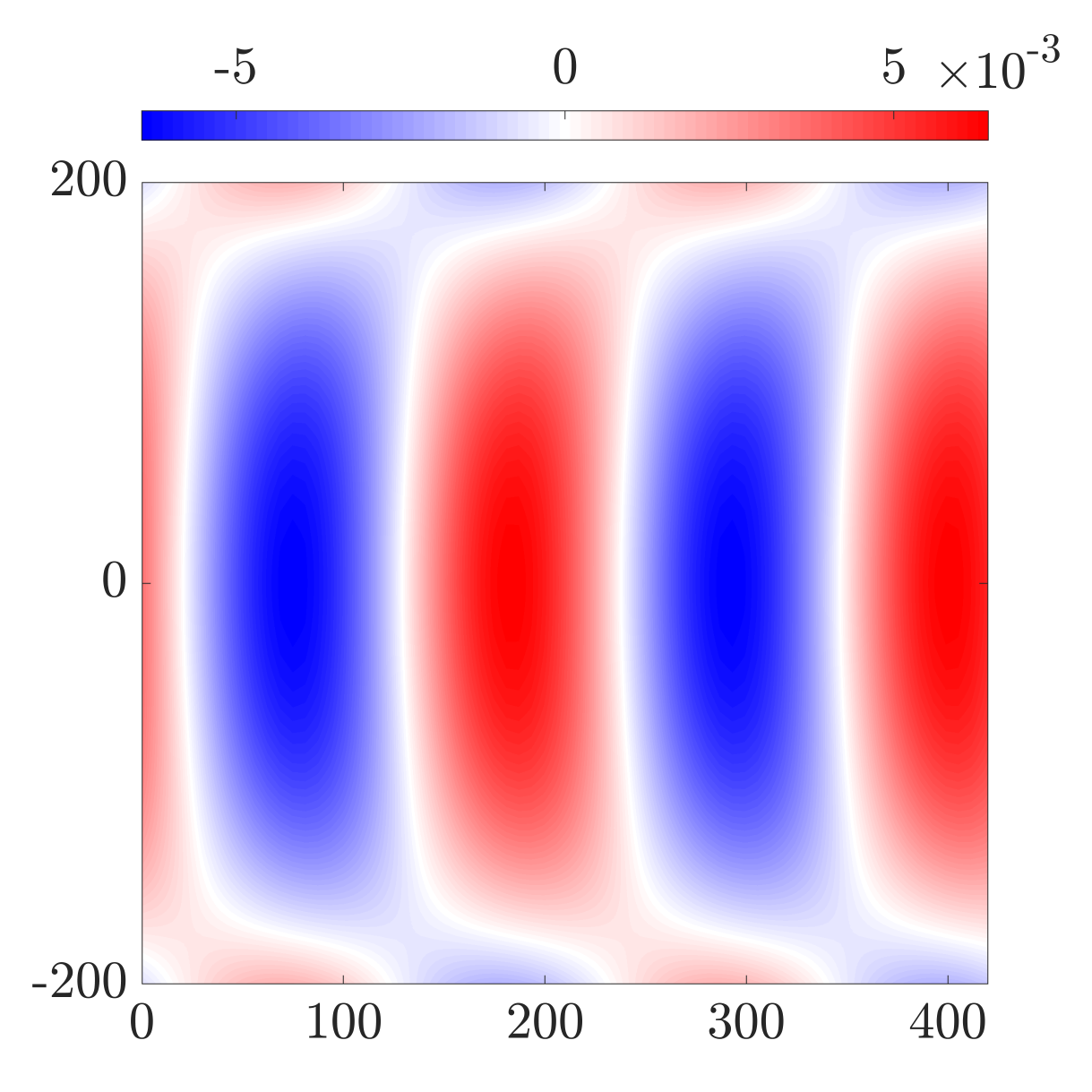}
       \\[-.1cm]
            \hspace{.2cm}
            $\tilde{x}^+$
       \end{tabular}
       \end{tabular}
       \end{center}
        \caption{{Velocity vectors ($u, v$) corresponding to spanwise vortices (top) and colorplots of the wall-normal velocity $v$ (bottom) in channel flow with (a) $Re_\tau=186$, (b) $Re_\tau=547$, (c) $Re_\tau=934$, and (d) $Re_\tau=2003$ over riblets with $\alpha/s=0.55$ and $l^+_g \approx 20$, which corresponds to spatial frequencies $\omega_z=40$, $115$, $200$, and $420$, respectively. Flow patterns result from a superposition of dominant eigenmodes of the covariance matrix $\bPhi_\theta(k_x)$ for the $\lambda^+_x$ corresponding to the the maximum riblet-induced amplification in figure~\ref{fig.E2vvRe} and $250 < \lambda^+_z < 1000$. The side-view ($\tilde{x}$-$\tilde{y}$) planes correspond to $\tilde{z}^+ = 0$ and the top view ($\tilde{x}$-$\tilde{z}$) planes correspond to $\tilde{y}^+ \approx 5$, i.e., one viscous unit above the crest of riblets.}}
        \label{fig.KHRe}
\end{figure}

Previous numerical studies reveal the presence of K-H rollers that arise from the amplification mechanisms described in prior figures using instantaneous visualizations of wall-normal velocity~\citep{garjim11b,garjim12}. The excitation of such flow structures can also be captured by analyzing the dominant eigenmode of the covariance matrices $\bPhi_\theta(k_x)$ for {$\lambda^+_x$ corresponding to the peak amplification in the premultiplied wall-normal energy spectrum (cf. figure~\ref{fig.E2vvRe}) and $250 < \lambda^+_z < 1000$. 
The lower threshold of $250$ on spanwise wavelengths offers a more conservative criterion for excluding structures associated with the near-wall cycle~\citep{endmodgarhutchu21} while better preserving the spanwise coherence of structures influenced by the K-H instability.}
As shown in figure~\ref{fig.KHRe}, the visualization of this eigenmode illustrates alternating patterns of downwash and upwash flow across several riblet grooves. The riblets considered in this figure are scalloped with $\alpha/s=0.55$ and $l^+_g \approx 20$ for $Re_\tau = 186$, $547$, $934$, and $2003$. Coherent regions of high and low wall-normal velocity {become wider in the spanwise direction and spanwise vortices becoming larger as the Reynolds number increases.} The intensity of high and low-speed regions also increases with the Reynolds number, which is attributed to the interaction of outer-layer structures with the spanwise rollers~\citep{garjim12} above the tips of large riblets. As a result, the wall-normal momentum transfer and turbulent mixing increases, which causes an increase in the drag. Finally, we observe a shift in the core of spanwise rollers from {$\tilde{y}^+ \approx 15$ at $Re_\tau = 186$ to $\tilde{y}^+ \approx 38$ at $Re_\tau = 2003$} with an extension to the bottom wall.

\subsubsection{Effect of Kelvin-Helmholtz instability on the shear stress}
\label{sec.KH-uv}

\begin{figure}
        \begin{center}
        \begin{tabular}{cccccc}
        \hspace{-.6cm}
        \subfigure[]{\label{fig.E12uv2DT919o179Re547}}
        &&
        \hspace{-.7cm}
        \subfigure[]{\label{fig.E12uv2DT950o460Re547}}
        &&
        \hspace{-.6cm}
        \subfigure[]{\label{fig.E12uv2DT321o163Re547}}
        &
        \\[-.5cm]\hspace{-.3cm}
	\begin{tabular}{c}
        \vspace{.2cm}
        {\small \rotatebox{0}{$\lambda^+_z$}}
       \end{tabular}
       &\hspace{-.3cm}
	\begin{tabular}{c}
       \includegraphics[width=4cm]{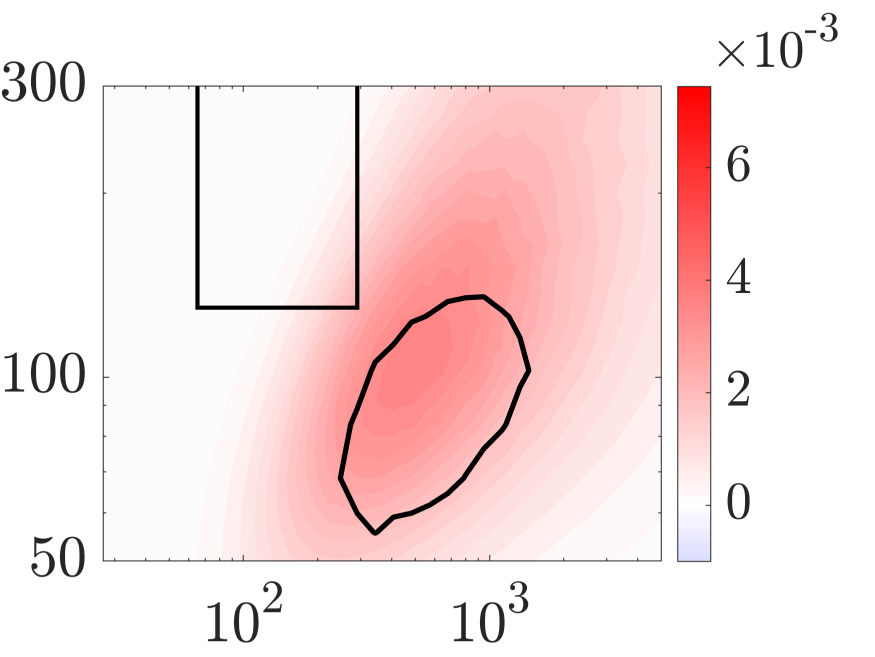}
       \\[-.1cm]
            \hspace{-.2cm}
            $\lambda^+_x$
       \end{tabular}
       &&\hspace{-.3cm}
    \begin{tabular}{c}
       \includegraphics[width=4cm]{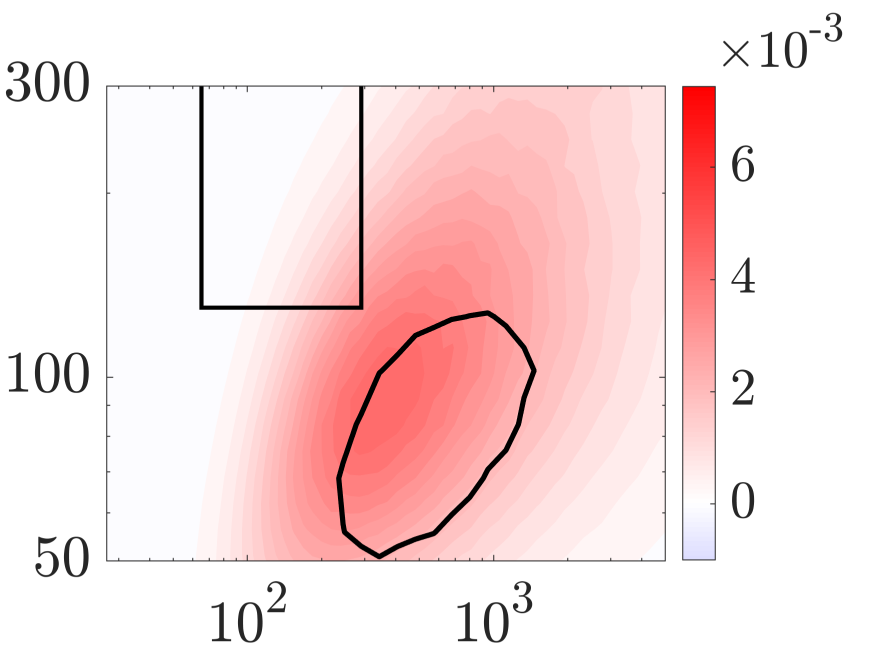}
       \\[-.1cm]
            \hspace{-.2cm}
            $\lambda^+_x$
       \end{tabular}
       &&\hspace{-.3cm}
    \begin{tabular}{c}
       \includegraphics[width=4cm]{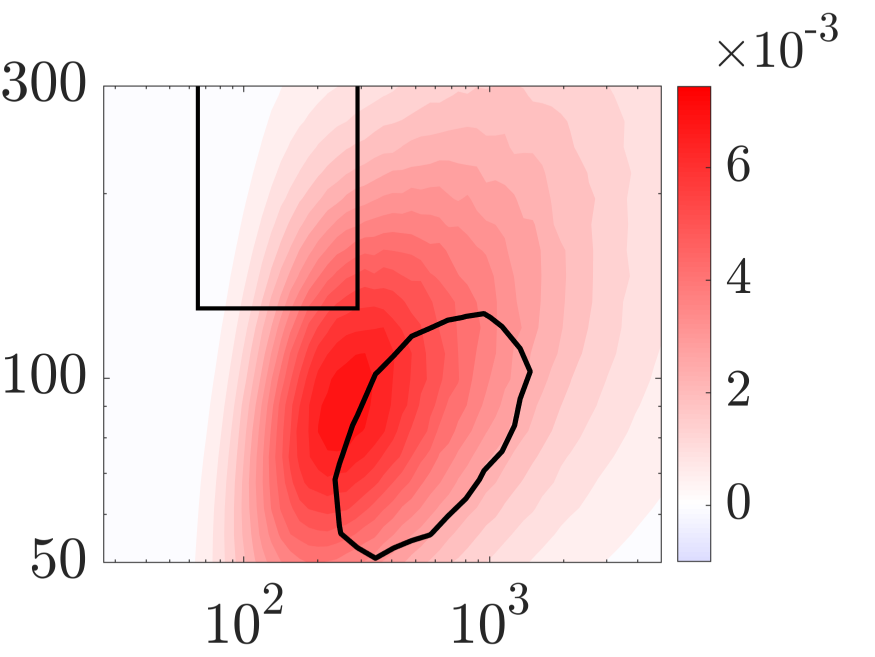}
       \\[-.1cm]
            \hspace{-.2cm}
            $\lambda^+_x$
       \end{tabular}
       \end{tabular}
       \end{center}
        \caption{{Premultiplied modifications to the energy spectrum of Reynolds shear stress,  $k_x \theta (\alpha\,E_{uv,1} + \alpha^2\,E_{uv,2})$, of turbulent channel flow with $Re_\tau=547$ one viscous unit above the crest of triangular riblets. (a) $\alpha/s=0.5$ and $l^+_g \approx 10$ ($\omega_z = 179$), (b) $\alpha/s=0.5$ and $l^+_g \approx 25$ ($\omega_z = 69$), and (c) $\alpha/s=1.87$ and $l^+_g \approx 20$ ($\omega_z = 163$). Black open boxes delimit the spectral window of K-H rollers according to~\cite{garjim11b} and black contour lines correspond to the $70\%$ contour level of the energy spectrum of smooth channel flow from the DNS of~\cite{deljimzanmos04}.}}
        \label{fig.E12uvTvalid}
\end{figure}

In addition to its imprint on the wall-normal energy spectrum, the K-H instability also results in an increase in the Reynolds shear stress, which, consequently, affects momentum transfer and skin-friction drag~\citep{garjim11b,gomgar19}. {Figure~\ref{fig.E12uvTvalid} shows the spectral evidence of K-H rollers in the premultiplied modifications to the Reynolds shear stress co-spectrum (i.e., $k_x \theta (\alpha\,E_{uv,1} + \alpha^2\,E_{uv,2})$) in channel flow with $Re_\tau = 547$ due to the presence of triangular riblets. The spectra are computed for a horizontal plane located 1 viscous unit above the riblets’ crest. These are the same riblet geometries as in figure~\ref{fig.E12vvTvalid}, which are known for their tendency to either permit or inhibit the K–H instability based on~\cite{endmodgarhutchu21}. As shown in figure~\ref{fig.E12uv2DT321o163Re547}, sharp triangular riblets have a more pronounced effect on the amplification of the premultiplied co-spectrum compared to their blunt counterparts (cf.~figures~\ref{fig.E12uv2DT919o179Re547} and~\ref{fig.E12uv2DT950o460Re547}). While the evidence of K-H instability is more clearly revealed in the wall-normal energy spectrum (figure~\ref{fig.E12vvTvalid}), its signature is also evident from the shear stress co-spectra as it attains more amplified values and extends into the spectral range corresponding to the K-H instability due to progressively larger and sharper triangular riblets. We note that since the case of $\alpha/s=0.5$ and $l_g^+\approx 25$ corresponds to a drag-increasing riblet configuration~\citep{endmodgarhutchu21}, the weak footprint of K-H modes may also be indicative of an alternative destructive mechanism, e.g., dispersive stresses induced by secondary flows~\citep{goltua98,modendhutchu21}, which we, however, do not account for in this study.}

\begin{figure}
        \begin{center}
        \begin{tabular}{cccccc}
        \hspace{-.6cm}
        \subfigure[]{\label{fig.E1uvas55O460ypRe547}}
        &&
        \hspace{-.7cm}
        \subfigure[]{\label{fig.E1uvas55O175ypRe547}}
        &&
        \hspace{-.6cm}
        \subfigure[]{\label{fig.E1uvas55O115ypRe547}}
        &
        \\[-.5cm]\hspace{-.3cm}
	\begin{tabular}{c}
        \vspace{-.2cm}
        {\small \rotatebox{0}{$\lambda^+_z$}}
       \end{tabular}
       &\hspace{-.45cm}
	\begin{tabular}{c}
        $l^+_g \approx 5$
        \\
       \includegraphics[width=4cm]{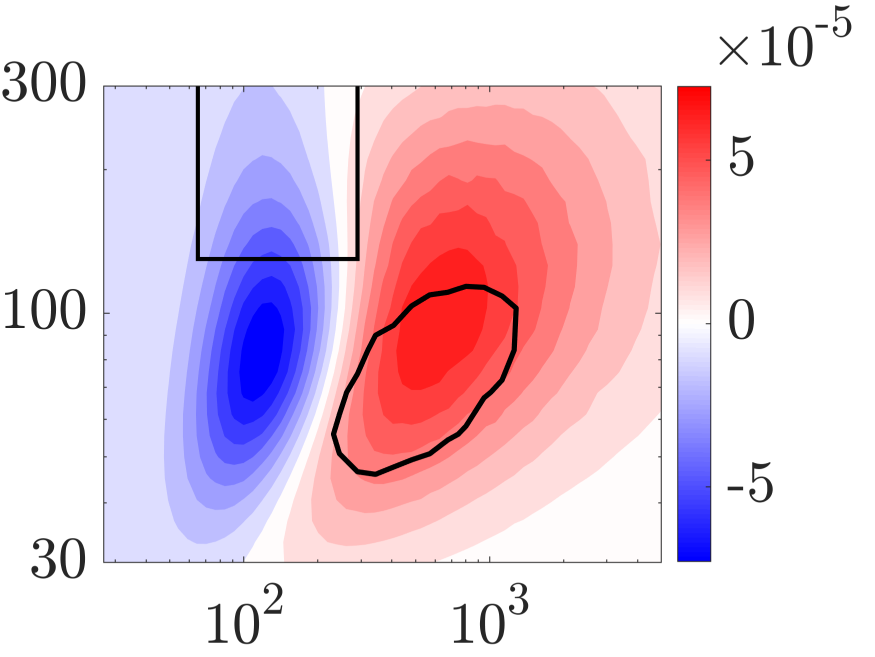}
       \end{tabular}
       &&\hspace{-.45cm}
    \begin{tabular}{c}
    $l^+_g \approx 10$
    \\
       \includegraphics[width=4cm]{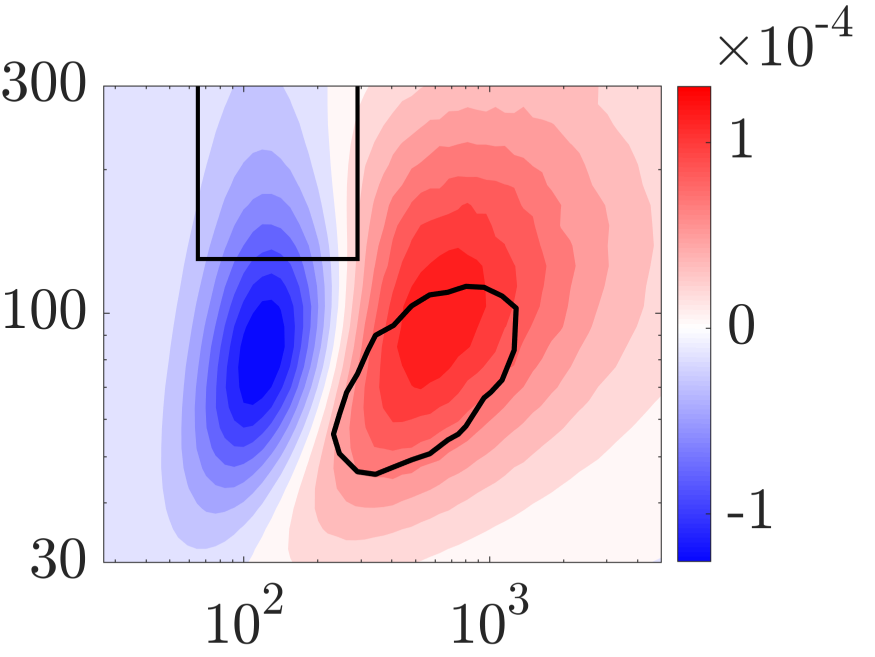}
       \end{tabular}
       &&\hspace{-.45cm}
    \begin{tabular}{c}
    $l^+_g \approx 20$
    \\
       \includegraphics[width=4cm]{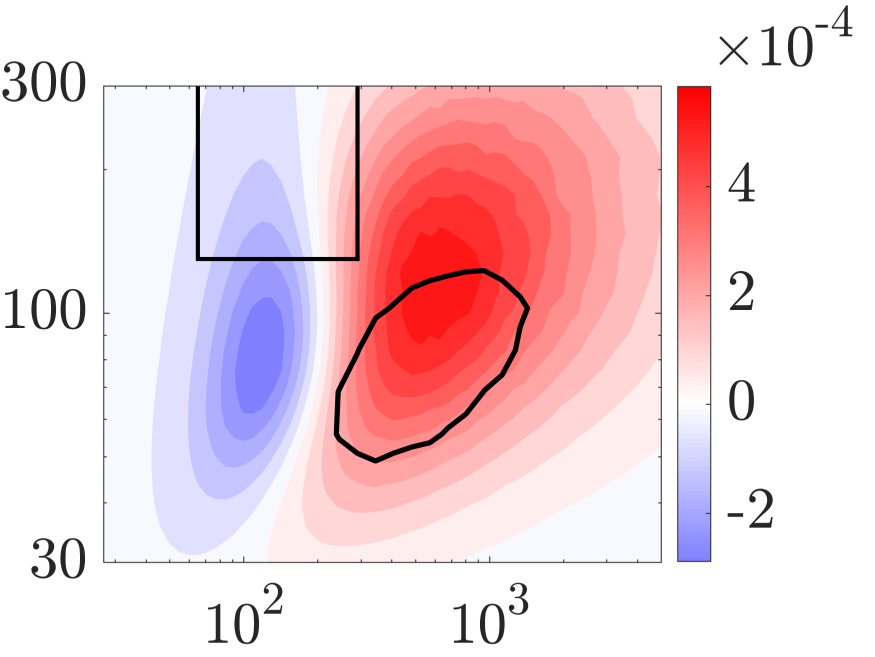}
       \end{tabular}
       \\[-0.1cm]
        \hspace{-.6cm}
        \subfigure[]{\label{fig.E2uvas55O460ypRe547}}
        &&
        \hspace{-.7cm}
        \subfigure[]{\label{fig.E2uvas55O175ypRe547}}
        &&
        \hspace{-.6cm}
        \subfigure[]{\label{fig.E2uvas55O115ypRe547}}
        &
        \\[-.5cm]\hspace{-.3cm}
	\begin{tabular}{c}
        \vspace{-.2cm}
        {\small \rotatebox{0}{$\lambda^+_z$}}
       \end{tabular}
       &\hspace{-.3cm}
	\begin{tabular}{c}
       \includegraphics[width=4cm]{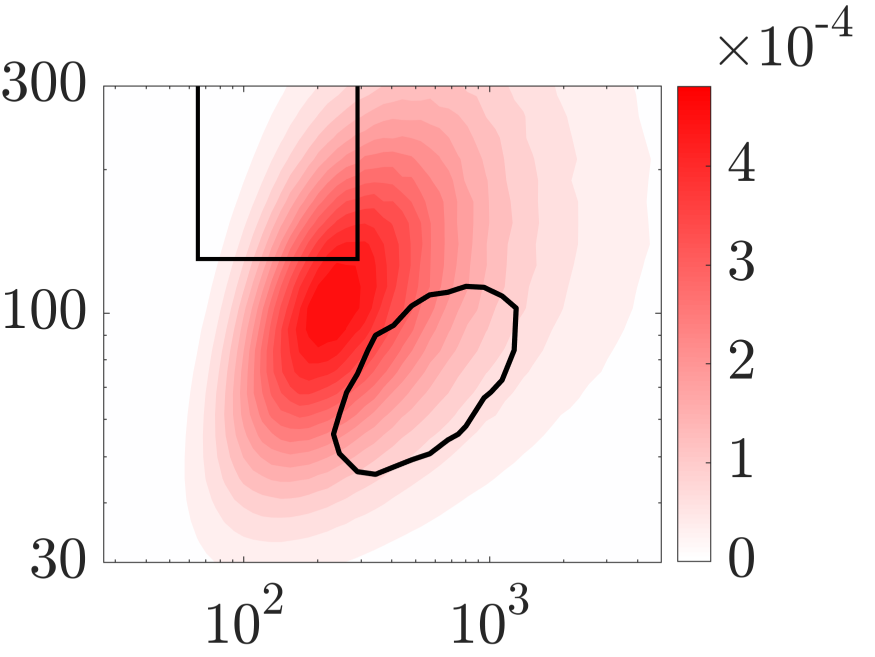}
       \end{tabular}
       &&\hspace{-.3cm}
    \begin{tabular}{c}
       \includegraphics[width=4cm]{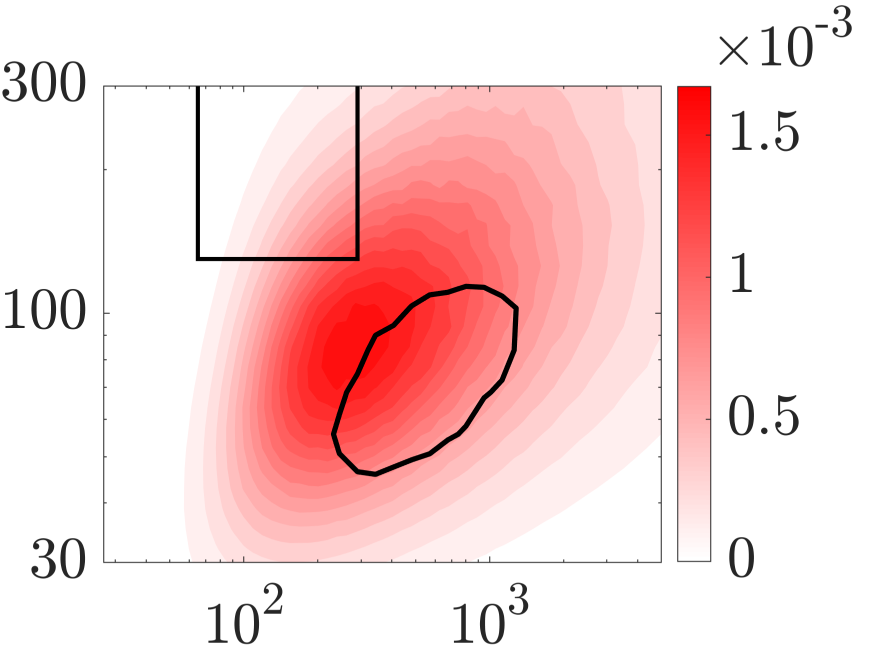}
       \end{tabular}
       &&\hspace{-.3cm}
    \begin{tabular}{c}
       \includegraphics[width=4cm]{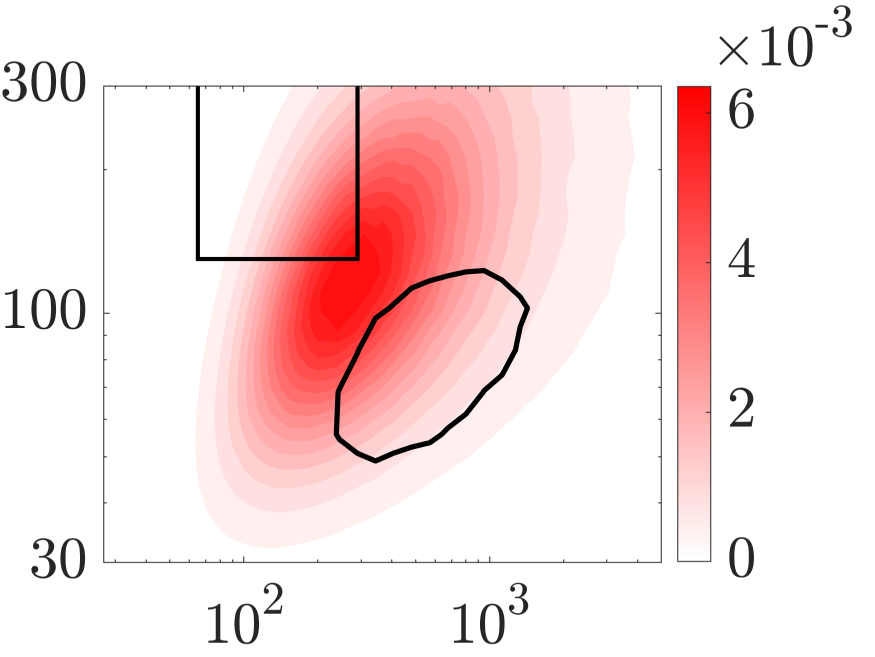}
       \end{tabular}
       \\[-0.1cm]
       \hspace{-.6cm}
        \subfigure[]{\label{fig.E12uvas55O460ypRe547}}
        &&
        \hspace{-.6cm}
        \subfigure[]{\label{fig.E12uvas55O175ypRe547}}
        &&
        \hspace{-.6cm}
        \subfigure[]{\label{fig.E12uvas55O115ypRe547}}
        &
        \\[-.5cm]\hspace{-.3cm}
	\begin{tabular}{c}
        \vspace{.2cm}
        {\small \rotatebox{0}{$\lambda^+_z$}}
       \end{tabular}
       &\hspace{-.3cm}
	\begin{tabular}{c}
       \includegraphics[width=4cm]{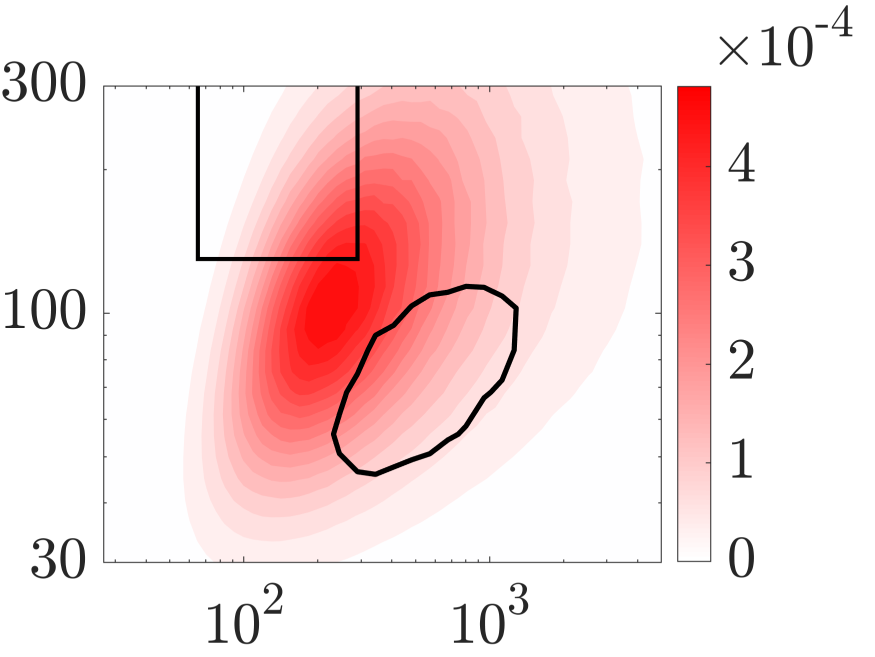}
        \\[-.1cm]
            \hspace{-.2cm}
            $\lambda^+_x$
       \end{tabular}
       &&\hspace{-.25cm}
    \begin{tabular}{c}
       \includegraphics[width=4cm]{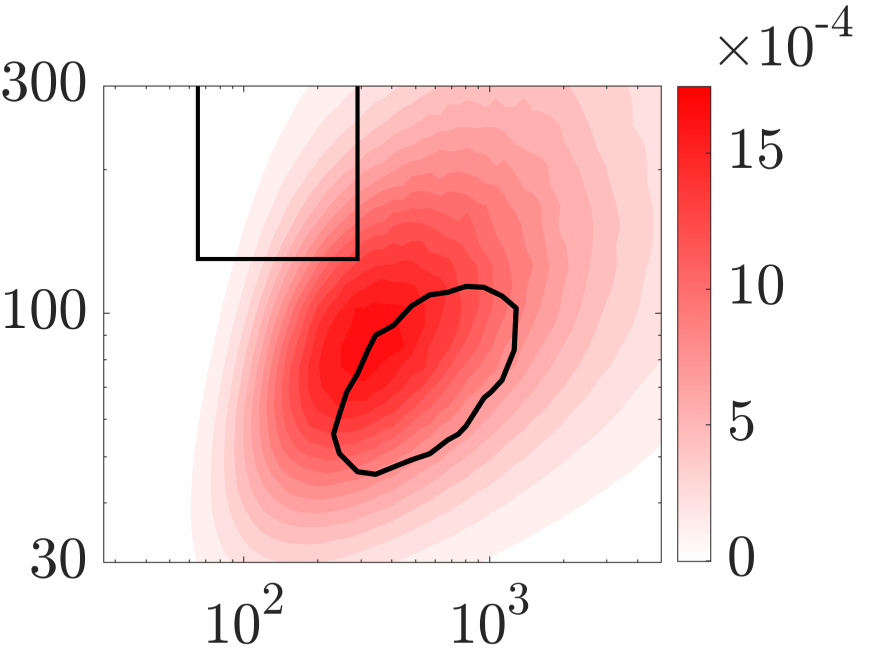}
       \\[-.1cm]
            \hspace{-.2cm}
            $\lambda^+_x$
       \end{tabular}
       &&\hspace{-.3cm}
    \begin{tabular}{c}
       \includegraphics[width=4cm]{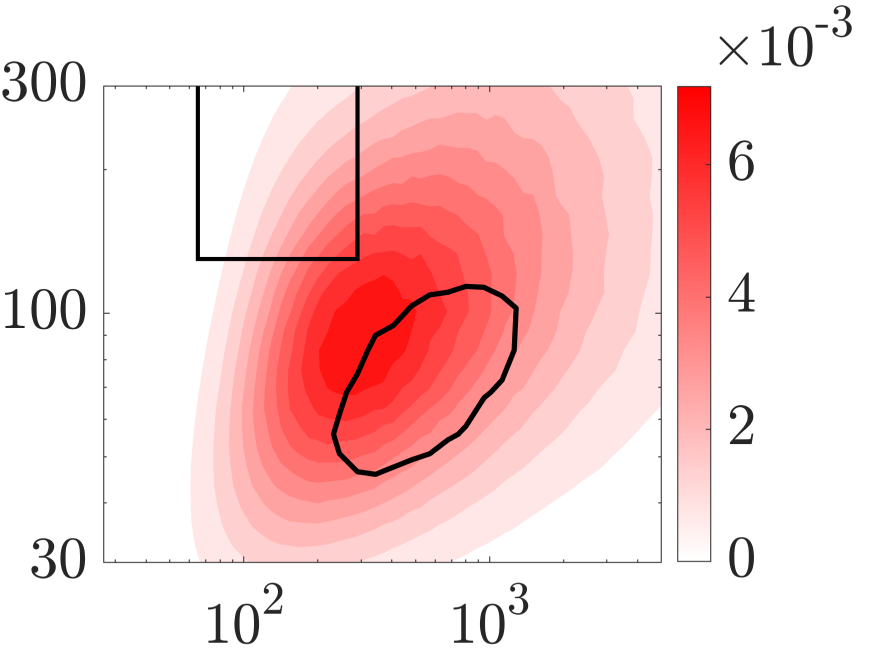}
       \\[-.1cm]
            \hspace{-.2cm}
            $\lambda^+_x$
       \end{tabular}
       \end{tabular}
       \end{center}
        \caption{{Premultiplied modifications to the energy spectrum of Reynolds shear stress in channel flow with $Re_\tau=547$ one viscous unit above the crest of scalloped riblets with $\alpha/s = 0.55$ and $l^+_g \approx 5$ ($\omega_z=460$) (left column), $l^+_g \approx 10$ ($\omega_z=230$) (middle column), and $l^+_g \approx 20$ ($\omega_z=115$) (right column). (a-c) $\alpha\,k_x \theta E_{uv,1}$; (d-f) $\alpha^2\,k_x \theta E_{uv,2}$; and (g-i) $k_x \theta (\alpha\,E_{uv,1} + \alpha^2\, E_{uv,2})$. Black open boxes delimit the spectral window of K-H rollers according to~\cite{garjim11b} and black contour lines correspond to the $70\%$ contour level of the energy spectrum of smooth channel flow from the DNS of~\cite{deljimzanmos04}.}}
    \label{fig.E2uvas087}
\end{figure}

\begin{figure}
        \begin{center}
        \begin{tabular}{cccccc}
        \hspace{-.6cm}
        \subfigure[]{\label{fig.E2uvas65O115ypRe547}}
        &&
        \hspace{-.7cm}
        \subfigure[]{\label{fig.E2uvas87O115ypRe547}}
        &&
        \hspace{-.6cm}
        \subfigure[]{\label{fig.E2uvas12O115ypRe547}}
        &
        \\[-.5cm]\hspace{-.3cm}
	\begin{tabular}{c}
        \vspace{.2cm}
        {\small \rotatebox{0}{$\lambda^+_z$}}
       \end{tabular}
       &\hspace{-.3cm}
	\begin{tabular}{c}
       \includegraphics[width=4cm]{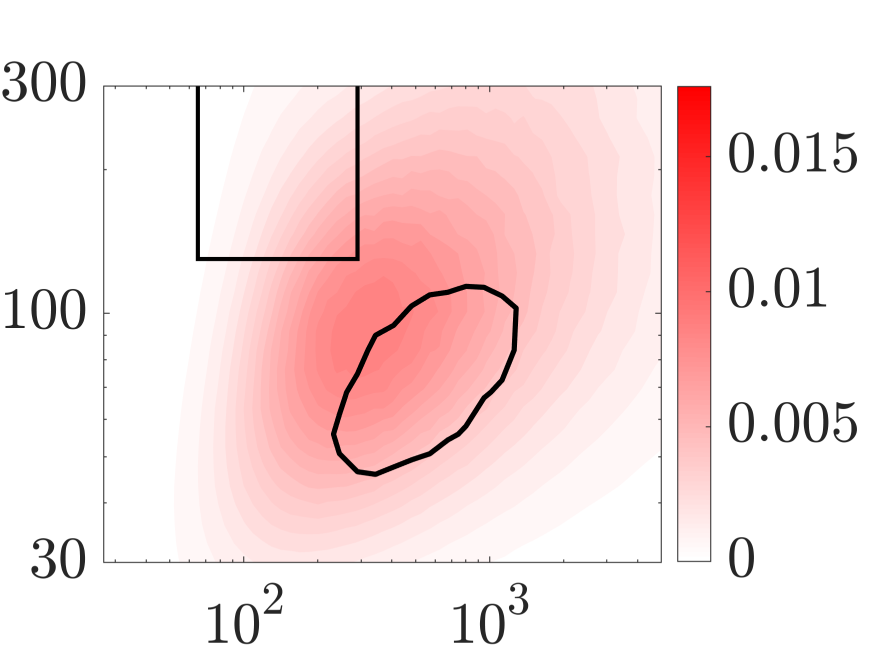}
       \\[-.1cm]
            \hspace{-.2cm}
            $\lambda^+_x$
       \end{tabular}
       &&\hspace{-.3cm}
    \begin{tabular}{c}
       \includegraphics[width=4cm]{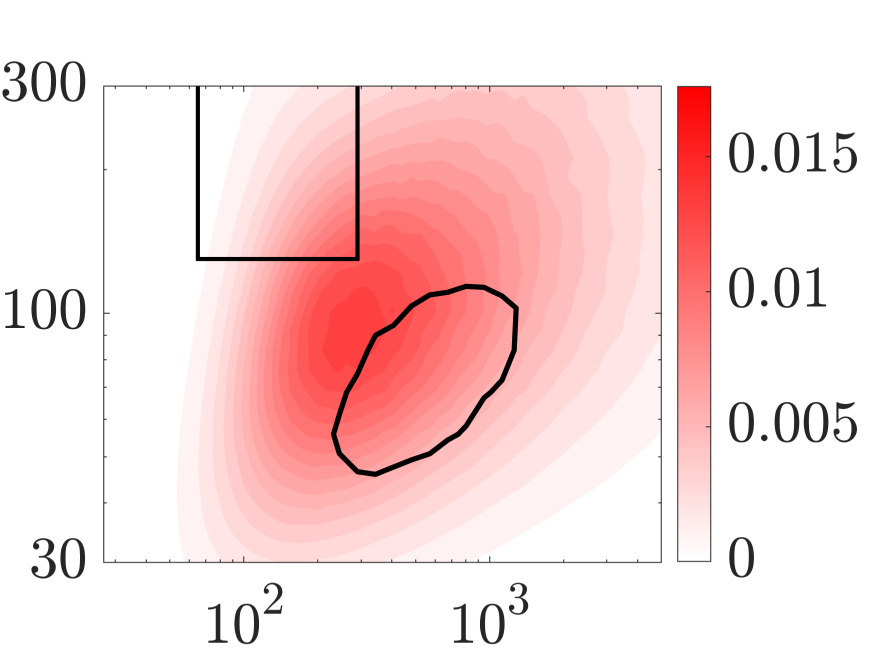}
       \\[-.1cm]
            \hspace{-.2cm}
            $\lambda^+_x$
       \end{tabular}
       &&\hspace{-.3cm}
    \begin{tabular}{c}
       \includegraphics[width=4cm]{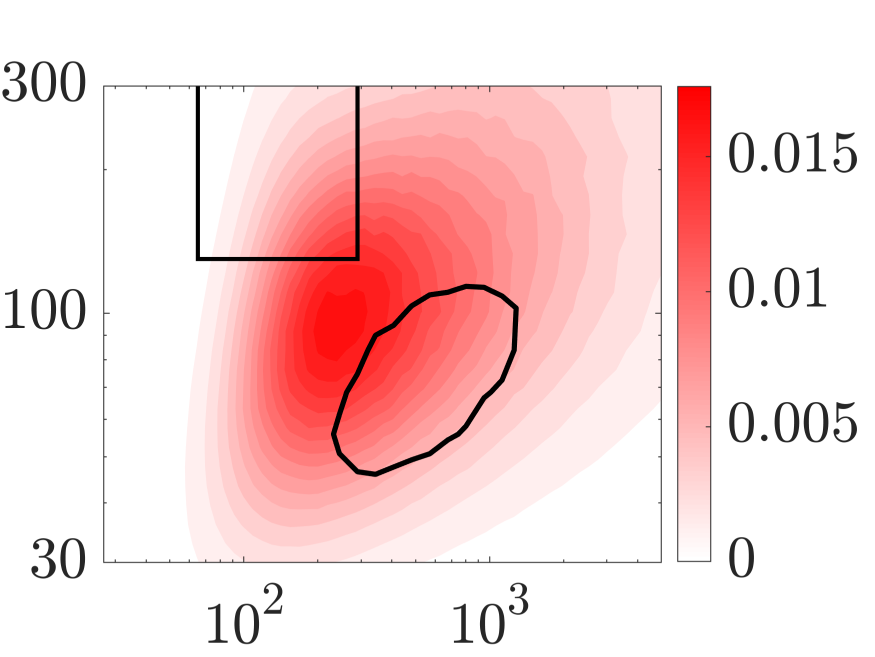}
       \\[-.1cm]
            \hspace{-.2cm}
            $\lambda^+_x$
       \end{tabular}
       \end{tabular}
       \end{center}
        \caption{{Premultiplied modifications to the energy spectrum of Reynolds shear stress,  $k_x \theta (\alpha\,E_{uv,1} + \alpha^2\,E_{uv,2})$, of turbulent channel flow with $Re_\tau=547$ one viscous unit above the crest of scalloped riblets with the same viscous spacing but different viscous height. (a) $\alpha/s=0.65$ and $l^+_g \approx 22$ ($\omega_z = 115$); (b) $\alpha/s=0.87$ and $l^+_g \approx 25$ ($\omega_z = 115$); and (c) $\alpha/s=1.2$ and $l^+_g \approx 29$ ($\omega_z = 115$). Black open boxes delimit the spectral window of K-H rollers according to~\cite{garjim11b} and black contour lines correspond to the $70\%$ contour level of the energy spectrum of smooth channel flow from the DNS of~\cite{deljimzanmos04}.}}
    \label{fig.E2uvas}
\end{figure}

\begin{figure}
        \begin{center}
        \begin{tabular}{cccccccc}
        \hspace{-.6cm}
        \subfigure[]{\label{fig.E2uvas55O45ypRe186}}
        &&
        \hspace{-.8cm}
        \subfigure[]{\label{fig.E12uvas55O115ypRe547_2}}
        &&
        \hspace{-.7cm}
        \subfigure[]{\label{fig.E2uvas55O200ypRe934}}
        &&
        \hspace{-.7cm}
        \subfigure[]{\label{fig.E2uvas55O420ypRe2003}}
        &
        \\[-.5cm]\hspace{-.3cm}
	\begin{tabular}{c}
        \vspace{.3cm}
        {\small \rotatebox{0}{$\lambda^+_z$}}
       \end{tabular}
       &\hspace{-.45cm}
	\begin{tabular}{c}
       \includegraphics[width=2.9cm]{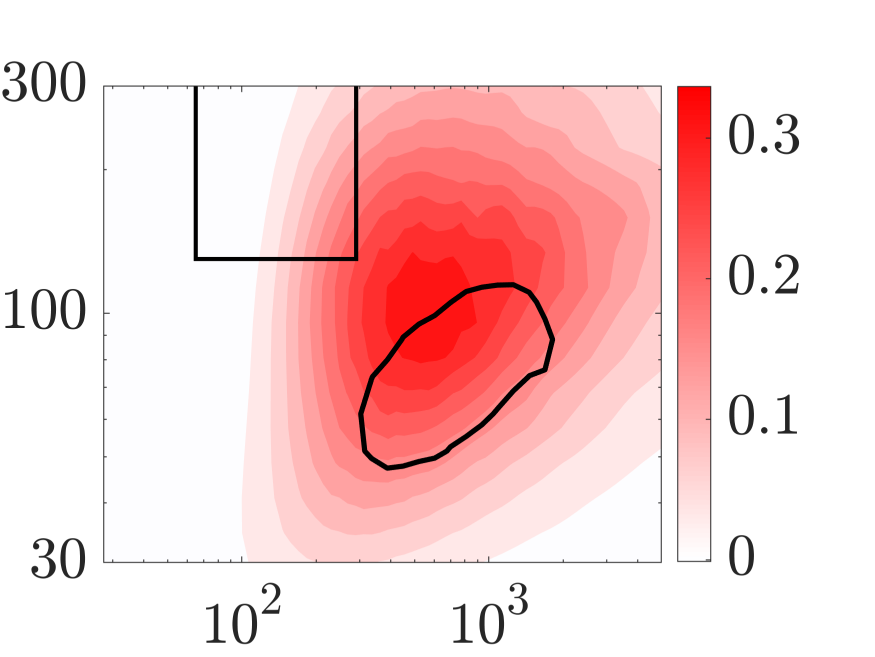}
        \\[-.1cm]
            \hspace{-.1cm}
            {\small $\lambda^+_x$}
       \end{tabular}
       &&\hspace{-.35cm}
    \begin{tabular}{c}
       \includegraphics[width=2.9cm]{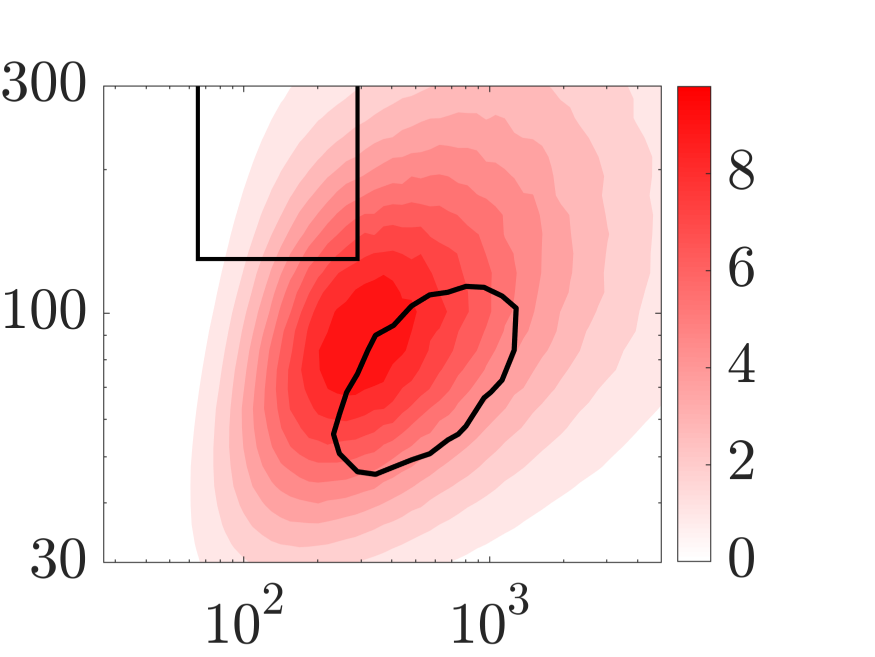}
       \\[-.1cm]
            \hspace{-.1cm}
            {\small $\lambda^+_x$}
       \end{tabular}
       &&\hspace{-.35cm}
    \begin{tabular}{c}
       \includegraphics[width=2.9cm]{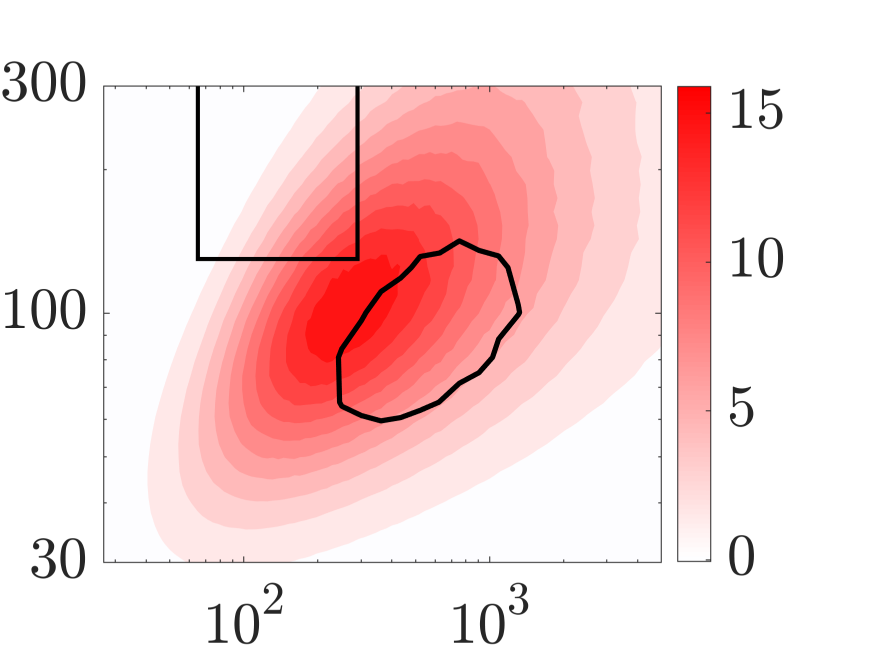}
       \\[-.1cm]
            \hspace{-.1cm}
            {\small $\lambda^+_x$}
       \end{tabular}
       &&\hspace{-.3cm}
    \begin{tabular}{c}
       \includegraphics[width=2.9cm]{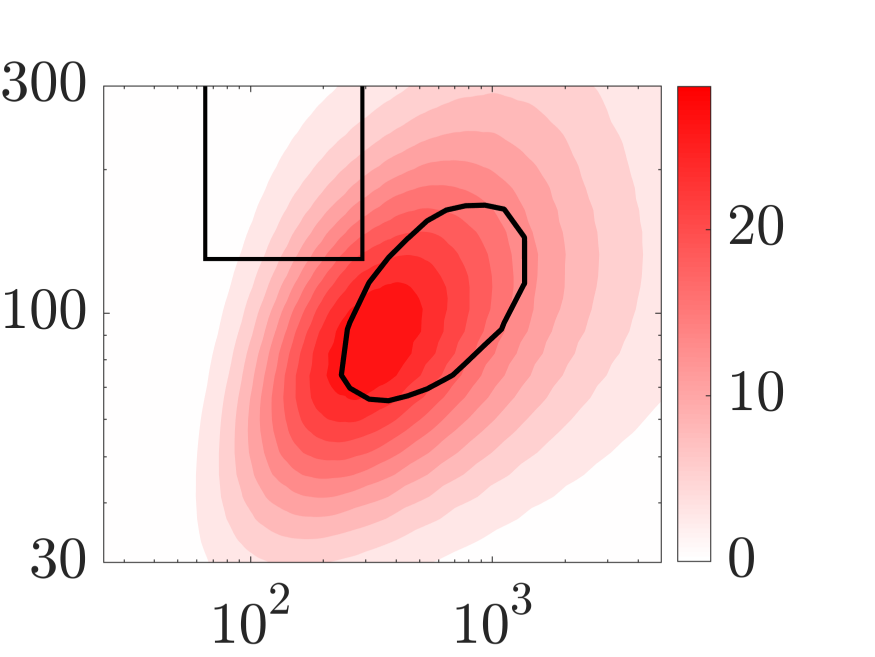}
       \\[-.1cm]
            \hspace{-.1cm}
            {\small $\lambda^+_x$}
       \end{tabular}
       \end{tabular}
       \end{center}
        \caption{{Riblet-induced modifications to the energy spectrum of Reynolds shear stress $k_x \theta (\alpha\,E_{uv,1} + \alpha^2\,E_{uv,2})$ for turbulent channel flow with (a) $Re_\tau=186$, (b) $Re_\tau=547$, (c) $Re_\tau=934$, and (d) $Re_\tau=2003$ one viscous unit above the crest of scalloped riblets with $\alpha/s=0.55$ and $l^+_g \approx 20$, which corresponds to spatial frequencies
        $\omega_z=40$, $115$, $200$, and $420$, respectively. Black open boxes delimit the spectral window of K-H rollers and black contour lines correspond to the $70\%$ contour level in the energy spectrum of smooth channel flow from the DNS of~\cite{deljim03,deljimzanmos04,hoyjim06}.}}
    \label{fig.E2uvRe}
\end{figure}

Figure~\ref{fig.E2uvas087} shows {spectral modifications at the same wall distance due to scalloped riblets of different size.
Similar to figure~\ref{fig.E2vvbeta75} we study such modifications for riblets with the same $\alpha/s$ but different $l_g^+$.}
As the riblet size increases, the co-spectra show more amplification at both $\alpha^1$ and $\alpha^2$ levels. These observations are similar to that of the wall-normal energy spectrum (figure~\ref{fig.E2vvbeta75}). To investigate the effect of larger riblets, figure~\ref{fig.E2uvas} includes the cases with $l^+_g \approx 22$, $25$, and $29$ ($\omega_z = 115$ for all three). The {modifications to the Reynolds shear stress co-spectra become stronger} for taller riblets. {Similar to the observation made in figure~\ref{fig.E2vvas}, we observe a uniform pattern in how the $-uv$ spectrum is amplified due to riblets with the same spanwise frequency $\omega_z$, which suggests a potential geometric scaling for this modification.} Finally, figure~\ref{fig.E2uvRe} compares the influence of riblets with $\alpha/s = 0.55$ and $l^+_g \approx 20$ on the {co-spectrum at} different Reynolds numbers. {Similar to figure~\ref{fig.E2vvRe}, the co-spectra have been normalized by the maximum $uv$ correlation of smooth channel flow at the same wall-parallel plane.
Although not centered in the spectral window associated with the K-H instability, the amplification in this region is nonetheless enhanced at higher Reynolds numbers—mirroring the trend observed in the wall-normal energy spectrum  (cf.~figure~\ref{fig.E2vvRe}).}

\begin{figure}
        \begin{center}
        \begin{tabular}{cccccc}
        \hspace{-.6cm}
        \subfigure[]{\label{fig.Euvs}}
        &&
        \hspace{-.7cm}
        \subfigure[]{\label{fig.E12uvas055O175Re547}}
        &&
        \hspace{-.72cm}
        \subfigure[]{\label{fig.E12uvas055O115Re547}}
        &
        \\[-.5cm]\hspace{-.3cm}
	\begin{tabular}{c}
        \vspace{.2cm}
        {\small \rotatebox{90}{$\tilde{y}^+$}}
       \end{tabular}
       &\hspace{-.3cm}
	\begin{tabular}{c}
       \includegraphics[width=3.9cm]{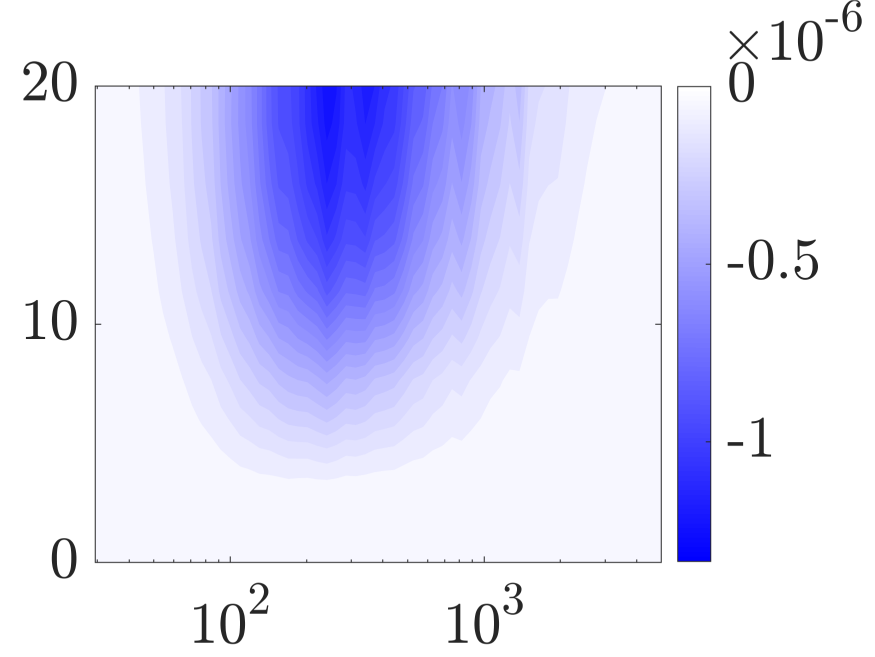}
       \end{tabular}
       &&\hspace{-.3cm}
    \begin{tabular}{c}
       \includegraphics[width=3.9cm]{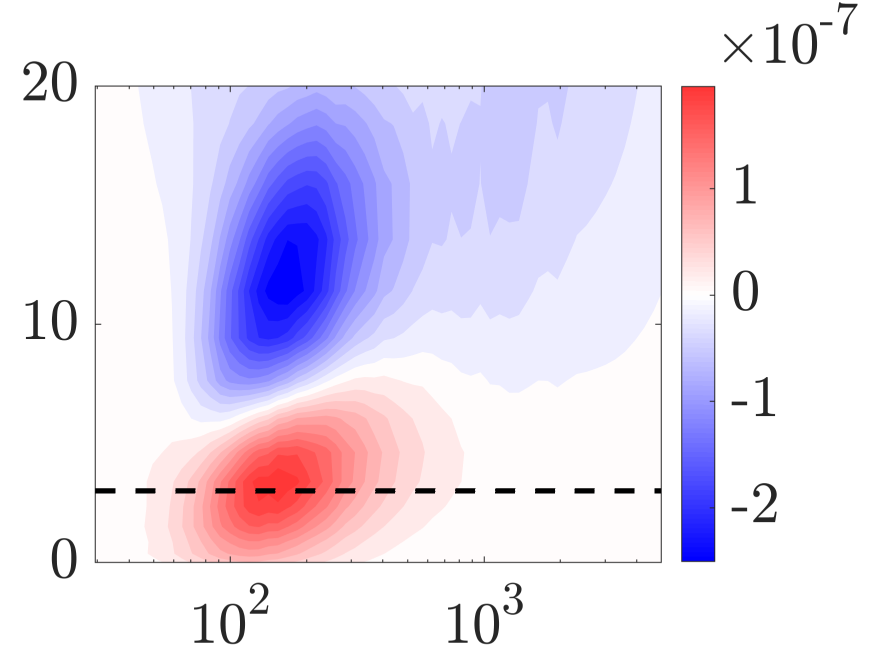}
       \end{tabular}
       &&\hspace{-.35cm}
    \begin{tabular}{c}
       \includegraphics[width=3.9cm]{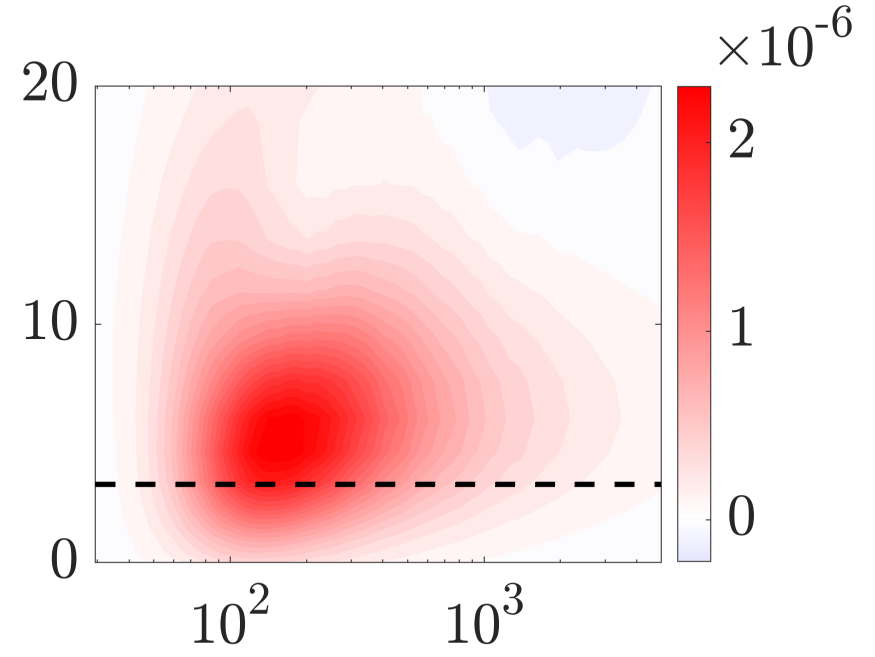}
       \end{tabular}
       \\[-0.1cm]
       \hspace{-.6cm}
        \subfigure[]{\label{fig.E12uvas065O115Re547}}
        &&
        \hspace{-.6cm}
        \subfigure[]{\label{fig.E12uvas087O115Re547}}
        &&
        \hspace{-.7cm}
        \subfigure[]{\label{fig.E12uvas12O115Re547}}
        &
        \\[-.5cm]\hspace{-.3cm}
	\begin{tabular}{c}
        \vspace{.4cm}
        {\small \rotatebox{90}{$\tilde{y}^+$}}
       \end{tabular}
       &\hspace{-.3cm}
	\begin{tabular}{c}
       \includegraphics[width=3.9cm]{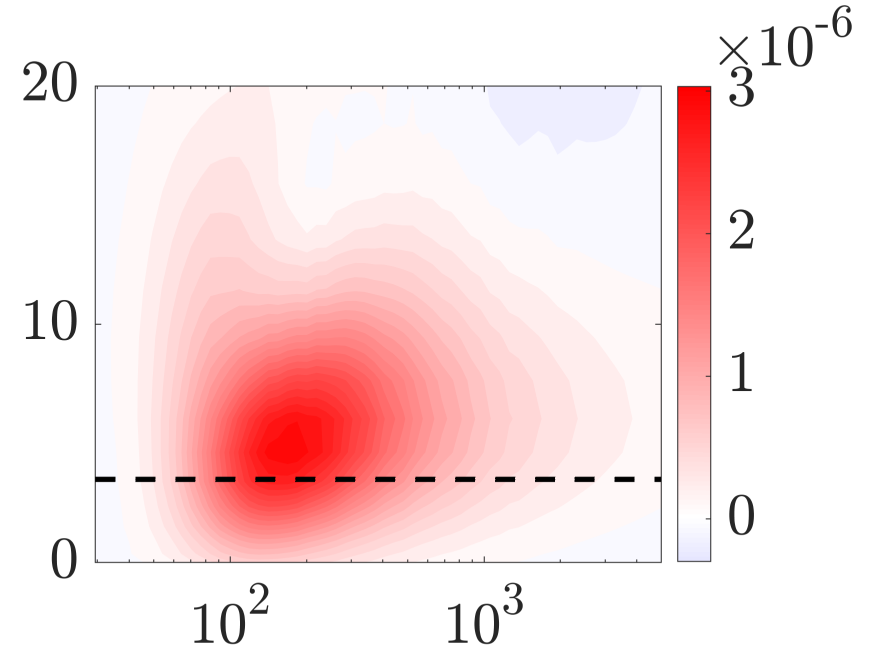}
        \\[-.1cm]
            \hspace{-.4cm}
            $\lambda^+_x$
       \end{tabular}
       &&\hspace{-.25cm}
    \begin{tabular}{c}
       \includegraphics[width=3.9cm]{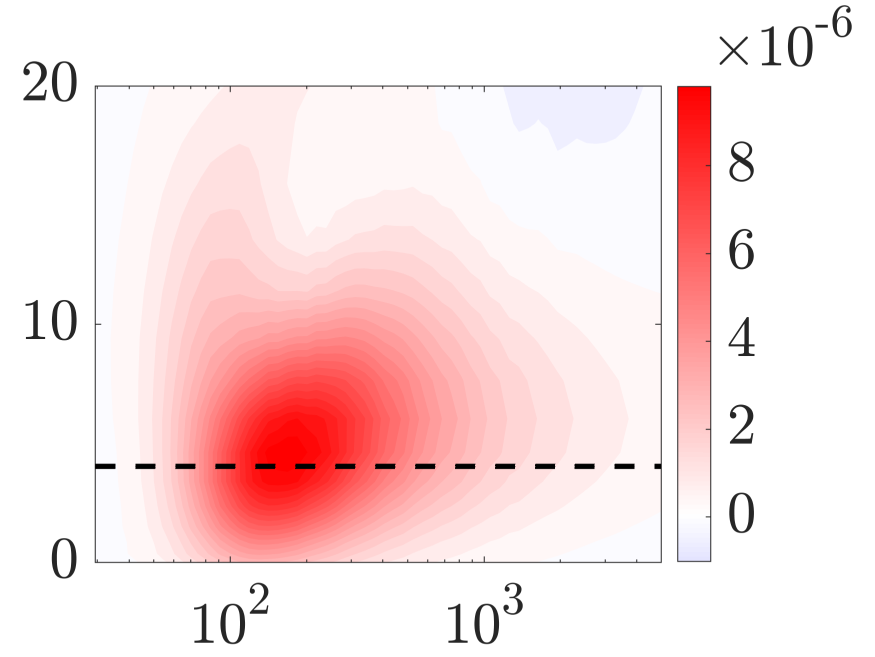}
       \\[-.1cm]
            \hspace{-.4cm}
            $\lambda^+_x$
       \end{tabular}
       &&\hspace{-.35cm}
    \begin{tabular}{c}
       \includegraphics[width=3.9cm]{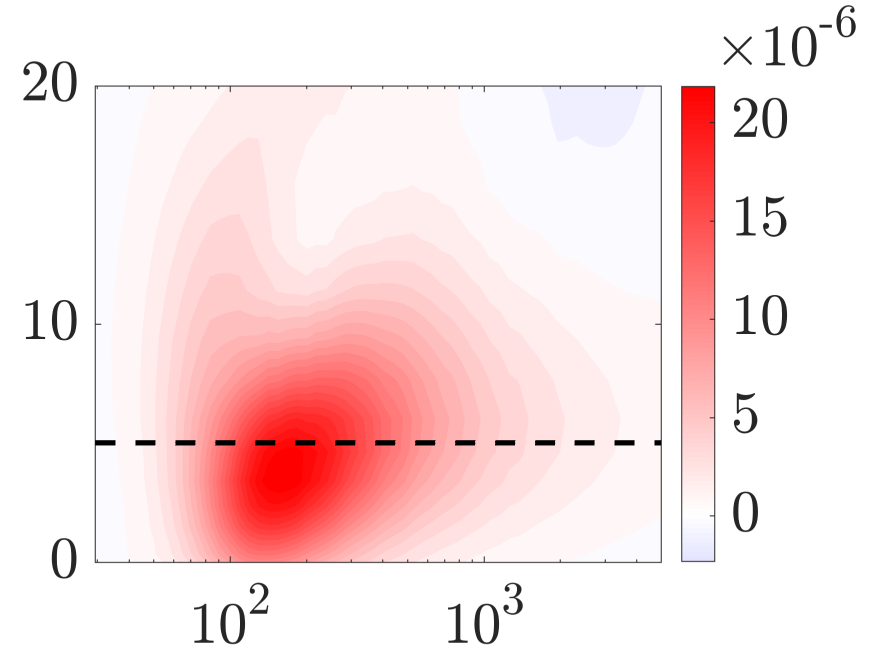}
       \\[-.1cm]
            \hspace{-.4cm}
            $\lambda^+_x$
       \end{tabular}
       \end{tabular}
       \end{center}
        \caption{{(a) Premultiplied one-dimensional energy spectrum of Reynolds shear stress, $-k_x E_{uv}$, integrated over spanwise wavelengths $\lambda^+_z>130$ for smooth channel flow  with $Re_\tau=547$ from the DNS of~\cite{deljimzanmos04}. (b, c) Premultiplied modifications up to $\alpha^2$, i.e., $-k_x (\alpha\,E_{uv,1} + \alpha^2\,E_{uv,2})$ and integrated over $\lambda^+_z>130$ due to scalloped riblets with $\alpha/s=0.55$ and $l^+_g \approx 10$ ($\omega_z = 230$) (b) and  $l^+_g \approx 20$ ($\omega_z = 115$) (c). (d-f) The same quantity for large scalloped riblets that share the same viscous spacing as the case in (c) but different height: (d) $\alpha/s=0.65$ and $l^+_g \approx 22$ ($\omega_z = 115$); (e) $\alpha/s=0.87$ and $l^+_g \approx 25$ ($\omega_z = 115$); and (f) $\alpha/s=1.2$ and $l^+_g \approx 29$ ($\omega_z = 115$). The dashed lines mark the location for the riblet tips.}}
    \label{fig.E12uvasRe547}
\end{figure}

{To exclusively study the effect of the K-H instability on the Reynolds shear stress we integrate the riblet-induced modifications over the length-scales corresponding to the such modes. 
Figure~\ref{fig.E12uvasRe547} compares the one-dimensional spectrum of Reynolds shear stress $-k_xE_{uv}$ in smooth turbulent channel flow with riblet-induced modifications $-k_x (\alpha\,E_{uv,1} + \alpha^2\,E_{uv,2})$ (up to $\alpha^2$). The shear stress spectrum is strengthened in the vicinity of the lower wall and surrounding the riblet tips ($\tilde{y}^+<10$). This amplification, which was also observed in the numerical study of~\cite{endmodgarhutchu21} corresponds to large spanwise-coherent motions and becomes stronger for larger riblets. Figures~\ref{fig.E12uvasRe547}(c-f) show that as the riblet height increases, this added stress protrudes farther into the riblet grooves while maintaining a dominant spectral footprint at $\lambda^+_x \approx 170$. The observed added stress within the grooves is consistent with the decline in drag reduction for these larger riblets.
This observation made for scalloped riblets} is in agreement with previous studies on K-H rollers in flows over filament canopies~\citep{shagar20} and riblets~\citep{endmodgarhutchu21}.

\begin{figure}
        \begin{center}
        \begin{tabular}{cccc}
       \hspace{-.2cm}
        \subfigure[]{\label{fig.EuvProfhs055Re547b}}
        &&
        \hspace{-0.8cm}
        \subfigure[]{\label{fig.DeltaKHhs055Re547}}
        &
        \\[-.5cm]
        \hspace{.2cm}
	\begin{tabular}{c}
        \vspace{.2cm}
        {\small \rotatebox{90}{{$-{uv}_\mathrm{KH}$}}}
       \end{tabular}
       &\hspace{-.4cm}
	\begin{tabular}{c}
       \includegraphics[width=0.4\textwidth]{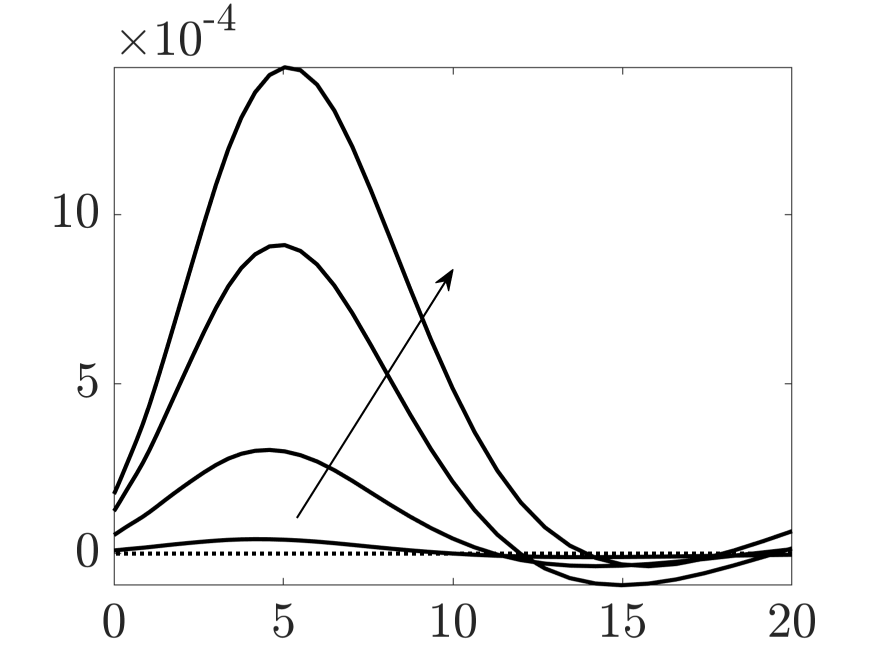}
        \\[-.1cm]
            \hspace{.2cm}
            $\tilde{y}^+$
       \end{tabular}
       &
       \hspace{-.3cm}
	\begin{tabular}{c}
        \vspace{.2cm}
        {\rotatebox{90}{$\Delta uv_{KH}$}}
       \end{tabular}
       &\hspace{-.4cm}
    \begin{tabular}{c}
       \includegraphics[width=0.4\textwidth]{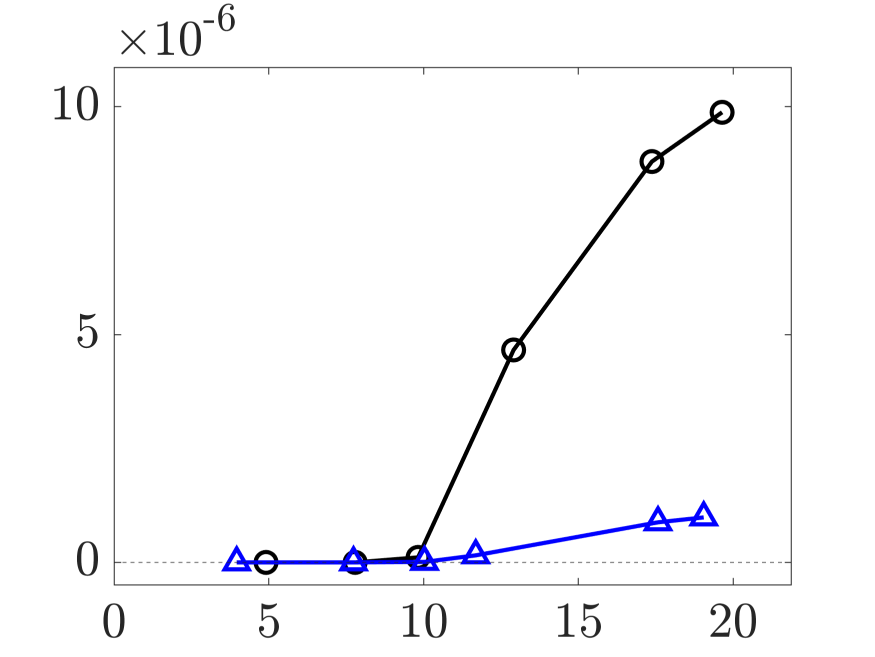}
       \\[-.1cm]
            \hspace{.2cm}
            $l_g^+$
       \end{tabular}
       \end{tabular}
       \end{center}
        \caption{{(a) Modifications to the shear stress $-{uv}_\mathrm{KH}$ resulting from the K-H modes in channel flow with $Re_\tau = 547$ over scalloped riblets with $\alpha/s=0.55$ and $l^+_g \in (10,20)$ ($\omega_z \in [115,230]$). The black dotted line corresponds to smooth channel flow and $l^+_g$ increases in the direction of the arrow. (b) Added shear stress due to K-H modes due to scalloped riblets with $\alpha/s = 0.55 (\bigcirc)$ and triangular riblets with $\alpha/s = 0.87 (\triangle)$.}}
        \label{fig.EuvProfhs055Re547}
\end{figure}

For channel flow with $Re_\tau = 547$ over optimal to large-size scalloped riblets with $\alpha/s=0.55$, {figure~\ref{fig.EuvProfhs055Re547b}} shows the wall-normal profiles of the modifications to the streamwise-wall-normal stress profile ${uv}_\mathrm{KH}$, which can be computed by integrating the riblet-induced modification $\alpha E_{uv,1} + \alpha^2 E_{uv,2}$ over the spectral region associated with the K-H modes (cf.~equation~\eqref{eq.vv-KH}). 
Modifications to the shear stress uniformly peak at $\tilde{y}^+ \approx 5$, which is slightly above the riblet tips, and increase for larger riblets. 
{The observation of the riblet-induced peak in $-uv_{\mathrm{KH}}$ below $\tilde{y}^+\approx 10$ is consistent with the results of~\cite{endmodgarhutchu21} for large and sharp triangular and blade riblets. By further integrating ${uv}_\mathrm{KH}$ over the wall-normal dimension, we quantify the total effect of the K-H instability on the shear stress as $\Delta uv_{KH} \DefinedAs \int_{-1}^{1} - {uv}_\mathrm{KH} \,\mrd \tilde{y}$. Figure~\ref{fig.DeltaKHhs055Re547} compares $\Delta uv_{KH}$ in flow over sharp scalloped and blunt triangular riblets. It is evident that shear stress is strongly influenced by the onset of K–H modes when the scalloped riblets exceed the optimal drag-reducing size of $l_g^+ \approx 10$. In contrast, $\Delta uv_{KH}$ remains consistently low regardless of the size of the triangular riblets. This again highlights the critical role of riblet geometry in the emergence of K–H instabilities~\citep{endmodgarhutchu21}.}

\subsection{Near-wall cycle}
\label{sec.NW}

\begin{figure}
        \begin{center}
        \begin{tabular}{cccccc}
        \hspace{-.6cm}
        \subfigure[]{\label{fig.E12uukxas55O460Re547}}
        &&
        \hspace{-.7cm}
        \subfigure[]{\label{fig.E12uukxas55O175Re547}}
        &&
        \hspace{-.7cm}
        \subfigure[]{\label{fig.E12uukxas55O115Re547}}
        &
        \\[-.5cm]\hspace{-.3cm}
	\begin{tabular}{c}
        \vspace{.3cm}
        {\small \rotatebox{90}{$\tilde{y}^+$}}
       \end{tabular}
       &\hspace{-.3cm}
	\begin{tabular}{c}
       \includegraphics[width=4cm]{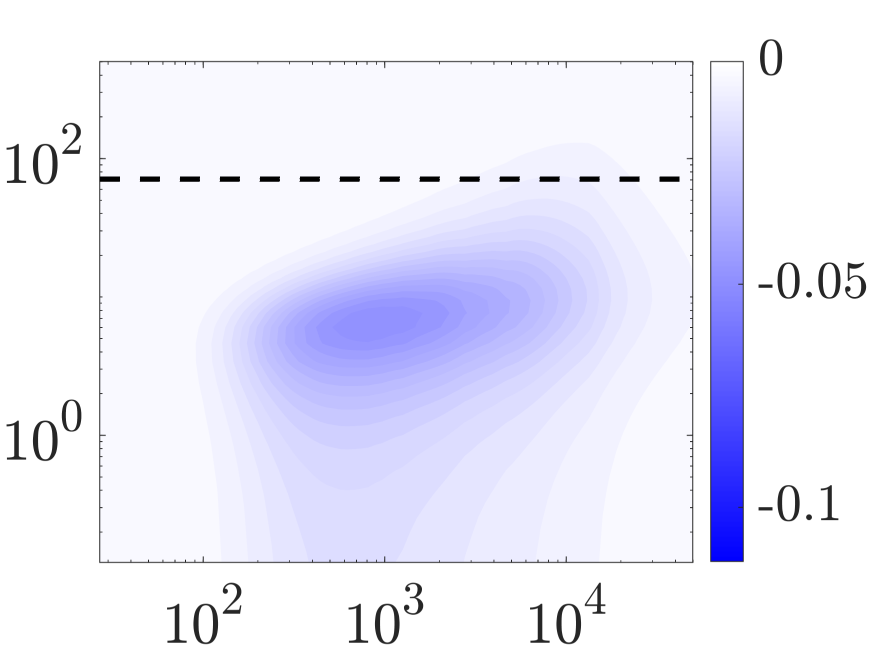}
       \end{tabular}
       &&\hspace{-.3cm}
    \begin{tabular}{c}
       \includegraphics[width=4cm]{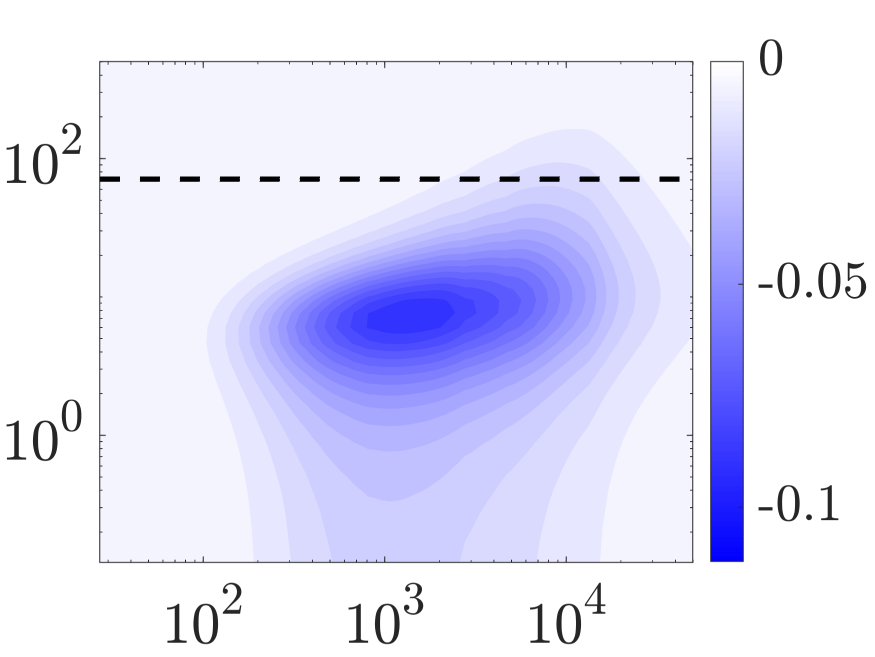}
       \end{tabular}
       &&\hspace{-.3cm}
    \begin{tabular}{c}
       \includegraphics[width=4cm]{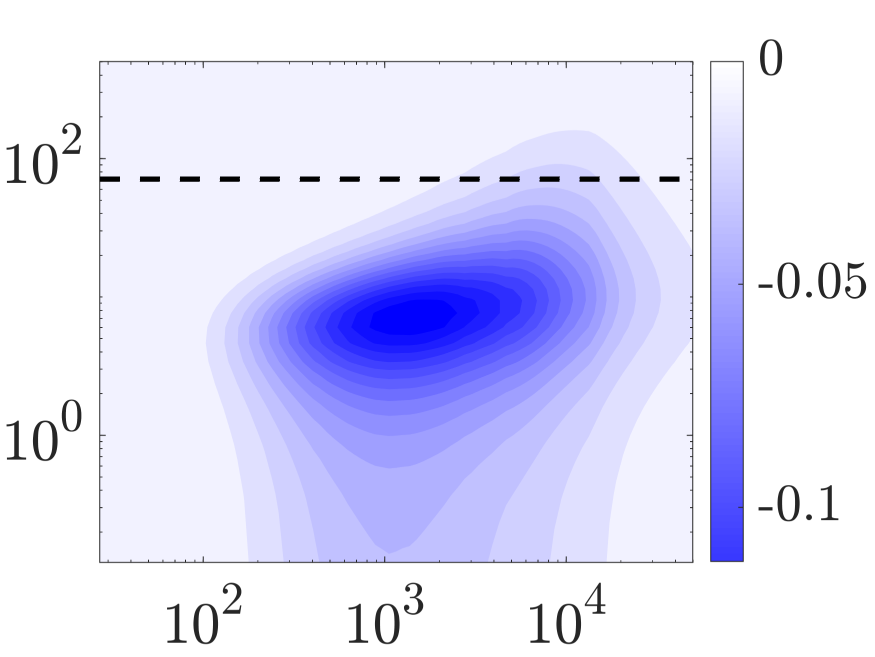}
       \end{tabular}
       \\
        \hspace{-.6cm}
        \subfigure[]{\label{fig.E12uukxas65O115Re547}}
        &&
        \hspace{-.7cm}
        \subfigure[]{\label{fig.E12uukxas87O115Re547}}
        &&
        \hspace{-.7cm}
        \subfigure[]{\label{fig.E12uukxas12O115Re547}}
        &
        \\[-.5cm]\hspace{-.3cm}
	\begin{tabular}{c}
        \vspace{.3cm}
        {\small \rotatebox{90}{$\tilde{y}^+$}}
       \end{tabular}
       &\hspace{-.3cm}
	\begin{tabular}{c}
       \includegraphics[width=4cm]{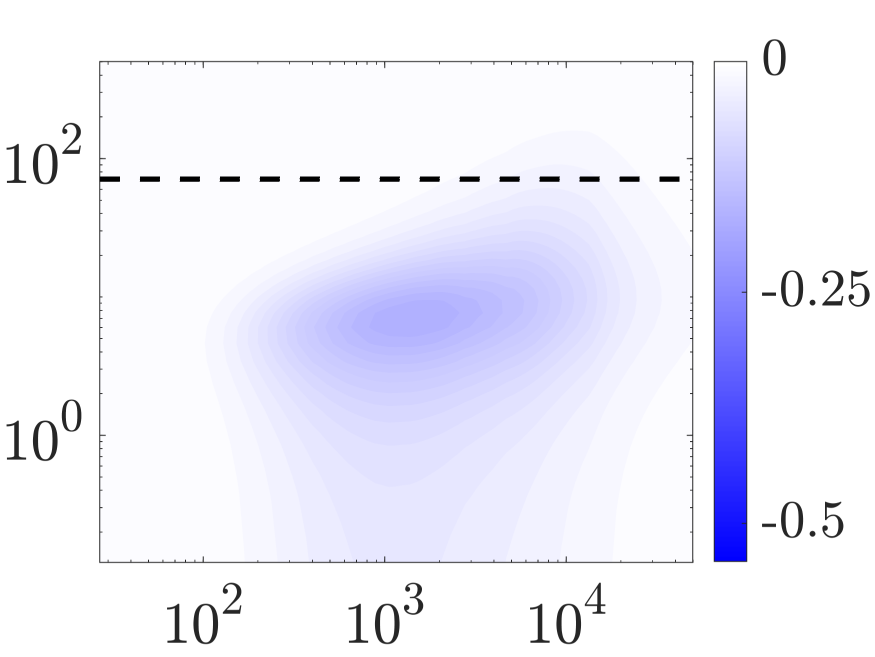}
       \\[-.1cm]
            \hspace{-.2cm}
            $\lambda^+_x$
       \end{tabular}
       &&\hspace{-.3cm}
    \begin{tabular}{c}
       \includegraphics[width=4cm]{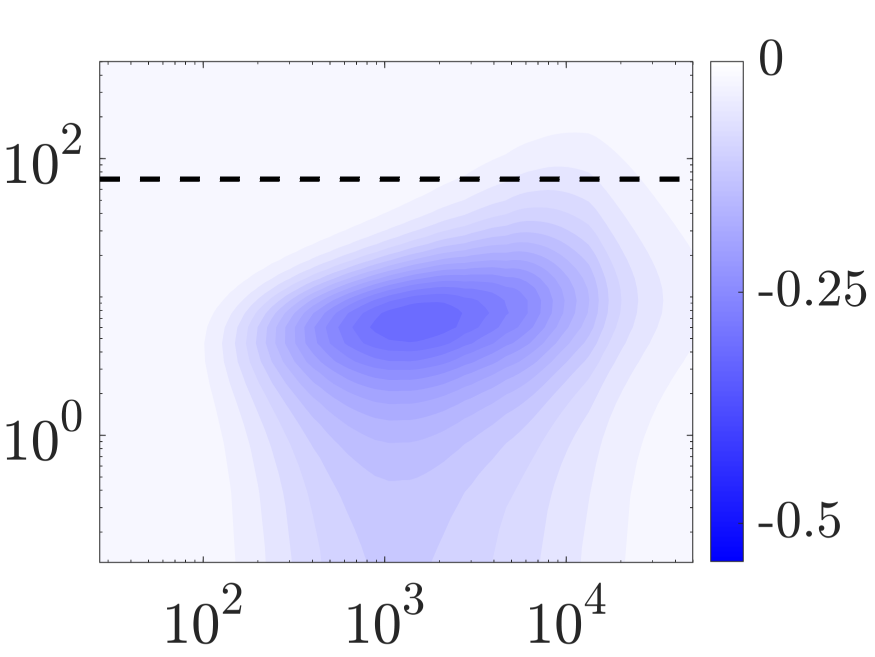}
       \\[-.1cm]
            \hspace{-.2cm}
            $\lambda^+_x$
       \end{tabular}
       &&\hspace{-.3cm}
    \begin{tabular}{c}
       \includegraphics[width=4cm]{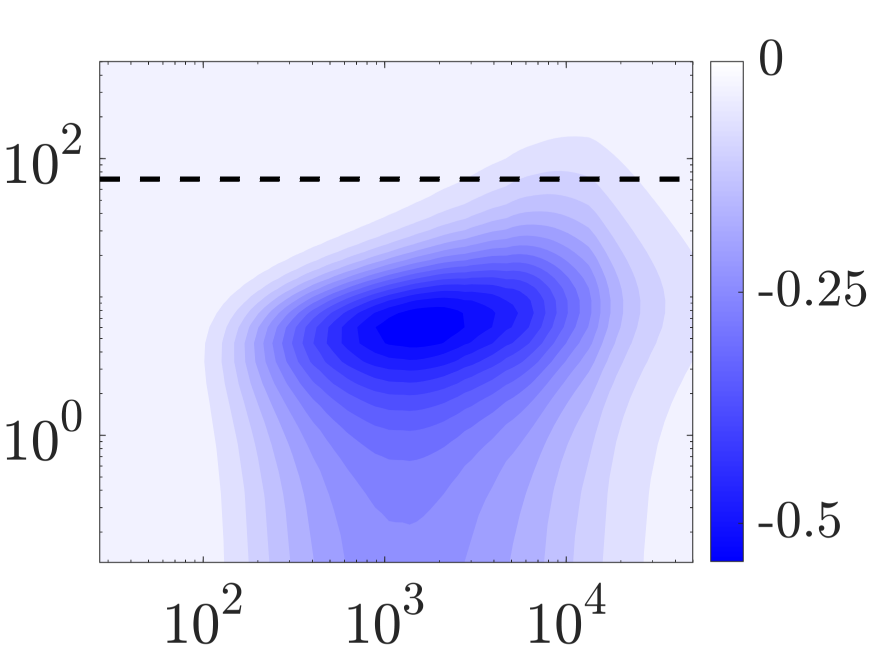}
       \\[-.1cm]
            \hspace{-.2cm}
            $\lambda^+_x$
       \end{tabular}
       \end{tabular}
       \end{center}
        \caption{{Premultiplied modifications to the one-dimensional energy spectrum of streamwise velocity, computed up to $\alpha^2$ as $k_x(\alpha\,E_{uu,1} + \alpha^2\, E_{uu,2})$ and integrated over all spanwise wavelengths in channel flow with $Re_\tau=547$ over scalloped riblets with $\alpha/s = 0.55$ and (a) $l^+_g \approx 5$ ($\omega_z = 460$); (b) $l^+_g \approx 10$ ($\omega_z = 230$); and (c) $l^+_g \approx 20$ ($\omega_z = 115$). (d-f) The same quantity is shown over scalloped riblets with the same viscous spacing but different viscous height: (d) $\alpha/s=0.65$ and $l^+_g \approx 22$ ($\omega_z = 115$); (e) $\alpha/s=0.87$ and $l^+_g \approx 25$ ($\omega_z = 115$); and (f) $\alpha/s=1.2$ and $l^+_g \approx 29$ ($\omega_z = 115$).
        The black dashed lines mark the beginning of the logarithmic layer.}}
    \label{fig.E12uuas055}
\end{figure}

The near-wall regeneration cycle involves the formation of streaks driven by the advection of the mean profile by streamwise vortices, which themselves arise from streak instabilities~\citep{hamkimwal95,jimpin99,schhus02,jim13,hwaben16}. The presence of large sharp riblets can affect the structure and energy of such energetic motions~\citep{endnewmodgarhutchu22}. In this section, we study the energy and spatial attributes of dominant motions that reside close to the corrugated surface. 
Figure~\ref{fig.E12uuas055} shows riblet-induced changes to the premultiplied one-dimensional streamwise energy spectrum $k_x E_{uu}$ integrated over all spanwise wavelengths in channel flow with $Re_\tau = 547$. As the riblet size becomes larger, the large streamwise wavelengths are suppressed in the near-wall region below the logarithmic layer~\citep{endnewmodgarhutchu22}. The similarity between {spectral modification patterns across the horizontal wavenumber space} suggests a potential geometric scaling over all riblet sizes. To study the {dependence on riblet height}, we further increase the riblet size but with a constant spacing in {figures~\ref{fig.E12uuas055}(d-f)}. The modifications to the one-dimensional streamwise energy spectra again expose the suppression of large wavelengths in the near-wall region but with a {deeper reach into the riblet grooves} in the case of taller riblets. In all cases, maximum attenuation happens for {$\lambda^+_x \approx 1500$} at $\tilde{y}^+ \approx 6$, which indicates the suppression of the near-wall portion of streamwise-elongated flow structures.

\begin{figure}
        \begin{center}
        \begin{tabular}{cccccccc}
        \hspace{-.6cm}
        \subfigure[]{\label{fig.E2uukxas55O40Re186}}
        &&
        \hspace{-.8cm}
        \subfigure[]{\label{fig.E2uukxas55O115Re547}}
        &&
        \hspace{-.7cm}
        \subfigure[]{\label{fig.E2uukxas55O200Re934}}
        &&
        \hspace{-.7cm}
        \subfigure[]{\label{fig.E2uukxas55O420Re2003}}
        &
        \\[-.5cm]\hspace{-.3cm}
	\begin{tabular}{c}
        \vspace{.3cm}
        {\scriptsize \rotatebox{90}{$\tilde{y}^+$}}
       \end{tabular}
       &\hspace{-.34cm}
	\begin{tabular}{c}
       \includegraphics[width=2.9cm]{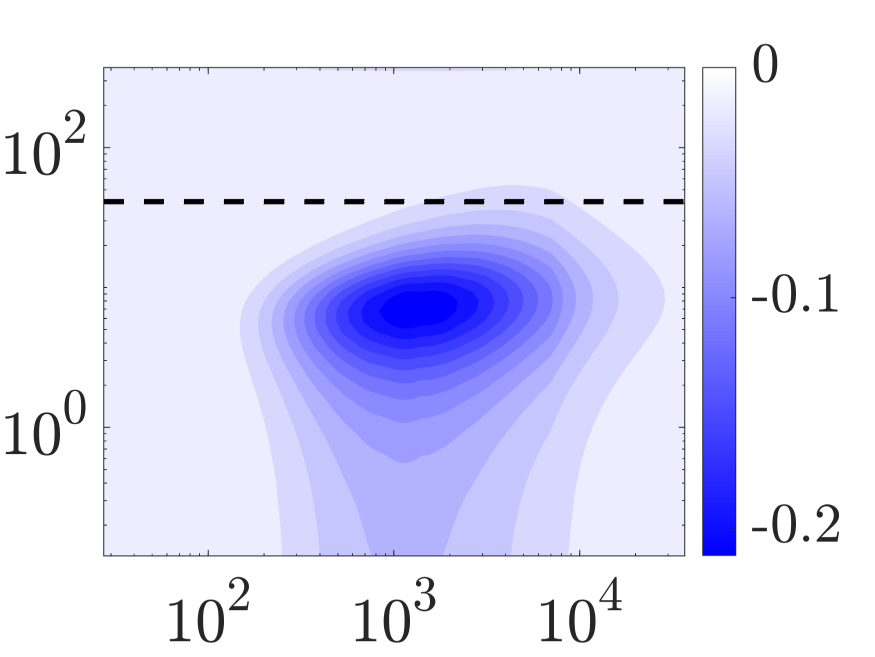}
        \\[-.1cm]
            \hspace{-.1cm}
            {\scriptsize $\lambda_x^+$}
       \end{tabular}
       &&\hspace{-.35cm}
    \begin{tabular}{c}
       \includegraphics[width=2.9cm]{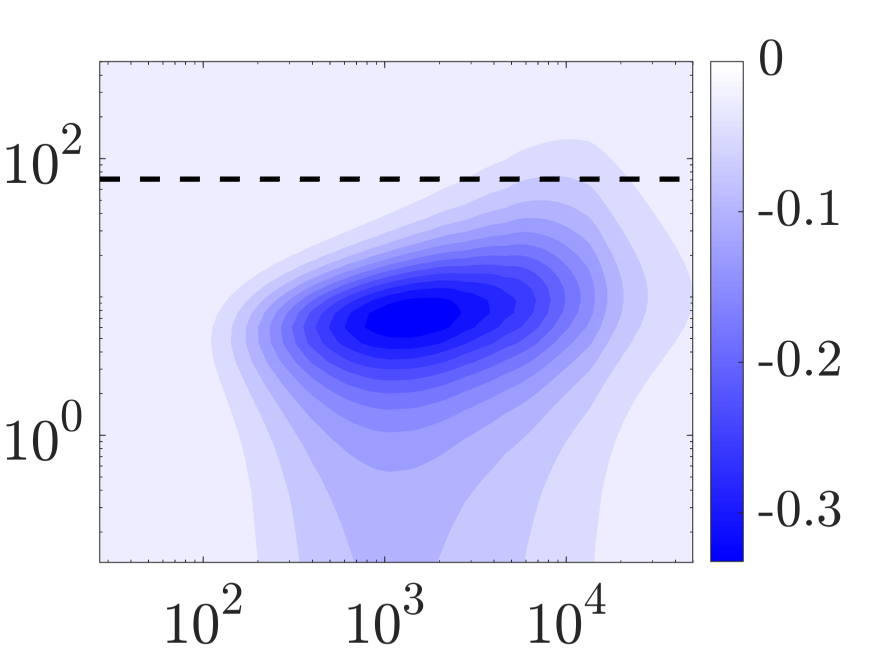}
       \\[-.1cm]
            \hspace{-.1cm}
            {\scriptsize $\lambda_x^+$}
       \end{tabular}
       &&\hspace{-.35cm}
    \begin{tabular}{c}
       \includegraphics[width=2.9cm]{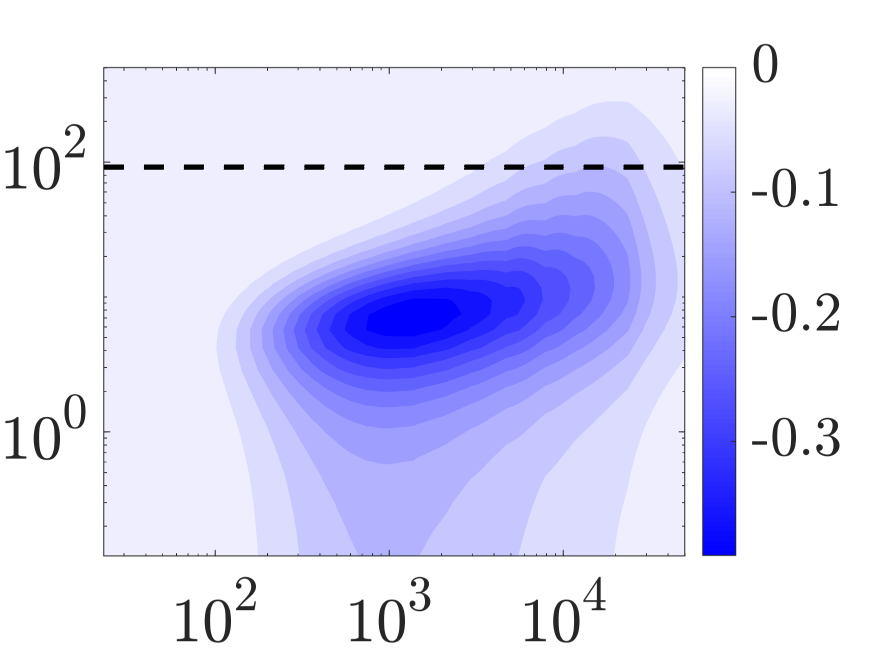}
       \\[-.1cm]
            \hspace{-.1cm}
            {\scriptsize $\lambda_x^+$}
       \end{tabular}
       &&\hspace{-.3cm}
    \begin{tabular}{c}
       \includegraphics[width=2.9cm]{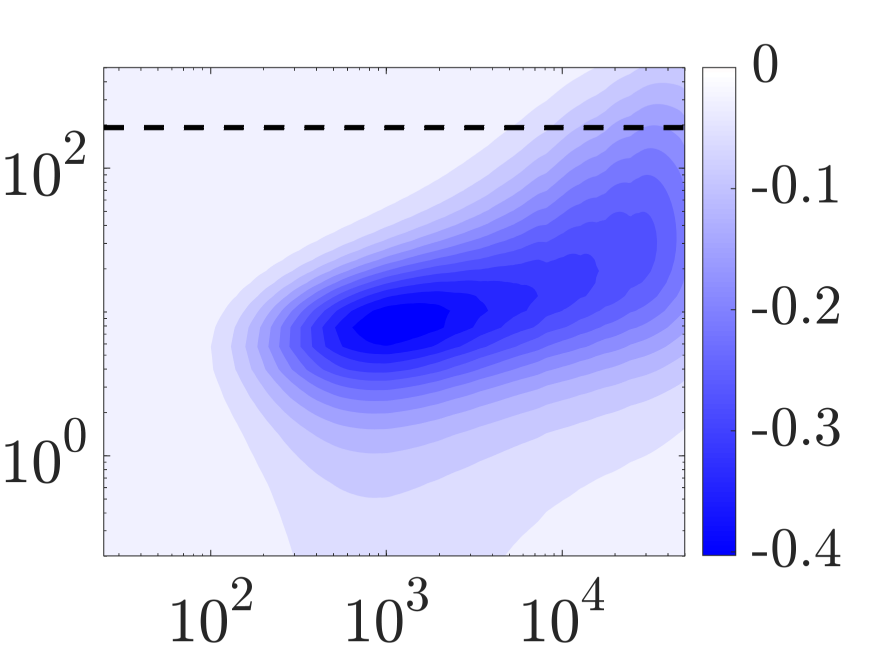}
       \\[-.1cm]
            \hspace{-.1cm}
            {\scriptsize $\lambda_x^+$}
       \end{tabular}
       \end{tabular}
       \end{center}
        \caption{{Premultiplied modifications to the one-dimensional energy spectrum of streamwise velocity, computed up to $\alpha^2$ as $k_x (\alpha\,E_{uu,1} + \alpha^2\,E_{uu,2})$ and integrated over all spanwise wavelengths in channel flow with (a) $Re_\tau=186$, (b) $Re_\tau=547$, (c) $Re_\tau=934$, and (d) $Re_\tau=2003$ over scalloped riblets of $\alpha/s=0.55$ and $l^+_g \approx 20$, which corresponds to spatial frequencies $\omega_z=40$, $115$, $200$, and $420$, respectively. The black dashed lines mark the beginning of the logarithmic layer.}}
    \label{fig.E2uuRe}
\end{figure}

Figure~\ref{fig.E2uuRe} shows the effect of lower-wall scalloped riblets with $\alpha/s = 0.55$ and $l^+_g \approx 20$ on the one-dimensional streamwise energy spectrum of turbulent channel flow with different Reynolds numbers. 
{The spectra have been normalized by their corresponding maxima in smooth channel flow to better understand the effect of riblets. While the streamwise-elongated structures become slightly more suppressed at higher Reynolds number relative to $Re_\tau=186$, the effect of riblets remains largely concentrated to the near-wall region. We note however that the effect of riblets slightly protrudes into the logarithmic layer by $Re_\tau = 2003$ (figure~\ref{fig.E2uukxas55O420Re2003}). The observed large-scale energy suppression is} in agreement with that of~\cite{endnewmodgarhutchu22}, in which DNS of minimal channel flow was combined with experiments to study the effect of large riblets in high-Reynolds-number wall-bounded flows up to $Re_\tau = 1000$. {This} study, however, identified the regions of missing energy at larger streamwise wavelengths ($\lambda_x^+ \approx 5000$).

\begin{figure}
        \begin{center}
        \begin{tabular}{cccc}
        \hspace{-.6cm}
        \subfigure[]{\label{fig.NWhs055O290XZRe547}}
        &&
        \hspace{-1cm}
        \subfigure[]{\label{fig.NWhs055O290YZRe547}}
        &
        \\[-.5cm]
        \hspace{-.3cm}
	\begin{tabular}{c}
        \vspace{.1cm}
        {\small \rotatebox{90}{$\tilde{z}^+$}}
       \end{tabular}
       &\hspace{-.3cm}
	\begin{tabular}{c}
       \includegraphics[width=0.45\textwidth]{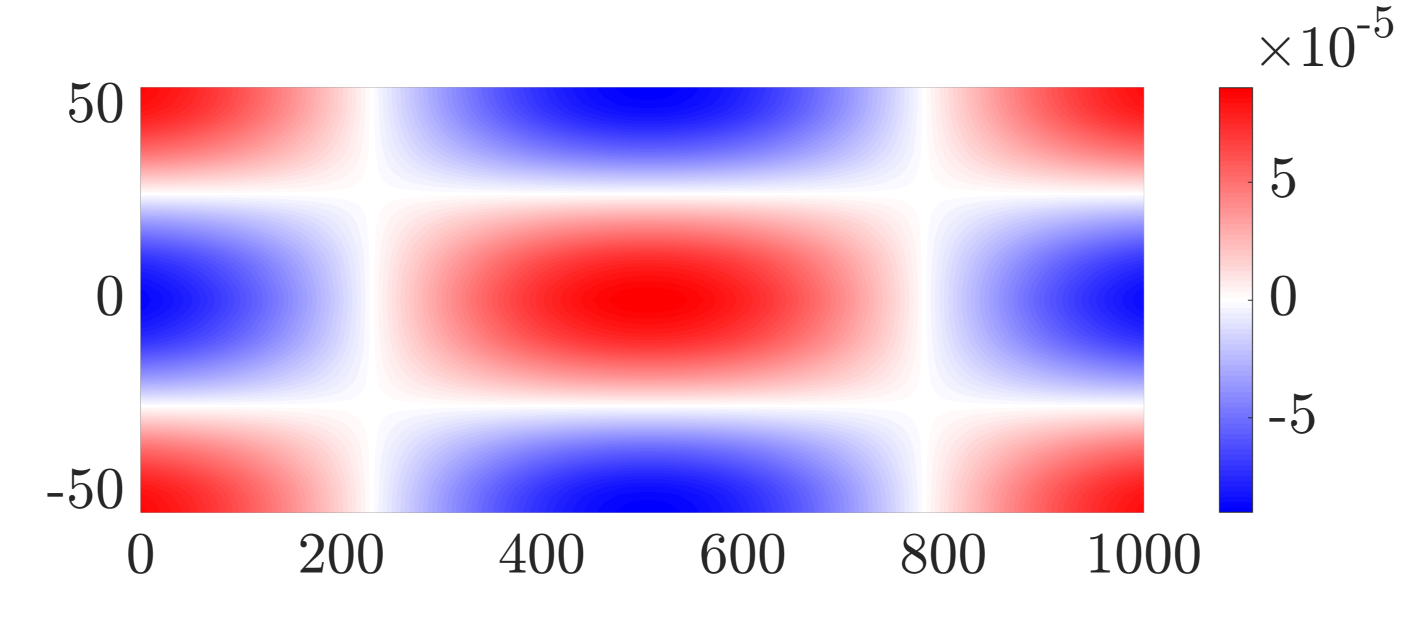}
       \end{tabular}
       &
       \hspace{-.5cm}
       \begin{tabular}{c}
        \vspace{-.2cm}
        {\small \rotatebox{90}{{$\tilde{y}^+$}}}
       \end{tabular}
       &\hspace{-.4cm}
    \begin{tabular}{c}
       \includegraphics[width=0.45\textwidth]{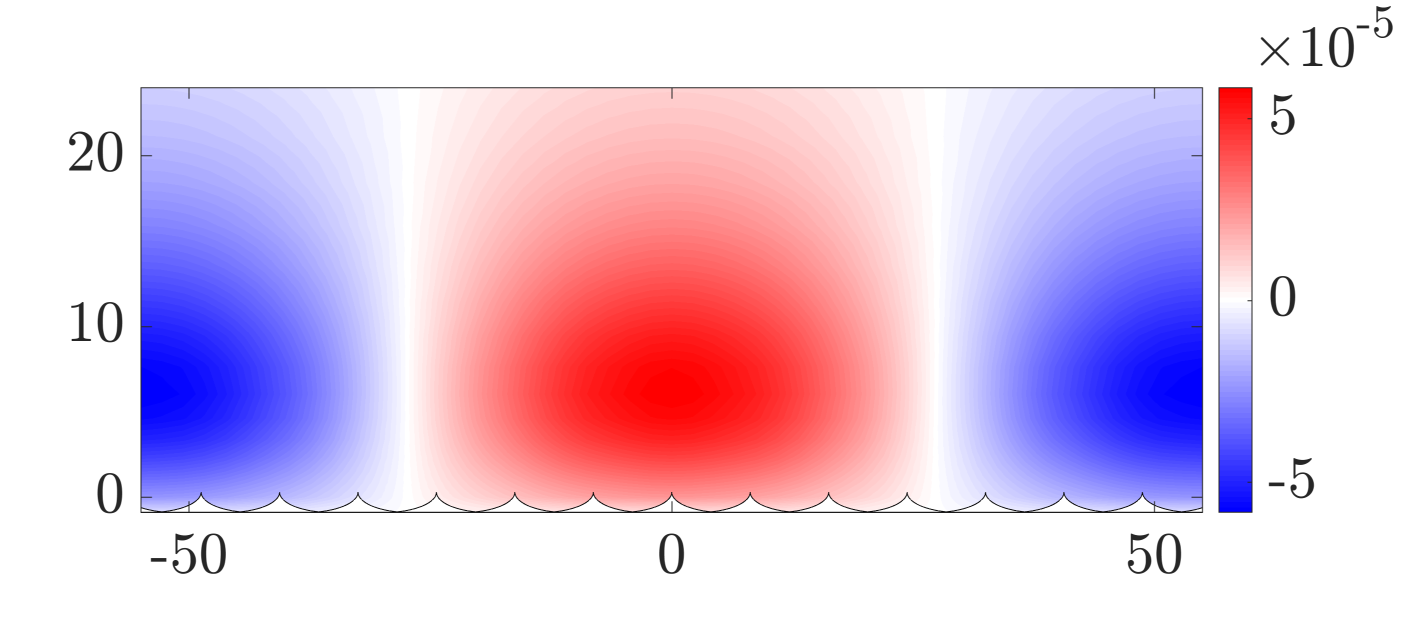}
       \end{tabular}
       \\[-0.1cm]
        \hspace{-.6cm}
        \subfigure[]{\label{fig.NWhs055O175XZRe547}}
        &&
        \hspace{-1cm}
        \subfigure[]{\label{fig.NWhs055O175YZRe547}}
        &
        \\[-.5cm]\hspace{-.3cm}
	\begin{tabular}{c}
        \vspace{.1cm}
        {\small \rotatebox{90}{$\tilde{z}^+$}}
       \end{tabular}
       &\hspace{-.3cm}
	\begin{tabular}{c}
       \includegraphics[width=0.45\textwidth]{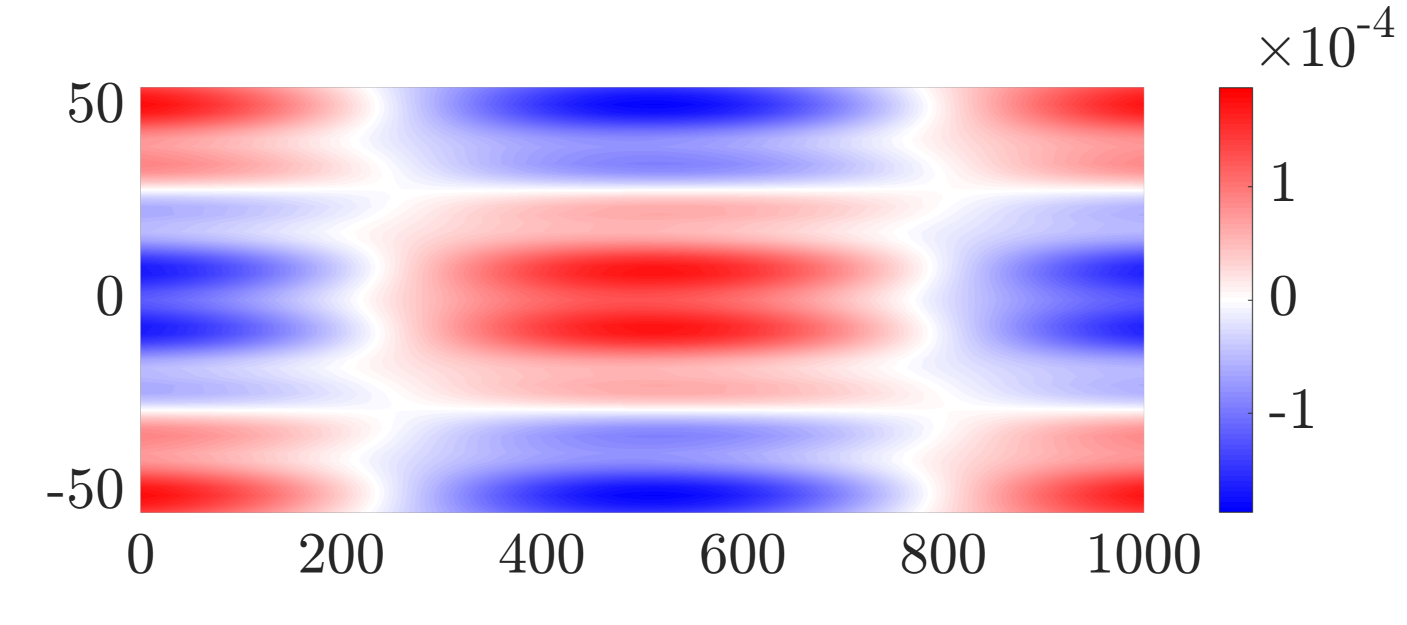}
       \end{tabular}
       &\hspace{-0.5cm}
       \begin{tabular}{c}
        \vspace{-.2cm}
        {\small \rotatebox{90}{{$\tilde{y}^+$}}}
       \end{tabular}
       &\hspace{-.4cm}
    \begin{tabular}{c}
       \includegraphics[width=0.45\textwidth]{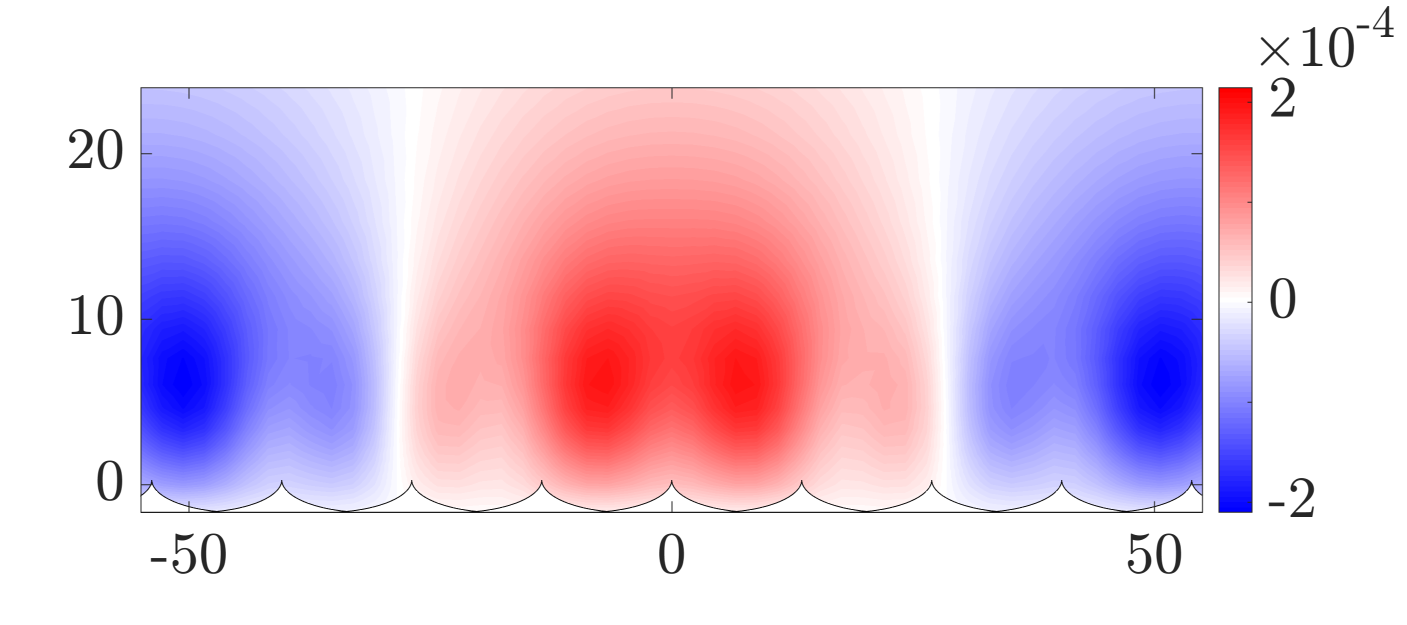}
       \end{tabular}
       \\[-0.1cm]
       \hspace{-.6cm}
        \subfigure[]{\label{fig.NWhs055O115XZRe547}}
        &&
        \hspace{-1cm}
        \subfigure[]{\label{fig.NWhs055O115YZRe547}}
        &
        \\[-.5cm]\hspace{-.3cm}
	\begin{tabular}{c}
        \vspace{.2cm}
        {\small \rotatebox{90}{$\tilde{z}^+$}}
       \end{tabular}
       &\hspace{-.3cm}
	\begin{tabular}{c}
       \includegraphics[width=0.45\textwidth]{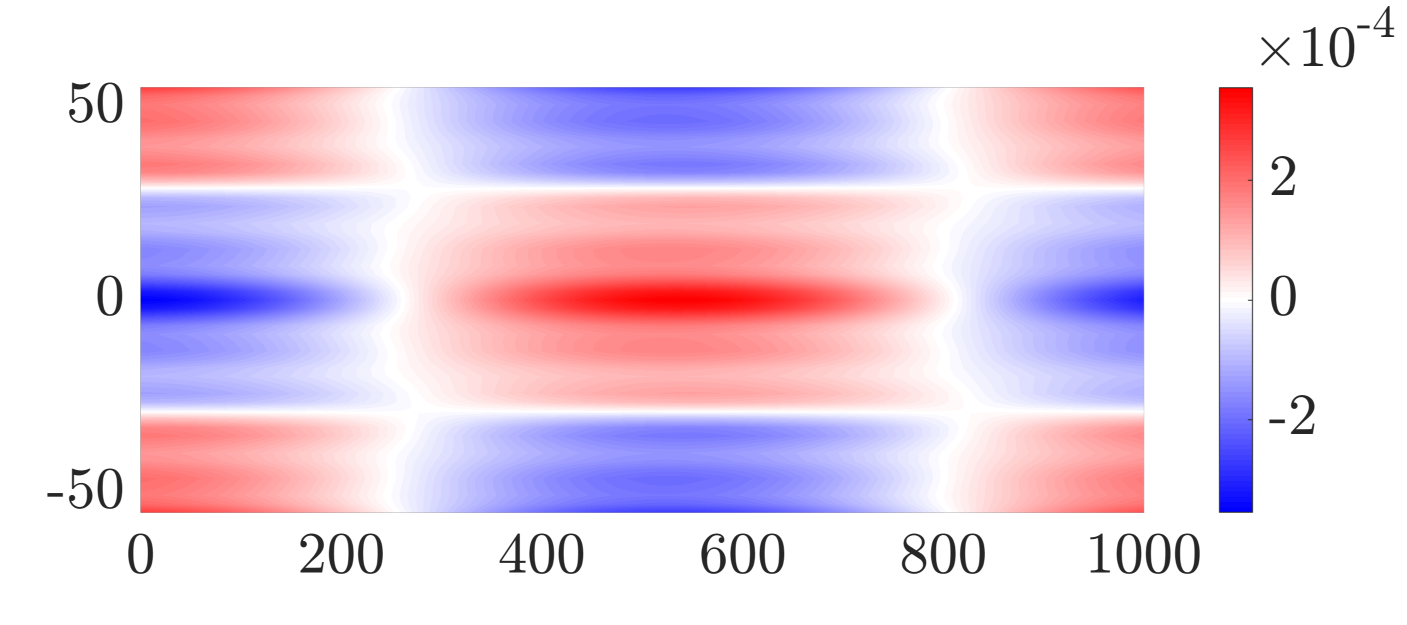}
        \\[-.1cm]
            \hspace{-.1cm}
            $\tilde{x}^+$
       \end{tabular}
       &
       \hspace{-.5cm}
       \begin{tabular}{c}
        \vspace{-.1cm}
        {\small \rotatebox{90}{{$\tilde{y}^+$}}}
       \end{tabular}
       &\hspace{-.4cm}
    \begin{tabular}{c}
       \includegraphics[width=0.45\textwidth]{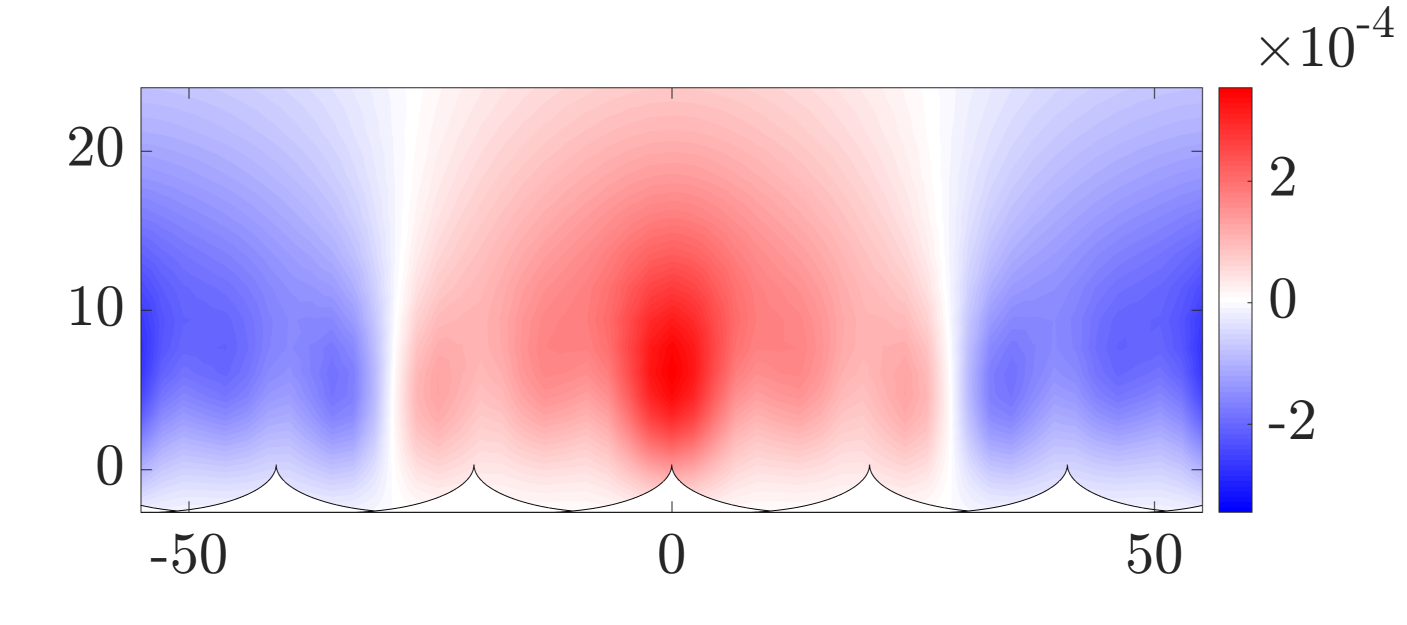}
       \\[-.1cm]
            \hspace{-.1cm}
            $\tilde{z}^+$
       \end{tabular}
       \end{tabular}
       \end{center}
        \caption{{Dominant near-wall flow structures in turbulent channel flow with $Re_\tau=547$ over scalloped riblets with $\alpha/s=0.55$ and (a,b) $l^+_g \approx 5$ $(\omega_z = 460)$; (c,d) $l^+_g \approx 10$ $(\omega_z = 230)$; (e,f) $l^+_g \approx 20$ $(\omega_z = 115)$. The first column shows the streamwise velocity $u$ from the top view ($\tilde{x}$-$\tilde{z}$ plane) one viscous unit above the crest of riblets, i.e., at (a) $\tilde{y}^+ \approx 2$, (c) $\tilde{y}^+ \approx 3$, and (e) $\tilde{y}^+ \approx 5$. The second column shows the same quantity from the cross-plane view ($\tilde{y}$-$\tilde{z}$ plane) at $\tilde{x}^+ = 500$. Here, $(\lambda^+_x,\lambda^+_z) \approx (1100,110)$ corresponds to typical scales of the near-wall cycle, which are extracted from the dominant eigenmode of the covariance matrix $\bPhi_\theta(k_x)$.}}
        \label{fig.NW}
\end{figure}

Figure~\ref{fig.NW} compares the dominant near-wall flow structures for turbulent channel flow with $Re_\tau=547$ over small ($l^+_g \approx 5$), optimal ($l^+_g \approx 10$), and large ($l^+_g \approx 20$) scalloped riblet configurations that share the height to spacing ratio of $\alpha/s=0.55$. The wavelengths of these flow structures, i.e.,  $(\lambda_x^+,\lambda_z^+) \approx (1100,110)$, correspond to the typical length scales that dominate the near-wall cycle~\citep{jimpin99}. The periodicity of the targeted flow structures in the horizontal plane is reflected in the {side} views of figure~\ref{fig.NW} as regions of high and low streamwise velocity that encompass the surface corrugation on the lower wall. As evident from figure~\ref{fig.NWhs055O290YZRe547}, small riblets have little to no effect on the near-wall structures, i.e., the flow structures resemble streamwise-elongated flow structures (streaks) over smooth walls. However, they do push the near-wall structures upward allowing them to interact with the riblet tips. The resulting increase in spanwise friction at the wall restricts the spanwise {meandering of streaks}~\citep{chomoikim93,jimpin99,leelee01} and, ultimately, weakens the quasi-streamwise vortices and reduces drag.
On the other hand, larger riblets distort the streamwise streaks by increasing their lateral span from {$54$} viscous units above small riblets ($l^+_g \approx 5$) to {$58$} viscous units above large ones ($l^+_g \approx 20$); see {figures~\ref{fig.NWhs055O290YZRe547}} and~\ref{fig.NWhs055O115YZRe547}. In the latter case, flow structures penetrate the riblet grooves (figure~\ref{fig.NWhs055O115YZRe547}), increasing drag by exposing a larger surface area to {faster stream above the riblet tips}~\citep{chomoikim93}. 
The reduction in large-scale streamwise energy near the wall in the presence of riblets (figure~\ref{fig.E12uuas055}) is attributed to the interaction between near-wall turbulence and the spanwise-periodic surface corrugation. This interaction induces secondary motions and amplifies cross-flow fluctuations in the vicinity of the riblets~\citep{goltua98,endnewmodgarhutchu22}.

\subsection{The absence of Kelvin-Helmholtz rollers over separated riblets}
\label{sec.separatedriblets}

\begin{figure}
  \centerline{\includegraphics[width=6.5cm]{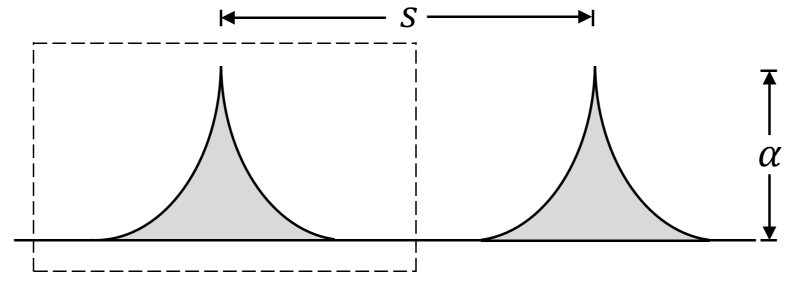}}
  \caption{Separated scalloped riblets of height $\alpha$ and peak to peak spacing $s$. The black dashed box delimits a $2\pi/\omega_z$ period of the riblets.}
\label{fig.separatedrib}
\end{figure}

\begin{figure}
\centering
        \begin{tabular}{ccc}
            \begin{tabular}{c}
            \vspace{0.5cm}
            \hspace{-1.35cm}
            \rotatebox{90}{$-\Delta D/m_l$}
            \end{tabular}
		    &
	        \hspace{-.9cm}
                \vspace{1.5cm}
    		\begin{tabular}{c}
    		        \includegraphics[width=0.45\textwidth]{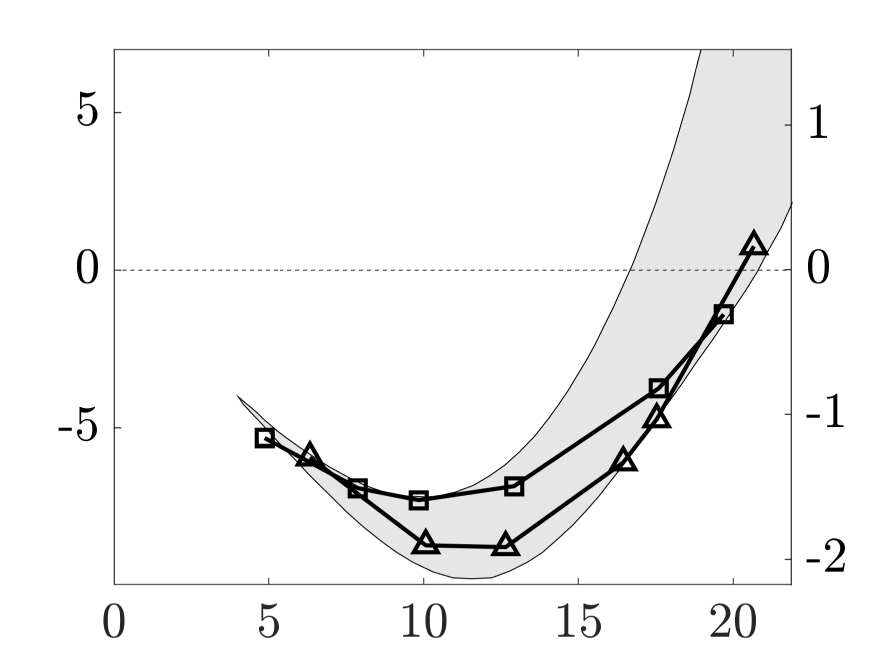}
    	        	\\
    		    \hspace{0.01cm}
    		     $l^+_g$
    	    \end{tabular}
                &
            \begin{tabular}{c}
                \vspace{0.5cm}
                \hspace{-0.75cm}
                \rotatebox{90}{$\Delta U$}
            \end{tabular}
            \end{tabular}
\vspace{-1.5cm}
	\caption{{Normalized drag reduction (left axis) and roughness function (right axis) due to the presence of connected $(\square)$ and separated $(\triangle)$  scalloped riblets with $\alpha/s = 0.55$ on the lower wall of a turbulent channel flow with $Re_\tau = 934$ as a function of $l^+_g$. The shaded region corresponds to the envelope of experimentally measured drag reduction levels from prior studies~\citep{becbruhaghoehop97,garjim11b}.}}
    \label{fig.DRSeparated}
\end{figure}

\begin{figure}
        \begin{center}
        \begin{tabular}{cccccc}
        \hspace{-.6cm}
        \subfigure[]{\label{fig.E12vvas55O780ypRe934spaced}}
        &&
        \hspace{-.75cm}
        \subfigure[]{\label{fig.E12vvas55O390ypRe934spaced}}
        &&
        \hspace{-.75cm}
        \subfigure[]{\label{fig.E12vvas55O200ypRe934spaced}}
        &
        \\[-.5cm]\hspace{-.3cm}
	\begin{tabular}{c}
        \vspace{.2cm}
        {\small \rotatebox{0}{$\lambda^+_z$}}
       \end{tabular}
       &\hspace{-.4cm}
	\begin{tabular}{c}
       \includegraphics[width=4cm]{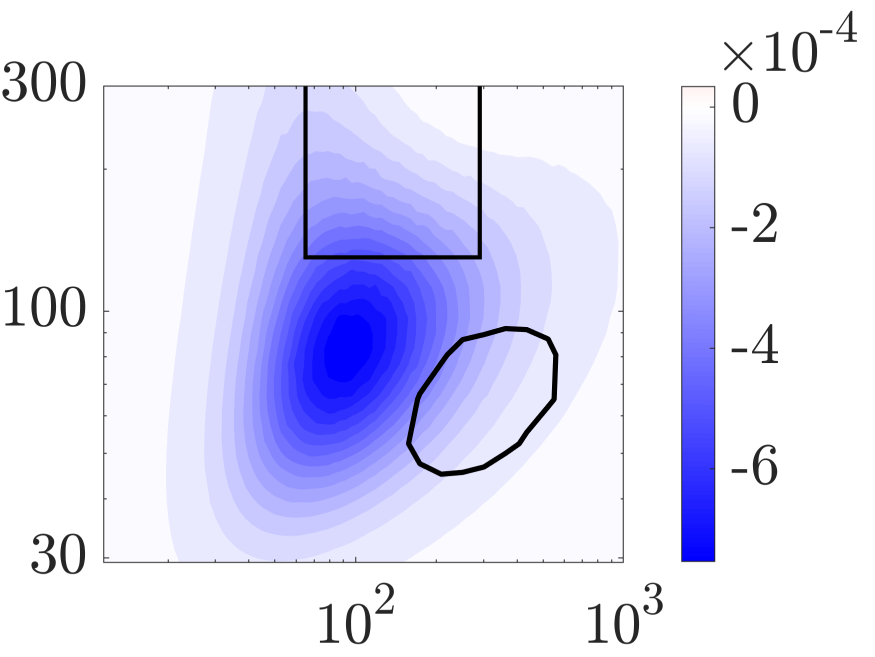}
       \\[-.2cm]
            \hspace{-.2cm}
            $\lambda^+_x$
       \end{tabular}
       &&\hspace{-.32cm}
    \begin{tabular}{c}
       \includegraphics[width=4cm]{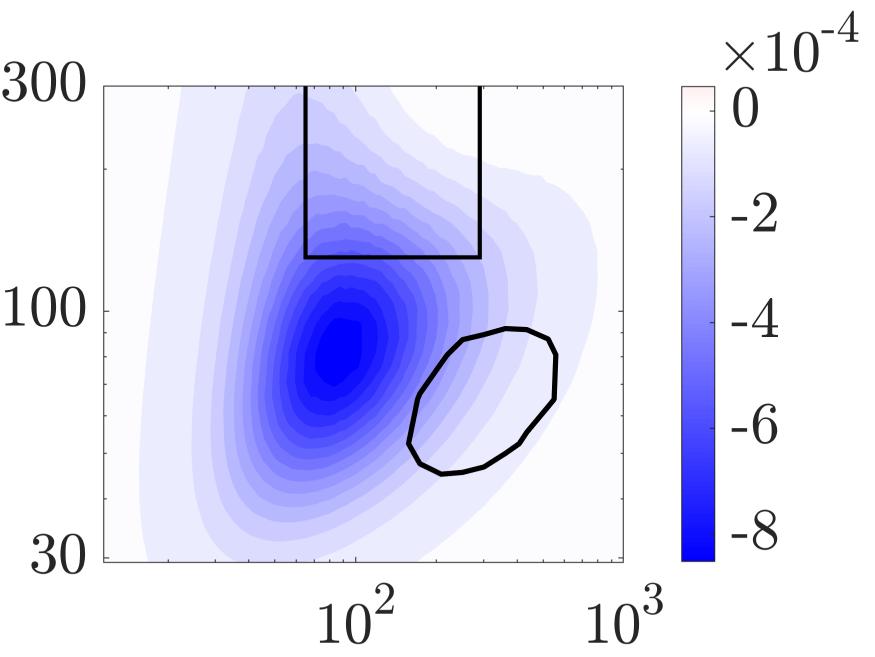}
       \\[-.2cm]
            \hspace{-.2cm}
            $\lambda^+_x$
       \end{tabular}
       &&\hspace{-.35cm}
    \begin{tabular}{c}
       \includegraphics[width=4cm]{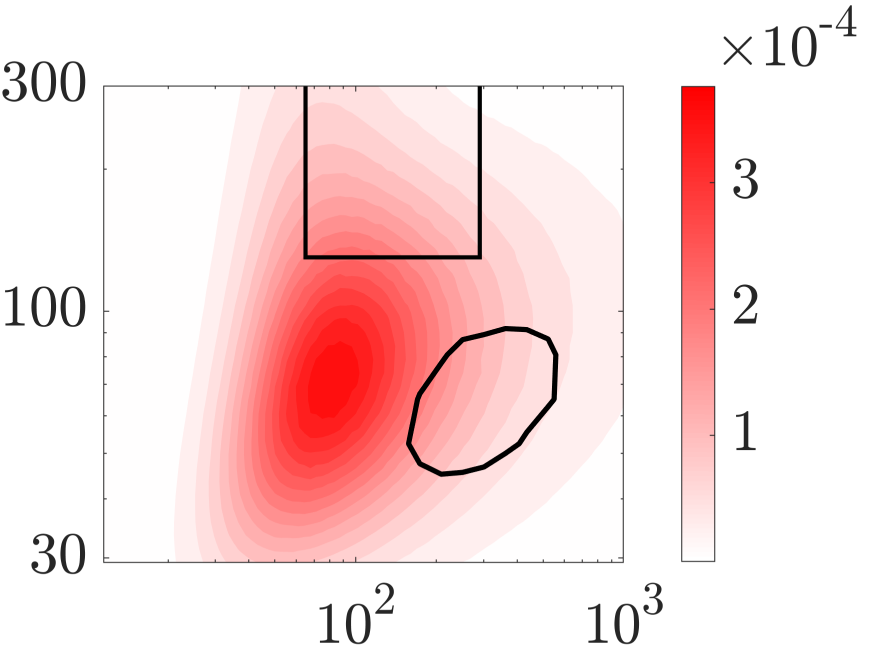}
       \\[-.2cm]
            \hspace{-.2cm}
            $\lambda^+_x$
       \end{tabular}
       \end{tabular}
       \end{center}
        \caption{{Premultiplied modifications to the wall-normal energy spectrum,  computed up to $\alpha^2$ as $k_x \theta (\alpha\,E_{vv,1} + \alpha^2\,E_{vv,2})$, of turbulent channel flow with $Re_\tau=934$ one viscous unit above the crest of separated scalloped riblets with $\alpha/s=0.55$ and (a) $l^+_g \approx 6$ $(\omega_z = 390)$, (b) $l^+_g \approx 12$ $(\omega_z = 195)$, and (c) $l^+_g \approx 20$ $(\omega_z = 100)$. Black open boxes delimit the spectral window of K-H rollers and black contour lines correspond to the $80\%$ contour level of the energy spectrum of smooth channel flow.}}
    \label{fig.E12vvspaced}
\end{figure}

\begin{figure}
\centering
    \begin{tabular}{cc}
        \begin{tabular}{c}
        \vspace{0.5cm}
        \hspace{-1.45cm}
        \rotatebox{90}{$\Delta uv_{KH}$}
        \end{tabular}
        &
        \hspace{-.9cm}
            \vspace{1.5cm}
        \begin{tabular}{c}
            \includegraphics[width=0.45\textwidth]{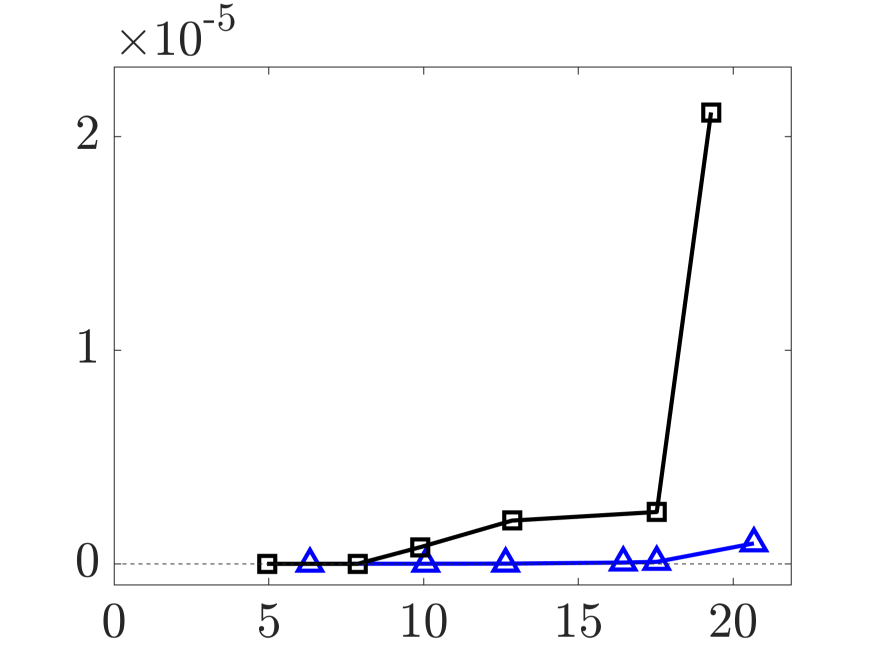}
                \\
            \hspace{0.01cm}
             $l^+_g$
        \end{tabular}
        \end{tabular}
\vspace{-1.5cm}
	\caption{{Added shear stress due to K-H modes in channel flow with $Re_\tau = 934$ over connected $(\square)$ and separated $(\triangle)$ scalloped riblets with $\alpha/s = 0.55$.}}
    \label{fig.DeltaKHhs055Re934}
\end{figure}

As demonstrated in \S~\ref{sec.KH}, large riblets generate spanwise rollers driven by the K-H instability. In this subsection, we analyze the spectral signature of these K-H rollers in the presence of separated scalloped riblets. 
To achieve separation, we reduce the spanwise coverage of the scalloped riblets within one period, creating periodic roughness elements separated by a spanwise distance of $\pi/\omega_z$; see figure~\ref{fig.separatedrib}. {As shown in figure~\ref{fig.DRSeparated}, the separated scalloped riblets provide greater drag reduction than the connected scalloped riblets, but they exhibit a similar breakdown of the linear viscous regime, leading to a decline in drag reduction performance.
Given that the most prominent signature for the onset of K–H instability is captured by the premultiplied wall-normal energy spectrum (cf. figure~\ref{fig.E2vvbeta75}), we examine this quantity to elucidate the role of K–H instability in the degradation of drag reduction.}
Figure~\ref{fig.E12vvspaced} shows the effect of separated scalloped riblets with $\alpha/s = 0.55$ and different sizes on the premultiplied wall-normal energy spectrum of channel flow with $Re_\tau = 934$ one viscous unit above the riblet crests. {Separated riblets of small and optimal size suppress the wall-normal energy across all wavelengths (figures~\ref{fig.E12vvas55O780ypRe934spaced} and~\ref{fig.E12vvas55O390ypRe934spaced}).}
In contrast to large connected riblets that amplify the wall-normal energy in the spectral range corresponding to spanwise rollers (figure~\ref{fig.E2vvas55O200ypRe934}), large separated riblets amplify the energy outside this range (figure~\ref{fig.E12vvas55O200ypRe934spaced}){, leaving the K-H spectral region largely untouched.}
This effect, which can be related to the penetration of turbulence in the riblet valleys {and weakening of the mixing layer at the riblet tips}, is in agreement with the findings of~\cite{endmodgarhutchu21}.
{We note that since the case of $\alpha/s=0.55$ and $l_g^+\approx 20$ corresponds to a drag-increasing riblet configuration (cf. figure~\ref{fig.DRSeparated}), the weak footprint of K-H modes is indicative of an alternative destructive mechanism, e.g., dispersive stresses~\citep{goltua98,modendhutchu21}. However, we do not account for such components of the velocity field in this study. Finally, we isolate the additional Reynolds shear stress due to K–H modes in the flow at $Re_\tau = 934$ over connected and separated scalloped riblets with $\alpha/s=0.55$ (figure~\ref{fig.DeltaKHhs055Re934}). As expected, the added shear stress due to separated riblets is insignificant.}

\section{Concluding remarks}
\label{sec.conclusion}

We develop a model-based framework for investigating the effects of surface corrugation{, in the form of streamwise-elongated riblets, on turbulent channel flow.} Our approach combines the turbulence modeling technique {of~\cite{moajovJFM12}} with a domain transformation that accurately represents {the surface geometry}. The turbulence modeling technique uses the second-order statistics of velocity fluctuations to modify the turbulent eddy-viscosity of smooth channel flow. As the domain transformation reflects the spatial periodicity of the boundary conditions onto the differential operators, it significantly increases the computational complexity of the harmonic equations that need to be solved for the statistical response of the linearized dynamics. To manage this complexity, we {employ the riblet} height as a small parameter, enabling a perturbation analysis of flow quantities. This approach breaks the dimensional complexity of the governing equations over smaller, more-manageable equations {that are of the same size as those in smooth channel flow,} facilitating the use of more collocation points near sharp {tips} in high-Reynolds-number flows.

Our perturbation analysis enables the calculation of riblet-induced modifications to the second-order statistics and, thus, the energy spectrum of turbulent channel flows, capturing the effect of riblets on various spatial wavelengths. By employing an appropriate turbulence model, we extend this analysis to the turbulent viscosity and the mean velocity, from which skin-friction drag can be computed. {We apply our framework to a range of triangular and scalloped riblet geometries, with particular emphasis on the more challenging scalloped shapes, leveraging the computational efficiency afforded by perturbation analysis. Our model-based} predictions closely align with drag-reduction trends reported by previous high-fidelity simulations and experiments. Furthermore, we analyze the primary flow mechanisms affecting drag reduction for large and sharp riblets. By examining riblet-induced changes to the energy spectrum, we assess how riblet size and height influence the amplification of K-H rollers at different Reynolds numbers. Our findings reveal that spectral evidence for these spanwise-coherent flow structures diminishes as the separation between roughness elements increases. Finally, we explore the effect of large riblets in suppressing the energy of streaks close to the wall. 

This work marks a significant step toward establishing a unified framework for low-complexity modeling of turbulent flows over rough surfaces. {Our framework relies on several simplifying assumptions: linear fluctuation dynamics, white-in-time representation of the stochastic effects of nonlinearity, the absence of turbulence ($\nu_T=0$) within riblet grooves below $\tilde{y}=-1$, the small riblet height condition required for perturbation analysis, and the absence of wall-normal and spanwise dispersive velocities from the linearized flow analysis. While such assumption are enabling, they can also limit the applicability of our framework in studying large riblets that allow turbulence to protrude deep into their grooves.}
Our ongoing efforts are directed at leveraging this model-based framework to predict the impact of riblet erosion, such as tip roundness and height loss~\citetext{{\citealp{leifeischfla23}}; \citealp*{pacdueadj24}}, on the performance of ariel vehicles~\citep{bilbelshywanbriquimcg24}. The integration of alternative turbulent viscosity models (e.g.,~\cite*{laszamgan24}), along with the incorporation of data from numerical simulations and experiments (e.g.,~\cite{zarchejovgeoTAC17,zarjovgeoJFM17,zargeojovARC20}), can further enhance the predictive accuracy of our framework and playing a crucial role in broadening its applicability across various engineering applications.

\section*{Acknowledgments}
Financial support from the Air Force Office of Scientific Research under award FA9550-23-1-0219 is gratefully acknowledged. The Office of Information Technology Cyberinfrastructure Research Computing (CIRC) at The University of Texas at Dallas and the Texas Advanced Computing Center are acknowledged for providing computing resources.

\section*{Declaration of Interests} The authors report no conflict of interest.

\section*{Author ORCIDs}Armin Zare, \hyperlink{https://orcid.org/0000-0002-3532-5767}{https://orcid.org/0000-0002-3532-5767},
Mohammadamin Naseri, \hyperlink{https://orcid.org/0000-0002-4646-2935}{https://orcid.org/0000-0002-4646-2935}

\appendix
\section{The operators {$\cA_\theta$}, $\cB_\theta$, and $\cC_\theta$ in equations~\eqref{eq.lnse}}\label{app.opABC}
The dynamical generator matrix $\cA_\theta(k_x)$ in evolution model~\eqref{eq.lnse} has the bi-infinite structure
\begin{align}
\label{eq.opA}
        \cA_\theta(k_x) 
        \, = \,
        \begin{bmatrix}
        \ddots & \vdots  & \vdots  & \vdots  & \iddots
        \\[.15cm]  
         \cdots & A_{0}(\theta_{n-1}) & A_{-1}(\theta_{n}) & A_{-2}(\theta_{n+1}) &\cdots 
         \\[.15cm] 
         \cdots & A_{1}(\theta_{n-1}) & A_{0}(\theta_{n}) & A_{-1}(\theta_{n+1}) &\cdots 
         \\[.15cm] 
         \cdots & A_{2}(\theta_{n-1}) & A_{1}(\theta_{n}) & A_{0}(\theta_{n+1}) &\cdots 
         \\[.15cm]  
        \iddots & \vdots  & \vdots  & \vdots  &\ddots
        \end{bmatrix},
\end{align}
where the block operator $A_{m}(\theta_n)$ accounts for the influence of the ($m+n$)th harmonic of the state, $\hat{\bpsi}_{m+n}$, on the dynamics of the $n$th harmonic of the state, $\hat{\bpsi}_{n}$. Each block operator takes the $2\times 2$ form:
\begin{align*}
    A_{m}(\theta_n)
        \, = \,
        \tbt{A_{m,1,1}(k_x,\theta_n)}{A_{m,1,2}(k_x,\theta_n)}{A_{m,2,1}(k_x,\theta_n)}{A_{m,2,2}(k_x,\theta_n)}.
\end{align*}
For the block operator on the main diagonal of $\cA_\theta(k_x)$ ($m=0$),
\begin{align}
\label{eq.opAmaindiag}
    \ba{rcl}
        A_{0,1,1}(k_x,\theta_n)
        &\!\!=\!\!&
        \dfrac{\Delta^{-1}_n}{Re_\tau} \Big[ (1\,+\,\nu_T)\Delta^2_{n}
        \,+\,
        2{\nu_{T}''}{\partial_{yy}}
        \,-\,
        \nu_{T}''\Delta_n
       \,+\,
        2\nu_{T}'\Delta_n \partial_{y} \Big] 
        \\[0.25cm]
        && ~ + \;
        \Delta^{-1}_{n}\Big[
        \mri k_x(U''_{0}\,-\,
         U_0\Delta_n)\Big] \,+\, \Gamma_{0,1,1},
        \\[0.25cm]
        A_{0,1,2}(k_x,\theta_n)
        &\!\!=\!\!&
        \Gamma_{0,1,2}, 
        \\[0.25cm]
        A_{0,2,1}(k_x,\theta_n)
        &\!\!=\!\!&
        -\, \mri \theta_n U'_{0} 
        \,+\,
        \Gamma_{0,2,1},
        \\[0.25cm]
        A_{0,2,2}(k_x,\theta_n)
        &\!\!=\!\!&
        \dfrac{1}{Re_\tau}  \left[
        (1\,+\,\nu_{T})\Delta_n
        \,+\,
        \nu_{T}'{\partial_y} \right]
        \,-\,
        \mri k_x U_0 
        \,+\,
        \Gamma_{0,2,2},
    \ea
\end{align}
and otherwise ($m\neq 0$),
\begin{align}
\label{eq.opAoffdiag}
    \ba{lllll}
        A_{m,1,1}(k_x,\theta_n)
        &\!\!=\!\!&
        \Gamma_{m,1,1}, \qquad
        A_{m,1,2}(k_x,\theta_n)
        &\!\!=\!\!&
        \Gamma_{m,1,2}, 
        \\[0.15cm]
        A_{m,2,1}(k_x,\theta_n)
        &\!\!=\!\!&
        \Gamma_{m,2,1}, \qquad
        A_{m,2,2}(k_x,\theta_n)
        &\!\!=\!\!&
        \Gamma_{m,2,2}.
    \ea
\end{align}
In equations~\eqref{eq.opAmaindiag} and~\eqref{eq.opAoffdiag},
{
\begin{align*}
    \ba{rcl}
        \Gamma_{m,1,1}
        &\!\!=\!\!&
        \dfrac{\Delta^{-1}_n}{Re_\tau} \Big[ F'_m * \big( 2\nu_{T}'\Delta_{m+n} \partial_{y} \,+\, 2\nu_{T}''(\Delta_{m+n} + \partial_{yy})
        \, - \,
        2\nu_{T}''' \partial_{yy} \big)
        \\[0.25cm]
        && ~ + \,
        \mri k_x (F'_m * U''_{m})
         +  2 F'_m * U'_m \partial_y) 
         + 
        \mri k_x\big(m^2\omega_z^2 U_m \,-\, U_m * \Delta_{m+n} \, - \, U''_m{\partial_{y}}\big) \Big],
        \\[0.25cm]
        \Gamma_{m,1,2}
        &\!\!=\!\!&
        \Delta^{-1}_n\Big[
        2 \dfrac{ \omega_z k_x}{k^2_{m+n}}( F'_m * (F_m * U''_m) + 2 F'_m * U'_m
         +  
        2 F'_m * (F_m * U'_m + U_m) \partial_{y})\Big],
        \\[0.45cm]
        \Gamma_{m,2,1}
        &\!\!=\!\!&
        -\,\mri \omega_z( F'_m * (F_m * U'_m) \partial_y + F'_m * (F_m * U''_m) + F'_m * U'_m) \,-\, \mri \theta_{m+n} F'_m * U'_m 
        \\[0.25cm]
        && ~ + \;
        \Big[
        \dfrac{-\mri \omega_z}{k_xk^2_{m+n}}(\mri F_m * (U'_m * F'_m) \partial_{yyy}
        \,-\,
         \theta_{m+n} F_m * (U'_m * F'_m) \partial_{yy}
         \\[0.45cm]
        && ~ - \;
         \omega_z F_m * (U_m * F'_m) \partial_{yyy}
        - \mri \theta_{m+n} \omega_z F_m * (F_m * F'_m) \partial_{yy} 
        \\[0.3cm]
        && ~ + \;
        \mri F_m * (U''_m * F'_m) \partial_{yy} - 2 \omega_z F_m * (U'_m * F'_m) \partial_{yy}
        +
        \mri\, \omega_z^2 U_m * (F_m * F'_m) {\partial_{yy}})
        \\[0.25cm]
        && ~ - \;
        \dfrac{\mri \omega_z k_x^2}{k^2_{m+n}}( F_m * (U'_m * F'_m) {\partial_{y}} 
        \,+\, U_m * F'_m {\partial_{y}}) \Big],
        \\[0.25cm]
        \Gamma_{m,2,2}
        &\!\!=\!\!&
        \dfrac{1}{Re_\tau}  \, \left[
        (1\,+\,\nu_{T})\Delta_{m+n}
        \,+\,
        F'_m * F'_m \,\nu_{T}'\,{\partial_y} \right]
        \,-\,
        \mri k_x U_m \,+\,
        \left(\mri m\,\omega_z F_m * (F_m * U'_m) \right.
        \\[0.25cm]
        && \left. ~ - \,
        \omega_z^2 F_m * U_m \right)\partial_{y}
        \,+\, \dfrac{\omega_z}{k^2_{m+n}}\left(\mri F_m * U'_m \,-\, \theta_{m+n} U_m \right.
        \\[0.45cm]
        && \left. ~ +\,
        \mri F_m * (F_m * U''_m) \,-\,
        2 \mri \omega_z F_m * U'_m \,-\, \omega_z U_m \,-\,
        \mri F_m * U'_m \,+\, \mri k_x U_m\right),
    \ea
\end{align*}
}
with $\theta_{m+n} = \theta + (m+n)\omega_z$, $k^2_{m+n} = k_x^2 + \theta_{m+n}^2$, and $\Delta_{m+n} = F'^2_m \partial_{yy} - k^2_{m+n}${, and $*$ denotes the convolution of two Fourier series expansions, e.g., $F_m * U_m$ corresponds to
\begin{align*}
        F_m * U_m 
            \;=\,
        \ds{\sum\limits_{m=-T}^{T} \sum\limits_{l=-T}^{T+m} l\,(m-l) {F}_l\, {U}_{m-l}},
\end{align*}
where $T$ denotes the truncation level in the Fourier expansions.}

The input and output matrices in equations~\eqref{eq.lnse} take the block diagonal forms $\cB_\theta(k_x) = \diag\left \{ B(k_x,\theta_n) \right \}_{n\in\bbZ}$ and $\cC_\theta(k_x) =\diag\left \{ C(k_x,\theta_n) \right \}_{n\in\bbZ}$, respectively, with
\begin{align}
\label{eq.Binveta}
    B(k_x,\theta_n) \;=\;
    \tbo{B_v}{B_\eta} \;=\;
    \begin{bmatrix}
     -\mri k_x \Delta^{-1}_n \partial_y& -\mri k^2_n \Delta^{-1}_n & -\mri \theta_n \Delta^{-1}_n \partial_y
     \\[0.2cm] 
     \mri \theta_n I& 0 & -\mri k_x I
    \end{bmatrix},
\end{align}
and
\begin{align}
\label{eq.Coutuvw}
    C(k_x,\theta_n) \;=\;
    \thbo{C_u}{C_v}{C_w} \;=\;
    \begin{bmatrix}
     \mri (k_x/k^2_n) \partial_y& -\mri (\theta_n/k^2_n) I
     \\[0.1cm] 
     I& 0
     \\[0.1cm] 
     \mri (\theta_n/k^2_n) \partial_y& \mri (k_x/k^2_n) I
    \end{bmatrix}.
\end{align}

\section{Perturbation analysis on the dynamic generator $\cA_\theta(k_x)$}
\label{app.Apert}

Following the structure of the dynamic generator $\cA_\theta(k_x)$ given in equation~\eqref{eq.opA}, the components of its perturbation expansion~\eqref{eq.Apert} take the following forms:
\begin{align}
\label{eq.A0-structure}
        \cA_{0,\theta} 
        &\; = \;
        \begin{bmatrix*}
        \ddots &  &  &  &\\ 
         & A_{0,0}(\theta_{n-1}) & &  &\\ 
         &  & A_{0,0}(\theta_{n}) &  &\\ 
         &  &  & A_{0,0}(\theta_{n+1}) &\\ 
         &  &  &  &\ddots
        \end{bmatrix*},
        \\[.15cm]
        \label{eq.A1-structure}
        \cA_{1,\theta} 
        &\; = \;
        \begin{bmatrix*}
        \ddots &  &  &  &\\ 
         & A_{1,0}(\theta_{n-1}) & A_{1,-1}(\theta_{n}) & A_{1,-2}(\theta_{n+1}) &\\ 
         & A_{1,1}(\theta_{n-1}) & A_{1,0}(\theta_{n}) & A_{1,-1}(\theta_{n+1}) &\\ 
         & A_{1,2}(\theta_{n-1}) & A_{1,1}(\theta_{n}) & A_{1,0}(\theta_{n+1}) &\\ 
         &  &  &  &\ddots
        \end{bmatrix*}, 
        \\[.15cm]
        \label{eq.A2-structure}
        \cA_{2,\theta}
        &\; = \;
        \begin{bmatrix*}
        \ddots &  &  &  &\\ 
         & A_{2,0}(\theta_{n-1}) & A_{2,-1}(\theta_{n}) & A_{2,-2}(\theta_{n+1}) &\\ 
         & A_{2,1}(\theta_{n-1}) & A_{2,0}(\theta_{n}) & A_{2,-1}(\theta_{n+1}) &\\ 
         & A_{2,2}(\theta_{n-1}) & A_{2,1}(\theta_{n}) & A_{2,0}(\theta_{n+1}) &\\ 
         &  &  &  &\ddots
        \end{bmatrix*}.
\end{align}
The block-diagonal operator-valued matrix $\cA_{0,\theta}(k_x)$ captures the dynamics of fluctuations in the absence of riblets and block-Toeplitz operators $\cA_{1,\theta}(k_x)$ and $\cA_{2,\theta}(k_x)$ capture the effect of the periodic surface corrugation at the levels of $\alpha^1$ and $\alpha^2$ of the perturbation expansion. Following the state evolution~\eqref{eq.lnse}, all operators take the following $2\times 2$ form:
\begin{align*}
    A_{l,m}(\theta_n)
        \, = \,
        \tbt{A_{l,m,1,1}(k_x,\theta_n)}{A_{l,m,1,2}(k_x,\theta_n)}{A_{l,m,2,1}(k_x,\theta_n)}{A_{l,m,2,2}(k_x,\theta_n)}.
\end{align*}
At the level of $\alpha^0$ the operator-valued submatrices are given as
\begin{align}
    \ba{rcl}
        A_{0,0,1,1}(k_x,\theta_n)
        &\!\!\!\!=\!\!\!\!&
        \dfrac{\Delta^{-1}_{0,n}}{Re_\tau} \left[ (1\,+\,\nu_T)\Delta^2_{0,n}
        \, + \,
        2{\nu_{T}''}{\partial_{yy}}
        \, - \,
        \nu_{T}''\Delta_{0,n}
        \, + \,
        2\nu_{T}'\Delta_{0,n} \partial_{y} \right] 
        \\[0.25cm]
        && ~ + \,
        \Delta^{-1}_{0,n}\left[
        \mri k_x(U''_{0}\, - \,
         U_0\Delta_{0,n})\right],
        \\[0.25cm]
        A_{0,0,1,2}(k_x,\theta_n)
        &\!\!\!\!=\!\!\!\!&
        0
        \\[.25cm]
        A_{0,0,2,1}(k_x,\theta_n)
        &\!\!\!\!=\!\!\!\!&
        -\mri \theta_n U'_{0},
        \\[.25cm]
        A_{0,0,2,2}(k_x,\theta_n)
        &\!\!\!\!=\!\!\!\!&
        \dfrac{1}{Re_\tau} \left[
        (1+\nu_{T})\Delta_{0,n}
        \, + \,
        \nu_{T}'{\partial_y} \right]
        \, - \,
        \mri k_x U_0.
    \ea
\end{align}
{The submatrices at the $\alpha^1$ perturbation level are given as
\begin{align*}
    \ba{rcl}
        A_{1,m,1,1}
        &\!\!\!=\!\!\!&
        \Delta^{-1}_{0,n} \Big[ -\mri k_x \big ( U''_{1,m} + U_{1,m} \Delta_{0,n} - m^2\omega_z^2 U_{1,m} \big ) 
        \,+\, 2 \mri m \omega_z U_{1,m} k_x \theta_n 
        \\[0.25cm]
        && ~ +\,
        \mri k_x  F'_{1,m} U'_0 \partial_y \,+\, \mri k_x  F'_{1,m} U''_0 \partial_y \,+\, 2 \mri k_x  F'_{1,m} U'_0 \partial_y
        \\[0.25cm]
        && ~ - \,
        \dfrac{1}{Re_\tau} \left[ 2 F'_{1,m} \nu_{T}'\Delta_{0,n} \partial_{y} \,+\, 2 F'_{1,m} \nu_{T}''(\Delta_{0,n} + \partial_{yy}) \,+\, 2 F'_{1,m} \nu_{T}''' \partial_{y} \right]
        \\[0.3cm]
        && ~ - \,
        k_x \theta_n \big( \big(\mri m \omega_z F_{1,m} U''_0
        \,+\,
        2\mri m \omega_z U'_{1,m} \big)/k^2 \partial_y 
        - \big(\mri m \omega_z F_{1,m} U'_0
        \\[0.25cm]
        && ~ +\,
        2\mri m \omega_z U_{1,m} \big)/k^2 \partial_{yy} \big) 
        \,-\,
        \dfrac{\Delta^{-1}_{0,n}}{Re_\tau} \Big[ \nu_{T}'' \big( -F'^2_{1,m} k_x^2/k^2 \partial_{yy} \,+\, 4 F'^2_{1,m} \partial_{yy} 
        \\[0.25cm]
        && ~ -\, 
        F'^2_{1,m} \theta_n^2/k^2 \partial_{yy} \,-\, m \omega_z F_{1,m} \theta_n \partial_y \big)
        \,+\, \nu_{T}' \big( -F'^2_{1,m} k_x^2/k^2 \partial_{yyy} \,+\, 2 F'^2_{1,m} \partial_{yyy} 
        \\[0.25cm]
        && ~ -\, 
        F'^2_{1,m} \theta_n^2/k^2 \partial_{yyy} - m \omega_z F_{1,m} \theta_n \partial_{yy} \big)
        + \nu_{T}''' 2 F'^2_{1,m} \partial_y \Big]
        - F'_{1,m} \Delta^{-1}_{0,n} \Big[ 2 \mri U'_0 k_x^3 \partial_y 
        \\[0.25cm]
        && ~ + \,
        \dfrac{1}{Re_\tau} \big( \nu_{T}' \big( -2 k_x^2 \Delta_{0,n} \partial_y -2 \theta_n^2 \Delta_{0,n} \partial_y \big) - \nu_{T}'' \big( 2 k_x^2 \partial_{yy} -2 k_x^2 \Delta_{0,n} -2 \theta_n^2 \partial_{yy} 
        \\[0.25cm]
        && ~ -\, 
        2 \Delta_{0,n} \theta_n^2 \big) \,-\, \nu_{T}''' \big( 2k_x^2 \partial_y -2\theta_n^2 \partial_y \big) \big) -2 \mri k_x \theta_n^2 \big( U'_{0} \partial_y - U''_0 \big) \Big]
        \\[0.25cm]
        && ~ +\, 
        \mri m \omega_z F_{1,m} \Delta^{-1}_{0,n} \Big[ 2k_x \theta_n U'_0 \partial_{yy} \,+\, \dfrac{1}{Re_\tau} \big( 2\mri \theta_n \nu_{T}' \Delta_{0,n} \partial_{yy} \,+\, 2\mri \theta_n \nu_{T}'' \partial_{yyy} 
        \\[0.25cm]
        && ~ +\, 
        2\mri \theta_n \nu_{T}''' \Delta_{0,n} \,+\, 2\mri \theta_n \nu_{T}'''' \partial_y \big) \Big]
%
        \\[0.25cm]
        A_{1,m,1,2}
        &\!\!\!=\!\!\!&
        \Delta^{-1}_{0,n} \Big[ - \mri m \omega_z k_x^2 \big( \big( F_{1,m} U''_0
        \,+\,
        2 U'_{1,m} \big)/k^2 
        - \big( F_{1,m} U'_0
        \, +\,
        2 U_{1,m} \big)/k^2 \partial_y \big) \Big]
        \\[0.25cm]
        A_{1,m,2,1}
        &\!\!\!=\!\!\!&
        -\mri \theta_n \big( F'_{1,m} U'_0 \,+\, U'_{1,m} \big) - \mri m \omega_z \big( U'_{1,m} \,-\, F_{1,m} U'_0 \,-\, \theta_n^2/k^2 U_{1,m} \partial_y \big)
        \\[0.2cm]
        && ~ +\, 
        \mri m^2 \omega_z^2 \theta_n/k^2 U_{1,m} \partial_y \,+\, 
        m \omega_z k_x/k^2 U_{1,m} \partial_y \,+\, \dfrac{1}{Re_\tau} \big( \nu_{T}' m \omega_z k_x F_{1,m} \partial_y \big)
        \\[0.2cm]
        &&  ~ +\,
        \mri m \omega_z F_{1,m} \Delta^{-1}_{0,n} \Big[ 2 k_x^2 U''_0 \,+\, 2 k_x^2 U'_0 \partial_y \,+\, \dfrac{1}{Re_\tau} \big( 2 \nu_{T}' \mri k_x \Delta_{0,n} \partial_y 
        \\[0.2cm]
        && ~ +\, 2 \mri k_x \nu_{T}'' \big(\Delta_{0,n} \,+\, \partial_{yy} \big) \,+\, 2 \mri k_x \nu_{T}''' \partial_y \big) \Big]
        \\[0.25cm]
        A_{1,m,2,2}
        &\!\!\!=\!\!\!&
        \mri m \omega_z k_x \theta_n/k^2 \big( F_{1,m} U'_0 \,+\, U_{1,m} \big) 
        \, + \,
        \mri m^2 \omega_z^2 k_x/k^2 U_{1,m} \,-\, \mri m \omega_z k_x \theta_n/k^2  U_{1,m}  
        \\[0.25cm]
        && ~ +\,
        \mri k_x U_{1,m} \,+\, \dfrac{1}{Re_\tau} \left[ \nu_{T}' F'^2_{1,m} \partial_y \right],
    \ea
\end{align*}
}
{and at the $\alpha^2$ perturbation level, the submatrices are given as}
{
\begin{align*}
    \ba{rcl}
        A_{2,m,1,1}
        &\!\!\!=\!\!\!&
        \Delta^{-1}_{0,n} \Big[ -\mri k_x \big( U''_{2,m} -m^2 \omega_z^2 U_{2,m} \,+\, \Delta_{0,n} U_{2,m} \,+\, 2 U'_{2,m} \partial_y \,-\, 2 m \omega_z \theta_n U_{2,m} \big)
        \\[0.2cm]
        && ~ +\,
        \Delta^{-1}_{0,n} \Big[ \mri k_x F'_{2,m} U''_0 \,+\, \mri k_x F'_{2,m} U'_0 \partial_y \,+\, \mri k_x U''_{2,m} \,+\, \mri k_x U'_{2,m} \partial_y 
        \\[0.2cm]
        && ~ +\,
        \mri k_x (F'_1 * U''_1)_m \,+\, \mri k_x (F'_1 * U'_1)_m \partial_y \,-\, \mri \omega_z k_x \theta_n/k^2 \big( m F_{2,m} U''_0 \partial_y 
        \\[0.2cm]
        && ~ -\,
        m F_{2,m} U'_0 \partial_{yy} - (F_1 * U''_1)_m \partial_y - (F_1 * U'_1)_m \partial_{yy} - m U'_{2,m} \partial_y - m U_{2,m} \partial_{yy} \big) 
        \\[0.2cm]
        && ~ -\,
        \mri k_x^3/k^2 U'_{2,m} \partial_y - \mri k_x^3/k^2 U_{2,m} \partial_{yy} +
        \mri k_x U''_{2,m} \,+\, 2\mri k_x U'_{2,m} \partial_y \,+\, \mri k_x U_{2,m} \partial_{yy}
        \\[0.2cm]
        && ~ -\,
        \dfrac{1}{Re_\tau} \left[ 2\nu_{T}' F'^2_{2,m} \partial_{yyy} + 2\nu_{T}'' F'^2_{2,m} \partial_{yy} +
        \nu_{T}''' F'^2_{2,m} \partial_{y} \right] - \mri k_x \theta_n/k^2 \big( m \omega_z U'_{2,m} \partial_y 
        \\[0.2cm]
        && ~ -\, 
        m \omega_z U_{2,m} \partial_{yy} \,-\,  U'_{2,m} \partial_y \,-\, U_{2,m} \partial_{yy} \big) \Big]
        \,+\,
        \dfrac{1}{Re_\tau} \left[ m \omega_z \theta_n \nu_{T}' F_{2,m} \partial_{yy} \right.
    \ea
\end{align*}
}
{
\begin{align*}
    \ba{rcl}
        && ~ +\, \left.
         m \omega_z \theta_n \nu_{T}'' F_{2,m} \partial_y \right] \,+\, \Delta^{-1}_{0,n} \Big[ \mri k_x \big( F'^2_{2,m}U''_0
        \,+\, 
        (F'_1 * U''_1)_m \,+\, F'^2_{2,m} U'_0 \big) 
        \\[0.2cm]
        && ~ +\, 
        2\,(F'_1 * U'_1)_m \partial_y \,-\, \mri \omega_z k_x \theta_n/k^2 \big( (F'_1 * F_1)_m U''_0
        \,+\, 
        2 (F'_1 * U'_1)_m \big) \partial_y 
        \\[0.2cm]
        && ~ -\,
        \mri \omega_z k_x \theta_n/k^2 \big( (F'_1 * F_1)_m U'_0 + 2 (F'_1 * U_1)_m \big) \partial_{yy} \Big] -
        F'^2_{2,m} \Delta^{-1}_{0,n} \Big[2\mri k_x U'_0 \partial_{yy} 
        \\[0.2cm]
        && ~ -\, 
        \dfrac{1}{Re_\tau} \left[ 2 \nu_{T}' \Delta_{0,n} \partial_{yy} + 2 \nu_{T}'' \partial_{yyy} + 2 \nu_{T}''' \Delta_{0,n} + 2 \nu_{T}'''' \partial_y \right] \Big]
        \,+\,
        (F'_1 * F_1)_m \Delta^{-1}_{0,n} \Big[
        \\[0.2cm]
        && ~ +\,
         2 k_x \theta_n U''_0 \,+\, 2 k_x \theta_n U'_0 \partial_y \,+\, \dfrac{1}{Re_\tau} \left[ 2 \mri \theta_n \nu_{T}' \Delta_{0,n} \partial_y  \,+\,
        2 \mri \theta_n \nu_{T}'' \big( \Delta_{0,n} \,+\, \partial_{yy} \big) \right.
        \\[0.2cm]
        && ~ +\, \left. 
        2 \mri \theta_n \nu_{T}''' \partial_y \right] \Big] \,-\, \dfrac{1}{Re_\tau} \left[ \nu_{T}' \big( F'^2_{2,m} k_x^2/k^2 \partial_{yyy} + 2 F'^2_{2,m} \partial_{yyy}
        \,+\, 
        F'^2_{2,m} \theta_n^2/k^2 \partial_{yyy} \right.
        \\[0.25cm]
        && ~ +\, \left.
        \mri \theta_n (F'_1 * F_1)_m \partial_{yy} \big)
        \,+\, \nu_{T}'' \big( F'^2_{2,m} k_x^2/k^2 \partial_{yy} \,+\, 4F'^2_{2,m} \partial_{yy} 
        \,+\, F'^2_{2,m} \theta_n^2/k^2 \partial_{yy} \right.
        \\[0.2cm]
        && ~ +\, \left. 
        \mri \theta_n (F'_1 * F_1)_m \partial_y \big) \,+\, 2\nu_{T}''' F'^2_{2,m} \partial_y \right] \,+\, F'_{2,m} \big( 2\mri k_x ( U''_0 + U'_0  \partial_y )
        \,-\, \dfrac{1}{Re_\tau} \left[  \right. 
        \\[0.2cm]
        && ~ +\, \left.
        \nu_{T}' \Delta_{0,n} \partial_y 
        \,+\,
        2\nu_{T}'' \big( \Delta_{0,n} \,+\, \partial_{yy} \big) \,+\, 2 \mri \theta_n \nu_{T}''' \partial_y \right]
        \big) \,+\, \dfrac{1}{Re_\tau} \left[ 2 \nu_{T}' F'^2_{2,m} \Delta_{0,n} \partial_y \right.
        \\[0.2cm]
        && ~ +\, \left. 
        4 \nu_{T}'' F'^2_{2,m} \partial_{yy} \,+\, 2 \nu_{T}''' F'^2_{2,m} \partial_y \right] \Big]
        \\[0.2cm]
        A_{2,m,1,2}
        &\!\!\!=\!\!\!&
        \Delta^{-1}_{0,n} \Big[ -\mri \omega_z k_x^2/k^2 \big( m F_{2,m} (U''_0 + U'_0 \partial_y) - (F_1 * U''_1)_m \partial_y + (F_1 * U'_1)_m \partial_{yy}
        \\[0.2cm]
        && ~ -\, 
        m U'_{2,m} \,-\, m U_{2,m} \partial_y \big) \,+\, \mri k_x^2 \theta_n/k^2 \big( U'_{2,m} \,+\, U_{2,m} \partial_y \big) \,-\, \mri m \omega_z k_x^2/k^2 \big( U'_{2,m}
        \\[0.2cm]
        && ~ +\, 
        U_{2,m} \partial_y \big) \,+\, \Delta^{-1}_{0,n} \Big[ -\omega_z^2 
        k_x^2 \theta_n/k^2 (F_1 * F_1)_m \big( U''_0 \,+\, U'_0 \partial_y \big)
        \\[0.2cm]
        && ~ -\, 2\mri \omega_z^2 k_x^2 \theta_n/k^2 (F_1 * U_1)_m \partial_y \Big] \Big],
        \\[0.2cm]
        A_{2,m,2,1}
        &\!\!\!=\!\!\!&
        -\mri \theta_n \big( F'_{2,m} U'_0 \,+\, U'_{2,m} \,-\, (F'_1 * U'_1)_m \big) + \mri \omega_z \big( m U'_{2,m} \,+\, (F'_1 * U'_1)_m \big)
        \\[0.2cm]
        && ~ +\, \mri \omega_z \theta_n^2/k^2 \big( m F_{2,m} U'_0 \,+\, (F_1 * U'_1)_m \,+\, U_{2,m} \big) \partial_y \,+\, \mri \omega_z^2 \theta_n/k^2 \big( (F'_1 * U'_1)_m 
        \\[0.2cm]
        && ~ +\, m^2 U_{2,m} \big) \partial_y \,+\, \mri m \omega_z k_x^2/k^2 U_{2,m} \partial_y \,+\, \dfrac{1}{Re_\tau} \left[ m \omega_z k_x F_{2,m} \nu_{T}' \partial_y \right],  
        \\[0.25cm]
        A_{2,m,2,2}
        &\!\!\!=\!\!\!&
        \mri \omega_z k_x \theta_n/k^2 \big( m F^2_{2,m} U'_0 \,+\, (F'_1 * U'_1)_m \,+\, m U_{2,m} \,-\, m U_{2,m} \big) \\[0.2cm]
        && ~ +\, \mri \omega_z^2 k_x/k^2 \big( (F'_1 * U'_1)_m \,+\, m^2 U_{2,m} \big) \,-\, \mri k_x U_{2,m} \,+\, \dfrac{1}{Re_\tau} \left[ \nu_{T}' F_{2,m} \partial_y \right].
    \ea
\end{align*}
}

\section{Perturbation analysis of Lyapunov equation~\eqref{eq.lyap}}
\label{app.Xpert}

For $\alpha\ll1$, Lyapunov equation~\eqref{eq.lyap} can be efficiently solved via perturbation analysis. Substituting~\eqref{eq.Apert} and~\eqref{eq.Xpert} into~\eqref{eq.lyap} and collecting equal powers of $\alpha$ yields the sequence of Lyapunov equations,
\begin{align}
\label{eq.Xpertset}
        \ba{lrcl}
            \alpha^0:\quad
            &
            \cA_{0,\theta}\,\cX_{0,\theta} \,+\, \cX_{0,\theta}\, \cA^* _{0,\theta} 
            &\!\!=\!\!& 
            -\,\cM_\theta
            \\
            \alpha^l:\quad
            &
            \cA_{0,\theta}\,\cX_{l,\theta} \,+\, \cX_{l,\theta}\,\cA^* _{0,\theta} 
            &\!\!=\!\!&
            -\,
            \ds{\sum_{p=1}^l \left(\cA_{p,\theta}\,\cX_{l-p,\theta} \,+\, \cX_{l-p,\theta}\,\cA^*_{p,\theta}\right)}
        \ea
\end{align}
where the first subscript denotes the perturbation index and the dependence of operators $\cA_{l,\theta}$, $\cX_{l,\theta}$, and $\cM_\theta$ on $k_x$ is suppressed in favor of brevity. For the level of $\alpha^0$, we follow~\cite{moajovJFM12} and select the block-diagonal operator $\cM_\theta(k_x)$ such that the energy spectrum of the stochastically forced linearized NS equations given by $\trace(\cX_{0,\theta}(k_x))$ matches that of a turbulent channel flow with smooth walls. This is done by scaling the block covariances of forcing as
$
    \cM(k_x,\theta_n)
    =
    \bar{E}_s(k_x,\theta_n)\,
    \cM_s(k_x,\theta_n)/\bar{E}_{s,0}(k_x,\theta_n)
$,
where, $\bar{E}_s(k_x,\theta_n)$ is the two-dimensional energy spectrum of smooth channel flow and $\bar{E}_{s,0}(k_x,\theta_n)$ is the energy spectrum resulting from the linearized dynamics~\eqref{eq.lnse} subject to white-in-time stochastic forcing of covariance
\begin{align*}
        \cM_s(k_x,\theta_n)
        \,=\,
       \begin{bmatrix}
        \sqrt{{E}_s(y,k_x,\theta_n)} \,I\!\!\! & 0\\ 
        0 & \!\!\!\sqrt{{E}_s(y,k_x,\theta_n)} \,I
        \end{bmatrix}\!\!\!
        \begin{bmatrix}
        \sqrt{{E}_s(y,k_x,\theta_n)} \,I\!\!\! & 0\\ 
        0 & \!\!\!\sqrt{{E}_s(y,k_x,\theta_n)} \,I
        \end{bmatrix}^*\!\!\!.
\end{align*}
{In this study, we compute the two-dimensional energy spectrum of smooth channel flow using a DNS-generated database (\url{https://torroja.dmt.upm.es/channels/data}) as $\bar{E}_s(k_x,\theta_n)=\int_{-1}^{1}{E}_s(y,k_x,\theta_n)dy$, where ${E}_s(y,k_x,\theta_n)$ is the energy spectrum resulting from DNS.}
Alternative forcing models (e.g.,~\citep{zarjovgeoACC14,zarchejovgeoTAC17,zarjovgeoJFM17,zargeojovARC20}) that may result in more accurate predictions of two-point correlations at the level of $\alpha^0$ could also be used here. Due to the block-diagonal structure of $\cA_{0,\theta}$ (equation~\eqref{eq.A0-structure}), the solution $\cX_{l,\theta}$ inherits the structure of the right-hand side operator, i.e., 
\begin{align*}
\cX_{0,\theta}(k_x) &= \diag\left \{ X_{0,0}(k_x,\theta_n) \right \}
\\[0.15cm]
\cX_{l,\theta}(k_x) &= \text{Toep}\left \{\cdots, X^*_{l,1}(k_x,\theta_n) ,\boxed{\!X_{l,0}(k_x,\theta_n)\!}, X_{l,1}(k_x,\theta_n), \cdots \right \}
\end{align*}
where the box denotes the element on the main diagonal of $\cX_{l,\theta}$. Each block can be computed by substituting for $\cA_{l,\theta}$ and $\cX_{l,\theta}$ into equations~\eqref{eq.Xpertset} to obtain a coupled system of Lyapunov equations as shown below up to $\alpha^2$:
\begin{align}
            \,\alpha^0:& \quad
            A_{0,0}(\theta_{n})X_{0,0}(\theta_{n}) \,+\, X_{0,0}(\theta_{n})A^* _{0,0}(\theta_{n}) 
            \;=\; 
            -M(\theta_{n})
            \label{eq.lyap-set}
            \\[0.2cm]
            \,\alpha^1:&\quad 
            A_{0,0}(\theta_{n-m})X_{1,m}(\theta_{n}) \,+\, X_{1,m}(\theta_{n})A^* _{0,0}(\theta_{n-m}) 
            \;=\; 
            -\big(A_{1,-m}(\theta_{n})X_{0,0}(\theta_{n}) 
            \non
            \\
            &\hspace{8.4cm} ~ + \,
            X_{0,0}(\theta_{n-m})A^* _{1,m}(\theta_{n-m})\big) 
            \non
            \\[0.2cm]
            \,\alpha^2:&\quad
            A_{0,0}(\theta_{n})X_{2,0}(\theta_{n}) \,+\, X_{2,0}(\theta_{n})A^* _{0,0}(\theta_{n}) 
            \;=\; 
            -\big(A_{2,0}(\theta_{n})X_{0,0}(\theta_{n}) \,+ X_{0,0}(\theta_{n})A^* _{2,0}(\theta_{n})
            \non
            \\
            &\hspace{8.cm} ~ + \,
            \ds{\sum_{m\in \mathbb{Z}} A_{1,m}(\theta_{n-m})X_{1,m}(\theta_{n})} 
            \non
            \\
            &\hspace{8cm} ~ +\,
            \ds{\sum_{m\in \mathbb{Z}} X_{1,m}(\theta_{n})A^* _{1,m}(\theta_{n-m})\big)}.
            \non
\end{align}

\section{Computing modifications at the level of $\alpha^1$ and $\alpha^2$ to $k$ and $\epsilon$ in equations~\eqref{eq.Pertexpansionkeps}}
\label{app.corrKEps}
The average effect of velocity fluctuations on $\alpha^1$- and $\alpha^2$-level modifications to $k$ and $\eps$ are obtained from autocorrelation operators $X_{1,0}(k_x, \theta_{n})$ and $X_{2,0}(k_x, \theta_{n})$, respectively, as
\begin{align*}
    \ba{rcl}
        k_l(y) 
        &\!\!=\!\!&
        \ds{\int_{0}^{\infty}\int_{0}^{\omega_z} \sum_{n \in \mathbb{Z}} K_{l,k}(y,k_x,\theta_n) \,\mrd \theta \,\mrd k_x}, 
        \\[0.4 cm]
        \epsilon_l(y) 
        &\!\!=\!\!&
        \ds{\int_{0}^{\infty}\int_{0}^{\omega_z} \sum_{n \in \mathbb{Z}} K_{l,\epsilon}(y,k_x,\theta_n) \,\mrd \theta \,\mrd k_x}.
    \ea
\end{align*}
Here, $K_{l,k}(y,k_x,\theta_n)$ and $K_{l,\epsilon}(y,k_x,\theta_n)$ are the kernel representation of operators ${\cal K}_{l,k}$ and ${\cal K}_{l,\epsilon}$, respectively, that 
can be computed from the central block (corresponding to the $0$th harmonic) of solution $\cX_{l,\theta}(k_x)$ of the respective Lyapunov equation at the $l$th perturbation level, i.e.,
\begin{align*}
    \ba{rcl}
        \hspace{-.1cm}{\cal K}_{l,k}(k_x,\theta_n) 
        &\!\!\!\!\!=\!\!\!\!\!&
        (C_uX_{l,0}C_u^* + C_vX_{l,0}C_v^* + C_wX_{l,0}C_w^*)/2, 
        \\[0.15 cm]
        \hspace{-.1cm}{\cal K}_{l,\eps}(k_x,\theta_n) 
        &\!\!\!\!\!=\!\!\!\!\!&
        2 \left(k_x^2C_uX_{l,0}C_u^* +\, \partial_y C_vX_{l,0}C_v^* \partial_y^* 
        +\,
        \theta_n^2 C_wX_{l,0}C_w^* 
        -\,
        \mri k_x \partial_y C_uX_{l,0}C_v^* \right.
        \\[0.15cm]
        && \left.  +\, 
        k_x \theta_n C_uX_{l,0}C_w^* 
        +\,
         \mri \theta_n C_vX_{l,0}C_w^* \partial_y^* \right) 
        +\,
        \partial_y C_uX_{l,0}C_u^* \partial_y^* 
        + 
        k_x^2 C_vX_{l,0}C_v^*
        \\[0.15cm]
        && +\, 
        \partial_y C_wX_{l,0}C_w^* \partial_y^*
        +\,
        k_x^2 C_wX_{l,0}C_w^* 
        +\,
        \theta_n^2 C_uX_{l,0}C_u^*
    \ea
\end{align*}
where, $C_u$, $C_v$, and $C_w$ are finite-dimensional representations of the output operators in appendix~\eqref{eq.Coutuvw} and covariance matrices $X_{l,0}$ are confined to the wall-normal range $y \in [-1,1]$ to provide appropriate comparison between the flows over smooth and corrugated surfaces. 
In figure~\ref{fig.kcorr}, we retrieve the dependence of the turbulent kinetic energy on the spanwise dimension via
\begin{align*}
    \ba{rcl}
    k_l(y,z) &\!\!=\!\!\!& \ds{\int_{0}^\infty \int_{0}^{\omega_z} \sum_{n\in \bbZ}}\, \mathrm{Re}\left(K_{l,k}(y,k_x,\theta_n) \,\mre^{\mri k_x x}\right) \cos(\theta_n z)\, \mrd \theta\, \mrd k_x.
    \ea
\end{align*}

\section{{Model-based predictions of} the effect of fluctuations on the mean velocity and skin-friction drag}
\label{app.Upert-Dpert}

The effect of fluctuations on the mean velocity are realized through the riblet induced modifications to the turbulent viscosity. By substituting expansions~\eqref{eq.Fpert},~\eqref{eq.Upert}, and~\eqref{eq.nuTpert} into the mean flow equations~\eqref{eq.meanVelT}, one can solve for zero-bulk perturbations to the mean velocity $U_l$ via
\begin{align*}
        U_l(y)
        \;=\;
        \bar{U}_{l,0}(y) \,-\, \bar{P}_{x,l}\, U_0(y)
\end{align*}
where $\bar{U}_{l,0}$ ($l=1$ and 2) are the $0$th harmonics of the solution to the coupled equations
\begin{align}
\label{eq.U-pert}
        \ba{rcl} 
             (1 + \nu_{T_0})\left [ \partial_{yy} + \partial_{zz}  \right ] \bar{U}_1
            \,+\, \nu_{T_0}'\, \partial_y\, \bar{U}_1
            &\!\!=\!\!&
            - \partial_y\, \big ( \nu_{T_1} \partial_y\, U_0 \big )
            -2 F_{\tilde{y}_1} \partial_y\, \big((1 + \nu_{T_0}) \, \partial_y\, U_0 \big)
            \\[0.25cm]
             (1 + \nu_{T_0})\left [\partial_{yy} + \partial_{zz}  \right ] \bar{U}_2
            \,+\, \nu_{T_0}' \,\partial_y\, \bar{U}_2 
            &\!\!=\!\!&
            2F_{\tilde{z}_1} (1 + \nu_{T_0})  \partial_{yz} \bar{U}_1
            \,-\,
            F_{\tilde{z}_1}^{2} \partial_y\, \big((1 + \nu_{T_0}) \, \partial_y\, U_0 \big)
            \\[0.15cm]
            && ~ -\,
            \partial_y\, \big(\nu_{T_1} \, \partial_y\, \bar{U}_1 \big)
            -2 F_{\tilde{y}_1} \partial_y\, \big((1 + \nu_{T_0}) 
            \, \partial_y\, \bar{U}_1 \big)
            \\[0.15cm]
            && ~ -\,
            2F_{\tilde{y}_1} \partial_y\, \big(\nu_{T_1} \, \partial_y\, U_0 \big)
            \,-\,
            \partial_y\, \big(\nu_{T_2} \, \partial_y\, U_0 \big)
            \\[0.15cm]
            && ~ -\,
            (F_{\tilde{y}_1}^{2} + 2F_{\tilde{y}_2}) \partial_y \big((1 + \nu_{T_0}) \partial_y\, U_0 \big)
        \ea
\end{align}
and $\bar{P}_{x,l} = (\int_{-1-\alpha r_p}^{1} \bar{U}_{l,0}(y)\, \mrd y ) / (2\,U_B)$. We use $25$ harmonics, i.e., $k=-12$, $\dots$,$12$ in equation~\eqref{eq.UFourier}, to capture riblet-induced modifications to the mean velocity at both perturbation levels. The inclusion of a sufficient number of harmonics is ensured by evaluating the incremental effect of additional harmonics on the overall results.

\refstepcounter{equation}


%

\end{document}